%% file: MainThesis.tex
\documentclass[final]{uqthesis}

\input{LaTexPackages.tex}

\title{Ultimate-rate quantum repeaters for quantum communications}

\author{Matthew Winnel}
\currentdegrees{BMusStudies BSc(Hons)(USyd)}

\orcid{\href{https://orcid.org/0000-0003-3457-4451}{https://orcid.org/0000-0003-3457-4451}}

\date{2023}
\submittedfor{Doctor}

\school{School of Mathematics and Physics}

\begin{document}

\frontmatter
\maketitle
\clearpage

\section{Abstract}

\normalfont

\input{./Abstract/Abstract.tex}

\clearpage
\section*{Declaration by author}
\input{./Authordeclaration.tex}

\clearpage
\input{./PreliminaryAndBackPages/Preliminary.tex}

\mainmatter


\input{./Chapter1/Introduction.tex}

\input{./Chapter2/Chapter2.tex}

\input{./chapter_min_leakage/chapter_min_leakage.tex}


\input{./chapter_scissors/chapter_scissors.tex}


\input{./chapter_CVrepeater/chapter_CVrepeater.tex}

\input{./chapter_simple_repeater/chapter_simple_repeater.tex}

\input{./chapter_purification/chapter_purification.tex}

\input{./Chapter7/Chapter7.tex}


\input{./Conclusion/Conclusion.tex}


\appendix

\include{./Appendix_minimum_leakage/Appendix_minimum_leakage}

\include{./Appendix_scissors/Appendix_scissors}

\include{./Appendix_CV_QR/Appendix_CV_QR}


\include{./Appendix_purification/Appendix_purification}








\nocite{apsrev41Control}
\bibliographystyle{apsrev4-1}





\include{bib_file2}

\end{document}

%% file: LaTexPackages.tex
\usepackage{cite}				 
\usepackage{pdfpages}			 
\usepackage{wrapfig}			 

\usepackage[breaklinks=true]{hyperref}
\hypersetup{
colorlinks   = true, 
urlcolor     = blue, 
linkcolor    = blue, 
citecolor    = red 
}

\usepackage{bm} 				 

\usepackage{upgreek}			 
\usepackage{dsfont}				 
\usepackage{simplewick}			 
\usepackage{mathtools}		     

\usepackage{framed}			     
\usepackage{microtype}			 
\usepackage{marvosym}			 
\usepackage{color}				 
\usepackage{transparent}	     
\usepackage{placeins}			 
\usepackage{mdframed,mdwlist}    
\usepackage{graphicx}            
\usepackage{float}               
\usepackage{longtable}           

\usepackage{mathdots}            

\usepackage{eucal}               

\usepackage{array}               

\usepackage{stmaryrd}            
\usepackage{amsthm}              
\usepackage{pifont}              

\usepackage{lipsum}              
\usepackage{enumerate}           
\usepackage[all]{xy}             
\usepackage{amsmath}             
\usepackage{amssymb}             
\usepackage[utf8]{inputenc}      

\usepackage{blindtext}           
\usepackage{tikz}                
\usepackage[figuresright]{rotating}	
\usetikzlibrary{shapes.geometric, arrows}

\usepackage{bm}
\usepackage{amsmath,amssymb}
\usepackage{amsfonts}

\DeclareMathAlphabet{\mathcal}{OMS}{cmsy}{m}{n}

\usepackage{bm}
\usepackage{amsmath,amssymb}
\usepackage{amsfonts}
\usepackage{layouts}
\usepackage{mathtools}
\usepackage{dsfont}
\usepackage{graphicx}
\usepackage{float}
\usepackage{verbatim}
\usepackage{amsfonts}
\usepackage{bbold}
\usepackage{dcolumn}
\usepackage{bm}
\usepackage{color}

\usepackage{gensymb}
\usepackage[hyphenbreaks]{breakurl}
\usepackage{enumitem}

\definecolor{blueprl}{RGB}{46,48,146}

\usepackage{graphics}
\usepackage{amsmath}
\usepackage{amssymb}
\usepackage{longtable}
\usepackage{epsfig}
\usepackage{latexsym}
\usepackage{bbm}
\usepackage{psfrag}%

\usepackage{mathrsfs}

\newcommand{\eg}{\textit{e.g.}} 
\newcommand{\etc}{\textit{etc.}} 

\makeatletter
\def\@bibdataout@aps{%
 \immediate\write\@bibdataout{%
  @CONTROL{%
   apsrev41Control,author="08",editor="1",pages="0",title="0",year="1",eprint="1"%
  }%
 }%
 \if@filesw
  \immediate\write\@auxout{\string\citation{apsrev41Control}}%
 \fi
}%
\makeatother 

\usepackage{gensymb}

\usepackage{comment}
\usepackage[normalem]{ulem}
\usepackage{xcolor}
\usepackage{graphicx}

\usepackage{cleveref}
\crefname{equation}{Eq.}{Eqs.}
\Crefname{equation}{Equation}{Equations}
\crefname{figure}{Fig.}{Figs.}
\Crefname{figure}{Figure}{Figures}
\crefname{figure}{Fig.}{Figs.}
\Crefname{figure}{Figure}{Figures}
\crefname{section}{Sec.}{Secs.}
\Crefname{section}{Section}{Sections}
\crefname{appendix}{Appendix}{Appendices}
\Crefname{appendix}{Appendix}{Appendices}
\crefname{table}{Table}{Tables}
\Crefname{table}{Table}{Tables}
\crefname{chapter}{Chapter}{Chapters}
\Crefname{Chapter}{Chapter}{Chapters}

\newcommand{\bra}[1]{\mbox{$\langle #1 |$}}
\newcommand{\ket}[1]{\mbox{$| #1 \rangle$}}
\def\ketbra#1#2{{\vert#1\rangle\!\langle#2\vert}}
\newcommand{\braket}[2]{\mbox{$\langle #1 | #2 \rangle$}}

\newcommand{\bblack}[1]{\textcolor{black}{#1}}

\newcommand{\Tr}{\text{Tr}}
\newcommand{\tr}{\text{tr}}

\newcommand{\Id}{\mathbb{1}}

\renewcommand{\Re}{\text{Re}}

\usepackage[sort&compress,numbers]{natbib}
\usepackage{doi}

\usepackage{tensor}



\newcommand{\oneinjectedscissor}{1\text{-scissor}}
\newcommand{\twoinjectedscissor}{2\text{-scissor}}
\newcommand{\oneinjectedscissors}{1\text{-scissors}}
\newcommand{\threeinjectedscissor}{3\text{-scissor}}

\newcommand{\Nscissors}{N\text{ 1-scissors}}
\newcommand{\twoscissors}{\text{two 1-scissors}}
\newcommand{\threescissors}{\text{three 1-scissors}}
\newcommand{\twocatrelay}{2\text{-cat tele-amplification}}

\newcommand{\fourcatrelay}{4\text{-cat tele-amplification}}

\usepackage{subfig}

\newcommand\NLAto{\stackrel{\mathclap{\footnotesize	\mbox{NLA}}}{\longmapsto}}

\newcommand\lossto{\stackrel{\mathclap{\footnotesize	\mbox{loss}}}{\longmapsto}}

%% file: Abstract/Abstract.tex








The field of quantum communications promises the faithful distribution of quantum information, quantum entanglement, and absolutely secret keys. However, the highest rates of these tasks are fundamentally limited by the transmission distances. Quantum repeaters, i.e., untrusted intermediate relay stations, are necessary to overcome the repeaterless bound which sets the fundamental rate-distance limit of any repeaterless quantum communication protocol. The ultimate end-to-end rates of quantum repeater networks are known to be achievable by an optimal entanglement-distillation protocol followed by quantum teleportation. In this thesis, we give physical repeater designs for this achievability. We also propose practical repeater designs with near-term potential for real-world practical applications. We focus on protocols with continuous variables since they enable the highest quantum communication rates and are compatible with existing network infrastructure.

First, we consider a repeaterless protocol for continuous-variable quantum key distribution with zero information leakage to an eavesdropper. This may simplify the required classical post-processing and increase the secret key rate. However, the protocol is strictly not a quantum repeater, and the rate is still fundamentally limited by the separation distance between the end users.

Next, we introduce a linear-optical technique called quantum scissors that can implement ideal noiseless linear amplification (NLA) and teleportation up to an energy cutoff. Our simple protocol consists of a beamsplitter and a generalised beamsplitter (an $(n+1)$-splitter), with $n$ resource photons and the detection of $n$ photons. NLA is a useful technique for many quantum repeater proposals. Remarkably, the quantum scissor can be used as a loss-tolerant quantum repeater for entanglement distillation without quantum memories, beating the repeaterless bound for $n=1$.

We then consider complete effective repeater designs. First, we introduce repeaters based on NLA and dual-homodyne detection with post-selection. These repeaters require quantum memories. Although the rates scale optimally, they fall short of the ultimate bounds.

After that, we introduce repeaters which can surprisingly beat the repeaterless bound without quantum memories. This is possible because our quantum scissors for NLA are, as we discovered, highly tolerant to loss. This loss tolerance significantly reduces the number of physical resources required to reliably transmit quantum information over a particular distance, effectively halving the total number of resources. These repeaters enjoy fast rates; however, they cannot saturate the ultimate bounds. Repeaters using quantum scissors are highly tolerant to additional loss and experimental imperfections, as well as being practical and scalable. They will form an important component of the first multi-node quantum network in the near future.

Then, we introduce repeaters using purification, where the quantum information is encoded in a quantum-error-detection code. This approach achieves the pure-loss channel's two-way (classically-assisted) quantum capacity. Our repeaters give the highest rates allowed by physics, saturating the ultimate end-to-end rates of quantum communication networks. Our protocol can be performed entirely using linear optics and photon-number measurements, retaining excellent rates. These repeaters will be implementable with developing technology.

Finally, we introduce repeaters with quantum error correction. While this approach is the most demanding experimentally, it requires only one-way communication, avoiding back-and-forth classical signalling. We employ the bosonic cat code for error correction. 

Explained in this thesis are physical repeater designs for achieving the fundamental limits of quantum communication networks. We explore scalable and practical approaches that both address the challenges of quantum error correction and that we expect to be implemented in the near term. When these approaches are scaled and combined, we are left with the blueprint for building extensive quantum networks, i.e., a highly efficient global quantum internet.



%% file: Authordeclaration.tex
%
\begin{instructional}
\textit{(All candidates to reproduce this section in their thesis verbatim)\\}
\end{instructional}

\noindent
This thesis is composed of my original work, and contains no material previously published or written by another person except where due reference has been made in the text. I have clearly stated the contribution by others to jointly-authored works that I have included in my thesis.\\

\noindent
I have clearly stated the contribution of others to my thesis as a whole, including statistical assistance, survey design, data analysis, significant technical procedures, professional editorial advice, financial support and any other original research work used or reported in my thesis. The content of my thesis is the result of work I have carried out since the commencement of my higher degree by research candidature and does not include a substantial part of work that has been submitted to qualify for the award of any other degree or diploma in any university or other tertiary institution. I have clearly stated which parts of my thesis, if any, have been submitted to qualify for another award.\\

\noindent
I acknowledge that an electronic copy of my thesis must be lodged with the University Library and, subject to the policy and procedures of The University of Queensland, the thesis be made available for research and study in accordance with the Copyright Act 1968 unless a period of embargo has been approved by the Dean of the Graduate School. \\

\noindent
I acknowledge that copyright of all material contained in my thesis resides with the copyright holder(s) of that material. Where appropriate I have obtained copyright permission from the copyright holder to reproduce material in this thesis and have sought permission from co-authors for any jointly authored works included in the thesis.

%% file: PreliminaryAndBackPages/Preliminary.tex


\section*{Publications included in this thesis}

\begin{instructional}

	Start this section on a new page [this template will automatically handle this].\\
	
	\noindent
	If you choose to include publications as part of your thesis as described in UQ policy (\href{http://ppl.app.uq.edu.au/content/4.60.07-alternate-thesis-format-options}{\color{blue}{PPL 4.60.07 Alternative Thesis Format Options}}) use this section to detail accepted or in press publication/s using the standard citation format for your discipline. \\
    
    \noindent
	Papers submitted for publication and awaiting review should appear in the next section, \textbf{Submitted manuscripts included in this thesis}.\\
    
    \noindent
	On the page immediately preceding the chapter that includes your publication, in no more than one (1) page, describe your contribution to the authorship if you are not a sole author. In describing your contribution, you must satisfy the University's authorship policy (\href{http://ppl.app.uq.edu.au/content/4.20.04-authorship}{\color{blue}{PPL 4.20.04 Authorship}}). Authorship is based on having made a substantive contribution to at least one, and usually more than one, of the following activities:
	\begin{enumerate}
		\item	conception and design of the project;
		\item	analysis and interpretation of the research data on which the publication is based;
		\item	drafting significant parts of the publication or critically reviewing it so as to contribute to the interpretation.
	\end{enumerate}
	
	\noindent
	As an author, you must have participated sufficiently in the publication to take public responsibility for at least that part of the work that you contributed.\\
    
    \noindent
	It may be useful to refer to specific parts of the methods, analyses, results, or discussion to illustrate your contribution to the paper.\\
    
    \noindent
	If you have not included any of your publications in the thesis then state ``No publications included''.\\
	
	\textbf{Example:}
	\begin{enumerate}

    \item \cite{DumyCitationKey} \textbf{Your Name}, Co-author 1, and Final Author, \href{linktoyourpaper}{Title of your paper}, \textit{Journal}, Issue, Number, Year

    \item \cite{DumyCitationKey} \textbf{Your Name}, Co-author 1, and Final Author, \href{linktoyourpaper}{Title of your paper}, \textit{Journal}, Issue, Number, Year

    \end{enumerate}
	
\end{instructional}


\begin{enumerate}

    \item \cite{PhysRevA.102.052425} J. Dias, \textbf{M. S. Winnel}, N. Hosseinidehaj, and T. C. Ralph, ``Quantum repeater for continuous-variable entanglement distribution,'' \href{http://dx.doi.org/10.1103/physreva.102.052425}{Phys. Rev. A \textbf{102}, 052425 (2020)}.

    \item \cite{PhysRevA.102.063715} \textbf{M. S. Winnel}, N. Hosseinidehaj, and T. C. Ralph, ``Generalized quantum scissors for noiseless linear amplification,'' \href{http://dx.doi.org/10.1103/PhysRevA.102.063715}{Phys. Rev. A \textbf{102}, 063715 (2020)}.

    \item \cite{PhysRevA.104.012411} \textbf{M. S. Winnel}, N. Hosseinidehaj, and T. C. Ralph, ``Minimization of information leakage in continuous-variable quantum key distribution,'' \href{http://dx.doi.org/10.1103/PhysRevA.104.012411}{Phys. Rev. A \textbf{104}, 012411 (2021)}.
    
    \item \cite{PhysRevLett.128.160501} J. J. Guanzon, \textbf{M. S. Winnel}, A. P. Lund, and T. C. Ralph, ``Ideal quantum teleamplification up to a selected energy cutoff using linear optics,'' \href{https://journals.aps.org/prl/abstract/10.1103/PhysRevLett.128.160501}{Phys. Rev. Lett. \textbf{128}, 160501 (2022)}.

        \item \cite{https://doi.org/10.48550/arxiv.2203.13924} \textbf{M. S. Winnel}, J. J. Guanzon, N. Hosseinidehaj, and T. C. Ralph, ``Achieving the ultimate end-to-end rates of lossy quantum communication networks,'' \href{https://www.nature.com/articles/s41534-022-00641-0}{npj Quantum Information \textbf{8}, 129 (2022)}.

\end{enumerate}


\section*{Submitted manuscripts included in this thesis}

\begin{instructional}
	List manuscript/s submitted for publication here. As described above for \textbf{Publications included in the thesis}, on the page immediately preceding the chapter that includes the submitted manuscript, in no more than one (1) page, detail your contribution to the authorship if you are not the sole author.\\
    
    \noindent
    If you have no submitted manuscripts from your candidature then state ``No manuscripts submitted for publication''.\\
    
    \textbf{Example:}
    \begin{enumerate}

    \item \cite{DumyCitationKey} \textbf{Your Name}, Co-author 1, and Final Author, Title of your paper, submitted to \textit{Journal} on 4th June 2018.

    \end{enumerate}
\end{instructional}


\begin{enumerate}

    \item \cite{winnel2021overcoming} \textbf{M. S. Winnel}, J. J. Guanzon, N. Hosseinidehaj, and T. C. Ralph, ``Overcoming the repeaterless bound in continuous-variable quantum communication without quantum memories,'' \href{https://arxiv.org/abs/2105.03586}{arXiv:2105.03586 (2021)}.

\end{enumerate}


\section*{Other publications during candidature}

\begin{instructional}
    List other publications arising during your candidature using the standard citation format for your discipline. Divide your publications into sub-sections as appropriate in your discipline \eg{} peer-reviewed papers, book chapters, conference abstracts. Papers submitted for publication and awaiting review are not considered publications and cannot be included in this section.\\
    
    \noindent
    If you have no publications from your candidature then state ``No other publications''.\\
    
    \textbf{Example:}
    \subsection*{Conference abstracts}

    \begin{enumerate}

    \item \cite{DumyCitationKey} \textbf{Your Name}, Co-author 1, and Final Author, Title of your conference paper, \textit{Proceedings of Conference}, other details.

    \end{enumerate}

    \subsection*{Book chapters}

    \begin{enumerate}

    \item \cite{DumyCitationKey} \textbf{Your Name}, Co-author 1, and Final Author, Title of your chapter, Book, editor, \etc{}.

    \end{enumerate}

\end{instructional}

\begin{enumerate}

    \item \cite{PhysRevA.105.032602} N. Hosseinidehaj, \textbf{M. S. Winnel}, and T. C. Ralph, ``Simple and loss-tolerant free-space quantum key distribution using a squeezed laser,'' \href{https://journals.aps.org/pra/abstract/10.1103/PhysRevA.105.032602}{Phys. Rev. A \textbf{105}, 032602 (2022)}.
    
        \item \cite{dias2019comparison} J. Dias, \textbf{M. S. Winnel}, W. J. Munro, T. C. Ralph, and K. Nemoto, ``Distributing entanglement in first-generation discrete- and continuous-variable quantum repeaters,'' \href{https://journals.aps.org/pra/abstract/10.1103/PhysRevA.106.052604}{Phys. Rev. A \textbf{106}, 052604 (2022).}

     \item \cite{Guanzon2022cat_relay} J. J. Guanzon, \textbf{M. S. Winnel}, A. P. Lund, and T. C. Ralph, ``Noiseless linear amplification and loss-tolerant quantum relay using coherent state superpositions.'' \href{https://arxiv.org/abs/2211.08035}{arXiv:2211.08035 (2022)}.
    
\end{enumerate}

\clearpage
\section*{Contributions by others to the thesis}

\begin{instructional}
	List the significant and substantial inputs made by others to the research, work and writing represented and/or reported in the thesis. These could include significant contributions to: the conception and design of the project; non-routine technical work; analysis and interpretation of research data; drafting significant parts of the work or critically revising it so as to contribute to the interpretation. \\
    
    \noindent
	If no one contributed significantly then state ``No contributions by others''.
\end{instructional}


I acknowledge the following contributions by my colleagues and co-authors to the results included in this thesis. Prof. Timothy C. Ralph and Nedasadat Hosseinidehaj have supervised the research discussed in this thesis and made significant contributions to the project. I acknowledge the substantial contributions by Joshua J. Guanzon in Refs.~\cite{PhysRevLett.128.160501,winnel2021overcoming,https://doi.org/10.48550/arxiv.2203.13924} who wrote and edited parts of the manuscripts and contributed to the mathematical details, especially Ref.~\cite{PhysRevLett.128.160501} where Joshua J. Guanzon is the lead author. I also acknowledge Austin P. Lund for contributions to Ref.~\cite{PhysRevLett.128.160501}. Finally, I acknowledge Josephine Dias for contributions to Ref.~\cite{PhysRevA.102.052425} where she is the lead author.


\section*{Statement of parts of the thesis submitted to qualify for the award of another degree}

\begin{instructional}
    The thesis must be comprised only of research undertaken while enrolled in the HDR program unless otherwise approved by the Dean, Graduate School in advance of submission.\\
    
    \noindent
    If you have been given permission to include your previous work that has been used towards another degree, you must list the relevant parts of the thesis that incorporates this work including, the degree name, year and institution, and the outcome of the submission of material. \\
    
    \noindent
    If no parts of the thesis have been submitted in this way then state ``No works submitted towards another degree have been included in this thesis''.
\end{instructional}


No works submitted towards another degree have been included in this thesis.


\section*{Research involving human or animal subjects}

\begin{instructional}
	All research involving human or animal subjects requires prior ethical review and approval by an independent review committee. At UQ, the relevant committee for research involving human subjects is the \href{http://www.uq.edu.au/research/integrity-compliance/human-ethics}{\color{blue}{Human Ethics Unit}} and the relevant committee for research involving animal subjects is the relevant \href{http://www.uq.edu.au/research/integrity-compliance/animal-welfare}{\color{blue}{Animal Ethics Committee}}.  Please provide details of any ethics approvals obtained including the ethics approval number and name of approving committees.  A copy of the ethics approval letter must be included in the thesis appendix.\\
    
    \noindent
	If no human or animal subjects were involved in this research please state: ``No animal or human subjects were involved in this research''.
\end{instructional}


No animal or human subjects were involved in this research.


\clearpage
\section*{Acknowledgements}

\begin{instructional}
    Start this section on a new page [the template will handle this for you].\\
    
    \noindent
    Acknowledgements recognise those who have been instrumental in the completion of the project.  Acknowledgements should include any professional editorial advice received including the name of the editor and a brief description of the service rendered.
\end{instructional}


I would like to thank and acknowledge the following people who made this thesis possible.

First, I would like to acknowledge my principal supervisor, Tim Ralph, for inspiration and guidance throughout my studies. Also, I would like to thank my associate supervisor, Neda Hosseinidehaj, for our wonderful discussions and her expertise.

Next, I would especially like to acknowledge my co-authors and colleagues, Josephine Dias and Josh Guanzon. My discussions with them resulted in many achievements in this thesis. They also helped prepare and refine our manuscripts for publication, always aiming for a high level of clarity and quality.

Then, I would also like to thank Jonas Neergaard-Nielsen, Ulrik Andersen, Iyad Suleiman, and Lucas Nunes Faria from the Technical University of Denmark. I also acknowledge Ozlem Erkilic, Sebastian Kish, Ping Koy Lam, and Syed Assad from the Australian National University for valuable discussions during my investigations.

I would like to especially thank Deepesh Singh for the delightful discussions about purification protocols, as well the recommendations for great books and literature. Finally, I would like to thank Kaerin Gardner and all the members of the Quantum Optics and Quantum Information group and CQC2T.


\clearpage
\section*{Financial support}

\begin{instructional}
    Start this section on a new page [the template will handle this for you].\\
    
    \noindent
    If you are the recipient of an Australian Government Research Training Program (RTP) scholarship, you are required to acknowledge this contribution.  Please include the text below:\\
    
    \noindent
    ``This research was supported by an Australian Government Research Training Program Scholarship''\\
    
    \noindent
    If you received any other financial support for your project, you are also required to acknowledge the funding body/bodies in this section.\\
    
    \noindent
    If no financial provided then state ``No financial support was provided to fund this research''.
\end{instructional}


This research was supported by an Australian Government Research Training Program Scholarship. The author acknowledges support from the Australian Research Council (ARC) under the Centre of Excellence for Quantum Computation and Communication Technology (Project No. CE170100012) and the Linkage Program (Project No. LP200100601).


\section*{Keywords}

\begin{instructional}
	Maximum 10 words; use lower case throughout, separating words/phrases with commas. For example: word, word word, word, word, word word
\end{instructional}

quantum communication, quantum error correction, quantum repeaters, entanglement distillation, teleportation


\section*{Australian and New Zealand Standard Research Classifications (ANZSRC)}

\begin{instructional}
    Provide data that links your thesis to the disciplines and discipline clusters in the Federal Government’s Excellence in Research for Australia (ERA) initiative.\\
    
    \noindent
    Please allocate the thesis a \textbf{maximum of 3} \href{http://www.abs.gov.au/Ausstats/abs@.nsf/Latestproducts/6BB427AB9696C225CA2574180004463E?opendocument}{\color{blue}{Australian and New Zealand Standard Research Classifications (ANZSRC) codes}} at the \textbf{6 digit level} and include the descriptor and a percent weighting for each code. Total percent must add to 100.\\

\textbf{Example:}\\

    ANZSRC code: 060101, Analytical Biochemistry, 60\% \\
    \indent ANZSRC code: 060104, Cell Metabolism, 20\% \\
    \indent ANZSRC code: 060199, Biochemistry and Cell Biology not elsewhere classified, 20\%
\end{instructional}

\noindent ANZSRC code: 510803, Quantum Information, Computation and Communication, 70\% \\
\noindent ANZSRC code: 510804, Quantum Optics and Quantum Optomechanics, 30\%


\section*{Fields of Research (FoR) Classification}

\begin{instructional}
    Allows for categorisation of the thesis according to the field of research. \\
    
    \noindent
    Please allocate the thesis a \textbf{maximum of 3} \href{http://www.abs.gov.au/Ausstats/abs@.nsf/Latestproducts/6BB427AB9696C225CA2574180004463E?opendocument}{\color{blue}{Fields of Research (FoR) Codes}} at the \textbf{4 digit level} and include the descriptor and a percent weighting for each code. Total percent must add to 100. \\

\textbf{Example:}\\

FoR code: 0601, Biochemistry and Cell Biology, 80\% \\
\indent FoR code: 0699, Other Biological Sciences, 20\%
\end{instructional}


\noindent FoR code: 5108, Quantum Physics, 100\%


\begin{instructional}
\section*{Order for the Remainder of the Thesis}
\noindent
    Remainder of the thesis should be in the following order

    \begin{itemize}
        \item Dedications (if applicable)
        \item Table of Contents
        \item List of Figures and Tables
        \item List of Abbreviations used in the thesis
        \item Main text of the thesis
        \item Bibliography or List of References
        \item Appendices
    \end{itemize}

\noindent
\textbf{Date of thesis template release:} 22 March 2019
\end{instructional}
\clearpage

	\rmfamily
	\normalfont

	\begin{vplace}[1]
		\begin{center}
			 This thesis is dedicated to my parents, who encouraged me to dream about science.
		\end{center}
	\end{vplace}

\clearpage
\pagestyle{headings}

\tableofcontents
	\clearpage
\listoffigures
	\clearpage
\newpage
\input{./PreliminaryAndBackPages/Symbols.tex} 


%% file: PreliminaryAndBackPages/Symbols.tex
%

\chapter[List of Abbreviations and Symbols]{List of Abbreviations and Symbols}


\begin{center}
	\small
	\begin{longtable}{ll}
	\toprule
	Abbreviations & {} \\
	\bottomrule
	\hypertarget{hyperlinkbacklabel_QKD}{\hyperlink{hyperlinklabel_QKD}{QKD}} & \hyperlink{hyperlinklabel_QKD}{Quantum key distribution}  \\
	\hypertarget{hyperlinkbacklabel_QR}{\hyperlink{hyperlinklabel_QR}{QR}} & \hyperlink{hyperlinklabel_QR}{Quantum repeater}  \\
	\hypertarget{hyperlinkbacklabel_PM}{\hyperlink{hyperlinklabel_PM}{PM}} & \hyperlink{hyperlinklabel_PM}{Prepare-and-measure}  \\
	\hypertarget{hyperlinkbacklabel_EB}{\hyperlink{hyperlinklabel_EB}{EB}} & \hyperlink{hyperlinklabel_EB}{Entanglement-based}  \\
	\hypertarget{hyperlinkbacklabel_MDI}{\hyperlink{hyperlinklabel_MDI}{MDI}} & \hyperlink{hyperlinklabel_MDI}{Measurement device independent}  \\
	\hypertarget{hyperlinkbacklabel_QEC}{\hyperlink{hyperlinklabel_QEC}{QEC}} & \hyperlink{hyperlinklabel_QEC}{Quantum error correction}  \\
    \hypertarget{hyperlinkbacklabel_QED}{\hyperlink{hyperlinklabel_QED}{QED}} & \hyperlink{hyperlinklabel_QED}{Quantum error detection}  \\
    \hypertarget{hyperlinkbacklabel_DV}{\hyperlink{hyperlinklabel_DV}{DV}} & \hyperlink{hyperlinklabel_DV}{Discrete variable}  \\
	\hypertarget{hyperlinkbacklabel_CV}{\hyperlink{hyperlinklabel_CV}{CV}} & \hyperlink{hyperlinklabel_CV}{Continuous variable}  \\
	\hypertarget{hyperlinkbacklabel_BSM}{\hyperlink{hyperlinklabel_BSM}{BSM}} & \hyperlink{hyperlinklabel_BSM}{Bell-state measurement}  \\
	\hypertarget{hyperlinkbacklabel_QND}{\hyperlink{hyperlinklabel_QND}{QND}} & \hyperlink{hyperlinklabel_QND}{Quantum non-demolition}  \\
	\hypertarget{hyperlinkbacklabel_QFT}{\hyperlink{hyperlinklabel_QFT}{QFT}} & \hyperlink{hyperlinklabel_QFT}{Quantum Fourier transform}  \\
	\hypertarget{hyperlinkbacklabel_NLA}{\hyperlink{hyperlinklabel_NLA}{NLA}} & \hyperlink{hyperlinklabel_NLA}{Noiseless linear amplification}  \\
	\hypertarget{hyperlinkbacklabel_TMSV}{\hyperlink{hyperlinklabel_TMSV}{TMSV}} & \hyperlink{hyperlinklabel_TMSV}{Two-mode squeezed vacuum state}  \\
	\hypertarget{hyperlinkbacklabel_EOF}{\hyperlink{hyperlinklabel_EOF}{EOF}} & \hyperlink{hyperlinklabel_EOF}{Entanglement of formation}  \\
	\hypertarget{hyperlinkbacklabel_GEOF}{\hyperlink{hyperlinklabel_GEOF}{GEOF}} & \hyperlink{hyperlinklabel_GEOF}{Gaussian entanglement of formation}  \\
	\hypertarget{hyperlinkbacklabel_RCI}{\hyperlink{hyperlinklabel_RCI}{RCI}} & \hyperlink{hyperlinklabel_RCI}{Reverse coherent information}  \\
	\hypertarget{hyperlinkbacklabel_GKP}{\hyperlink{hyperlinklabel_GKP}{GKP}} & \hyperlink{hyperlinklabel_GKP}{Gottesman-Kitaev-Preskill}  \\
	\hypertarget{hyperlinkbacklabel_PLOB}{\hyperlink{hyperlinklabel_PLOB}{PLOB}} & \hyperlink{hyperlinklabel_PLOB}{Pirandola-Laurenza-Ottaviani-Banchi}  \\
 \hypertarget{hyperlinkbacklabel_ebit}{\hyperlink{hyperlinklabel_ebit}{ebit}} & \hyperlink{hyperlinklabel_ebit}{Entanglement bit}  \\
	\hypertarget{hyperlinkbacklabel_BS}{\hyperlink{hyperlinklabel_BS}{BS}} & \hyperlink{hyperlinklabel_BS}{Beamsplitter}  \\
	\hypertarget{hyperlinkbacklabel_CSUM}{\hyperlink{hyperlinklabel_CSUM}{CSUM}} & \hyperlink{hyperlinklabel_CSUM}{Controlled-SUM quantum gate}  \\
	\hypertarget{hyperlinkbacklabel_SNU}{\hyperlink{hyperlinklabel_SNU}{SNU}} & \hyperlink{hyperlinklabel_SNU}{Shot-noise units}  \\
 \hypertarget{hyperlinkbacklabel_SDP}{\hyperlink{hyperlinklabel_SDP}{SDP}} & \hyperlink{hyperlinklabel_SDP}{Semidefinite program}  \\
	\hline
	\end{longtable}
\end{center}

\clearpage

\begin{center}
	\small
	\begin{longtable}{ll}
	\toprule
	Symbols & {} \\
	\bottomrule
	\hypertarget{hyperlinkbacklabel_a}{\hyperlink{hyperlinklabel_a}{$\hat{a},\hat{a}^\dagger$}} & \hyperlink{hyperlinklabel_a}{Annihilation and creation operators}  \\
	\hypertarget{hyperlinkbacklabel_q}{\hyperlink{hyperlinklabel_q}{$\hat{q},\hat{p}$}} & \hyperlink{hyperlinklabel_a}{Quadrature operators}  \\
	\hypertarget{hyperlinkbacklabel_r}{\hyperlink{hyperlinklabel_r}{$r$}} & \hyperlink{hyperlinklabel_r}{Squeezing parameter}  \\
	\hypertarget{hyperlinkbacklabel_alpha}{\hyperlink{hyperlinklabel_alpha}{$\ket{\alpha}$}} & \hyperlink{hyperlinklabel_alpha}{Coherent state}  \\
	\hypertarget{hyperlinkbacklabel_sqz}{\hyperlink{hyperlinklabel_sqz}{$\ket{\alpha, r}$}} & \hyperlink{hyperlinklabel_sqz}{Single-mode squeezed state}  \\
	\hypertarget{hyperlinkbacklabel_chi}{\hyperlink{hyperlinklabel_chi}{$\ket{\chi}$}} & \hyperlink{hyperlinklabel_chi}{Two-mode squeezed vacuum state}  \\
	\hypertarget{hyperlinkbacklabel_n}{\hyperlink{hyperlinklabel_n}{$\ket{n}$}} & \hyperlink{hyperlinklabel_n}{Fock state}  \\
	\hypertarget{hyperlinkbacklabel_ij}{\hyperlink{hyperlinklabel_ij}{$\ket{i,j}$}} & \hyperlink{hyperlinklabel_ij}{Shorthand for tensor product of Fock states, $\ket{i}\otimes\ket{j}$}  \\
 \hypertarget{hyperlinkbacklabel_rho}{\hyperlink{hyperlinklabel_rho}{$\hat{\rho}$}} & \hyperlink{hyperlinklabel_rho}{Density operator}  \\
 \hypertarget{hyperlinkbacklabel_thermal}{\hyperlink{hyperlinklabel_thermal}{$\hat{\rho}_\text{th}$}} & \hyperlink{hyperlinklabel_thermal}{Thermal state}  \\
	\hypertarget{hyperlinkbacklabel_g}{\hyperlink{hyperlinklabel_g}{$g$}} & \hyperlink{hyperlinklabel_g}{Gain of noiseless linear amplification}  \\
	\hypertarget{hyperlinkbacklabel_G}{\hyperlink{hyperlinklabel_G}{$G$}} & \hyperlink{hyperlinklabel_G}{Gain of quantum-limited amplification}  \\
	\hypertarget{hyperlinkbacklabel_eta}{\hyperlink{hyperlinklabel_eta}{$\eta$}} & \hyperlink{hyperlinklabel_eta}{Transmissivity of lossy channel}  \\
	\hypertarget{hyperlinkbacklabel_xi}{\hyperlink{hyperlinklabel_xi}{$\xi$}} & \hyperlink{hyperlinklabel_xi}{Excess noise}  \\
 \hypertarget{hyperlinkbacklabel_barn}{\hyperlink{hyperlinklabel_barn}{$\bar{n}$}} & \hyperlink{hyperlinklabel_barn}{Mean photon number}  \\
	\hypertarget{hyperlinkbacklabel_E_D}{\hyperlink{hyperlinklabel_E_D}{$E_\text{D}$}} & \hyperlink{hyperlinklabel_E_D}{Rate of entanglement distillation}  \\
	\hypertarget{hyperlinkbacklabel_E}{\hyperlink{hyperlinklabel_E}{$E$}} & \hyperlink{hyperlinklabel_E}{Rate of entanglement purification}  \\
	\hypertarget{hyperlinkbacklabel_C}{\hyperlink{hyperlinklabel_C}{$C$}} & \hyperlink{hyperlinklabel_C}{Two-way assisted quantum capacity }  \\
	\hypertarget{hyperlinkbacklabel_K}{\hyperlink{hyperlinklabel_K}{$K$}} & \hyperlink{hyperlinklabel_K}{Secret key rate}  \\
	\hypertarget{hyperlinkbacklabel_R}{\hyperlink{hyperlinklabel_R}{$R$}} & \hyperlink{hyperlinklabel_R}{Reverse coherent information}  \\
	\hypertarget{hyperlinkbacklabel_von_Neumann}{\hyperlink{hyperlinklabel_von_Neumann}{$S(\hat{\rho})$}} & \hyperlink{hyperlinklabel_von_Neumann}{Von Neumann entropy of $\hat{\rho}$}  \\
	\hypertarget{hyperlinkbacklabel_Shannon}{\hyperlink{hyperlinklabel_Shannon}{$I_{AB}$}} & \hyperlink{hyperlinklabel_Shannon}{Mutual information between Alice and Bob}  \\
	\hypertarget{hyperlinkbacklabel_Holevo}{\hyperlink{hyperlinklabel_Holevo}{$\chi_{EB}$}} & \hyperlink{hyperlinklabel_Holevo}{Holevo bound with Bob}  \\
	\hline
	\end{longtable}
\end{center}


%% file: Chapter1/Introduction.tex
\chapter{Introduction}
\label{Chap:Intro}






\section{Introduction}


Quantum physics describes matter and energy at a most fundamental level. It revolutionised our descriptions of natural phenomena and allowed the development of remarkable technologies that operate on a scale where quantum effects matter, for example, chemistry, optics, superconducting magnets, light-emitting diodes, lasers, semiconductors, and electronics, as well as imaging for medicine and research. These advances, the first quantum revolution, changed our understanding of nature and led to the technological revolution of the 20th century, based on technologies such as the transistor, laser, and atomic clock, which are critical to computers, optical communications, and global positioning systems, respectively.


Such advances are undeniably magnificent and have had major impacts on science and our world. However, information-processing tasks traditionally involve the manipulation of classical information, such as a single binary value, ``0'' or ``1''. That is, until recently, the manipulation and transmission of information was only performed by computers and machines fully describable using classical mechanics which is not a full description of nature. Undoubtedly, this restriction barely scratches the surface of what is possible.


Machines following the laws of quantum mechanics could allow for exponential speed ups in computation time~\cite{Arute_2019} and absolute security of communication~\cite{RevModPhys.81.1301}. This is known as the second quantum revolution~\cite{doi:10.1098/rsta.2003.1227}. To many companies, investors, and government agencies, these new quantum technologies are indistinguishable from magic.

Classical computing has progressed from early computing machines and mainframes to the modern internet age of globally connected high-speed computers, phones, and other devices. Indeed, the internet was ground-breaking, leading to remarkable information-processing technologies and many of the applications we observe today were initially regarded as impossible. 

The second quantum revolution will provide technologies such as quantum computers, sensors, and simulators, designed to outperform classical technologies for certain tasks. It will be required that these technologies are connected in a network, especially to realise their full potential.


The goal of quantum communication~\cite{Gisin_2007} is to enable the reliable exchange of quantum information between quantum devices at high rates and often at long distances, i.e., the faithful transmission of quantum entanglement~\cite{PhysRev.47.777,Yin_2013} and quantum bits (qubits) from one place to another. Entanglement is the phenomenon in which physically separated systems are inherently connected and cannot be described without reference to the other~\cite{schrödinger_1935,PhysRevA.40.4277}. The most important applications of quantum communication are quantum key distribution~\cite{RevModPhys.81.1301,Pirandola_2020,Xu_2020,Curty_2004}, distributed quantum computing~\cite{Cirac_1999,jiang2007distributed,7562346} and enhanced sensing and metrology~\cite{PhysRevA.97.032329,PhysRevLett.121.043604}. Quantum key distribution is the most mature and accessible quantum technology. Its goal is to distribute a secret random key between two physically-separated parties. Its unconditional security relies on quantum physics, a promising solution to the vulnerability of current classical cryptosystems.

The faithful transmission of quantum information between parts of a network, or between components of a quantum computer or other quantum devices (for example, if the quantum computer is too large to be contained in a single refrigerator) is no easy task since quantum information differs a great deal from classical information in that different components can exist in superposition and be entangled with one another. This means quantum communication requires a different network infrastructure. It must be able to generate, distribute, and store quantum information and quantum entanglement. Existing classical networks do not have these capabilities.

Quantum information is encoded in a physical quantum system, and quantum communication involves physically moving a quantum system from one place to another. Light is a great choice for encoding quantum information since photons travel at the speed of light and light does not interact with the environment in a vacuum. In this thesis, we often imagine quantum communication as sending light signals.


Quantum teleportation is a hallmark protocol for quantum communication designed to transfer quantum information from one system or particle to another via classical communication~\cite{bennett_teleporting_1993,Pirandola_2015}. The laws of physics forbid copying of quantum information, but teleportation works since the original is destroyed and (ideally) no information about the state is revealed by the classical channel. Teleportation relies on quantum entanglement. 






Sending quantum information large distances across space or through optical fibres introduces loss and noise. Since quantum states are very fragile, extreme care must be taken; otherwise, they will scramble into a classical state, and the quantum nature of the state will be lost. Fundamentally, the achievable highest rates of the tasks in quantum communications are limited by the transmission distance, suffering from an exponential rate-distance scaling~\cite{takeoka2014}. The achievable highest rates are proved in Ref.~\cite{Pirandola_2017}.

The dominant source of noise for quantum communication comes in the form of losses~\cite{cerf2007quantum}, for instance, scattering and absorption loss in fibres, beam wandering and scattering in atmospheric links, and beam spreading in free space. While quantum communication does not need to distribute entanglement, the transfer needs to be able to preserve entanglement. The job of quantum repeaters is to solve the problem of loss and noise and allow the faithful transfer of quantum information between two nodes of a network separated by some distance.


Classical repeaters solve the problem of distance by amplifying the classical information at repeater stations to combat loss and noise. The laws of quantum physics forbid such a simple solution. Quantum repeaters are needed, exploiting clever ideas such as quantum-error-correcting codes, which are designed to detect and correct errors, and other quantum techniques, such as, exploiting spooky-quantum phenomena such as teleportation and entanglement. Quantum repeaters are much more exotic than optical amplifiers.

Quantum repeaters allow quantum computers, quantum sensors, and other quantum devices to communicate with each other over long distances, forming a quantum version of the conventional internet~\cite{kimble2008,referencE}. While the current internet is composed of computers and devices connected by classical communication channels, the quantum internet will be composed of nodes connected by quantum channels. Quantum repeaters are currently an active field of research~\cite{sangouard2011quantum,munro2015inside,Muralidharan2016}; however, most quantum repeater programs are at their early stages of development. 


Encodings of quantum information can roughly be characterised as discrete variable (DV) and continuous variable (CV)~\cite{Braunstein_2005,yonezawa2008continuousvariable,cerf2007quantum,Weedbrook_2012,Serafini2017QuantumCV}. The difference is related to the way quantisation manifests itself, as a discrete or continuous measurement spectrum. Examples of DV encoding include the polarisation of a photon, the time-bin of a photon, the photon number of an optical mode, or the spin of a nucleus or electron. CV encodings exploit the infinite-dimensional Hilbert space of an oscillator, i.e., encode information into the CV quadratures of the light, such as amplitude $\hat{q}$ and phase $\hat{p}$ which have a continuous spectrum.  Examples of CV states include coherent states, squeezed states, and thermal states. There are also states which use a carefully selected subset of the full infinite-dimensional Hilbert space of a CV mode to encode a DV system, for instance, cat states~\cite{Mirrahimi_2014,PhysRevA.94.042332} and Gottesman-Kitaev-Preskill (GKP) states~\cite{GKP2001}. These states have error-correcting properties which can be used to protect against noise in a quantum device such as a quantum computer or a quantum repeater node.

Considering the CVs phase and momentum is only one way to describe light. Light can also be described as superpositions of photon numbers, and so light can be thought of in terms of DVs as well. CV technologies may be more versatile than DV technologies since CV teleportation allows any state to be teleported (not just a quantum state which is ``matched'' to the teleporter). This is unique to CV teleportation. If we can distribute CV entanglement throughout a network, then we can use CV teleportation to teleport any state. In principle, CV protocols offer easier state manipulation~\cite{Braunstein_2005} and compatibility with existing optical telecom infrastructure~\cite{Kumar_2015}.

Recently, a variety of quantum repeater designs have been proposed for CV encodings~\cite{dias2020quantum,dias2017,dias2018quantum,PhysRevA.98.032335,PhysRevResearch.2.013310,Ghalaii_2020,PhysRevA.102.052425,winnel2021overcoming}. This is a promising way to encode quantum information; the infinite-dimensional Hilbert space of the light can enable higher quantum communication rates compared with DV encodings; however, the rates fall short of the ultimate achievable rates allowed by physics~\cite{Pirandola_2019}. In this thesis, we present a repeater architecture which indeed in principle achieves the highest possible rates~\cite{https://doi.org/10.48550/arxiv.2203.13924}; see \cref{Chap:purification}.

In this thesis, we identify proof-of-principle experiments for demonstrating soon practical quantum repeaters for quantum communications, exploiting in some way the CV nature of light. We present practical protocols which are also stepping stones to the first multi-node networks and a full-scale quantum internet. We also identify physical architectures to achieve the highest rates in quantum communications. Our protocols can realise the full potential of the quantum internet while also being practical and can be implemented with existing and developing technology. 

Our repeater designs are based on quantum light sources, detectors, and memories. Although compatible with large-scale quantum communication networks, our designs are also useful for point-to-point quantum communications without an intermediate repeater node; i.e., our techniques for building large-scale networks can be used even if the network consists of only two users, for example, for point-to-point quantum key distribution.

\begin{figure}
        \centering
        \includegraphics[width=1\textwidth]{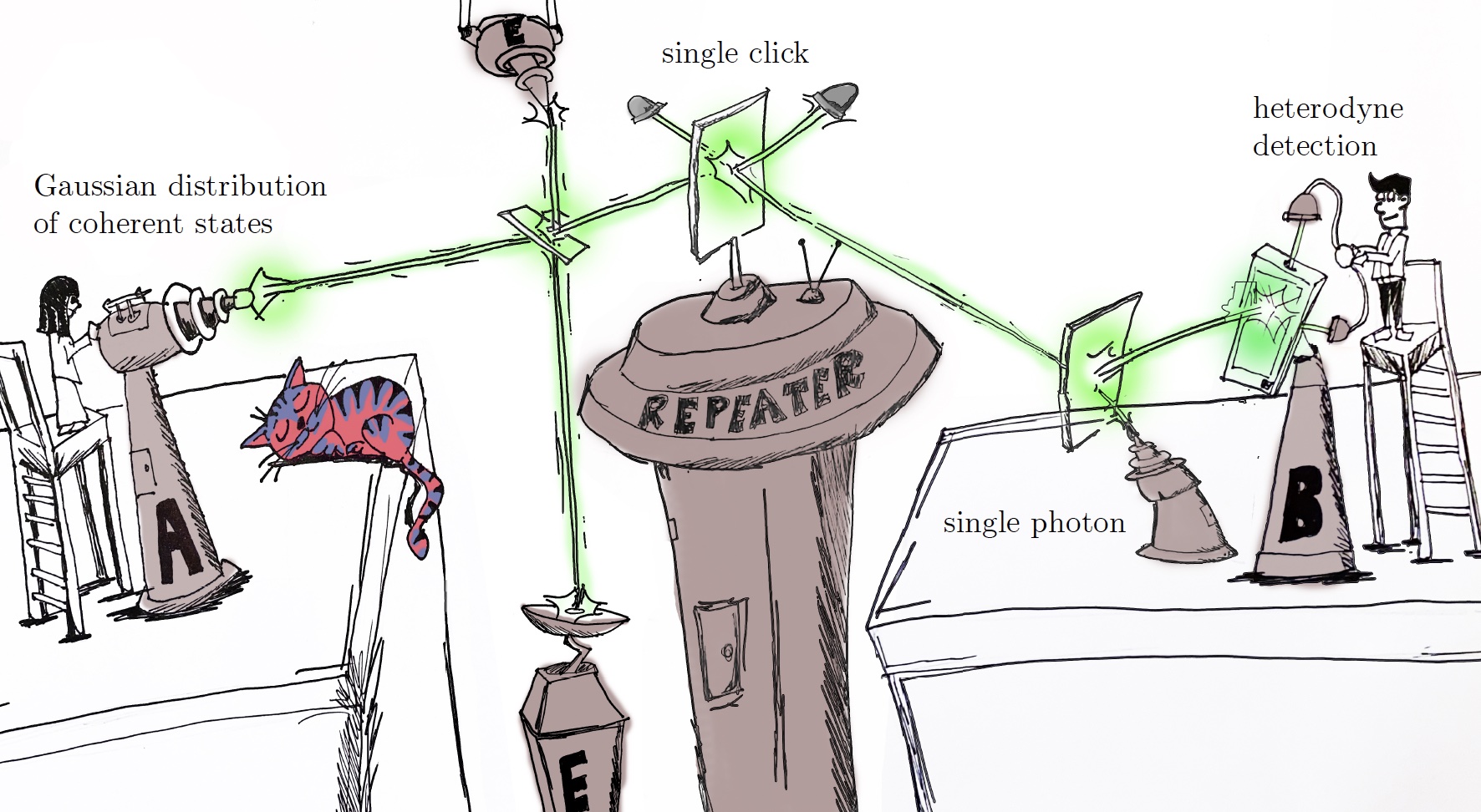}
\caption[Illustration of our simple quantum repeater experiment for continuous-variable quantum key distribution, introduced in~\cref{Chap:simple_repeater}.]{Illustration of our simple quantum repeater experiment for continuous-variable quantum key distribution, introduced in~\cref{Chap:simple_repeater}~\cite{Dave_Winnel,winnel2021overcoming}. The goal is for two trusted, neighbouring parties, Alice ($A$) and Bob ($B$), separated by a large distance, to distribute a secret random key. Quantum physics ensures a potential eavesdropper, Eve ($E$), learns nothing about the final key despite her best attempts (she is shown attacking one of the links). Our untrusted, quantum repeater overcomes the limitation of distance. It works as follows. Alice encodes information into the amplitude and phase of a laser (i.e., a Gaussian distribution of coherent states) and transmits the states to the repeater. Meanwhile, Bob fires a single photon at a mirror, preparing single-photon entanglement, and shares one output mode with the repeater. The untrusted repeater performs a measurement and when a single click is observed, the quantum states are teleported and noiselessly amplified to Bob. Finally, Bob measures the light, and Alice and Bob can distil a secret random key from their data. Schr\"{o}dinger's cat (from~\cref{Chap:QEC}) is peacefully sleeping in Alice's station.\label{Chap1fig:Matt_Poster_FNL_full_colour_annotated}}
\end{figure}

Recently, Ref.~\cite{slussarenko_quantum_2022} performed a nice experiment where the noisy channel was improved above the ambient level in a heralding way, thus, beating direct transmission. This is necessary but not sufficient towards building a quantum repeater. The theoretical results in~\cref{Chap:simple_repeater} of this thesis will lead to the first full demonstration of a CV repeater, and an experimental collaboration aimed at demonstrating the device is underway~\cite{winnel2021overcoming}. A cartoon illustration of the experiment is shown in~\cref{Chap1fig:Matt_Poster_FNL_full_colour_annotated}. This simple repeater will be concatenated using quantum memories into larger networks for quantum communications over intercontinental distances.

We also introduce other practical optimal and efficient designs for exciting cutting-edge demonstrations. In particular, we present novel noiseless linear amplification and teleportation devices for quantum states of light~\cite{PhysRevA.102.063715,PhysRevLett.128.160501} in~\cref{Chap:scissors}, optimal repeater designs with the fastest possible rates for long-distance quantum communication~\cite{https://doi.org/10.48550/arxiv.2203.13924} in~\cref{Chap:purification}, and a technique for error-corrected quantum repeaters based on bosonic codes in~\cref{Chap:QEC}.




\section{Structure of this thesis}

The structure of this thesis is detailed as follows.

\subsubsection{\cref{Chap:math_background}}

\noindent
We present a brief review of quantum information and then we describe the components necessary for a quantum repeater.


\subsubsection{\cref{Chap:other}}

\noindent
We consider a novel repeaterless protocol for continuous-variable quantum key distribution where information leakage to an eavesdropper is eliminated.

\subsubsection{\cref{Chap:scissors}}

\noindent
We generalise the linear-optics technique called quantum scissors to higher-order photon numbers and describe its use in performing noiseless linear amplification (NLA).


\subsubsection{\cref{Chap:CV_QR}}

\noindent
We consider a quantum repeater based on NLA and CV entanglement swapping.

\subsubsection{\cref{Chap:simple_repeater}}

\noindent
We show that our quantum scissors can be used as a simple quantum repeater for CV quantum key distribution without requiring quantum memories. We also introduce a method for optimising entanglement-swapping protocols.

\subsubsection{\cref{Chap:purification}}

\noindent
We introduce repeaters that achieve the ultimate limits and can purify completely from pure loss.



\subsubsection{\cref{Chap:QEC}}

\noindent
We consider repeaters with quantum error correction.


\subsubsection{\cref{Chap:Conclusion}}

\noindent
We summarise the results of this thesis and provide an outlook in this last chapter.

%% file: Chapter2/Chapter2.tex
\chapter{Background}
\label{Chap:math_background}	
\pagestyle{headings}

To begin, we briefly review the mathematical concepts that are required to understand this thesis. We discuss the fundamental limits of quantum communications, and we describe the components necessary for effective quantum repeaters. These sections will be very relevant to the material covered in later chapters.


\section{Quantum information theory}
\label{Chap2Sec:QI_theory}	

This thesis is presented from the angle of quantum optics. In quantum optics, Gaussian states are pervasive. For instance, a laser outputs a coherent state of light which makes it a good candidate for transmitting quantum information, and the generation of entangled two-mode Gaussian states of light is a prominent method for distributing continuous-variable entanglement between two users of a network. To our toolbox of Gaussian quantum information elements, we also add photon-number states and measurements. We show later in this thesis that photon-number measurements are extremely powerful for designing a quantum repeater. 

We begin with a summary of bosonic systems and some elements from quantum optics. Our aim is not to include a full review of Gaussian quantum information and we refer the reader instead to, for example, Refs.~\cite{Weedbrook_2012,Serafini2017QuantumCV}.

\subsection{Summary of bosonic systems}

A quantum state of a physical system is described by a density matrix, obtained from the density operator by a choice of basis. The density operator (or density matrix), \hypertarget{hyperlinklabel_rho}{\hyperlink{hyperlinkbacklabel_rho}{$\hat{\rho}$}} is defined
\begin{align}
    \hat{\rho} &= \sum_j p_j \ketbra{\psi_j}{\psi_j},
\end{align}
representing the state of an ensemble where a pure state $\ket{\psi_j}$ is prepared with probability $p_j$. The density operator is Hermitian, positive semi-definite, and has trace one.  Note that we will often denote operators and density matrices without the hat for simplicity.

$N$ bosonic modes, i.e., $N$ quantum harmonic oscillators, represent a prototypical continuous-variable system, associated with a tensor-product Hilbert space $\mathcal{H}^{\otimes N}$ and corresponding $N$ pairs of creation and annihilation operators $\{\hat{a}_k, \hat{a}_k^\dagger\}_{k=1}^N$. We arrange these operators into the vector:
\begin{align}
    \mathbf{\hat{b}} &\equiv (\hat{a}_1, \hat{a}^\dagger_1, \hat{a}_2, \hat{a}^\dagger_2, \dots ,\hat{a}_N, \hat{a}^\dagger_N)^\text{T},
\end{align}
which must obey the commutation relations:
\begin{align}
    [ \hat{b}_i, \hat{b}_j ] &= \Omega_{ij},
\end{align}
where $\Omega_{ij}$ is an element of $\Omega \equiv \bigoplus_{i=1}^{N} \begin{bmatrix}  0 & 1 \\ -1 & 0 \end{bmatrix}$.

The Hilbert space of $N$-mode systems is infinite-dimensional and separable. The single-mode infinite-dimensional Hilbert space is spanned by the Fock or photon-number basis, $\{ \hypertarget{hyperlinklabel_n}{\hyperlink{hyperlinkbacklabel_n}{\ket{n}}} \}_{n=0}^\infty$, composed by the eigenstates of the photon-number operator $\hat{n}\equiv \hat{a}^\dagger \hat{a}$, i.e. $\hat{n}\ket{n}=n \ket{n}$. We have
\begin{align}
    \hat{a}\ket{0}=0, \; \hat{a}\ket{n}=\sqrt{n}\ket{n-1} \text{ for } n\geq 1,
\end{align}
and
\begin{align}
    \hat{a}^\dagger\ket{n}= \sqrt{n+1}\ket{n+1} \text{ for } n\geq 0.
\end{align}
We use the following shorthand for the tensor product of Fock states: $\ket{i}\otimes\ket{j} = $ \hypertarget{hyperlinklabel_ij}{\hyperlink{hyperlinkbacklabel_ij}{\ket{i,j}}}.

We make use of the following definition of the quadrature operators \hypertarget{hyperlinklabel_q}{\hyperlink{hyperlinkbacklabel_q}{$\hat{q}$, $\hat{p}$}} for the position and momentum, respectively:
\begin{align}
    \hat{q} \equiv (\hat{a}+\hat{a}^\dagger),\; \hat{p} \equiv {i}(\hat{a}^\dagger-\hat{a}).
\end{align}
We arrange the quadrature operators in the vector: 
\begin{align}
    \mathbf{\hat{x}} &\equiv (q_1, p_1, q_2, p_2, \dots ,q_N, p_N)^\text{T}.
\end{align}
They satisfy the canonical commutation relation $[\hat{x}_i,\hat{x}_j] =2 i \Omega_{ij}$, in units $\hslash = 2$.

The quadrature operators are observables with eigenstates:
\begin{align}
    \hat{q} \ket{q} = q \ket{q},\; \hat{p}\ket{p} = p \ket{p},
\end{align}
with continuous real eigenvalues. The eigenstates $\ket{q}$ and $\ket{p}$ are improper eigenstates since they are nonnormalisable.

The two bases identified by the sets of eigenstates $\{\ket{q}\}_{q\in \mathbb{R}}$ and $\{\ket{p}\}_{p\in \mathbb{R}}$ are related by a Fourier transform
\begin{align}
    \ket{q} &= \frac{1}{2\sqrt{\pi}} \int \text{d}p \; e^{-iqp/2} \ket{p},\\
    \ket{p} &= \frac{1}{2\sqrt{\pi}} \int \text{d}q \; e^{iqp/2} \ket{q}.
\end{align}
The quadrature eigenvalues are continuous variables that can fully describe the $N$-mode bosonic system.

This is all conveniently expressed with the notion of phase-space representation. Any density matrix $\hat{\rho}$ has an equivalent representation in terms of a  quasi-probability distribution called the Wigner function, defined over phase space. The main quantities that characterise the Wigner representation are the statistical moments of the quantum state: the first moment or the mean value, ${\bar{x}} = \Tr(\mathbf{\hat{x}}\hat{\rho})$, and the second moment or covariance matrix, $V$. The elements of the covariance matrix are
\begin{align}
    V_{ij} &\equiv \frac{1}{2} \langle \{ \Delta \hat{x}_i, \Delta \hat{x}_j \} \rangle,
\end{align}
where $\Delta \hat{x}_i \equiv \hat{x}_i - \langle \hat{x}_i \rangle$, and $\{,\}$ is the anticommutator.

The covariance matrix must satisfy the uncertainty principle:
\begin{align}
    V+i\Omega &\geq 0, \; \Omega \equiv \bigoplus_{i=1}^{N} \begin{bmatrix}  0 & 1 \\ -1 & 0 \end{bmatrix}, \label{eq:uncertainty_principle}
\end{align}
and one can derive the Heisenberg uncertainty principle for position and momentum, $V(\hat{q})V(\hat{p})\geq 1$. Gaussian states are fully characterised by the first and second moments. Further, a pure state (i.e., where $\tr(\hat{\rho}^2)=1$) is Gaussian if and only if its Wigner function is non-negative~\cite{HUDSON1974249}.

A quantum state undergoes a transformation called a quantum operation which is a linear completely-positive (CP) trace-non-increasing map. A quantum operation is called a quantum channel when it is trace-preserving (TP) (i.e., a CPTP map). A quantum operation is Gaussian when it transforms Gaussian states into Gaussian states~\cite{De_Palma_2015}.

A quantum measurement is described by a set of operators $\{E_i\}$ satisfying the completeness relation $\sum_i E_i^\dagger E_i = \Id$. Outcome $i$ has probability $P_i=\tr(\hat{\rho}E_i^\dagger E_i)$ and the state is projected onto $\hat{\rho}_i = E_i\hat{\rho}E_i^\dagger /P_i $. For CV systems, measurements are described by continuous outcomes $i \in \mathbb{R}$, so that $P_i$ is a probability density. A measurement is Gaussian if when applied to Gaussian states it provides outcomes that are Gaussian-distributed. In practice, Gaussian measurements are performed using homodyne detection, ancilla Gaussian states, and linear optics.

The most important Gaussian states are listed as follows: vacuum state $\ket{0}$, thermal state \hypertarget{hyperlinklabel_thermal}{\hyperlink{hyperlinkbacklabel_thermal}{$\hat{\rho}_\text{th}$}}, coherent state \hypertarget{hyperlinklabel_alpha}{\hyperlink{hyperlinkbacklabel_alpha}{$\ket{\alpha}$}}, single-mode squeezed state \hypertarget{hyperlinklabel_sqz}{\hyperlink{hyperlinkbacklabel_sqz}{$\ket{\alpha,r}$}}, and two-mode squeezed vacuum (TMSV) state $\ket{\chi}$. The most important Gaussian unitaries are as follows: displacement, single-mode squeezing parameterised by squeezing parameter \hypertarget{hyperlinklabel_r}{\hyperlink{hyperlinkbacklabel_r}{r}}, two-mode squeezing, phase rotation, and the beamsplitter. In additional to these Gaussian states and transformations, photon-number (Fock) states $\ket{n}$ and photon-number measurements are useful non-Gaussian resources for many protocols.  Descriptions of all these elements are readily available in the literature~\cite{Weedbrook_2012}. Finally, superpositions of Gaussian states, for example, coherent-state superpositions (i.e., cat states) and squeezed-state superpositions (i.e., approximate Gottesman-Kitaev-Preskill (\hypertarget{hyperlinklabel_GKP}{\hyperlink{hyperlinkbacklabel_GKP}{GKP}}) states~\cite{GKP2001}) are useful non-Gaussian states for error correction~\cite{PhysRevA.97.032346}, considered in~\cref{Chap:QEC}.

In the following subsections, we define the beamsplitter transformation and noiseless linear amplification.

\subsection{Beamsplitter}\label{sec:beamsplitter}

A mirror, or beamsplitter (\hypertarget{hyperlinklabel_BS}{\hyperlink{hyperlinkbacklabel_BS}{BS}}), is a common optical component. It is described by a Gaussian unitary transformation, and it is the simplest example of an interferometer for the case of two bosonic modes. For every input beam, it has two output beams, one reflected and one transmitted. The properties of a beamsplitter are described by the transmissivity, $T$.

The beamsplitter unitary transformation is
\begin{align}
    {B}(\theta) &= e^{{\theta}(\hat{a}^\dagger \hat{b} - \hat{b} \hat{a}^\dagger)},
\end{align}
where $\hat{a}$ and $\hat{b}$ are the annihilation operators of the two modes, and the transmissivity is determined by $T = \cos^2{\theta} \in [0,1]$.

\subsection{Noiseless linear amplification}\label{sec:NLA}

Amplifying channels describe optical processes, such as phase-insensitive amplification where the input signals are amplified deterministically with the addition of thermal noise. Thermal noise is introduced due to the no-cloning theorem and the uncertainty principle.

We can get around these limitations if the device is allowed to be probabilistic. This is called noiseless linear amplification (\hypertarget{hyperlinklabel_NLA}{\hyperlink{hyperlinkbacklabel_NLA}{NLA}}) which performs the transformation $\sum_n c_n \ket{n} \to \sum_n g^n c_n \ket{n}$. The ideal transformation has zero success probability, so to be useful it is common to introduce an energy bound or a restriction on the allowed states.

Optimal NLA with gain, \hypertarget{hyperlinklabel_g}{\hyperlink{hyperlinkbacklabel_g}{$g$}}, is ideal up to the $n$th Fock component. It performs the transformation~\cite{PhysRevA.89.023846}
\begin{align}
    M &= \frac{1}{g^N} \sum_{n=0}^{N} g^n \ketbra{n}{n} + \sum_{n=N+1}^{\infty} \ketbra{n}{n}.
\end{align}
This can be shown by acting $M$ on a state in the Fock basis:
\begin{align}
    M \sum_n c_n \ket{n} &= \frac{1}{g^N} \sum_{n=0}^{N} g^n \ketbra{n}{n} \sum_n c_n \ket{n} + \sum_{n=N+1}^{\infty} \ketbra{n}{n} \sum_n c_n \ket{n} \\
    &= \frac{1}{g^N} \sum_{n=0}^{N} g^n  c_n \ket{n} + \sum_{n=N+1}^{\infty} c_n \ket{n}.
\end{align}
This is the optimal transformation with success probability $\sim 1/g^{2N}$ which is $\sim 1/g^2$ for small input states.


\section{Noise models}

Bosonic lossy channels are the most important channels for fibre and free-space quantum communications. We introduce the pure-loss and thermal-loss channels for optical fibre, and the fading channel for free-space. 

Other important channels not considered in this thesis are the qudit dephasing channel~\cite{devetak2004capacity}, and bosonic dephasing channels as considered in Ref.~\cite{Lami2023} (see also references therein).

The noise acts differently on the quantum information encoded in different ways. For instance, loss on polarisation qubits (encoded in definite-energy states of a single photon) becomes erasure errors, while the effect of loss on coherent states is to reduce the amplitude. On cat states, loss induces phase-flip errors. Against field qubits, loss becomes amplitude-damping errors.

\subsection{Pure loss}

Bosonic pure loss is the dominant source of noise for many quantum communication tasks. The pure-loss channel is equivalent to introducing a vacuum mode and mixing the data state on a beamsplitter with transmissivity \hypertarget{hyperlinklabel_eta}{\hyperlink{hyperlinkbacklabel_eta}{$\eta$}} $ \in [0,1]$. The Kraus-operator representation of the single-mode pure-loss channel is~\cite{PhysRevA.56.1114,PhysRevA.84.042311}
\begin{align}
    \mathcal{L} (\rho) &= \sum_{l=0}^{\infty} \hat{A}_l \rho \hat{A}_l^\dagger,\label{Chap2eq:pure_loss_Kraus_represenation}
\end{align}
with Kraus operators $\hat{A}_l  = \sqrt{\frac{(1-\eta)^l}{l!}}\eta^{\hat{\frac{n}{2}}}\hat{a}^l$ associated with losing $l$ photons to the environment, where $\hat{a}$ and $\hat{a}^\dagger$ are the single-mode annihilation and creation operators, respectively, and $\hat{n}=\hat{a}^\dagger \hat{a}$ is the photon-number operator. 

Throughout this thesis, we assume optical fibre with loss of 0.2 dB/km.

\subsection{Thermal-loss channel\label{sec:thermal_noise}}

The thermal-loss channel consists of pure loss with additional thermal noise. It is equivalent to combining a thermal state $\rho_\text{th}$ with mean photon number \hypertarget{hyperlinklabel_barn}{\hyperlink{hyperlinkbacklabel_barn}{$\bar{n}$}} and variance $V=2\bar{n}+1=\frac{\eta \xi}{ 1-\eta} + 1$, with excess noise $\xi$, with the input mode on a beamsplitter of transmissivity $\eta$.

It can be decomposed into pure loss of transmissivity $\tau = \eta/G$ followed by quantum limited amplification with gain \hypertarget{hyperlinklabel_G}{\hyperlink{hyperlinkbacklabel_G}{$G$}} $= 1+(1-\eta)\bar{n}= \frac{\eta\xi}{2} + 1$~\cite{Caruso_2006,PhysRevA.84.042311,PhysRevA.96.062306}. The Kraus representation is
\begin{align}
    \mathcal{N} (\rho) &= \sum_{k,l=0}^{\infty} \hat{B}_k \hat{A}_l \rho \hat{A}_l^\dagger \hat{B}_k^\dagger,\label{eq:loss_thermal_noise}
\end{align}
with Kraus operators
\begin{align}
    \hat{A}_l  &= \sqrt{\frac{(1-\tau)^l}{l!}}\tau^{\hat{\frac{n}{2}}}\hat{a}^l\\
    \hat{B}_k &= \sqrt{\frac{1}{k!}\frac{1}{G}\left(\frac{G-1}{G}\right)^k} \hat{a}^{\dagger k} G^{{-}\hat{\frac{n}{2}}}.
\end{align}
$\hat{A}_l$ are the Kraus operators of the pure-loss channel and $\hat{B}_k$ are the Kraus operators of the quantum limited amplifier.

Note that the opposite decomposition is possible whenever the thermal channel is not entanglement breaking: see Theorem 31 of Ref.~\cite{Sharma_2018}

\subsection{Fading channel}

The transmissivity of a free-space channel fluctuates in time due to atmospheric turbulence, called a fading channel. This contrasts a fibre link with a fixed transmissivity. Fading channels can be characterised by a probability distribution $p(\eta)$~\cite{PhysRevA.80.021802}, and decomposed into a set of sub-channels. The transmissivity for each sub-channel, $\eta_i$, is relatively stable and occurs with probability $p_i$, where $\sum_i p_i = 1$.

\section{Fundamental limits of quantum communications}\label{sec:fun_limits}

The great challenge for quantum communication~\cite{Gisin_2007} is how to overcome loss~\cite{cerf2007quantum}, the dominant source of noise through free space and telecom fibres. Many applications~\cite{PhysRevLett.120.080501,PhysRevLett.121.043604,PhysRevA.97.032329,7562346,DANOS200773}, including quantum key distribution (\hypertarget{hyperlinklabel_QKD}{\hyperlink{hyperlinkbacklabel_QKD}{QKD}})~\cite{RevModPhys.81.1301,Pirandola_2020,Xu_2020} (i.e., the task of sharing a secret random key between two distant parties), suffer from an \bblack{exponential rate-distance scaling~\cite{takeoka2014}}. Determining the most efficient protocols for distributing quantum information, entanglement, and secure keys is of vital importance to realise the full capability of the quantum internet~\cite{kimble2008}.

The rate of quantum information transmission is bounded by the maximum achievable rate of the so-called coherent information~\cite{Horodecki_2000}. The following papers are important in defining and interpreting what follows~\cite{Horodecki_2000,Devetak_2005,Devetak_2006}.

\bblack{It is known that the reverse coherent information (\hypertarget{hyperlinklabel_RCI}{\hyperlink{hyperlinkbacklabel_RCI}{RCI}})~\cite{PhysRevLett.102.210501}, \hypertarget{hyperlinklabel_R}{\hyperlink{hyperlinkbacklabel_R}{$R$}}, is an achievable rate for entanglement distillation, $E_D$, in units of ebits per use of the channel, via an implicit optimal protocol based on \textit{one-way}
    classical communication. An \hypertarget{hyperlinklabel_ebit}{\hyperlink{hyperlinkbacklabel_ebit}{ebit}} is the amount of entanglement in one maximally-entangled qubit (Bell pair). The RCI is defined in~\cref{sec:RCI}. For the bosonic pure-loss channel, this rate is $R = -\log_2{(1-\eta)}$~\cite{PhysRevLett.102.050503,Pirandola_2017}, where $\eta \in [0,1]$ is the channel transmissivity. This is an achievable rate for entanglement distillation, \hypertarget{hyperlinklabel_E_D}{\hyperlink{hyperlinkbacklabel_E_D}{$E_{\text{D}}$}}, over the lossy channel and is also an achievable rate for QKD, \hypertarget{hyperlinklabel_K}{\hyperlink{hyperlinkbacklabel_K}{$K$}}, since an ebit is a specific form of secret key bit. To summarise, we have $K \geq E_{\text{D}} \geq R = -\log_2{(1-\eta)}$. Ref.~\cite{Pirandola_2017} proved the upper bound, the so-called repeaterless or Pirandola–Laurenza–Ottaviani–Banchi (\hypertarget{hyperlinklabel_PLOB}{\hyperlink{hyperlinkbacklabel_PLOB}{PLOB}}) bound, that is, $K \leq  -\log_2{(1-\eta)}$; see also~\cite{Wilde_2017}. This, together with the lower bound, $R$, from Ref.~\cite{PhysRevLett.102.050503}, establishes $K = E_{\text{D}} = R = -\log_2{(1-\eta)} =$ \hypertarget{hyperlinklabel_C}{\hyperlink{hyperlinkbacklabel_C}{$C$}}, the two-way assisted entanglement distribution capacity and secret key distribution capacity of the pure-loss channel.}

Likewise, there are fundamental limits to the highest end-to-end rates of arbitrary quantum communication networks~\cite{Pirandola_2019}, where untrusted nodes divide the total distances into shorter quantum channels (links). The untrusted node is a special-purpose quantum processor called a quantum repeater (\hypertarget{hyperlinklabel_QR}{\hyperlink{hyperlinkbacklabel_QR}{QR}}), designed to overcome noise and distribute entanglement effectively. Quantum repeaters are strictly required to beat the repeaterless (PLOB) bound~\cite{munro2015inside,Muralidharan2016}. For a linear repeater chain, it is optimal to place repeaters equidistantly, then the ultimate end-to-end rate is given by $-\log_2{(1-\eta)}$~\cite{Pirandola_2019}, where $\eta$ now refers to the transmissivity of each link. For a multiband network, consisting of $m$ generally-entangled channels in parallel, the rate is additive, $-m\log_2{(1-\eta)}$~\cite{Pirandola_2019}. \bblack{These ultimate repeater bounds are achievable by using an optimal entanglement distillation protocol followed by quantum teleportation (entanglement swapping), while ideal quantum memories are most-likely required to achieve the highest rates.}  

\section{Quantifying rates}

 This thesis makes extensive use of the RCI and the EOF, defined next.

\subsection{Reverse coherent information\label{sec:RCI}}

Take a maximally entangled state of two systems $A$ and $B$. Propagating the $B$ system through the quantum channel defines the Choi state of the channel. Then the reverse coherent information (RCI) represents a lower bound for the distillable entanglement and for the optimal secret key rate. The RCI of a state $\rho_{AB}$ is defined as~\cite{Horodecki_2000,Devetak_2005,Devetak_2006,PhysRevLett.102.210501}
\begin{align}
R(\rho_{AB}) &= S(\rho_A) - S(\rho_{AB}),
\end{align}
where $S(\rho_A)$ and $S(\rho_{AB})$ are von Neumann entropies of $\rho_A = \tr_B(\rho_{AB})$ and $\rho_{AB}$ respectively. The von Neumann entropy of $\rho$ is \hypertarget{hyperlinklabel_von_Neumann}{\hyperlink{hyperlinkbacklabel_von_Neumann}{$S({\rho})$}} = $-\tr(\rho \log_2 \rho)$.

The Gaussian RCI is calculated from the covariance matrix of $\rho_{AB}$:
\begin{align}
\text{Gaussian RCI} &= g(\vec{\nu}_A)-g(\vec{\nu}_{AB}),
\end{align}
where $g(x) = \frac{x+1}{2}\text{log}_2(\frac{x+1}{2}) - \frac{x-1}{2}\text{log}_2(\frac{x-1}{2})$, and $\vec{\nu}_A$ and $\vec{\nu}_{AB}$ are the symplectic eigenvalues of the covariance matrices of $\rho_A$ and $\rho_{AB}$ respectively.

\subsection{Entanglement of formation}

The entanglement of formation (\hypertarget{hyperlinklabel_EOF}{\hyperlink{hyperlinkbacklabel_EOF}{EOF}}) is a useful measure of entanglement~\cite{PhysRevA.54.3824}. Compared with the EOF, the Gaussian entanglement of formation (\hypertarget{hyperlinklabel_GEOF}{\hyperlink{hyperlinkbacklabel_GEOF}{GEOF}}) is straightforward to calculate~\cite{Wolf_2004}. The GEOF quantifies the amount of two-mode squeezing required to prepare an entangled state from a classical state~\cite{PhysRevA.96.062338}, which is an upper bound on the EOF. When required in this thesis, the GEOF is calculated following Ref.~\cite{PhysRevA.96.062338} and using results from Ref.~\cite{PhysRevLett.84.2722,PhysRevLett.84.2726}.

\section{Quantum key distribution}

QKD~\cite{RevModPhys.81.1301,Pirandola_2020,Xu_2020} is the task of distributing a secret random key between two distant parties. Its provable security relies on quantum physics, and it is therefore a promising solution to the vulnerability of current classical cryptosystems. The first QKD protocols were based on discrete variables (\hypertarget{hyperlinklabel_DV}{\hyperlink{hyperlinkbacklabel_DV}{DV}}), such as BB84~\cite{BB84}, where the fundamental component is a qubit. DV QKD generally uses qubit encodings such as the polarisation of a photon. The original BB84 protocol~\cite{BB84} is the most widely studied and there are many variations and demonstrations.

QKD has been extended to continuous-variable (\hypertarget{hyperlinklabel_CV}{\hyperlink{hyperlinkbacklabel_CV}{CV}})~\cite{cerf2007quantum, Weedbrook_2012} systems, utilising the infinite-dimensional space of modes of light, with the benefit of using simpler experimental setups and off-the-shelf optical communication devices such as homodyne detectors.

There is a zoo of CV-QKD protocols~\cite{Weedbrook_2004,gehring2015single, usenko2018unidimensional,usenko2019generalized,Grosshans_2002,PhysRevA.63.052311,gottesman2001secure,Pirandola2015,PhysRevA.89.052301,Zhang_2014}. For instance, either coherent states~\cite{Grosshans_2002,Weedbrook_2004} or squeezed states~\cite{PhysRevA.63.052311,gottesman2001secure} may be distributed between the trusted parties, and the ensemble may be symmetric or asymmetric in phase space (i.e., with respect to the quadratures of light). See Ref.~\cite{Pirandola_2020} for a review. QKD protocols must be secure against the most general eavesdropping attacks allowed by the laws of quantum physics. For advanced, state-of-the-art CV security proofs, see Ref~\cite{leverrier2017security, furrer2011continuous}.

CV protocols give the highest rates since the states are infinite-dimensional. For instance, the maximum rate for BB84 without a repeater assuming a pure-loss channel is $K_\text{BB84} = \eta$, where $\eta$ is the transmissivity of the channel. Whereas, for CV QKD the maximum rate without a repeater is $K = \log_2{(1-\eta)} = C$, the capacity of the pure-loss channel, as we saw in~\cref{sec:fun_limits}.

In what follows, we briefly review the state-of-the-art security of CV QKD, then we discuss the achievable rates of CV-QKD protocols, and finally, we take a na\"ive attempt at designing a repeater using Gaussian states and operations only. This attempt of course fails, driving the need for powerful quantum repeaters, discussed in~\cref{sec:QR_intro}, and motivating the main content of this thesis.

\subsection{State-of-the-art security of CV-QKD protocols\label{Chap8sec:security}}

The goal of security analysis is to provide a useful upper bound on Eve's information using correlations between Alice and Bob, leading to an expression for the secret key rate. Unfortunately, proving security is problematic in CV QKD because the Hilbert space is infinite-dimensional, the measurement operators are unbounded, and channel parameters must be estimated of noisy quantum channels. Compare this with DV QKD where the Hilbert space is finite-dimensional, and loss can be dealt with by simply discarding the state. It is known that Gaussian collective attacks are asymptotically optimal for fully-Gaussian CV-QKD protocols~\cite{garcia-patron2006unconditional,navascues2006optimality}, i.e. in the asymptotic case of infinitely long keys. 

To go beyond asymptotic analysis, only a few Gaussian CV-QKD protocols are known to be composably secure against general attacks and include finite-size effects, for instance, the no-switching protocol based on Gaussian modulated coherent states in the $q$ and $p$ quadratures and heterodyne detection~\cite{leverrier2017security,Ghorai_2019}, and the switching protocol based on squeezed states with a Gaussian modulation in the $q$ or $p$ quadratures and homodyne detection such that the overall state prepared by Alice is a Gaussian thermal state ~\cite{furrer2011continuous,furrer2014reverse}. In this thesis, we restrict our attention to the asymptotic regime, though we caution that finite-size effects may be significant.

\subsection{Achievable rates of CV-QKD protocols}\label{sec:CVQKD_introduction}

In CV QKD, the quantum information is encoded into the quadratures of light, i.e., by preparing squeezed states or coherent states and measuring via homodyne or heterodyne detection.

\begin{figure}
        \centering
        \includegraphics[scale=1.3]{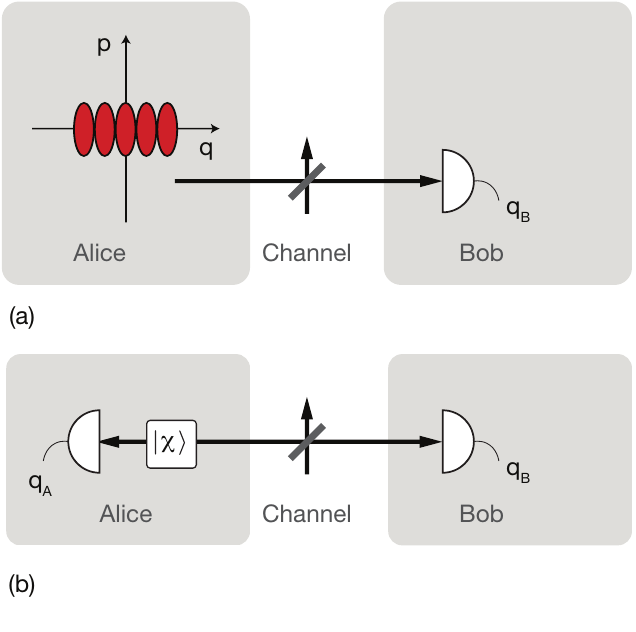}
    \caption[CV-QKD protocol based on squeezed states and homodyne detection]{CV-QKD protocol based on squeezed states and homodyne detection. (a) Prepare-and-measure (PM) version and (b) Entanglement-based (EB) version.}\label{Chap2fig:sqz_hom_QKD_protocol}
\end{figure}

For example, the repeaterless CV-QKD protocol based on squeezed states and homodyne detection is shown in~\cref{Chap2fig:sqz_hom_QKD_protocol}. In the prepare-and-measure (\hypertarget{hyperlinklabel_PM}{\hyperlink{hyperlinkbacklabel_PM}{PM}}) version shown in~\cref{Chap2fig:sqz_hom_QKD_protocol}a, squeezed states modulated in either the $q$ or $p$ quadrature, selected at random, are prepared by Alice and sent to Bob. Bob performs homodyne detection in a random quadrature, either $q$ or $p$. After classical communication, keeping only the times they coincidentally chose the same quadrature, Alice and Bob can generate a secret key from the data after classical error correction and privacy amplification.

During the protocol, correlations exist between all three parties, and the trusted parties, Alice, and Bob, must suppress their correlations with the eavesdropper, known as Eve, to ensure the final key is secret. This can be done via error correction, privacy amplification, or for an entanglement-based protocol, via entanglement distillation or purification. 

Security analysis is performed in the equivalent entanglement-based (\hypertarget{hyperlinklabel_EB}{\hyperlink{hyperlinkbacklabel_EB}{EB}}) version shown in~\cref{Chap2fig:sqz_hom_QKD_protocol}b, where Alice prepares a two-mode-squeezed-vacuum (\hypertarget{hyperlinklabel_TMSV}{\hyperlink{hyperlinkbacklabel_TMSV}{TMSV}}) state and keeps one mode and sends the other to Bob. The TMSV state is
\begin{equation}
\hypertarget{hyperlinklabel_chi}{\hyperlink{hyperlinkbacklabel_chi}{\ket{\chi}}} = \sqrt{1-\chi^2} \sum \chi^n \ket{n} \ket{n},
\end{equation}
where $0<\chi<1$ is the two-mode squeezing parameter. After homodyne detection on her mode, the PM and EB versions are equivalent.

The lower bound on the asymptotic secret key rate in the case of reverse reconciliation for collective attacks is given by the Devetak-Winter rate~\cite{devetak2005distillation} (note the discussion in Section VI-A (page 24) of Ref.~\cite{khatri2021secondorder})
\begin{equation}
K_{DW} = \beta I_{AB} - \chi_{EB},\label{eq:DW}
\end{equation}
where \hypertarget{hyperlinklabel_Shannon}{\hyperlink{hyperlinkbacklabel_Shannon}{$I_{AB}$}} is the classical (Shannon) mutual information between the trusted end users \hypertarget{hyperlinklabel_Holevo}{\hyperlink{hyperlinkbacklabel_Holevo}{$\chi_{EB}$}} is the Holevo quantity, the maximal quantum mutual information between Eve and Bob (the reference side of the information reconciliation), and $\beta$ is the reconciliation efficiency. 

The upper bound on the information extractable by Eve is given by the Holevo quantity~\cite{holevo1973bounds}
\begin{equation}
\chi_{EB} = S(\rho_E) - S(\rho_{E|b}),\label{eq:HOLEVO}
\end{equation}
where $S(\rho_E)$ is the von Neumann entropy of Eve's state, and $S(\rho_{E|b})$ is the von Neumann entropy of Eve's state conditioned on Bob's homodyne (or heterodyne) measurement. 

The Holevo information is obtained by allowing Eve access to the purification of the state shared between the end-users $\rho_{AB}$. The global state $\rho_{ABE}$ is pure and we can use the self-duality property of the von Neumann entropy to write $S(\rho_E)=S(\rho_{AB})$ and $S(\rho_{E|b})=S(\rho_{A|b})$, where $\rho_{A|b}$ is Alice's mode conditioned on a measurement on Bob's mode. 

For Gaussian states, $\chi_{EB}$ may be calculated from the symplectic eigenvalues of the covariance matrices of $\rho_{AB}$ and $\rho_{A|b}$ in the EB CV-QKD protocol. Thus, Eve's information is given by
\begin{equation}
\chi_{EB} = S(\rho_{AB}) - S(\rho_{A|b}),\label{eq:HOLEVO_pur}
\end{equation}
i.e., the Holevo quantity is given in terms of von Neumann entropies which can be calculated using the symplectic eigenvalues $\nu_k$ of the covariance matrix of the state~\cite{PhysRevA.59.1820}, i.e., $V_{AB}$ and $V_{A|b}$, via the relation $S(\rho)=\sum_{k=1}^N \frac{\nu_k+1}{2}\log_2 \frac{\nu_k+1}{2} - \frac{\nu_k-1}{2}\log_2 \frac{\nu_k-1}{2},$ where $N$ is the number of modes.

 We refer the reader to Ref.~\cite{Sanchez2007QuantumIW} for more details and for how to calculate Alice and Bob's mutual information $I_{AB}$. Furthermore, the aforementioned method to compute the asymptotic secret key rate works for many other CV-QKD protocols~\cite{Sanchez2007QuantumIW}; in particular, for the no-switching coherent-state CV-QKD protocol, where Alice and Bob instead do heterodyne detection in the EB version. 

It can be shown that the ideal performance of the repeaterless squeezed-state CV-QKD protocol in a pure-loss channel achieves the repeaterless bound~\cite{Pirandola_2017} up to a factor of $1/2$ to account for sifting. Indeed, by using quantum memories, Alice and Bob can ensure they always choose the same quadrature and remove the sifting factor $1/2$~\cite{PhysRevLett.102.050503}. While this protocol is optimal for QKD without a repeater, it does not suggest how to design a quantum repeater architecture for use in a multi-node network. We will detail this point in the next subsection.

\subsubsection{CV-MDI QKD}\label{sec:MDI_intro}

Recall that the main goal of this thesis is to present high-performance architectures which are effective quantum repeaters. To attempt to find such an architecture, let us place an untrusted node between the trusted users. We will show that this na\"ive design is not a quantum repeater for Gaussian states. We need something much more powerful to effectively repeat the quantum information.

\begin{figure}
\includegraphics[width=0.75\linewidth]{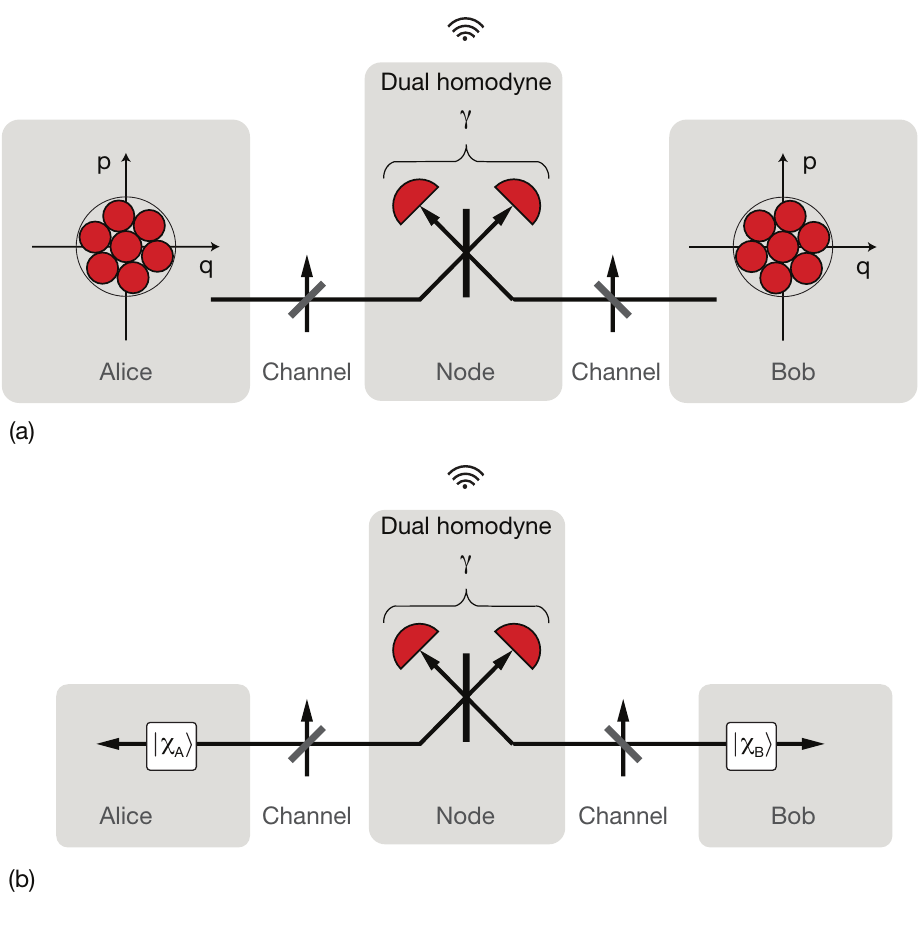}
\centering
\caption[CV-MDI QKD protocol]{The CV-MDI QKD protocol. (a) The PM version and (b) the EB version. The central node is untrusted. While useful for CV quantum communications, this setup is not an effective quantum repeater.}
\label{fig:CV_MDI_protocol}
\end{figure}

We consider the CV measurement-device-independent (\hypertarget{hyperlinklabel_MDI}{\hyperlink{hyperlinkbacklabel_MDI}{MDI}}) protocol based on dual-homodyne detection at the untrusted node~\cite{Ma_2014,Zhang_2014,Ottaviani_2015,Pirandola2015}. The CV-MDI protocol is shown in~\cref{fig:CV_MDI_protocol}. The PM version is shown in~\cref{fig:CV_MDI_protocol}a and the EB version in~\cref{fig:CV_MDI_protocol}b. The PM protocol works by Alice and Bob each preparing a coherent state (or a squeezed state) selected from a Gaussian distribution and transmitting the state to the untrusted node. Dual-homodyne detection is performed at the node. Dual-homodyne detection consists of a $50:50$ beamsplitter followed by orthogonal quadrature measurements. Outcome $\gamma$ is broadcast to all users, allowing Alice and Bob to generate a key from the data. This is possible since while the trusted users know information about the quantum states they each sent, Eve only knows information about $\gamma$. Since all measurements in the PM protocol are untrusted, we say that the protocol is measurement device independent.

The PM version shown in~\cref{fig:CV_MDI_protocol}a is equivalent to the EB version in~\cref{fig:CV_MDI_protocol}b, which is similar to CV entanglement swapping, discussed in~\cref{Chap2CVrepeatercomponents}. CV entanglement swapping is a useful technique to distribute entanglement which can then be used as a resource for quantum teleportation between nodes of a network.

\begin{figure}
\includegraphics[width=0.65\linewidth]{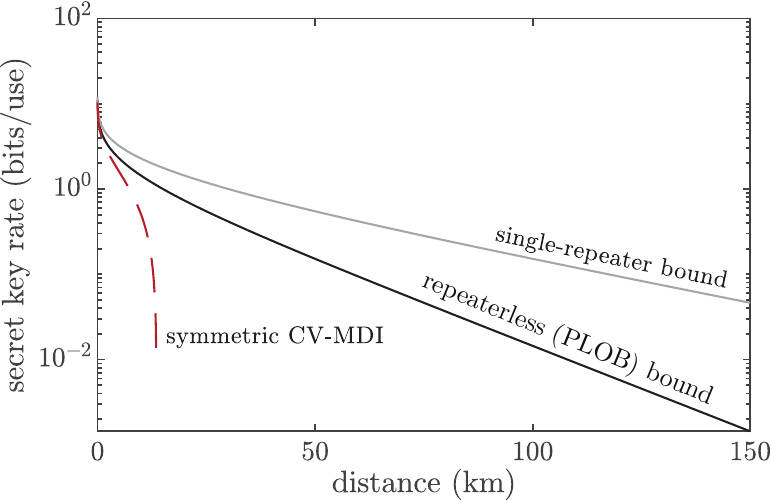}
\centering
\caption[RCI for the symmetric, Gaussian CV-MDI protocol as a function of the total distance for the pure-loss channel]{Achievable secret key rate given by the RCI for the symmetric, Gaussian CV-MDI protocol as a function of the total distance for the pure-loss channel. The ideal rate of the symmetric protocol optimises for infinite squeezing. The ultimate repeaterless (PLOB) bound~\cite{Pirandola_2017} and single-repeater bound~\cite{Pirandola_2019} are shown for comparison. The CV-MDI protocol cannot beat the repeaterless bound despite the node. Thus, the node is not effective at repeating Gaussian quantum information. The maximum node spacing is only about 13 km for a symmetric placement of the node. The protocol works best extremely asymmetrically, with the node placed in Bob's station with reverse reconciliation, then it is equivalent to a point-to-point protocol and achieves the repeaterless bound, but it cannot overcome it.}
\label{fig:CV_MDI_protocol_rate}
\end{figure}

We compute the highest asymptotic secret key rate of CV-MDI QKD (or equivalently CV entanglement swapping) using~\cref{eq:DW}, for the node placed symmetrically between Alice and Bob, shown in~\cref{fig:CV_MDI_protocol_rate}. For more details about the security analysis, see Ref.~\cite{Pirandola2015}.
 
We see that the RCI of CV-MDI QKD (or CV entanglement swapping) does not surpass the repeaterless (PLOB) bound~\cite{Pirandola_2017}, so the relay is not a repeater for Gaussian states and Gaussian channels. This unfortunate result is because it is impossible to distil entanglement using Gaussian states and operations only~\cite{PhysRevA.66.032316,PhysRevLett.89.137903}. The Gaussian protocol gives high rates at metropolitan distances, but the RCI falls to zero at large distances. 

Unfortunately, the highest rates are for an extremely-asymmetric configuration with the central node placed very close to one of the trusted parties. In the extremely-asymmetric configuration, the protocol is equivalent to direct transmission (i.e., the highest rate achieves the repeaterless bound but can never overcome it). In~\cref{Chap:simple_repeater}, we introduce for the first time possible CV-MDI QKD protocols with positive key rates at long distances by using a quantum repeater without quantum memories. Remarkably, the rates of our CV-MDI QKD repeater protocols are highest with the repeater node about \textit{half-way} between the trusted parties, so we do not require an asymmetric configuration, thus fixing the troublesome directional problem in CV QKD.

\section{Quantum repeaters}\label{sec:QR_intro}

Quantum repeaters are not just optical amplifiers, but rather exotic purpose-built quantum processors designed to protect and distribute quantum information and entanglement between users of a network. They overcome loss and noise in the quantum channel, extending the range~\cite{briegel1998quantum}. The first quantum repeater protocol used multiple rounds of entanglement swapping~\cite{ziukowski1993event} in order to connect entangled pairs and share entanglement between ends of a long-distance channel while entanglement purification~\cite{Bennett_1996} corrected against operational errors. 

There have been significant theoretical advancements in repeater protocols and experimental progress with repeater components~\cite{munro2015inside,Muralidharan2016, sangouard2011quantum}. Twin-field QKD~\cite{Lucamarini_2018} is repeater-like and can distribute a secret key at rates better than the repeaterless bound (unlike CV-MDI QKD); however, twin-field QKD cannot be used to communicate general quantum states. In~\cref{Chap:simple_repeater}, we introduce a simple design that can take quantum states as input, teleport and amplify those states across the long-distance channel, and beat the repeaterless bound. We are collaborating with experimentalists who are actively building this device~\cite{winnel2021overcoming}.

Quantum repeaters have previously been categorised into three generations depending on how they combat loss and other sources of noise~\cite{munro2015inside,Muralidharan2016}. We discuss the first and third generations in what follows (the second generation is somewhat outdated).

\subsection{Components of first-generation CV quantum repeaters}\label{Chap2CVrepeatercomponents}

With respect to CV systems, the first two generations remove loss via teleportation-based techniques, for instance, entanglement swapping and/or noiseless linear amplification. They are based on three essential ingredients: entanglement distribution, entanglement distillation, and entanglement swapping. We now discuss the operations of first-generation CV quantum repeaters.

First, entangled states, usually in the form of TMSV states, $\ket{\chi}$, are distributed between neighbouring nodes. Second, entanglement is distilled non-deterministically, which overcomes loss and noise in the links. Quantum memories are required to hold onto the quantum states while neighbouring links succeed in distilling their entanglement. Third, joint measurements are performed on some of the modes, heralding entanglement between non-neighbouring stations. Additional rounds of distillation and swapping can entangle stations separated over greater distances.

\begin{figure}
\centering
\includegraphics[width=1\linewidth]{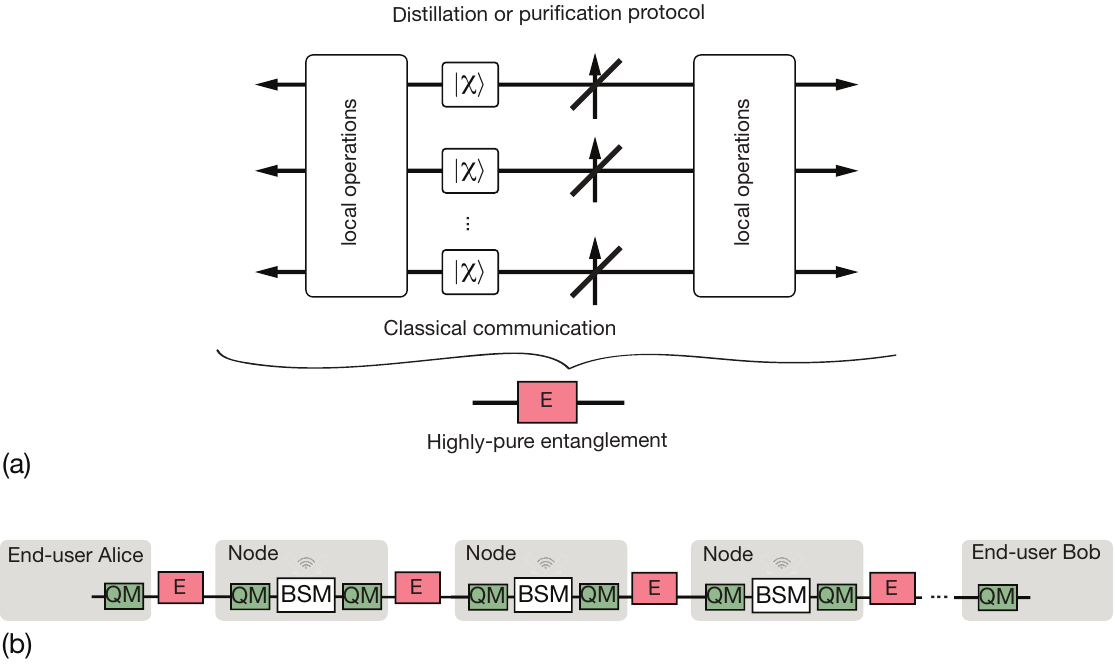}
\centering
\caption[Quantum networks using distillation or purification between nodes.]{Quantum networks using distillation or purification between nodes. \textbf{a} Distillation/purification protocol operating between neighbouring nodes. \textbf{b} Simple quantum network consisting of a linear quantum repeater chain. First, entanglement is distilled or purified between all neighbouring nodes of the network (red). The pure entanglement is held in quantum memories (QM, green). Finally, Bell-state measurements (BSM) and classical communication perform entanglement swapping such that eventually the end parties share highly-pure entanglement. The BSMs can be DV or CV depending on the mode of operation.}
\label{fig:purification_repeater_protocol}
\end{figure}

Let us consider a specific example. Consider a linear quantum network for simplicity where $N-1$ untrusted quantum repeaters divide the total distance between the end users into $N$ links. Neighbouring nodes perform entanglement distillation or purification shown in~\cref{fig:purification_repeater_protocol}a and store the improved entanglement in ideal quantum memories. The quantum memories allow the probabilistic distillation or purification schemes of each link to succeed independently from each other. Once this step is complete, highly-pure entanglement is shared between all neighbouring nodes of the network, as shown in~\cref{fig:purification_repeater_protocol}b. To distribute entanglement between the end users the nodes simply perform entanglement swapping via Bell-state measurements (\hypertarget{hyperlinklabel_BSM}{\hyperlink{hyperlinkbacklabel_BSM}{BSM}}) as shown in~\cref{fig:purification_repeater_protocol}b. (For now, let us assume that the swapping happens in parallel, and that many copies of the protocol are available.) 

The BSM step may be DV or CV, depending on whether DV BSMs or CV dual-homodyne measurements are performed, respectively. These two approaches are quite different. We now analyse in detail DV and CV entanglement swapping in the following.

\subsubsection{DV entanglement swapping}

If the distilled entangled states with dimension $d$ shared between neighbouring nodes of the network are maximally entangled, then the DV BSM at the node projects the state onto one of the known $d^2$ Bell states with success probability $1/d^2$. Hence, with a success probability of $1$, a maximally entangled state is heralded between non-neighbouring nodes. The nodes communicate classically which BSM outcome was obtained and based on this information the non-neighbouring nodes can perform the required correction. Eventually, the end users share maximal entanglement. If the initial entangled states are not maximally entangled, then the entanglement will decrease after the swapping.

\subsubsection{CV entanglement swapping}

We consider optimal Gaussian entanglement swapping, similarly described in Ref.~\cite{PhysRevA.83.012319}. This involves dual-homodyne detection and sending classical signals to both end users and performing displacements on both modes, unlike other protocols such as teleportation where operations are only performed on one mode. This ensures that the resulting entangled state remains pure after swapping pure entanglement. We will show next that the amount of entanglement decreases, in contrast to DV entanglement swapping. 

We assume that copies of pure TMSV states are shared between all neighbouring nodes of the network. For now, we assume these states are ideal but there may be additional truncation and noise. We will show in this thesis that our NLA devices, our simple repeater, and our purification protocol can all be used to distribute highly-pure TMSV states between any two neighbouring nodes of a lossy quantum network.

Then, we consider a deterministic CV BSM in the relay consisting of dual homodyne, i.e., mixing the two incoming modes on a $50:50$ beamsplitter and performing orthogonal quadrature measurements. The nodes communicate the outcomes of the dual-homodyne measurements so that the required displacement-correction operations can be performed. Since the state is Gaussian, the covariance matrix is all that is required to calculate the rate of distilled entanglement or estimate the secret key rate. The covariance matrix does not depend on the particular outcome of the dual-homodyne measurement, and it can be calculated easily via the symplectic transformations for the dual-homodyne measurements as follows.

First, assume all neighbouring parties in the network share copies of a Gaussian TMSV state, $\ket{\chi}$. This is sometimes difficult to achieve because recall that we need some non-Gaussian elements to distil entanglement from Gaussian states in the first place, so usually the purified pairs are slightly non-Gaussian; for example, often there is some truncation noise introduced by the chosen purification protocol or the NLA. So, if the purified states are not exactly Gaussian, we can at least require that the output states are sufficiently Gaussian by choosing a large enough truncation dimension cut-off for a given $\chi$. This then conveniently allows us to use the covariance-matrix formalism, simplifying our estimations of the secret key rate for CV QKD. Then the output states are approximately Gaussian and equal to copies of TMSV states $\ket{\chi}$, assuming the dimension of truncation is sufficiently large.

The Gaussian TMSV state $\ket{\chi}$ is fully characterised by the following covariance matrix~\cite{Weedbrook_2012}:
\begin{align}
   V &= \begin{bmatrix}
        \nu & 0 & \sqrt{\nu^2-1} & 0\\
        0 & \nu & 0 & -\sqrt{\nu^2-1}\\
        \sqrt{\nu^2-1} & 0 & \nu & 0\\
        0 & -\sqrt{\nu^2-1} & 0 & \nu\\
    \end{bmatrix}.\label{eq:CM_chi}
\end{align}
where $\nu=\cosh{2r}$ with $\chi=\tanh{r} \in [0,1]$.

We start with two copies of this state, i.e., two links connected by a single repeater. Mode 1 is entangled with mode 2 and mode 3 is entangled with mode 4. Recall that dual-homodyne detection consists of a $50:50$ beamsplitter followed by orthogonal quadrature measurements in $q$ and $p$. Let modes 1 and 4 be the two modes input to the quantum repeater and modes 2 and 3 be non-neighbouring nodes in the network. Modes 2 and 3 will remain and be entangled after the entanglement swapping operation. Modes 1 and 4 are input on a $50:50$ beamsplitter transforms the four-mode covariance matrix like~\cite{Weedbrook_2012}
\begin{align}
   \begin{bmatrix} V & \\ & V \end{bmatrix} &\to S_B \begin{bmatrix} V & \\ & V \end{bmatrix} S_B^T,
\end{align}
where
\begin{align}
   S_B &= \begin{bmatrix}
        \frac{1}{\sqrt{2}} & 0 & 0 & 0 & 0& 0 & \frac{1}{\sqrt{2}} & 0\\
        0 & \frac{1}{\sqrt{2}} & 0 & 0 & 0 & 0& 0 & \frac{1}{\sqrt{2}}\\
         0 & 0 & 1 & 0 & 0 & 0 & 0& 0\\
         0 & 0 & 0 & 1 & 0 & 0 & 0& 0\\
         0 & 0 & 0 & 0 & 1 & 0 & 0& 0\\
         0 & 0 & 0 & 0 & 0 & 1 & 0& 0\\
        -\frac{1}{\sqrt{2}} & 0 & 0 & 0& 0 & 0  & \frac{1}{\sqrt{2}} & 0\\
        0 & -\frac{1}{\sqrt{2}} & 0 & 0& 0 & 0 & 0 & \frac{1}{\sqrt{2}}\\
    \end{bmatrix}.
\end{align}

After the beamsplitter, the four-mode quantum system becomes
\begin{align}
  S_B \begin{bmatrix} V & \\ & V \end{bmatrix} S_B^T  &=  \left[ \begin{smallmatrix}  
                              \nu   &                                0   &     \frac{ \sqrt{2(\nu^2-1)}}{2}   &                                0   &    \frac{ \sqrt{2(\nu^2-1)}}{2}   &                                0   &                                0   &                                0\\
                            0   &                               \nu   &                                0   &    -\frac{ \sqrt{2(\nu^2-1)}}{2}   &                               0   &    -\frac{ \sqrt{2(\nu^2-1)}}{2}   &                                0   &                                0\\
 \frac{ \sqrt{2(\nu^2-1)}}{2}   &                                0   &                               \nu   &                                0   &                               0   &                                0   &    -\frac{ \sqrt{2(\nu^2-1)}}{2}   &                                0\\
                            0   &    -\frac{ \sqrt{2(\nu^2-1)}}{2}   &                                0   &                               \nu   &                               0   &                                0   &                                0   &     \frac{ \sqrt{2(\nu^2-1)}}{2}\\
 \frac{ \sqrt{2(\nu^2-1)}}{2}   &                                0   &                                0   &                                0   &                              \nu   &                                0   &     \frac{ \sqrt{2(\nu^2-1)}}{2}   &                                0\\
                            0   &    -\frac{ \sqrt{2(\nu^2-1)}}{2}   &                                0   &                                0   &                               0   &                               \nu   &                                0   &    -\frac{ \sqrt{2(\nu^2-1)}}{2}\\
                            0   &                                0   &    -\frac{ \sqrt{2(\nu^2-1)}}{2}   &                                0   &    \frac{ \sqrt{2(\nu^2-1)}}{2}   &                                0   &                               \nu   &                                0\\
                            0   &                                0   &                                0   &     \frac{ \sqrt{2(\nu^2-1)}}{2}   &                               0   &    -\frac{ \sqrt{2(\nu^2-1)}}{2}   &                                0   &                               \nu\\
                  \end{smallmatrix} \right] .
\end{align}

For a matrix written in block form:
\begin{align}
    \Gamma = \begin{bmatrix} \Gamma_X & \sigma \\ \sigma^T & \Gamma_Y \end{bmatrix},
\end{align}
measuring a quadrature of subsystem $\Gamma_Y$ via homodyne detection, where $\Gamma_Y$ is a $2\times2$ real matrix, transforms the covariance matrix of subsystem $\Gamma_X$ as follows~\cite{Weedbrook_2012}:
\begin{align}
    \Gamma &\to \Gamma_X-\sigma(\Pi \;  \Gamma_Y \; \Pi)^{-1} \sigma^T,\label{eq:hom_CM}
\end{align}
where $\Pi = \text{diag}(1,0)$ for $\hat{q}$ quadrature and $\Pi = \text{diag}(0,1)$ for $\hat{p}$ quadrature, and $(\Pi \;  \Gamma_Y \; \Pi)^{-1}$ is a pseudoinverse. Note that the output covariance matrix does not depend on the dual-homodyne measurement outcome.

For our four-mode system, we homodyne modes 1 and 4 in the repeater, one in $\hat{q}$ quadrature and one in $\hat{p}$ quadrature. We use~\cref{eq:hom_CM} twice since we perform two measurements. Performing homodyne in $\hat{q}$ quadrature of mode 1 and $\hat{p}$ quadrature of mode 4, the resulting covariance matrix is
\begin{align} 
&\to
 \begin{bmatrix}
      \frac{\nu^2+1}{2\nu}  &                     0  &     -\frac{\nu^2-1}{2\nu}  &                     0\\
                  0  &     \frac{\nu^2+1}{2\nu}  &                      0  &     \frac{\nu^2-1}{2\nu}\\
 -\frac{\nu^2-1}{2\nu}  &                     0  &      \frac{\nu^2+1}{2\nu}  &                     0\\
                  0  &     \frac{\nu^2-1}{2\nu}  &                      0  &     \frac{\nu^2+1}{2\nu}\\
                  \end{bmatrix},
\end{align}
so we can see that the output state is a TMSV state with reduced squeezing, quantified by the reduced noise variance $\tilde{\nu} = \frac{\nu^2+1}{2\nu}$. That is, where previously two non-neighbouring nodes were not entangled, now they share \textit{deterministically} a pure TMSV state with reduced squeezing. We keep doing this over all repeater nodes such that eventually the end users share deterministically a TMSV state with a covariance matrix of the same form with a new noise variance which depends on the number of repeater nodes and initial squeezing. This covariance matrix shared between the end users allows us to estimate the asymptotic secret key rate using~\cref{eq:DW}.

\bblack{Thus, entanglement is lost when distributing approximate TMSV states, captured by the covariance matrix transformation in~\cref{eq:hom_CM} and by the secret key rate formula in~\cref{eq:DW}.} However, the advantages of the CV scheme over the DV scheme is that CV entanglement swapping is ``easy''. The ``difficult'' parts of the protocol like counting many photons and preparing Fock states concern the entanglement distillation or purification steps only. Finally, to distribute a secret key from the distilled and distributed CV entanglement is also straightforward in CV since the end users simply do quadrature measurements on their final shared state and can distil a secret key after classical error correction and privacy amplification.

\subsection{Third-generation quantum repeaters}

First-generation quantum repeaters required probabilistic elements, quantum memories, and back-and-forth signalling. One can restore determinism in the repeaters and remove the requirement for quantum memories by using quantum error correction (\hypertarget{hyperlinklabel_QEC}{\hyperlink{hyperlinkbacklabel_QEC}{QEC}}) where the quantum information is redundantly encoded into a quantum-error-correction code and a recovery operation is applied after the noise to correct errors on the logical qubit (or quantum digit, or qudit). No classical communication is necessary, so the rates are bounded by the unassisted quantum capacity of the pure-loss channel~\cite{PhysRevLett.98.130501,PhysRevA.86.062306,PhysRevA.95.012339,Wilde_2018} which cannot withstand more than $50\; \%$ loss and is strictly less than the two-way assisted quantum capacity. Despite this limitation, repeaters using QEC avoid back-and-forth classical communication which can be quite slow. We consider repeaters using bosonic codes in~\cref{Chap:QEC}. 

We speculate that a hybrid approach, i.e., nondeterminstic error-corrected repeater, could enjoy benefits from both first and third-generation approaches.

\section{Main results of this thesis}

In this thesis, we investigate the performance and design of quantum repeaters for continuous variables. In~\cref{Chap:other}, we first consider a protocol for CV QKD which is repeaterless and the rate cannot overcome the limitation of distance; however, the novelty is that all information leakage to an eavesdropper is eliminated in a pure-loss channel.

Next, we consider protocols that can effectively improve the entanglement and overcome the destructive effect of loss. Before introducing our complete repeater designs, we first introduce quantum scissors for noiseless linear amplification (NLA) in~\cref{Chap:scissors}, an important component for the repeaters considered in \cref{Chap:CV_QR,Chap:simple_repeater}. It is a basic building block for a quantum repeater and can be implemented with existing technology.

In later chapters, we introduce effective repeaters, useful for long-distance quantum communication in complex quantum networks. In other words, we introduce repeater designs which are \textit{effective}, that is, the rate of these repeaters can surpass the highest possible point-to-point rates without a repeater, given by the repeaterless bound~\cite{Pirandola_2017}. The most practical repeaters are considered first, in~\cref{Chap:CV_QR,Chap:simple_repeater}. They are unable to saturate the ultimate limits; however, they are the most practical. Repeaters without memories, considered in~\cref{Chap:simple_repeater}, beat the repeaterless bound while being very efficient in the use of resources. A global quantum internet can be built using results from~\cref{Chap:simple_repeater} using just half the total resources as most other designs. Thus, we expect the repeaters considered in ~\cref{Chap:simple_repeater} to be the first to implement a working multi-node quantum repeater network in the near future.

Our quantum repeaters using purification (quantum error detection), introduced in~\cref{Chap:purification}, give the highest rates per use of the channel and can saturate the ultimate limits allowed by physics. In~\cref{Chap:QEC}, we consider repeaters with quantum error correction. They are effective so long as the repeater spacing is less than about 15 km.

Finally, we conclude and provide an outlook in~\cref{Chap:Conclusion}.


%% file: chapter_min_leakage/chapter_min_leakage.tex
\chapter{Minimisation of information leakage}
\label{Chap:other}	
\pagestyle{headings}

\noindent
The results of this chapter have been published in the following.\\

\noindent
1.~\cite{PhysRevA.104.012411} \textbf{M. S. Winnel}, N. Hosseinidehaj, and T. C. Ralph, ``Minimization of information leakage in continuous-variable quantum key distribution,'' \href{http://dx.doi.org/10.1103/PhysRevA.104.012411}{Phys. Rev. A \textbf{104}, 012411 (2021)}.\\

\noindent
A quantum communication protocol based on a Gaussian modulation of squeezed states in a single quadrature and measured via homodyne detection can completely eliminate information leakage to an eavesdropper in a pure-loss channel~\cite{jacobsen2018complete}. However, the asymmetry of the protocol with respect to the quadratures of light presents security issues and the eavesdropper's information is not necessarily minimised for general asymmetric attacks. In this chapter, we perform asymptotic security analysis of the asymmetric protocol against general asymmetric collective attacks and bound the eavesdropper's information via the Heisenberg uncertainty principle. The bound is not tight and therefore, we symmetrise the protocol in a heralding way, discarding the issues of asymmetry altogether. Our proposed heralding protocol asymptotically eliminates information leakage in a pure-loss channel and minimises leakage in a noisy channel. While this protocol is not a repeater since it does not improve the rate-distance scaling, it does simplify the classical post-processing part of the protocol which can speed up the secret key rate.

\section{\label{Chap8sec:intro}Introduction}

A CV analogue of BB84~\cite{BB84} is the protocol based on Gaussian modulated squeezed states with switching (i.e., randomly
choosing to squeeze and modulate either the $x$ or $p$ quadrature) and measured via homodyne detection~\cite{PhysRevA.63.052311}, discussed in~\cref{sec:CVQKD_introduction}. However, even in a pure-loss channel, information is inevitably leaked to an eavesdropper, i.e., Eve's maximal information $\chi_{EB}>0$ in~\cref{eq:DW}. During the protocol, Alice and Bob must suppress their correlations with Eve to ensure the final key is secret. 

Recently, a CV-QKD protocol has been devised which takes a different approach. By designing the alphabet of input states in a certain way, information leakage to Eve is completely and deterministically eliminated in a pure-loss channel and minimised in a symmetric noisy channel~\cite{jacobsen2018complete}. Information is encoded in a single quadrature via a Gaussian modulation of squeezed states, which are squeezed in the modulation direction such that the overall variance of the ensemble is shot noise. The motivation for minimising information leakage is that this will simplify classical post-processing and ultimately speed up the secret key rate. Another motivation is that zero information leakage in a pure-loss channel is analogous to discarded photons in DV QKD. However, since the protocol is Gaussian, it cannot be used to distil entanglement or to design a quantum repeater.

It is vital that QKD protocols can be proved secure against the most general eavesdropping attacks allowed by the laws of quantum physics. Indeed, currently, composable security for finite key lengths has been proved against general attacks for only several CV-QKD protocols, such as the no-switching protocol based on Gaussian modulated coherent states and heterodyne detection~\cite{leverrier2017security}, and the Gaussian modulated squeezed-state protocol with switching and homodyne detection~\cite{furrer2011continuous}. Recently, a discrete modulation protocol has been proved secure including finite-size effects~\cite{matsuura2020finitesize}. In Ref.~\cite{jacobsen2018complete}, the authors considered a symmetric Gaussian attack which is not asymptotically optimal since the protocol is asymmetric and the channel parameters in the unmodulated quadrature are unknown.

In this chapter, we deal with the asymmetry of the asymmetric minimum-leakage protocol from Ref.~\cite{jacobsen2018complete}. We first extend their security analysis to include general asymmetric Gaussian attacks. We show that Eve's information is not necessarily minimised for asymmetric attacks. Further, we introduce a new protocol that is strictly symmetric with respect to the quadratures of light. This forces Eve to implement symmetric attacks thereby avoiding the issues of asymmetry. Conditioned on a homodyne measurement at Alice's station, our protocol heralds squeezed states modulated in a single quadrature. We show that for this new protocol information leakage is asymptotically eliminated for pure-loss channels, and minimised for noisy channels.

In~\cref{Chap8sec:security}, we revisited state-of-the-art security for CV QKD. In~\cref{Chap8sec:asymmetric}, the original asymmetric protocol from Ref.~\cite{jacobsen2018complete} with squeezed states is analysed against collective attacks in the asymptotic limit. Our symmetric heralding protocol, discussed in~\cref{Chap8sec:herald}, is Gaussian and symmetric; however, current composable security proofs including finite-size effects still fail due to Alice's homodyne detection.

The outline of this chapter is as follows. In \cref{Chap8sec:asymmetric}, we recall the asymmetric protocol from Ref.~\cite{jacobsen2018complete} and extend their analysis to include general asymmetric attacks. In \cref{Chap8sec:herald}, we introduce a symmetric heralding protocol. In \cref{Chap8sec:discussion}, we discuss our results and conclude.

\section{\label{Chap8sec:asymmetric}Asymmetric protocol}

\begin{figure}
\centering
\subfloat[]{%
 \includegraphics[width=0.65\linewidth]{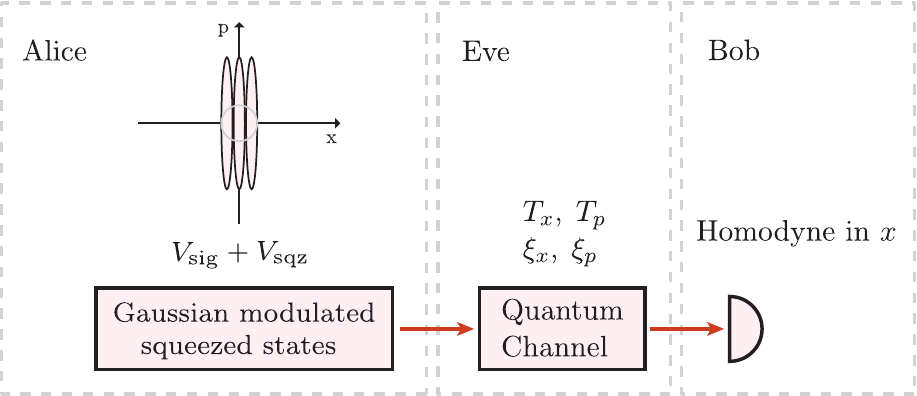} \label{Chap8fig:protocol_PM}%
}

\subfloat[]{%
  \includegraphics[width=0.65\linewidth]{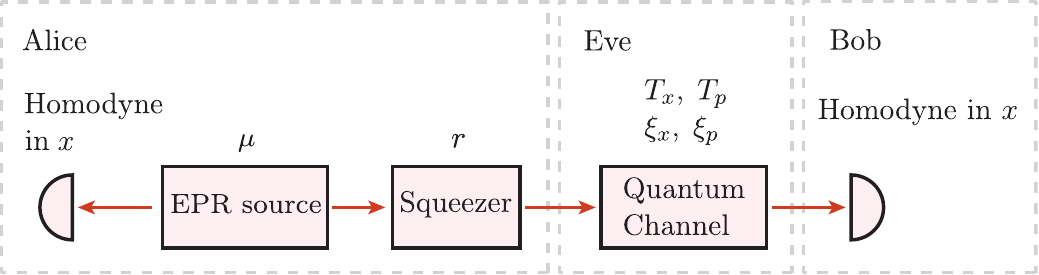}
  \label{Chap8fig:protocol_EB}%
  }
\caption[Equivalent schemes of the asymmetric minimum-leakage protocol.]{Equivalent schemes of the asymmetric minimum-leakage protocol from Ref.~\cite{jacobsen2018complete} \protect\subref{Chap8fig:protocol_PM} PM version and \protect\subref{Chap8fig:protocol_EB} EB version. The asymmetric protocol consists of squeezed states modulated and squeezed in the squeezing direction. When the variance of the overall ensemble is equal to shot noise and the channel is a pure-loss channel, the information leakage is equal to zero.}\label{Chap8fig:QKDschemes}
\end{figure}

The asymmetric minimum-leakage protocol from Ref.~\cite{jacobsen2018complete} is shown in~\cref{Chap8fig:QKDschemes}. It is based on squeezed states modulated in a single direction with an overall Gaussian modulation and squeezing chosen so that the Holevo information $\chi_{EB}$ is eliminated or minimised while the key rate is non-zero. The modulation forms an asymmetric ``blob'' in phase space. We do not expect symmetric attacks to be optimal since the protocol is asymmetric. Let us assume that the modulated quadrature is chosen to be the amplitude quadrature $q$. In this chapter, we denote it as the $x$ quadrature since it matches the notation of Ref.~\cite{PhysRevA.104.012411}. In a pure-loss channel, the requirement for complete elimination of $\chi_{EB}$ is for the ensemble of squeezed states to have an overall amplitude quadrature noise variance equal to that of the vacuum.

In the PM version of the protocol (the version Alice and Bob implement in practice), Alice prepares a Gaussian modulation of squeezed states of light to send to Bob. The PM scheme of the protocol is given in~\cref{Chap8fig:protocol_PM}. Alice prepares $x$ squeezed states with amplitude quadrature variance $V_{\text{sqz}}$ and applies modulation in the $x$ quadrature, displacing each squeezed state according to a random Gaussian variable with variance $V_{\text{sig}}$. The states are then sent to Bob through an asymmetric noisy channel with transmittance $T_x,\;T_p$ and excess noise $\xi_x,\;\xi_p$ in the $x$  and $p$ quadratures respectively. Bob performs homodyne measurements of the modulated $x$ quadrature, but sometimes also measures the unmodulated $p$ quadrature for estimating the properties of the channel in the $p$ quadrature. Alice and Bob extract a secret key from the $x$ quadrature data using a reverse-reconciliation procedure.

Security is analysed using the equivalent EB version of the protocol, shown in~\cref{Chap8fig:protocol_EB} which goes as follows. Alice prepares a two-mode squeezed vacuum state of variance $\mu$, keeps one of the modes, and squeezes the $x$ quadrature of the second mode with squeezing parameter $r$ before sending it to Bob. Alice performs homodyne measurements in the $x$ quadrature to project Bob's mode onto an ensemble of squeezed states with a Gaussian modulation. For instance, if Alice homodynes $x$ then effectively she has sent a Gaussian modulation of $x$ squeezed states in the $x$ direction to Bob, equivalent to the PM version.

As shown in Ref.~\cite{jacobsen2018complete}, the PM condition which completely decouples the eavesdropper in a pure-loss channel is $V_\text{sig}+V_\text{sqz} = 1$, an overall $x$ quadrature noise variance of vacuum. One can arrive at this condition by calculating Eve's maximal information via the covariance matrix shared between Alice and Bob in the EB version, and we show this in a moment, following Ref.~\cite{jacobsen2018complete}. A similar relation minimises the Holevo information under the assumption of a symmetric noisy channel but a little more squeezing is required because of the noise. By symmetric channel we mean $T_x = T_p$ and $\xi_x = \xi_p$, which is a restricted eavesdropping attack since the input state to the channel is asymmetric. The dashed curve in~\cref{Chap8fig:uniprotocol} shows that Eve's information is minimised in a symmetric noisy channel for an appropriate choice of $V_\text{sqz}$ given that $V_\text{sig}=0.5$.

We write down the covariance matrix of the asymmetric minimum-leakage protocol from Ref.~\cite{jacobsen2018complete} in the EB version (shown in~\cref{Chap8fig:protocol_EB}). This will allow us to calculate the Holevo bound after symmetric and asymmetric Gaussian noisy channels. The covariance matrix before the channel shared between Alice and Bob is
\begin{align*}
\Gamma_{\text{initial}} &= \left[ \begin{smallmatrix} 
                       \mu &                        0 & e^{-r}\sqrt{\mu^2 - 1} &                        0 \\
                        0 &                       \mu &                        0 & -e^{r}\sqrt{\mu^2 - 1} \\
                       e^{-r}\sqrt{\mu^2 - 1} &                        0 &             \mu e^{-2r} &                        0 \\
                        0 & -e^{r}\sqrt{\mu^2 - 1} &                        0 &              \mu e^{2r} \end{smallmatrix} \right],
\end{align*}
where $\mu$ is the strength of the TMSV state and $r$ is the amount of the squeezing in the $x$ direction.

If the EB squeezing parameter is $r=-\text{ln}\sqrt{\mu}$ (the minus sign means squeezing in the $p$ direction), then Alice has effectively sent coherent states to Bob in the PM version, modulated in the $x$ direction. For other amounts of squeezing, Alice has conditionally prepared squeezed states in the PM version. Specifically, for the squeezing parameter $r=\text{ln}\sqrt{\mu}$ (squeezing in the $x$ direction), the modulation variance in the $x$ quadrature is shot noise and this is the PM condition $V_\text{sig}+V_\text{sqz}=1$ which eliminates Eve's information for a pure-loss channel. The initial covariance matrix in the EB version for zero Holevo information (conditioned on Bob's homodyne in the $x$ quadrature) in a pure-loss channel is
\begin{align}\label{eq:shotnoiseCM}
\Gamma_{\text{initial}} &= \left[ \begin{smallmatrix} \mu & 0 & \sqrt{\frac{\mu^2-1}{\mu}} & 0\\ 0 & \mu & 0 & -\sqrt{\mu(\mu^2-1)} \\ \sqrt{\frac{\mu^2-1}{\mu}}   & 0 & 1 & 0 \\ 0 & -\sqrt{\mu(\mu^2-1)} & 0 & \mu^2\end{smallmatrix} \right].
\end{align}
There is a symmetry between the zero-information-leakage squeezed-state protocol and the coherent-state single-quadrature protocols. It arises because we squeeze to the shot noise, and coherent states have an intensity noise at the shot noise level. Given the covariance matrix in \cref{eq:shotnoiseCM}, if Alice chooses to homodyne in $x$ then $x$ modulated squeezed states are conditionally prepared at Bob. If Alice homodynes in $p$, then $p$ modulated coherent states are prepared at Bob. The Holevo quantity is eliminated only if Bob chooses to homodyne in the quadrature at the shot noise; i.e., Alice and Bob homodyne the same quadrature in the EB version. 

Midway between the two EB protocols (i.e. midway between coherent state and shot noise $r=\pm{}\text{ln}\sqrt{\mu}$), Alice performs no squeezing on Bob's mode in the EB version, $r=0$, then the EB protocol is equivalent to the original squeezed-state protocol with switching which is symmetric between the two quadratures \cite{PhysRevA.63.052311}.

\subsection{\label{Chap8app:sec:appendix:symmetricattack}Symmetric attack}

Our goal is to calculate the Holevo bound assuming the channel is symmetric. We propagate Bob's mode of the initial state $\Gamma_\text{initial}$ down a symmetric channel of transmissivity $T$ and with excess noise \hypertarget{hyperlinklabel_xi}{\hyperlink{hyperlinkbacklabel_xi}{$\xi$}}. After the symmetric channel, the covariance matrix shared between Alice and Bob is
\begin{align*}
\Gamma_{AB \text{, symmetric channel}} &= \left[ \begin{matrix}  
                               \mu &                                0 & c_x &                                0 \\
                                0 &                               \mu &                                0 & c_p \\
c_x &                                0 &     v_x^B &                                0 \\
                                0 & c_p &                                0 &  v_p^B \\
 \end{matrix} \right],
 \end{align*}
 with 
 \begin{align*}
 c_x = e^{-r}\sqrt{T(\mu^2 - 1)}\\
  c_p = -e^{r}\sqrt{T(\mu^2 - 1)}\\
 v_x^B = T(e^{-2r}\mu +\xi) + 1 - T\\
v_p^B = T(e^{2r}\mu +\xi) + 1 - T,
 \end{align*}
where $T$ is the transmissivity and $\xi$ is excess noise, symmetric in both the $x$ and $p$ quadratures (see the Supplemental Material of Ref.~\cite{jacobsen2018complete}).

The covariance matrix for Alice's mode conditioned on Bob's homodyne measurement, which is assumed to be in the $x$ quadrature, is~\cite{PhysRevLett.89.137903,PhysRevLett.89.137904}
 \begin{align*}
\Gamma_{A|b} &= \Gamma_{A} - \sigma (\Pi \;\Gamma_B \; \Pi)^{-1} \sigma^T,
\end{align*}
where $\Gamma_{A}$, $\Gamma_{B}$ and $\sigma$ are given by $\Gamma_{AB} = \left[ \begin{smallmatrix}  \Gamma_A   &    \sigma \\ \sigma^T &    \Gamma_B \end{smallmatrix} \right]$, $\Pi = \text{diag}(1,0)$, and $(\Pi \;\Gamma_B \; \Pi)^{-1}$ is a pseudoinverse. Thus, we have
\begin{align*}
\Gamma_{A|b} &= \left[ \begin{smallmatrix} 
 \mu - \frac{Te^{-2r}(\mu^2 - 1)}{T(e^{-2r}\mu +\xi) + 1 - T} &  0 \\
                                                             0 & \mu 
\end{smallmatrix} \right].
\end{align*}

The Holevo quantity is
\begin{equation*}
\chi_{EB} = S(\rho_{AB}) - S(\rho_{A|b}),
\end{equation*}
where $S(\cdot)$ is the von Neumann entropy which for Gaussian states can be determined numerically from the symplectic eigenvalues $\nu_k$ of the covariance matrix.

As $r\to\ln{\sqrt{\mu}}$ then $\chi_{EB}\to0$ in a pure-loss channel. In the next section, we go beyond symmetric attacks.

\subsection{General asymmetric attacks}

In this section, we consider channels where the noise and loss may be asymmetric in each of the quadratures. Recall that during the protocol, Alice and Bob perform parameter estimation to derive the covariance matrix shared between them. Usually, both quadratures are modulated allowing the full covariance matrix to be estimated; however, although only one quadrature is modulated, Bob can still estimate the variance of the unmodulated quadrature and use the physicality of the quantum state to bound Eve's information~\cite{usenko2018unidimensional}, i.e., via the Heisenberg uncertainty principle.

\subsubsection{\label{Chap8sec:Heisenberg}Bounding Eve using the Heisenberg Uncertainty Principle}

We use the EB version shown in~\cref{Chap8fig:protocol_EB} and bound Eve using the Heisenberg uncertainty principle. The initial covariance matrix in the EB version is
\begin{align}
\Gamma_{\text{initial}} &= \left[ \begin{smallmatrix} 
                       \mu &                        0 & e^{-r}\sqrt{\mu^2 - 1} &                        0 \\
                        0 &                       \mu &                        0 & -e^{r}\sqrt{\mu^2 - 1} \\
 e^{-r}\sqrt{\mu^2 - 1} &                        0 &             \mu e^{-2r} &                        0 \\
                        0 & -e^{r}\sqrt{\mu^2 - 1} &                        0 &              \mu e^{2r} \\
\end{smallmatrix} \right],
\end{align}
where $\mu$ is the TMSV variance and $V=e^{-2r}$ is the strength of the squeezing on the outgoing mode with squeezing parameter $r$. The EB parameters ($\mu$ and $r$) are related to the PM parameters ($V_\text{sig}$ and $V_\text{sqz}$) such that (see the Supplemental Material of Ref.~\cite{jacobsen2018complete})
\begin{align}
\begin{split}
    \text{Bob's variance of } x\text{ quadrature} &= \mu e^{-2r} =V_\text{sig}+V_\text{sqz},\\
    \text{Bob's variance of } p\text{ quadrature} &= \mu e^{2r}=\frac{1}{V_\text{sqz}}.\label{Chap8eq:EBparamaters1}
\end{split}
\end{align}
Explicitly, solving for $\mu$ and $r$ in terms of the PM parameters, we have
\begin{align}
\begin{split}
    \mu&=\sqrt{1+\frac{V_\text{sig}}{V_\text{sqz}}}\\
    r&=-\frac{1}{2}\ln(\sqrt{V_\text{sqz}(V_\text{sqz}+V_\text{sig})}).\label{Chap8eq:EBparamaters2}
\end{split}
\end{align}

Our goal is to bound Eve's information for general asymmetric channels. After the asymmetric channel, the covariance matrix shared by Alice and Bob is:
\begin{align*}
\Gamma_{AB \text{, asymmetric channel}} &= \left[ \begin{smallmatrix} \mu & 0 & \sqrt{T_x e^{-2r}(\mu^2-1)} & 0\\ 0 & \mu & 0 & -\sqrt{T_p e^{2r}(\mu^2-1)} \\ \sqrt{T_x e^{-2r}(\mu^2-1)}  & 0 & T_x (e^{-2r}\mu+\xi_x)+1-T_x & 0 \\ 0 & -\sqrt{T_p e^{2r}(\mu^2-1)} & 0 &T_p (e^{2r}\mu+\xi_p)+1-T_p\end{smallmatrix} \right],
\end{align*}
where $r$ is the squeezing parameter in the EB version.  $T_x$ and $\xi_x$ are, respectively, the channel transmittance and excess noise, estimated by Alice and Bob in the $x$ quadrature. However, since the $p$ quadrature is not modulated, $T_p$ and $\xi_p$ are unknown. We can bound Eve's information using the Heisenberg uncertainty relation.

Then, the covariance matrix shared between Alice and Bob after Eve's asymmetric attack is given by
\begin{align}
\Gamma_{AB} &= \left[ \begin{smallmatrix} \mu & 0 & e^{-r}\sqrt{T_x (\mu^2-1)} & 0\\ 0 & \mu & 0 & c_p \\ e^{-r}\sqrt{T_x (\mu^2-1)}  & 0 & T_x (e^{-2r}\mu+\xi_x)+1-T_x & 0 \\ 0 & c_p & 0 &v_p^B\end{smallmatrix} \right],
\end{align}
where $v_p^B$ and $c_p$ are unknown since the $p$ quadrature is unmodulated. Fixing $T_x$ and $\xi_x$, then given Bob's $p$ quadrature variance  $v_p^B$ measured during an experiment (which he only does sometimes since he is mostly performing a homodyne measurement of the $x$ quadrature), we bound the unknown correlation parameter $c_p$ by the physical requirement of the state given by~\cref{eq:uncertainty_principle}, i.e., $\Gamma_{AB} + i\Omega \geq 0$, where $\Omega$ is the symplectic form $\Omega = \bigoplus_{i=1}^{n} \omega,\;\omega = \left( \begin{smallmatrix} 0 & 1 \\ -1 & 0 \end{smallmatrix} \right)$.

To calculate a general upper bound on Eve's information we simulate an experiment and give Bob's variance $v_p^B$ the value it would have if the channel were symmetric (i.e., in our simulated experiment $v_p^B$ is calculated using $T_p=T_x$ and $\xi_p=\xi_x$), and then we maximise Eve's information $\chi_{EB}$ by going over all physical covariance matrices.  This is plotted in \cref{Chap8fig:uniprotocol} (solid). Also plotted is for a symmetric channel (dashed), which is considered in Ref.~\cite{jacobsen2018complete}. Note that Alice prepares a Gaussian modulation of coherent states for no squeezing $V_\text{sqz}=1$. For coherent states, bounding Eve's information in this way is not very pessimistic, as can be seen in the plot since the gap between symmetric and upper bound is not significant. But as squeezing is increased, the gap between the symmetric channel and the general upper bound becomes very significant. To provide tighter bounds, it is better to measure all terms of the covariance matrix. Thus, we need to do some estimation of the unmodulated quadrature.

\subsubsection{\label{Chap8sec:estimatep}Estimating the unmodulated quadrature}

\begin{figure}
        \centering
        \includegraphics[width=0.65\linewidth]{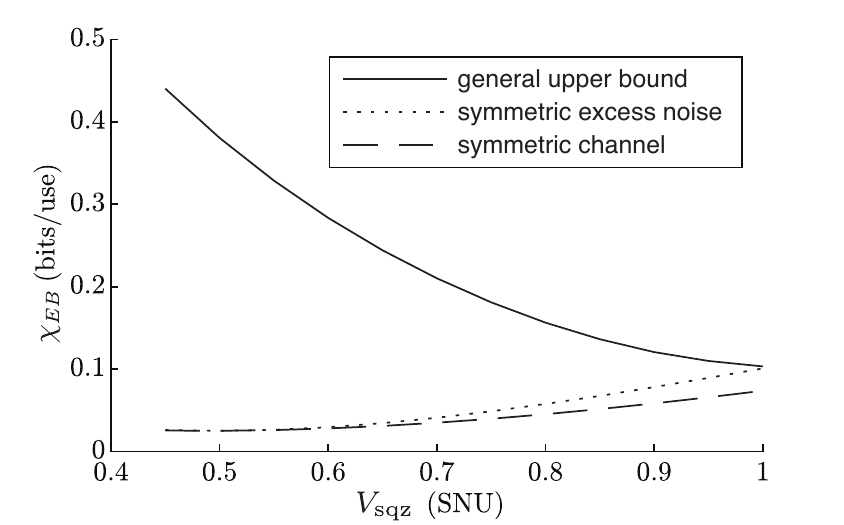}
    \caption[Bound on Eve's information versus squeezing in shot-noise units (SNU) for the asymmetric protocol with transmittance $T_x=0.5$, excess noise $\xi_x=0.01$ SNU, and $V_\text{sig}=0.5$ SNU]{Bound on Eve's information versus squeezing in shot-noise units (\hypertarget{hyperlinklabel_SNU}{\hyperlink{hyperlinkbacklabel_SNU}{SNU}}) for the asymmetric protocol with transmittance $T_x=0.5$, excess noise $\xi_x=0.01$ SNU, and $V_\text{sig}=0.5$ SNU. The upper bound becomes very loose with lower squeezing. Shown are symmetric lossy thermal channel (dashed), general upper bound (solid), and equal excess noise in both quadratures (dotted).}\label{Chap8fig:uniprotocol}
\end{figure}

One should perform some estimation of the unmodulated quadrature in an experiment to obtain the full covariance matrix; however, it is difficult to estimate the unmodulated quadrature since it is antisqueezed. For illustration purposes, we assume that some estimation of the $p$ quadrature is performed and that the excess noise is equal in both quadratures, $\xi_p = \xi_x$, and we again bound Eve's information using the physicality of the covariance matrix and plot this in~\cref{Chap8fig:uniprotocol} (dotted). We do not have to simulate Bob's variance in this case because there is only one unknown parameter, $T_p$. Compared to the general upper bound, assuming equal excess noise in both quadratures does much better. This shows that it is important to estimate the noise in the unmodulated quadrature; however, equal excess noise in both quadratures is not a realistic assumption because in an experiment the noise associated with the unmodulated quadrature is expected to be worse than the modulated quadrature. Therefore, one can expect Eve's information to lie between the symmetric channel and the upper bound. 

In summary, Eve's information is not necessarily minimised when considering stronger attacks beyond symmetric ones (and note that we have not yet considered finite-size effects). This motivates exploration of symmetric protocols with the goal of minimising Eve's information.

\section{\label{Chap8sec:herald}Heralding protocol}

\begin{figure}
\centering
\subfloat[]{%
  \includegraphics[width=0.65\linewidth]{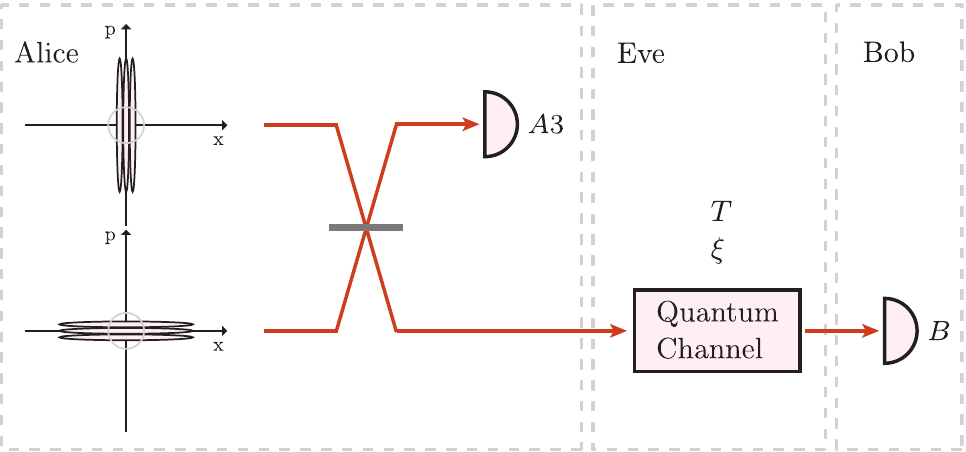}
  \label{Chap8fig:heralding_PM}
  }

\subfloat[]{%
  \includegraphics[width=0.65\linewidth]{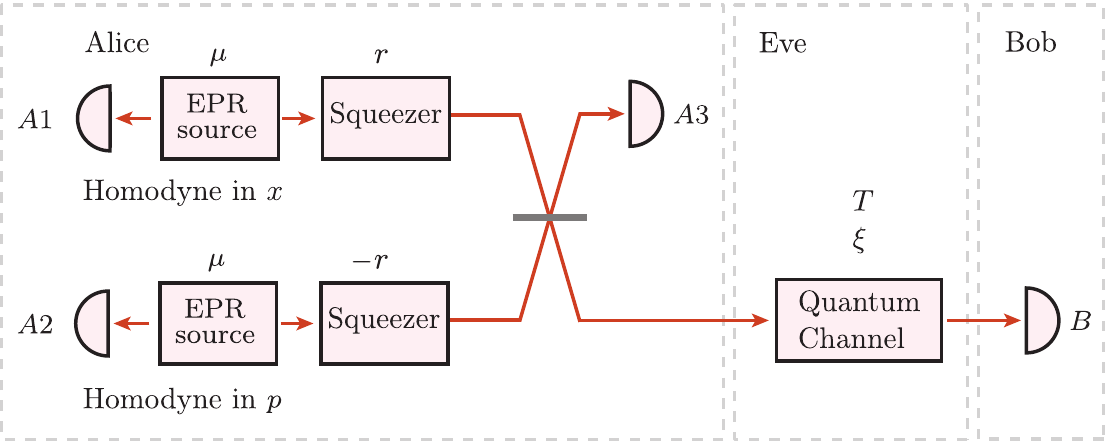}
  \label{Chap8fig:heralding_EB}
  }
\caption[Equivalent schemes of the heralding protocol]{Equivalent schemes of the heralding protocol \protect\subref{Chap8fig:heralding_PM} PM version and \protect\subref{Chap8fig:heralding_EB} EB version. Alice and Bob randomly and independently homodyne either $x$ or $p$ at both modes $A3$ and $B$ and later sift the results. The protocol consists of all Gaussian elements and is symmetric with respect to the quadratures, and eliminates information leakage in a pure-loss channel when~\cref{Chap8eq:heralding} is satisfied. See~\cite{Madsen2012} for a proof-of-principle experiment.}\label{Chap8fig:QKDschemesheralding}
\end{figure}

In this section, we symmetrise the asymmetric protocol of Ref.~\cite{jacobsen2018complete} in a heralding way to avoid worrying about asymmetric attacks in the security analysis. We present our heralding protocol in~\cref{Chap8fig:QKDschemesheralding}. In the PM version, Alice prepares two modes; two independent ensembles of squeezed states with a Gaussian modulation in a single direction, one mode modulated in the $x$ direction and the other mode in the $p$ direction. Note that here again we use $V_\text{sqz}$ for the squeezing variance and $V_\text{sig}$ for the modulation variance. These are combined on a beamsplitter (see~\cref{sec:beamsplitter}) and Alice homodynes one of the outputs (labelled $A3$) in either $x$ or $p$ at random while the other mode is sent to Bob (labelled $B$). The protocol heralds squeezed states modulated in a single quadrature, conditioned on Alice's homodyne measurement at $A3$. Bob independently and randomly homodynes in either $x$ or $p$ and, during classical post-processing, Alice and Bob sift their results, where they only keep the data for which they have used the same quadrature for the measurement.

In the EB version, Alice's ensembles are replaced with TMSV states plus auxiliary squeezing on one of the modes. Mode $B$ is squeezed in the $x$ quadrature with squeezing parameter $r$ and mode $A3$ is squeezed in the $p$ direction by an equal amount. The EB version is equivalent to the PM version if Alice homodynes $A1$ in $x$ and $A2$ in $p$.

Our heralding protocol consists of all Gaussian elements and is symmetric, so we can analyse security against Gaussian collective attacks in the asymptotic limit. For our heralding protocol we do know that in the asymptotic limit Gaussian attacks are optimal, and, unlike the asymmetric protocol, we no longer must worry about asymmetric attacks since the protocol is symmetric. We still have the issue that we do not know Eve's optimal attack in the finite-size regime.

Now we consider the EB version of the heralding protocol. Before going through the channel, the mode that will go to Bob is symmetric and is given by
\begin{align}
\Gamma_B &= \left[  \begin{smallmatrix} \mu  \cosh{2r} &0 \\ 0 & \mu \cosh{2r}  \end{smallmatrix}  \right].
\end{align}

After the channel, homodyne detection is performed on modes $A3$ and $B$ independently and randomly in $x$ and $p$. At the end Alice and Bob sift, and here, without loss of generality, we assume this sifting is in the $x$ quadrature (sifted results in the $p$ quadrature will behave similarly). After the quantum channel and Alice's heralding measurement (which we assume is homodyne $A3$ in $x$), the covariance matrix of Bob's mode is
\begin{align}
\Gamma_{B|A3_x} &= \left[  \begin{smallmatrix}   \frac{e^{4r} - T + T\xi - Te^{4r} + T\xi e^{4r} + 2T\mu e^{2r} + 1}{e^{4r} + 1} &0 \\ 0 & T\xi - T + T\mu \cosh{2r} + 1  \end{smallmatrix} \right].
\end{align}
The full covariance matrix is given in~\cref{Chap8app:sec:appendix:heralding}.

For zero information leakage in a pure-loss channel we require that the variance of the $x$ quadrature is equal to the shot noise. Solving for the PM parameters defined in~\cref{Chap8eq:EBparamaters2}, we find the following condition
\begin{align}
V_\text{sig} &= \frac{{V_\text{sqz}}^2 -2 V_\text{sqz} +1}{2 - V_\text{sqz}}.
\label{Chap8eq:heralding}
\end{align}

For example, if $V_\text{sig}=0.3$, then this condition implies that there are two solutions that eliminate the Holevo quantity: squeezing $V_\text{sqz} = 0.2821$ or antisqueezing $V_\text{sqz}=1.4179$. The squeezing solution maximises the key rate of our heralding protocol and not the antisqueezing solution since mutual information $I_{AB}$ decreases with antisqueezing. This is not inconsistent with Ref.~\cite{usenko2018unidimensional} where the authors show better performance when the modulation of the unidimensional protocol is in the antisqueezed quadrature. The reason that our protocol is different from theirs is that in that we choose the modulation $V_\text{sig}$ such that the Holevo information is zero in a pure-loss channel. Note that $V_\text{sqz}=0$ for infinite squeezing and $V_\text{sqz}=1$ for zero squeezing. 

\begin{figure}
\centering
\subfloat[]{%
 \includegraphics[width=0.65\linewidth]{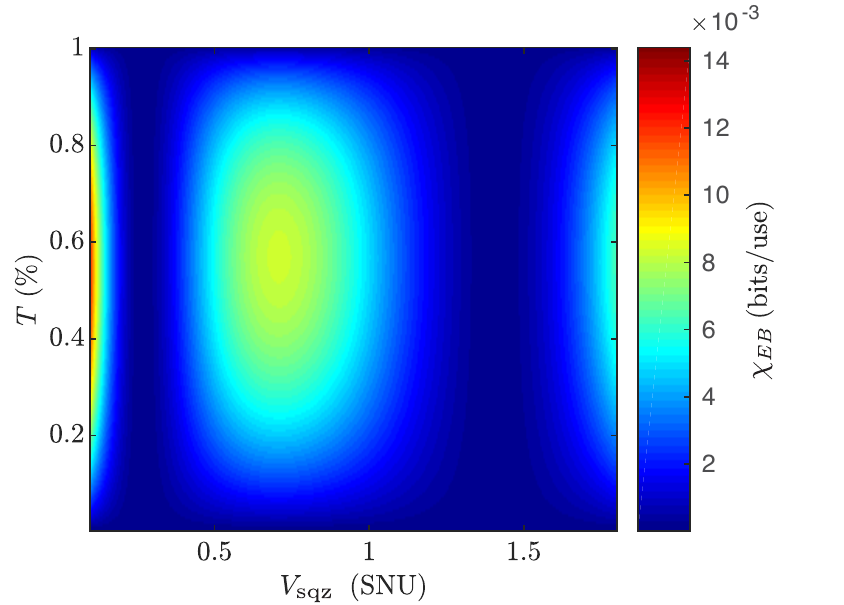} \label{Chap8fig:heraldingcontour_pure}
}

\subfloat[]{%
  \includegraphics[width=0.65\linewidth]{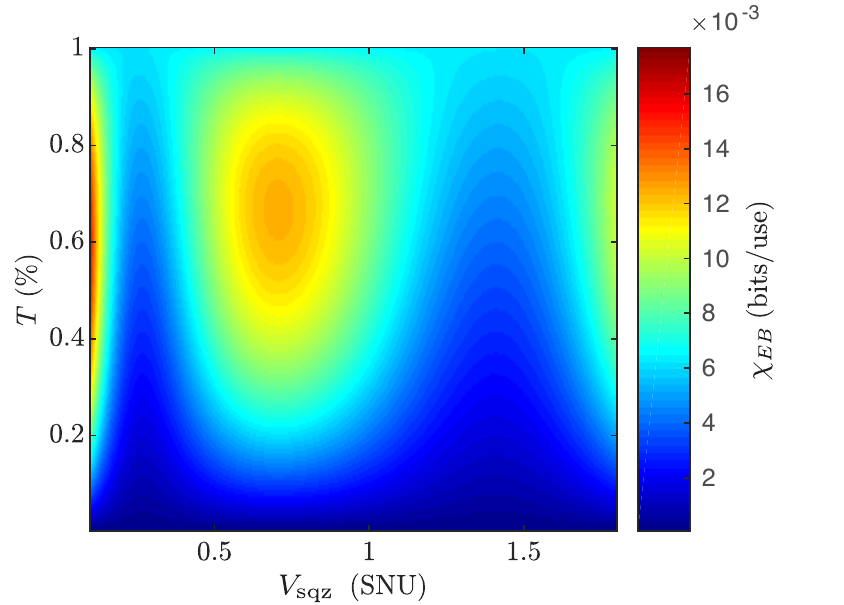}
  \label{Chap8fig:heraldingcontour_noise}
  }
\caption[Eve's information for the heralding protocol as a function of squeezing variance $V_\text{sqz}$ and transmissivity $T$ with fixed modulation variance $V_\text{sig}=0.3$]{Eve's information for the heralding protocol as a function of squeezing variance $V_\text{sqz}$ and transmissivity $T$ with fixed modulation variance $V_\text{sig}=0.3$: \protect\subref{Chap8fig:heraldingcontour_pure} no excess noise, \protect\subref{Chap8fig:heraldingcontour_noise} excess noise $\xi = 0.001$ SNU. In a pure-loss channel Eve's information can be zero, and with added thermal noise Eve's information is minimised. The region of interest is when $V_\text{sqz}<1$.} 
\label{Chap8fig:heraldingcontour}
\end{figure}

\cref{Chap8fig:heraldingcontour} shows Eve's information as a function of transmissivity $T$ and squeezing variance $V_\text{sqz}$ for \protect\subref{Chap8fig:heraldingcontour_pure} pure-loss and \protect\subref{Chap8fig:heraldingcontour_noise} excess-noise $\xi = 0.001$. Eve's information is eliminated in a pure-loss channel when~\cref{Chap8eq:heralding} is satisfied. For the noisy case,~\cref{Chap8fig:heraldingcontour_noise} shows that when~\cref{Chap8eq:heralding} is approximately satisfied and the Holevo information is minimised, the Holevo information actually decreases with loss. In other words, the Holevo information goes to zero as the transmissivity of the channel decreases. That Eve's information is less the more loss there is may be counter-intuitive (since she steals more but gets less) but makes sense when one remembers that Alice and Bob's mutual information is also less for more loss.  

\begin{figure}
    \centering
        \includegraphics[width=0.65\linewidth]{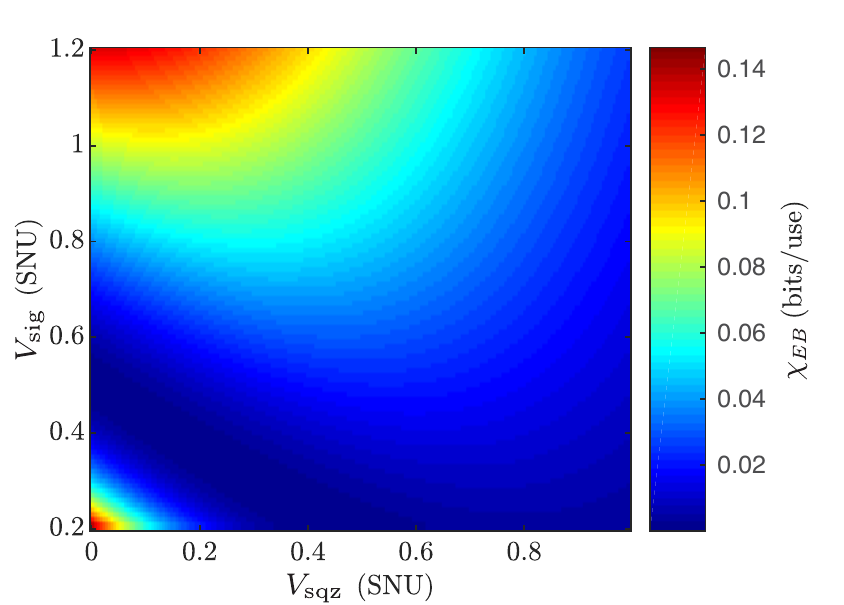}
\caption[Eve's information for the heralding protocol as a function of squeezing variance $V_\text{sqz}$ and signal variance $V_\text{sig}$ in a pure-loss channel with fixed transmissivity $T=0.5$]{Eve's information for the heralding protocol as a function of squeezing variance $V_\text{sqz}$ and signal variance $V_\text{sig}$ in a pure-loss channel with fixed transmissivity $T=0.5$. As $V_\text{sqz}\to0$ Eve's information is zero when $V_\text{sig}\to0.5$. Coherent states are prepared at Alice when $V_\text{sqz}{=}1$ and Eve's information cannot be zero in this case for any choice of modulation.}
\label{Chap8fig:heraldingcontour2}
\end{figure}

In~\cref{Chap8fig:heraldingcontour2}, we plot Eve's information as a function of signal modulation and squeezing in a pure-loss channel with fixed transmissivity $T=0.5$. This shows that as squeezing is increased from $V_\text{sqz}=1$ (coherent states) to $V_\text{sqz}\to0$ (infinite squeezing), the modulation variance $V_\text{sig}$ must also be increased in order to minimise information leakage.

\begin{figure}
    \centering
        \centering
        \includegraphics[width=0.65\linewidth]{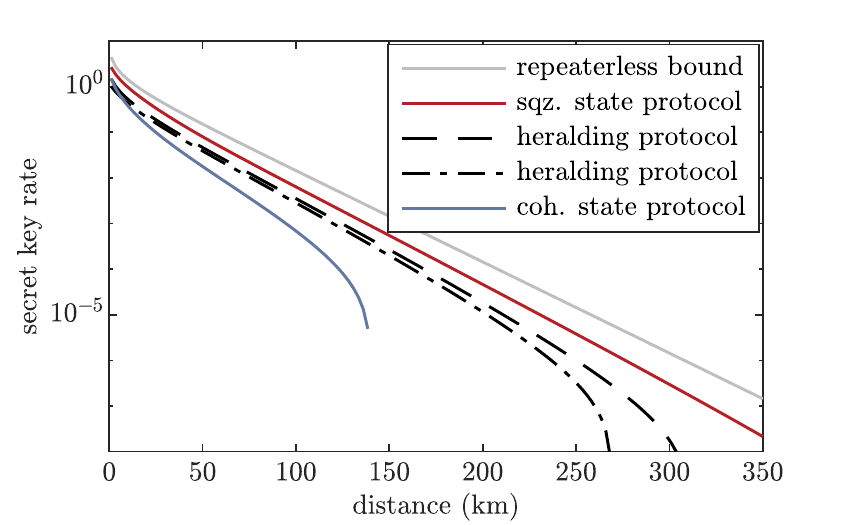}
\caption[Secret key rate of the heralding protocol as a function of distance for lossy thermal channel]{Secret key rate of the heralding protocol as a function of distance for lossy thermal channel. The dashed and dot-dashed lines (black) show the heralding protocol with infinite and 10 dB of finite squeezing respectively. The heralding protocol optimises at very large squeezing, i.e., $V_\text{sqz}{\to}0,\;V_\text{sig}{\to}0.5$. For these plotted rates, the condition \cref{Chap8eq:heralding} is satisfied at all distances, hence, Eve's information is zero at all distances for a pure-loss channel and small for a noisy channel. Shown for comparison are optimised rates for the original squeezed-state protocol with switching and homodyne detection~\cite{PhysRevA.63.052311} and no-switching coherent state protocol with heterodyne detection~\cite{Weedbrook_2004}. The excess noise is $\xi = 0.05$ for all protocols, the reconciliation efficiency is $\beta{=}0.95$ and we have considered optical fibre with loss of 0.2 dB/km. The repeaterless bound \cite{Pirandola_2017} is also plotted in the figure (grey).}
\label{Chap8fig:heraldingrates}
\end{figure}

In~\cref{Chap8fig:heraldingrates}, we show the secret key rate of the heralding protocol as a function of distance for a thermal-loss channel with excess noise $\xi = 0.05$. Alice and Bob's mutual information is calculated assuming modes $A1$, $A3$ and $B$ are all homodyned in $x$, and the parameters of the protocol are fixed such that~\cref{Chap8eq:heralding} is satisfied, and Eve's information is eliminated or approximately minimised at all distances. Alice and Bob's mutual information is calculated after Alice's heralding measurement of mode $A3$ in the $x$ quadrature and is given by
\begin{align}
I_{AB} = \frac{1}{2} \log_2 \frac{V_{B_x}}{V_{B_x|A1_{\text{hom}x}}},
\end{align}
where $V_{B_x}$ is the variance of Bob's $x$ quadrature, since this is the quadrature we assume he uses for the key, and ${V_{B_x|A1_{\text{hom}x}}}$ is the variance of Bob's $x$ quadrature conditioned on Alice's homodyne measurement in $x$ of mode $A1$.

The heralding protocol optimises for very large squeezing shown as the dashed line in~\cref{Chap8fig:heraldingrates}. Infinite squeezing is $V_\text{sqz}{\to}0$ and the zero-leakage condition~\cref{Chap8eq:heralding} means that $V_\text{sig}{\to}0.5$. Also shown is the key rate for the heralding protocol with 10 dB of finite squeezing (dot-dashed). For comparison, we plot the optimised key rate for Gaussian-modulated squeezed-state protocol and measured via homodyne detection~\cite{PhysRevA.63.052311} (red) and Gaussian-modulated coherent state protocol and heterodyne detection~\cite{Weedbrook_2004} (blue). The optimised squeezed-state protocol with homodyne detection outperforms our heralding protocol. However, since our heralding protocol likewise is based on squeezed states and homodyne detection it can outperform the coherent state protocol with heterodyne detection.

We note here that rates for the heralding protocol are numerically identical to those for the asymmetric protocol against a symmetric thermal-loss channel. Indeed, the rate for the heralding protocol for a pure-loss channel, infinite squeezing, and reconciliation efficiency $\beta{=}1$ reaches half of the fundamental repeaterless (PLOB) bound, which is given by $C = {-}\log_2{(1-\eta)}$ \cite{Pirandola_2017}. This result for the asymmetric protocol was indeed mentioned in~\cite{jacobsen2018complete}. So, it is not a repeater in a quantum sense, although it can simplify classical post-processing.

\section{Discussion and conclusion\label{Chap8sec:discussion}}

In this chapter, we have investigated CV-QKD protocols designed to eliminate information leakage to Eve. We extended the security analysis of the original asymmetric minimum-leakage protocol by considering general asymmetric channels by demanding the physicality of the quantum states. We also introduced a new protocol by symmetrising in a heralding way.

The main results from this chapter are the following: if the unmodulated quadrature can be properly estimated, the asymmetric protocol minimises Eve's information and is secure against Gaussian attacks in the asymptotic regime. However, using the Heisenberg uncertainty principle for security analysis can give Eve a lot of information. Our heralding protocol minimises Eve's information in a heralding way despite being symmetric with respect to the quadratures, but with added experimental complexity.

Another way to symmetrise the asymmetric minimum-leakage protocol is by switching, i.e., randomly sending $x$-squeezed or $p$-squeezed states with an overall modulation of shot noise in each quadrature. However, the overall ensemble prepared by Alice will be non-Gaussian (specifically, it will not be a Gaussian thermal state as in the squeezed-state protocol with homodyne detection from Ref.~\cite{PhysRevA.63.052311}). The non-Gaussianity complicates the security analysis, potentially giving more information to the eavesdropper, departing from our goal of minimum-leakage. We leave the investigation of minimum-leakage switching protocols for future work.

In conclusion, we have introduced a new CV-QKD protocol that minimises the amount of information leaked to an eavesdropper in a noisy channel and eliminates information leakage in a pure-loss channel, meaning that the classical part of the protocol will be computationally less complex, potentially leading to an overall increase in the secret key rate in situations of practical interest.

The protocol introduced in this chapter cannot be used as a repeater since it cannot overcome the repeaterless bound. The highest rate of the protocol achieves half the repeaterless bound. In the following chapters, we consider repeating techniques, dealing with the quantum part of the protocol.

%% file: chapter_scissors/chapter_scissors.tex
\chapter{Generalised quantum scissors}
\label{Chap:scissors}	
\pagestyle{headings}

\noindent
The results of this chapter have been published in the following.\\

\noindent
1.~\cite{PhysRevA.102.063715} \textbf{M. S. Winnel}, N. Hosseinidehaj, and T. C. Ralph, ``Generalized quantum scissors for noiseless linear amplification,'' \href{http://dx.doi.org/10.1103/PhysRevA.102.063715}{Phys. Rev. A \textbf{102}, 063715 (2020)}.\\

\noindent
2.~\cite{PhysRevLett.128.160501} J. J. Guanzon, \textbf{M. S. Winnel}, A. P. Lund, and T. C. Ralph, ``Ideal quantum teleamplification up to a selected energy cutoff using linear optics,'' \href{https://journals.aps.org/prl/abstract/10.1103/PhysRevLett.128.160501}{Phys. Rev. Lett. \textbf{128}, 160501 (2022)}.\\

\noindent
The author of this thesis, \textbf{Matthew Winnel}, discovered the three-photon quantum scissor and the seven-photon quantum scissor in Ref.~\cite{PhysRevA.102.063715}. Soon after, in Ref.~\cite{PhysRevLett.128.160501}, Joshua Guanzon generalised to all positive integers and showed the technique can work in reverse using single photons instead of bunched photons. The authors recently became aware of a new related work which investigated noiseless quantum tele-amplifiers from a different angle~\cite{Fiur_ek_2022}, based on the continuous-variable teleportation protocol~\cite{PhysRevLett.80.869}.\\

\noindent
In this chapter, we discuss how noiseless linear amplification (NLA) (see~\cref{{sec:NLA}}) is a powerful technique for designing a quantum repeater and can be implemented using linear optics. We generalise the concept of optical state truncation and NLA to enable truncation of the Fock-state expansion of an optical state to higher order and to simultaneously amplify it using linear optics. The resulting generalised quantum scissors are more efficient for NLA than employing multiple scissors in parallel and are experimentally practical. Advantages can be demonstrated in terms of fidelity with a target state, probability of success, distillable entanglement, and the amount of non-Gaussianity and truncation noise introduced.

We can implement ideal NLA up to the $n^\mathrm{th}$ Fock state, where $n$ can be any positive integer. Our protocol consists of a beamsplitter and an $(n+1)$-splitter, with $n$ ancillary photons and the detection of $n$ photons. Our protocol can be used as a loss-tolerant quantum relay for entanglement distribution and distillation. Our first-order device acts as a quantum repeater without quantum memories, as we will see in~\cref{Chap:simple_repeater}.

\section{Introduction}


The no-cloning theorem~\cite{wootters1982single} forbids the deterministic, linear (i.e., phase insensitive) amplification of a quantum state. Hence, all deterministic linear amplifiers must introduce noise~\cite{caves1981quantum-mechanical}. Nevertheless, non-deterministic noiseless linear amplification is possible if the amplifier is allowed to operate in a probabilistic, but heralded way, and the alphabet of states has an energy bound \cite{ralph2009nondeterministic,xiang2010heralded}.

Noiseless linear amplification (NLA) has proven a very useful technique in quantum optics with numerous experimental demonstrations \cite{barbieri2011nondeterministic} such as distillation of entanglement \cite{xiang2010heralded, PhysRevA.100.022315}, purification of entanglement \cite{Ulanov15}, amplification of qubits \cite{Kosis13} and enhanced metrology \cite{Usuga10}. Proposed applications include continuous-variable error correction \cite{Ralph11}, quantum key distribution~\cite{Blandino2012CV-QKD, Ghalaii_2020, ghalaii2020discretemodulation}, quantum repeaters~\cite{dias2018quantum, PhysRevResearch.2.013310}, and discrete-variable Bell inequalities~\cite{osorio2012heralded}. 

The non-deterministic quantum scissor introduced by Pegg et al.~\cite{pegg1998optical} truncates an input optical field to first order, retaining only the vacuum and one-photon components of the input state. 
In their original proposal, Ralph and Lund introduced a modified scissor device that truncates an input state to first order and simultaneously amplifies it by increasing the amplitude of the one-photon component relative to the vacuum component~\cite{ralph2009nondeterministic}. For input states of small amplitude, the modified scissor acts as an ideal NLA. 

In order to go beyond small input amplitudes, Ralph and Lund proposed employing multiple quantum scissors in parallel~\cite{ralph2009nondeterministic}. For a large finite number of scissors, the setup acts as an ideal NLA; however, with a vanishing probability of success. If a small number of scissors are used, the amplification is ``distorted'', that is, the Fock coefficients of the state obtained are multiplied by different constants than those required for ideal linear amplification. Other methods for NLA do not truncate the state but still distort Fock components higher than one~\cite{zavatta2010high-fidelity,Ulanov15}.

To improve the one-photon scissor-based NLA, there have been attempts to generalise the modified quantum scissor to two photons~\cite{jeffers2010nondeterministic}; however, the resulting device imposes an undesired non-linear sign change to the two-photon component of the amplified state. A generalisation of the original quantum scissor to higher order has also been made but this does not allow for amplification~\cite{koniorczyk2000general,
villas-boas2001recurrence,miranowicz2005optical-state}. Alternatively, a protocol that could truncate and amplify without distortion was described in Ref.~\cite{PhysRevA.89.023846}; however, this solution is impractical as it requires a massive high-order optical non-linearity. 
As a result, no demonstration of NLA without distortion of the higher order Fock state components has been achieved, thus seriously limiting future applications.

Here, we generalise the concept of optical state truncation and amplification and propose a practical linear optical device which can correctly amplify the input state up to any order $n$, where $n$ is a positive integer. 
The device naturally performs noiseless linear amplification without distorting the amplified Fock coefficients.

In this chapter, we first describe our $n$-scissor protocol, including its operation and probability of success. {The output fidelities are orders of magnitude better than previous linear-optical NLA protocols.} The NLA can be used for teleportation, as a loss-tolerant quantum repeater, and is tolerant to experimental imperfections. 

In this thesis, we focus on the first-order and third-order devices, since they were discovered first~\cite{PhysRevA.102.063715}, remembering that our scissor NLA protocol generalises to all positive integers~\cite{PhysRevLett.128.160501}.

\begin{figure}
\centering
\subfloat[]{%
  \includegraphics[scale=1.4]{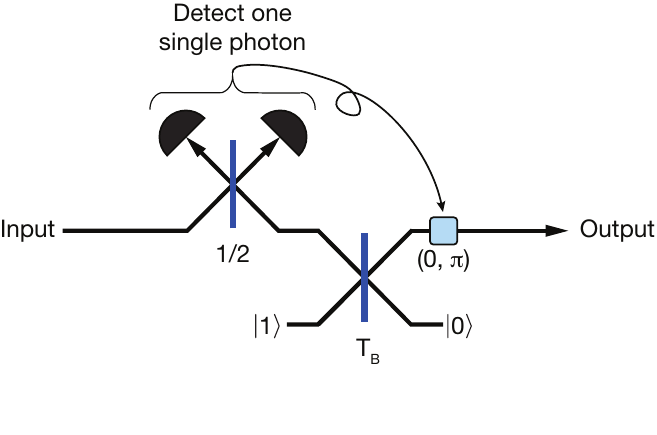} \label{Chap3fig:scissor}
}

\subfloat[]{%
  \includegraphics[scale=1.4]{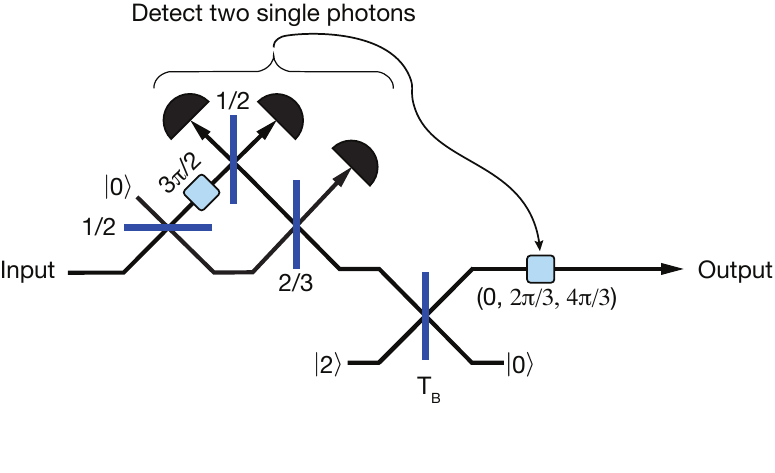}
  \label{Chap3fig:2scissor}
  }

\subfloat[]{%
  \includegraphics[scale=1.4]{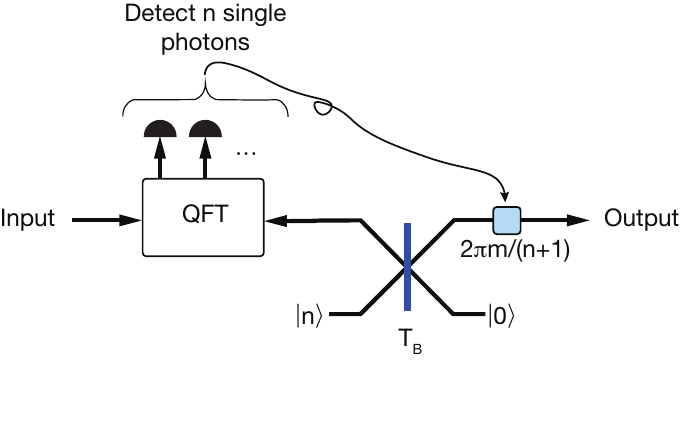}
  \label{Chap3fig:n_scissors}
  }
\caption[Quantum scissors]{Our scalable $n$-photon quantum scissors, which implements noiseless linear amplification (NLA) or de-amplification of an arbitrary state $|\psi\rangle \rightarrow g^{\hat{a}^\dagger \hat{a}}|\psi_n\rangle = |g\psi_n\rangle$, up to the $n^\mathrm{th}$ Fock state. \protect\subref{Chap3fig:scissor} Single-photon quantum scissor. \protect\subref{Chap3fig:2scissor} Two-photon quantum scissor. \protect\subref{Chap3fig:n_scissors} $n$-photon quantum scissor. The gain $g\in(0,\infty)$ is chosen by modifying the transmissivity $T_B = g^2/(1+g^2)\in(0,1)$ of the beamsplitter. This $n$-photon quantum scissors require an $n$ Fock state as a resource, but can also work with $n$ single photons as a resource~\cite{PhysRevLett.128.160501}.}
\label{Chap3fig:nQS}
\end{figure}

\section{Quantum scissors up to the $n$th Fock state}

In this section, we introduce our scalable linear-optical technique that can implement NLA up to the $n^\mathrm{th}$ Fock state, where $n$ can be any positive integer. NLA is a basic component of the quantum repeater considered in~\cref{Chap:CV_QR}, and can also be used as a loss-tolerant quantum relay introduced in~\cref{Chap:simple_repeater}.

Suppose the input state in the Fock basis is $\ket{\psi}{=}\sum_{n=0}^\infty c_n |n\rangle.$ As discussed in~\cref{sec:NLA}, an ideal NLA performs the transformation $\hat{T}_\text{ideal}\ket{\psi}{\to}\sum_{n=0}^\infty g^n c_n |n\rangle$, where $g$ is the gain of the NLA. The success probability is zero for any device that achieves this transformation perfectly \cite{Pandey_2013}.

The modified single-photon quantum scissors~\cite{pegg1998optical, ralph2009nondeterministic} truncate and amplify an optical state in Fock space and is shown in~\cref{Chap3fig:scissor}. It performs the transformation
\begin{align}\label{eq:1scissor}
\hat{T}_1 \ket{\psi} &= \sqrt{\frac{1}{2(g^2+1)}} (c_0 |0\rangle \pm g c_1 |1\rangle),
\end{align}
where the gain is $g{=}\sqrt{T_B/(1{-}T_B)}$. The passive phase flip depends on which detector fires.

This device is called a quantum scissor since all Fock components greater than one are truncated. For the device to operate as an ideal NLA, the two photon component must be negligible, that is $|g^2c_2|{\ll}|gc_1|$. The one-photon scissor acts as an ideal NLA only for small input states and the effect of the truncation is severe for large input states.

Shown in~\cref{Chap3fig:2scissor} is the second-order device $(n=2$), i.e., the two-photon quantum scissor~\cite{PhysRevLett.128.160501}. Two photons enter the device as a resource, two single-photon detectors register single photons, and one detector registers no photon. There are three possible patterns for getting two ``clicks'' and one ``no click.'' These click patterns lead to a heralded passive phase shift of the state, but the magnitudes of the Fock components are unchanged. Thus, the success probability is effectively increased by a factor of three.

The two-photon scissor truncates and ideally amplifies the input state to second order. The device operates as an ideal NLA as long as $|g^3c_3|{\ll}|g^2c_2|$, which is a good improvement from the single-photon scissor, at the cost of a reduced probability of success.

The general protocol is shown in~\cref{Chap3fig:n_scissors}, where the quantum Fourier transform (\hypertarget{hyperlinklabel_QFT}{\hyperlink{hyperlinkbacklabel_QFT}{QFT}}) is a coherent $(n+1)$-splitter which can be performed using linear optics. The $n$-scissor operates on an arbitrary bosonic state $|\psi\rangle$ and truncates the Fock components higher than $n$ and performs ideal NLA $g^{\hat{a}^\dagger \hat{a}}$:
\begin{align}
    |\psi\rangle &\equiv \sum_{j=0}^{\infty} c_j|j\rangle \xrightarrow{n\text{-scissor}} |g\psi_n\rangle = N \sum_{j=0}^{n} g^j c_j|j\rangle. \label{Chap3eq:nQS}
\end{align}
This is implemented via~\cref{Chap3fig:n_scissors} using a beamsplitter and a QFT, with $n$ extra resource photons and $n$ single-photon detections. The amount of amplification or de-amplification gain $g\in(0,\infty)$ can be chosen by setting the beamsplitter transmissivity to $T_B=g^2/(1+g^2)$. The $n$-scissor operation only occurs if the correct outcomes are obtained.

The overall action of the $n$-scissor is
\begin{align}
    |g\psi_n\rangle &= \frac{\sqrt{n!}}{(n+1)^{n/2}} \frac{1}{(g^2+1)^{n/2}} \sum_{j=0}^n g^j c_j |j\rangle, \label{Chap3eq:gpsi}
\end{align}
and has a success probability of $P = \braket{g\psi_n}{g\psi_n}$. The $n$-scissor applies ideal $g^{a^\dagger a}$ up to the $n^\mathrm{th}$ Fock state. The complete proof is presented in Ref.~\cite{PhysRevLett.128.160501}.

\section{Quantum scissors are cat-state tele-amplifiers\label{sec:appendixscissors}}

\begin{figure}
    \begin{center}
        \includegraphics[scale=1.3]{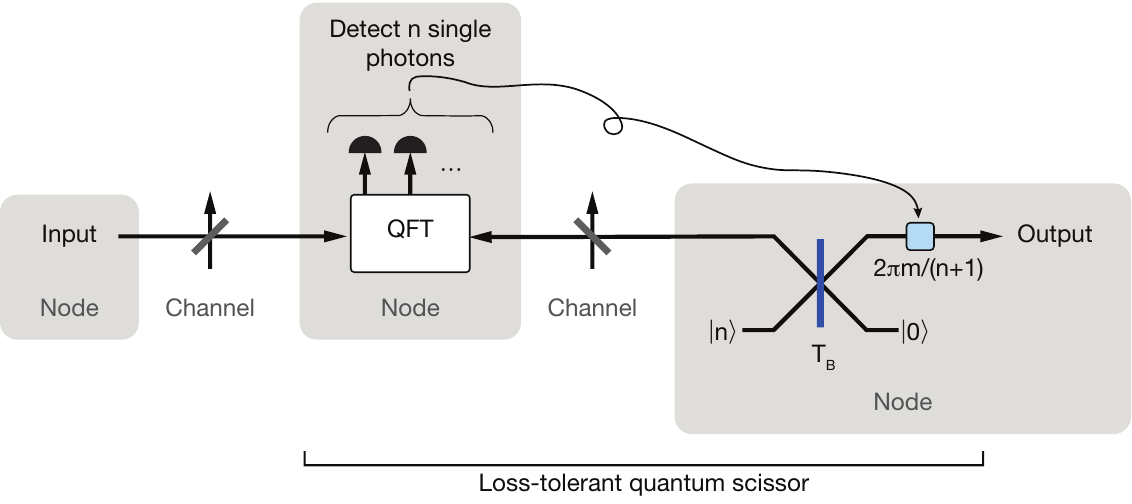}
        \caption[Our scalable $n$-scissor as a relay]{\label{Chap3fig:lossy_scissors}We consider a lossy $n$-scissor protocol with an arbitrary state input where Alice's channel has $\eta_A$ transmissivity and Bob's channel has $\eta_B$ transmissivity.}
    \end{center}
\end{figure}

Remarkably, when the input state consists of small constellations of Gaussian states, our generalised quantum scissors are tolerant to loss on the mode between the beamsplitter and the QFT, as can be seen in~\cref{Chap3fig:lossy_scissors}. Gaussian ensembles are exactly the ensembles of states transferred during the usual Gaussian CV-QKD protocols, discussed in~\cref{{sec:CVQKD_introduction}}. Intuitively, the reason the devices are tolerant to loss can be seen by considering a related class of protocols which contain the scissors in the limit of zero amplitude; quantum tele-amplification of cat states~\cite{neergaard-nielsen2013quantum,Guanzon2022cat_relay}.

Tele-amplification devices teleport states consisting of superpositions of coherent states on a ring in phase space (cat states). If there is loss between the beamsplitter and the QFT, the teleportation is still ideal if the input state is a mixture of coherent states (rather than a superposition). As the amplitude of the coherent state mixture goes to zero, we discover that our quantum scissors are very good at transferring small Gaussian ensembles of states despite loss on Bob's arm. They are loss-tolerant quantum relays. The first-order device ($n=1$) can beat the repeaterless bound, so it is an effective quantum repeater in this sense. We introduce a simple, memoryless repeater protocol for CV QKD in~\cref{Chap:simple_repeater}.

By taking the small-amplitude limit of tele-amplification, we thus derive the single-photon quantum scissor and the three-photon quantum scissor in what follows, which correspond to two-lobed and four-lobed cat state tele-amplification protocols. Cat states have application in DV quantum error correction and quantum computation, so we expect tele-amplification protocols to play an important role~\cite{neergaard-nielsen2013quantum,Guanzon2022cat_relay}.

For the rest of the chapter, we will refer to the scissors as $\oneinjectedscissor$, $2$-scissor, $\threeinjectedscissor$, etc. The prefix in our notation is the number of photons entering the device as a resource, the number of detectors that register single photons, and the Fock state truncation of the output state. See~\cref{sec:appendixscissors} for novel derivations of the $1$-scissor and our $3$-scissor.

\subsection{\label{sec:appendix1-scissor}Single-photon quantum scissor}

\begin{figure}
        \includegraphics[width=0.5\linewidth]{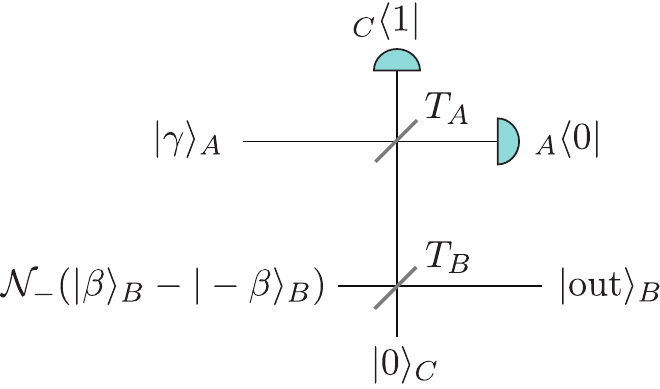}
        \centering
          \caption[$\twocatrelay$ with two-component cat state as a resource and two-port detection scheme.]{$\twocatrelay$ with two-component cat state as a resource and two-port detection scheme~\cite{neergaard-nielsen2013quantum}.}
          \label{fig:2cat_scissor}
          \end{figure}

We first reconsider the modified single-photon quantum scissor ($\oneinjectedscissor$). For simplicity, we will consider how an arbitrary coherent state $\ket{\gamma}$ transforms. It is convenient and intuitive to consider the tele-amplification procedure from~\cite{neergaard-nielsen2013quantum}, where a two-component cat state is used as the resource (replacing the single-photon resource). This setup is shown in \cref{fig:2cat_scissor} and we will refer to it as $\twocatrelay$. Further motivation for considering the 4-cat device, is that 4-lobed cat states have 4-fold rotation symmetry, the simplest cat code with error-correcting properties~\cite{Mirrahimi_2014} which can correct single loss errors. In the limit that $\beta\to0$, the $\oneinjectedscissor$ is equivalent to $\twocatrelay$ and we show this for coherent states in what follows.

Let's assume that the initial input state of mode A is a coherent state $\ket{\gamma}_A$. The resource state of mode B is an odd (minus) 2-cat state, and the initial state of mode $C$ is the vacuum state. The three-mode state after the action of the two beamsplitters $\hat{V}_B$ and $\hat{V}_A$, with transmissivity $T_B$ and $T_A$ respectively, is
\begin{equation}
\begin{split}
\ket{\psi}_{ABC} & = \mathcal{N}_{-} \hat{V}_A \hat{V}_B \ket{\gamma}_A ( \ket{\beta}_B{-}\ket{{-}\beta}_B )  \ket{0}_C\\
& = \mathcal{N}_{-}  \ket{\sqrt{T_A} \gamma - \sqrt{(1-T_A)(1-T_B)} \beta}_A \ket{\sqrt{T_B}\beta}_B \ket{{-}\sqrt{1-T_A} \gamma - \sqrt{T_A(1-T_B)}\beta}_C\\
& - \mathcal{N}_{-} \ket{\sqrt{T_A} \gamma + \sqrt{(1-T_A)(1-T_B)} \beta}_A \ket{{-}\sqrt{T_B}\beta}_B \ket{{-}\sqrt{1-T_A} \gamma + \sqrt{T_A(1-T_B)}\beta}_C,
\end{split}
\end{equation}
where $\mathcal{N}_{-}=[2(1{-}e^{-2|\beta|^2})]^{-1/2}$ is the normalisation constant for the cat-state resource. To simplify the calculation, we introduce a new complex variable $\alpha$, and impose the condition $\beta = \alpha\sqrt{T_A/[(1-T_A)(1-T_B)]}$. To simplify things further, we also set $T_A = 0.5$ (but it may be advantageous to alter the transmissivity of beamsplitter $A$ in other scenarios). Single-photon detection is performed on modes $A$ and $C$. Conditioned on no photon detection at $A$ and single-photon detection at $C$, the final state of mode $B$ is
\begin{align}
{}_C\bra{1}{}_A\braket{0}{\psi}_{ABC}&= \mathcal{N} ( (\alpha+\gamma) \ket{g\alpha} + (\alpha-\gamma) \ket{{-}g\alpha}),
\end{align}
where $g = \sqrt{{T_A T_B}/{(1-T_A)(1-T_B)}}$, and $\mathcal{N}$ is given by
\begin{align}
\mathcal{N} &= \frac{- \mathcal{N_{-}} e^{-(|\gamma|^2+|\alpha|^2)/2}}{\sqrt{2}}.
\end{align}
We expand in the Fock basis to first order in $\alpha$ and we get
\begin{align}
&= \mathcal{N} e^{-|g\alpha|^2/2}  ( (\alpha{+}\gamma) (\ket{0} {+} g\alpha \ket{1}) {+} (\alpha-\gamma) (\ket{0} {-} g\alpha \ket{1})\\
&= \mathcal{N} e^{-|g\alpha|^2/2}  ( 2\alpha \ket{0} + 2g\alpha\gamma \ket{1})\\
&= \mathcal{N} \;  2\alpha e^{-|g\alpha|^2/2}  (  \ket{0} + g\gamma \ket{1}).
\end{align}
Expanding and throwing away higher order $\alpha$ we find that $\mathcal{N}_{-}=1/(2|\beta|)$, also removing the global phase minus sign, we have
\begin{align}
&=  \sqrt{\frac{1-T_B}{2}} e^{-|\gamma|^2/2}  (  \ket{0} + g\gamma \ket{1})\\
&=  \sqrt{\frac{1}{2(g^2+1)}} e^{-|\gamma|^2/2} (  \ket{0} + g\gamma \ket{1}),
\end{align}
which is the known result for a $\oneinjectedscissor$. The setup also works for the opposite measurement outcome: no photon detection at $C$ and single-photon detection at $A$; however, there is a passive phase flip. The success probability is doubled since we keep both measurement outcomes: ${}_C\bra{1}{}_A\bra{0}$ and ${}_C\bra{0}{}_A\bra{1}$.

\subsection{\label{sec:appendix3-scissor}Three-photon quantum scissor}

\begin{figure}
      \makebox[\linewidth][c]{\includegraphics[width=0.5\linewidth]{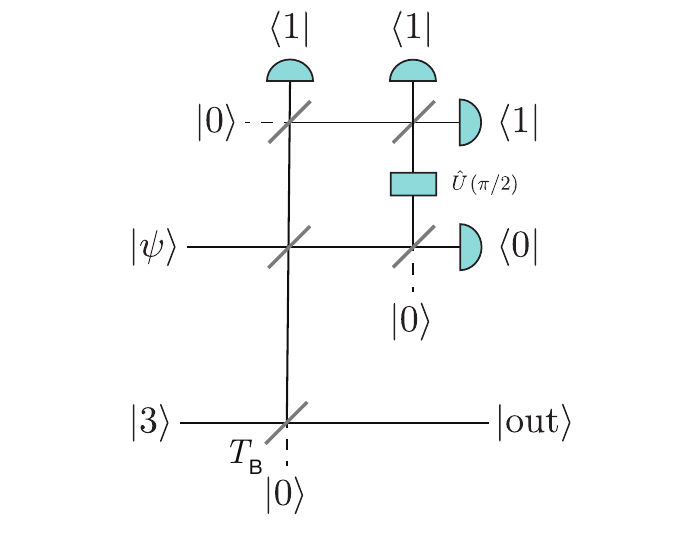}}
      \centering
          \caption[Our generalised three-photon quantum scissor]{Our generalised three-photon quantum scissor ($\threeinjectedscissor$) with the injection of a three photon state and four-port single-photon measurement scheme, with three ports detecting single photons and the remaining port detecting no photons. The upper four beamsplitters are all set to have transmissivity $1/2$. The lower beamsplitter has transmissivity $T_B$ which sets the gain of the NLA. There is a vital $\pi/2$ phase shift on one of the paths.}
          \label{fig:3scissor}
\end{figure}

We generalised the single-photon quantum scissor to all positive integers. As another practical example, consider our three-photon ($n=3$) quantum scissor shown in~\cref{fig:3scissor} (it was discovered before the $2$-scissor, which is why we focus on it here). The device can amplify input states of even larger amplitude than the $1$-scissor and the $2$-scissor. From~\cref{Chap3eq:gpsi}, we see that it performs the transformation
\begin{align}\label{eq:3scissor}
\hat{T}_3 \ket{\psi} &= \frac{\sqrt{6}}{8} \left(\frac{1}{g^2+1}\right)^\frac{3}{2}  \; (c_0 |0\rangle + g c_1 |1\rangle + g^2 c_2 |2\rangle + g^3 c_3 |3\rangle),
\end{align}
where the gain is $g{=}\sqrt{T_B/(1{-}T_B)}$ and the success probability is
\begin{align}
P_3 &= 4 \; | \bra{\psi }\hat{T}^\dagger_3 \hat{T}_3 \ket{\psi} |^2.
\end{align}
The success probability is effectively increased by a factor of four to account for the four successful combinations of obtaining three ``clicks''.

The three-photon scissor truncates and ideally amplifies the input state to third order. The device operates as an ideal NLA as long as $|g^4c_4|{\ll}|g^3c_3|$, which is a major improvement from the single photon scissor, at the cost of a reduced probability of success.

\begin{figure}
        \includegraphics[width=0.6\linewidth]{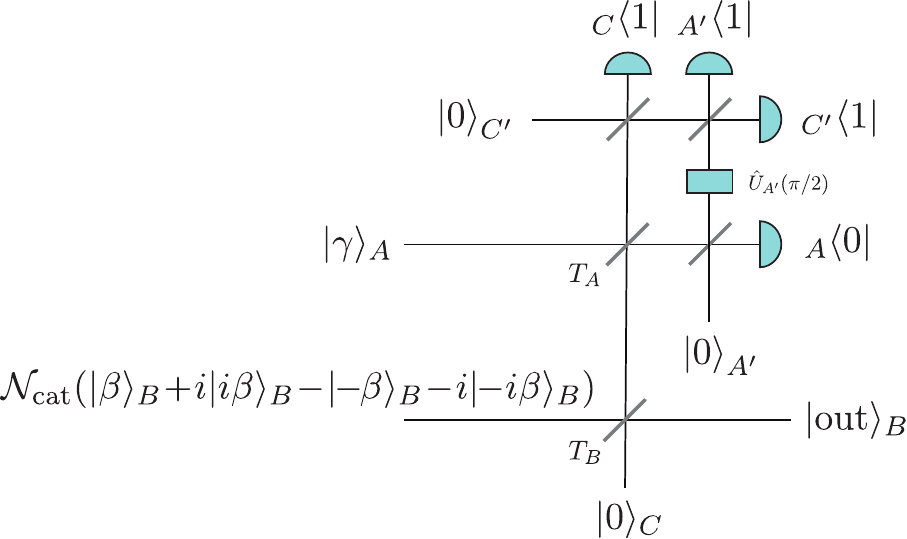}
        \centering
          \caption[$\fourcatrelay$ with four-component cat state as a resource and four-port detection scheme.]{$\fourcatrelay$ with four-component cat state as a resource and four-port detection scheme~\cite{neergaard-nielsen2013quantum}.}
          \label{fig:4cat_scissor}
\end{figure}

Tele-amplification can be scaled up: cat states with more components are used as the resource, the detection procedure detects multiple single photons, and the device can tele-amplify coherent states from a larger alphabet. Another simple case is the $\fourcatrelay$, for example, and is shown in \cref{fig:4cat_scissor}. In the limit that $\beta\to0$, $\fourcatrelay$ is equivalent to our $\threeinjectedscissor$. In the following, we show how the $\threeinjectedscissor$ transforms coherent states by again considering tele-amplification.

Consider~\cref{fig:4cat_scissor}. To simplify the calculation, we introduce $\alpha$ and replace $\beta {=} \alpha\sqrt{T_A/[(1-T_A)(1-T_B)]}$. The gain is $g{=} \sqrt{{T_A T_B}/{(1-T_A)(1-T_B)}}$, and we set $T_A{=}0.5$ also at this stage. The entire state after propagating through the circuit, but before the measurement, is given by
\begin{align}
\ket{\psi}_{BACA'C'} &= \mathcal{N}_\text{cat} \sum_{k=0}^3 i^k \ket{g \alpha i^k}_B \otimes \ket{\frac{\gamma - \alpha i^k}{2}}_A \otimes \ket{\frac{-(\gamma + \alpha i^k)}{2}}_C  \otimes \ket{\frac{\gamma(1-i)+\alpha(i^k+i^{k+1})}{2\sqrt{2}}}_{A'} \otimes \ket{\frac{\gamma(1+i)+\alpha(i^k-i^{k+1})}{2\sqrt{2}}}_{C'},
\end{align}
where $\mathcal{N}_\text{cat} = 1/\sqrt{8 e^{-\beta^2}(\sinh{\beta^2}-\sin{\beta^2})}$ is the normalisation factor from the four-lobed cat state.

After the measurement ${}_C\bra{1}{}_{A'}\bra{1}{}_{C'}\bra{1}{}_A\bra{0}$ we get
\begin{align}
{}_C\bra{1}{}_{A'}\bra{1}{}_{C'}\bra{1}{}_A\braket{0}{\psi}_{BACA'C'} &= \mathcal{N}_\text{cat} \sum_{k=0}^3 i^k   e^{-\frac{|A|^2}{2}} e^{-\frac{|A'|^2}{2}} e^{-\frac{|C|^2}{2}} e^{-\frac{|C'|^2}{2}}  A' C C' \ket{g \alpha i^k}_B,
\end{align}
where
\begin{align}
A &= \frac{\gamma - \alpha i^k}{2}\\
A' &= \frac{\gamma(1-i)+\alpha(i^k+i^{k+1})}{2\sqrt{2}}\\
C &= \frac{-(\gamma + \alpha i^k)}{2}\\
C' &= \frac{\gamma(1+i)+\alpha(i^k-i^{k+1})}{2\sqrt{2}}.
\end{align}
Simplifying, we get
\begin{align}
{}_C\bra{1}{}_{A'}\bra{1}{}_{C'}\bra{1}{}_A\braket{0}{\psi}_{BACA'C'} &= \mathcal{N}_\text{cat} \sum_{k=0}^3 i^k  e^{-(|\gamma|^2+|\alpha|^2)/2}  \frac{1}{8} (\alpha i^k - \gamma i)(\gamma i +\alpha i^k)(\gamma + \alpha i^k) \ket{g \alpha i^k}_B.\label{eq:cat4relayder}
\end{align}
We expand in the Fock basis to first order in $\alpha$
\begin{align}
&= \mathcal{N}_\text{cat} e^{-(|\gamma|^2+|\alpha|^2)/2}\frac{1}{2} \alpha^3 e^{-|g\alpha|^2/2}  \sum_{n=0}^3 \frac{g^n \gamma^n}{\sqrt{n!}} \ket{n}.
\end{align}
For small $\alpha$ note that $\mathcal{N}_\text{cat} = \sqrt{3}/(2\sqrt{2}\beta^3)$. Therefore, we have
\begin{align}
&= \frac{\sqrt{6}}{8 } \frac{\alpha^3}{\beta^3} e^{-(|\gamma|^2+|\alpha|^2)/2}  e^{-|g\alpha|^2/2}  \sum_{n=0}^3 \frac{g^n \gamma^n}{\sqrt{n!}} \ket{n}.
\end{align}
Recall that $\alpha/\beta = (1/(g^2+1))^{1/2}$, and throwing away all left-over high order terms in $\alpha$ we get
\begin{align}
&= \frac{\sqrt{6}}{8} \left(\frac{1}{g^2+1}\right)^\frac{3}{2} e^{-|\gamma|^2/2} \sum_{n=0}^3 \frac{g^n \gamma^n}{\sqrt{n!}} \ket{n}.
\end{align}
In the limit that $\alpha\to0$, the $\fourcatrelay$ is a generalised quantum scissor ideal NLA, truncated at Fock number 3. The generalised scissors truncate and amplify arbitrary input states (not only coherent states as shown above).

The different click patterns herald passive phase shifts on the state of 0, $\pi/2$, $\pi$, or $3\pi/2$. It can be corrected for by simply delaying the beam (i.e., a linear phase shift operation. For some applications the phase shift may simply be tracked in software. Explicitly, depending on the result at the four ports $(A, A’, C, C’)$, the transformation is given by:
\begin{align}
\ket{\psi}&\to \hat{T}_\text{3}\ket{\psi}\text{, for }(0,1,1,1)\\
\ket{\psi}&\to e^{\frac{\pi}{2}\hat{a}^\dagger\hat{a}}\hat{T}_\text{3}\ket{\psi}\text{, for }(1,0,1,1)\\
\ket{\psi}&\to e^{\pi\hat{a}^\dagger\hat{a}}\hat{T}_\text{3}\ket{\psi}\text{, for }(1,1,0,1)\\
\ket{\psi}&\to e^{\frac{3\pi}{2}\hat{a}^\dagger\hat{a}}\hat{T}_\text{3}\ket{\psi}\text{, for }(1,1,1,0)
\end{align}
where $\hat{T}_\text{3}$ is the $\threeinjectedscissor$ transformation.

Cat states with $n+1$ lobes may be used~\cite{Guanzon2022cat_relay} which is equivalent to injecting $n$ photons into the scissor (and checking for $n$ photons at the detector). In the limit of a large number of injected photons, generalised scissors perform as an ideal NLA without truncation, but with a vanishing success probability.

\subsection{\label{sec:appendix2-scissor}Lossy resource state}

We found an interesting result when there is loss at the input Fock state. Let us consider again $\fourcatrelay$. It requires three single-photon detections. This device actually works with other Fock-shifted 4-cat resources, not just the ``correct'' cat state which is Fock state $|3\rangle$ in the limit of zero amplitude. When one injects a two-photon state $\ket{2}$ instead of a three photon state $\ket{3}$, coherent states are transformed as
\begin{align}
\hat{T}_{\twoinjectedscissor}\ket{\gamma} &= \gamma \frac{\sqrt{2}}{8} \left(\frac{1}{g^2+1}\right) e^{-|\gamma|^2/2}  \;  (  |0\rangle + g \gamma |1\rangle + \frac{g^2 \gamma^2}{\sqrt{2}} |2\rangle),\label{eq:2scissor}
\end{align}
thus acting as an ideal NLA for coherent-state inputs of sufficiently small amplitude.

The success probability goes as $\gamma^2/(g^2+1)^2$. This result has surprisingly advantageous consequences for the performance of real experiments. For coherent state inputs, the scissors device allows for loss in the resource mode, i.e., preparation of a Fock state with non-perfect efficiency. The device will still successfully truncate and amplify the coherent state input, although now truncating at $\ket{n-l}$, where $l$ are the number of lost photons at the resource.

The reason it works for coherent states is that $\gamma$ is pulled out of all terms and can be absorbed into the normalisation constant. Seemingly magically, the factors $g$ and $1/\sqrt{n!}$ are in the correct place. This is because the factors $g$ and $1/\sqrt{n!}$ come from cat components $\ket{g\alpha i^k}_B$ and are not scrambled, whereas $\gamma$ comes from the input state and is scrambled. For coherent states, this scrambling is okay. For other input states, the coefficients on all the Fock states in superposition will be scrambled.

\section{Amplification of coherent states}

The fidelity $F$ of the output state with the ideally-amplified state (i.e., the target state) is useful as a measure of how well the input state has been amplified. When considering the performance of probabilistic amplifiers, it is also important to consider the success probability.

\begin{figure*}
\centering
\includegraphics[width=1\textwidth]{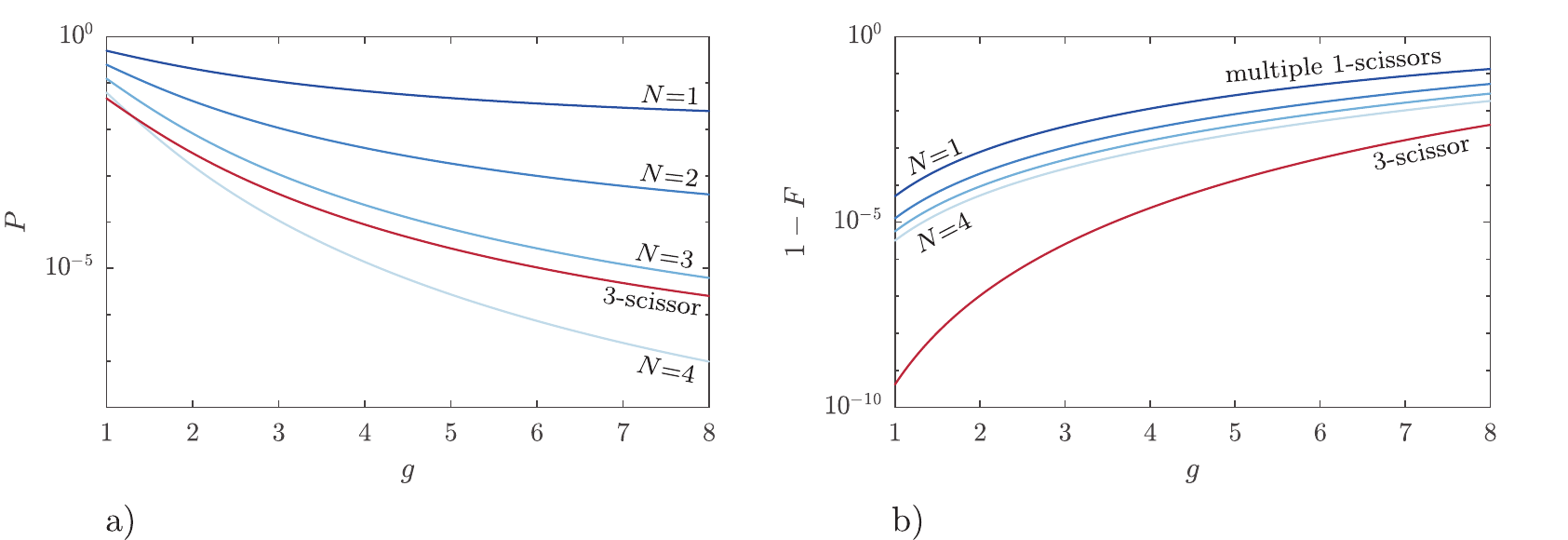}
\caption[a) Probability of success $P$ and b) infidelity $1-F$ versus gain $g$ comparing multiple $\oneinjectedscissors$ in an NLA with the $\threeinjectedscissor$.]{a) Probability of success $P$ and b) infidelity $1-F$ versus gain $g$ comparing multiple $\oneinjectedscissors$ in an NLA with the $\threeinjectedscissor$. The input state is a coherent state $\ket{\gamma}$ with magnitude $\gamma=0.1$. The number of $\oneinjectedscissors$ in the NLA is $N=1$ (darkest blue), $N=2$, $N=3$, and $N=4$ (lightest blue). There is a trade-off between fidelity and success probability; however, the $\threeinjectedscissor$ outperforms the $N=3$ device both in terms of fidelity and success probability. For more plots and a complete analysis of the full class of $n$-scissors devices, see Ref.~\cite{PhysRevLett.128.160501}.\label{fig:gammapoint1}}
\end{figure*}

In \cref{fig:gammapoint1} we plot the success probability and infidelity $(1{-}F)$ of the output state with the target state $\ket{g\gamma}$ as a function of the gain $g$ for a fixed input coherent state with amplitude $\gamma = 0.1$, and compare generalised scissors with NLA based on multiple $\oneinjectedscissors$ in parallel (see~\cref{sec:appendixPsuc,sec:appendixmultiplescissors} for details). Despite the trade-off between fidelity and success probability we see that our higher-order scissors simultaneously achieves higher fidelity \textit{and} success probability than the NLA based on several $\oneinjectedscissors$.

\section{Entanglement distillation with generalised quantum scissors}

A two-mode squeezed vacuum (TMSV) state with squeezing parameter $r$ is $\ket{\chi} {=}  \sqrt{1{-}\chi^2} \sum_{n=0}^\infty \chi^n \ket{nn}$, where $\chi{=}\tanh{r}$, and the mean photon number is $\bar{n}{=}\sinh^2{r}$. The notation $\ket{n,n}$ is shorthand for the two mode Fock state $\ket{n}{\otimes}\ket{n}$.

Placing a $\oneinjectedscissor$ on one arm transforms the TMSV state according to
\begin{align}
|\chi\rangle &\to  \sqrt{\frac{1-\chi^2}{2(g^2+1)}} ( \ket{0,0} + g \chi \ket{1,1} ).
\end{align}

Placing the $\threeinjectedscissor$ on one arm performs the transformation
\begin{align}
|\chi\rangle &\to   \frac{\sqrt{6}}{8} \sqrt{\frac{1-\chi^2}{(g^2+1)^3} } \sum_{n=0}^3 (g \chi)^n \ket{n,n}.
\end{align}

Both scissors herald states that have the form of a truncated TMSV state but with an effective increase in $\chi\to g\chi$. Therefore, the scissors are useful for entanglement concentration (when going from pure entanglement to more entangled pure entanglement). These protocols generalise to TMSV states distributed through loss, allowing purification of entanglement distributed over long distances~\cite{Ralph11,Ulanov15}.

For such protocols, the $\oneinjectedscissor$ usually works best when limited to small $\chi$ and large loss \cite{dias2018quantum}. The $\threeinjectedscissor$ allows distillation protocols to operate in regimes of higher squeezing and less loss, at the cost of a reduced probability of success, and it also introduces less non-Gaussianity. To demonstrate this, we calculate the entanglement of formation and reverse coherent information~\cite{PhysRevLett.102.050503} in the following.

Consider transmitting the second mode of a TMSV state through a pure loss channel of transmissivity $T$. The global state is instead given by
\begin{align}
|\chi\rangle_{AA'E} &\to  \sqrt{1-\chi^2} \sum_{n=0}^\infty \chi^n \sum_{k=0}^n \sqrt{{n \choose k}}  \; (1-T)^{k/2} T^{(n-k)/2}|n\rangle_{A} |n-k\rangle_{A'} |k\rangle_{E},
\end{align}
and the trace is taken over mode E so the state is mixed. 
Following the pure loss channel by a $\oneinjectedscissor$ results in a state with density operator
\begin{align}
\rho_{AB} &= \sum_{k=0}^{\infty} |\psi_k \rangle \langle \psi_k |,
\end{align}
where 
\begin{align}
|\psi_k \rangle &= \sqrt{\frac{1-\chi^2}{2(g^2+1)}} (1-T)^\frac{k}{2}  \; [(-\chi)^k |k\rangle |0\rangle + \sqrt{T(k+1)} (-\chi)^{k+1} g |k+1\rangle |1\rangle].
\end{align}
Similarly, following the pure loss channel by a $\threeinjectedscissor$ gives
\begin{align}
\rho_{AB} &= \sum_{k=0}^{\infty} |\psi_k \rangle \langle \psi_k |,
\end{align}
where
\begin{align}
|\psi_k \rangle &= \frac{\sqrt{6}}{8} \sqrt{\frac{1-\chi^2}{(g^2+1)^3} } \sum_{n=k}^{n+3} \chi^n g^{n-k} \sqrt{{n \choose k}}  \; (1-T)^{k/2} T^{(n-k)/2} \ket{n}\ket{n-k}.
\end{align}

After transmission of one mode of a TMSV state through a pure loss channel of transmissivity $T$ followed by either a $\oneinjectedscissor$ or a $\threeinjectedscissor$, we calculate the Gaussian entanglement of formation (GEOF) as an entanglement measure to evaluate the performance of each scissor. The GEOF quantifies the amount of two-mode squeezing required to prepare an entangled state from a classical state~\cite{PhysRevA.96.062338}, which is an upper bound on the entanglement of formation.

\begin{figure}
\includegraphics[width=0.6\linewidth]{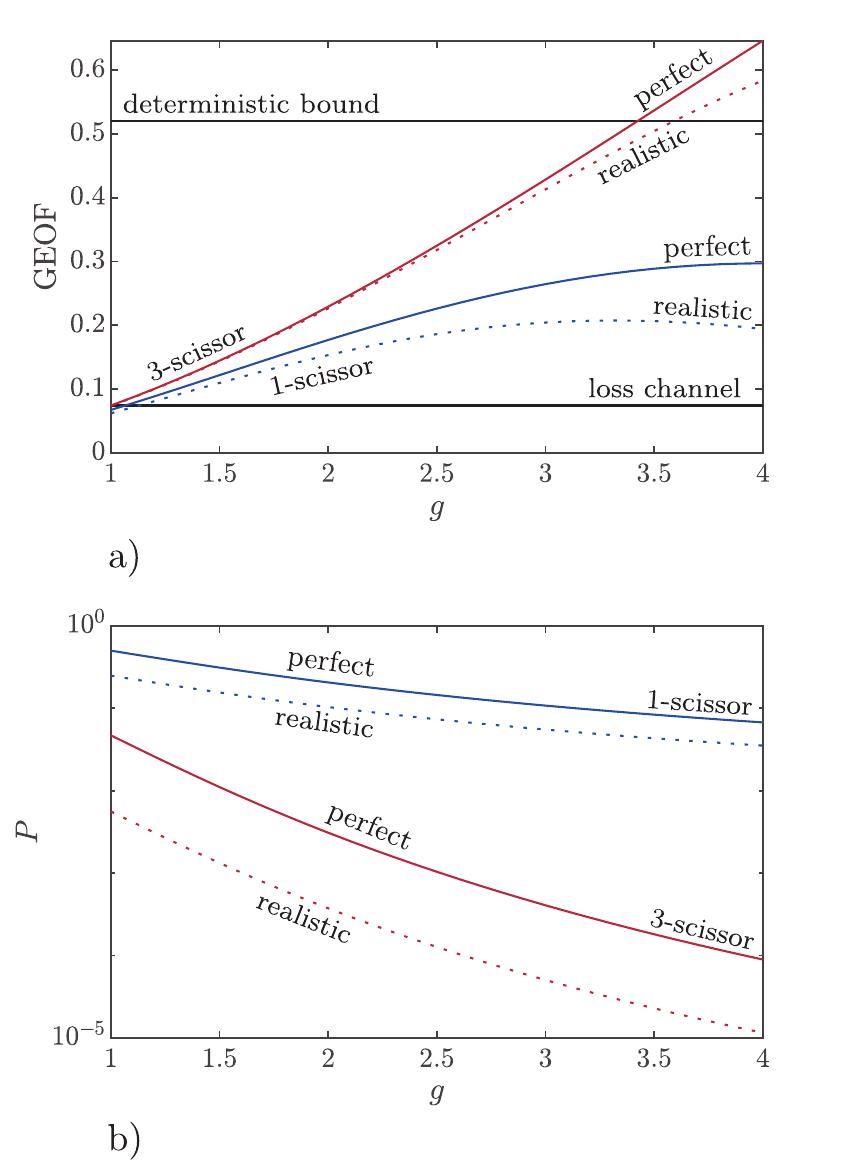}
\centering
\caption[a) Gaussian entanglement of formation (GEOF) and b) probability of success $P$ as a function of gain $g$ for a TMSV state, with one arm propagated through a lossy channel followed by a $\oneinjectedscissor$ or a $\threeinjectedscissor$.]{a) Gaussian entanglement of formation (GEOF) and b) probability of success $P$ as a function of gain $g$ for a TMSV state, with one arm propagated through a lossy channel followed by a $\oneinjectedscissor$ or a $\threeinjectedscissor$ for the perfect setup, and for non-ideal realistic resource and detectors $(\tau_\text{s} = \tau_\text{d} = 0.7)$. Channel transmissivity is $T=0.1$ and TMSV parameter is $\chi=0.3$. The ``loss channel'' is GEOF calculated for direct transmission, i.e., no quantum scissor. The ``deterministic bound'' is the amount of entanglement given that an infinitely squeezed state has been sent through the channel~\cite{PhysRevA.96.062338}. This plot shows a situation for which the crossing point of the deterministic bound, the minimum requirement for error correction \cite{dias2018quantum}, can be reached by the $\threeinjectedscissor$, even for a realistic experimental setup, but is unobtainable by the $\oneinjectedscissor$. Given \cref{fig:gammapoint1} has already demonstrated that the three-photon scissor has an improved fidelity and success probability over three or four scissors in parallel, we do not deem it necessary to further investigate entanglement distillation using scissors in parallel.\label{fig:GEOF}}
\end{figure}

The GEOF is calculated following Ref.~\cite{PhysRevA.96.062338} and using results from Ref.~\cite{PhysRevLett.84.2722,PhysRevLett.84.2726}. \Cref{fig:GEOF} shows the GEOF as a function of $g$ for the $\oneinjectedscissor$ and our $\threeinjectedscissor$ given TMSV parameter $\chi = 0.3$ and channel transmissivity $T=0.1$. Also shown is the amount of entanglement for the same TMSV state and loss channel but with no quantum scissor. The deterministic bound assumes an infinitely squeezed TMSV state sent through the same loss channel and no quantum scissor. Crossing the deterministic bound is a necessary condition for the distillation to be useful in error correction or repeater protocols \cite{SpyrosandJosephine}.
The $\threeinjectedscissor$ has a higher GEOF than the $\oneinjectedscissor$ for the same gain. For these parameters, the $\threeinjectedscissor$ crosses the deterministic bound whilst the $\oneinjectedscissor$ is unable to cross this bound.

To demonstrate the non-Gaussian effect of the scissors, we calculate the total reverse coherent information (RCI)~\cite{PhysRevLett.102.050503}, and compare it with the Gaussian RCI, i.e., the total RCI calculated for a Gaussian state with the same covariance matrix. The RCI gives a lower bound on the distillable entanglement~\cite{PhysRevLett.102.210501} as outlined in~\cref{sec:RCI}.

\begin{figure}
\includegraphics[width=0.5\linewidth]{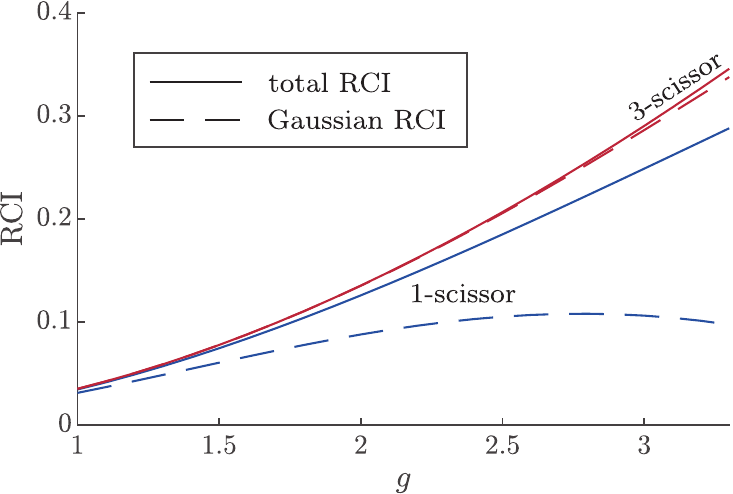}
\centering
\caption[Total reverse coherent information (RCI) and Gaussian RCI as a function of gain $g$ for a TMSV state, with one arm propagated through a lossy channel followed by a $\oneinjectedscissor$ or a $\threeinjectedscissor$.]{Total reverse coherent information (RCI) and Gaussian RCI as a function of gain $g$ for a TMSV state, with one arm propagated through a lossy channel followed by a $\oneinjectedscissor$ or a $\threeinjectedscissor$. Channel transmissivity is $T=0.1$ and TMSV parameter is $\chi=0.3$. Our $\threeinjectedscissor$ introduces less non-Gaussianity than the $\oneinjectedscissor$.\label{fig:RCI}}
\end{figure}

We plot the total RCI and Gaussian RCI as a function of gain in \cref{fig:RCI} for TMSV parameter $\chi=0.3$ and channel transmissivity $T=0.1$. This plot demonstrates that for these parameters, the non-Gaussian entangled state heralded after the $\oneinjectedscissor$ suffers severely from unwanted non-Gaussianity, whereas the non-Gaussianity introduced by the $\threeinjectedscissor$ is not so harsh.

\section{Imperfect operations}

An important consideration is how the performance of the 3-scissor is affected by experimental imperfections. 
Single-photon detectors with quantum efficiency $\tau_\text{d}$ can be modelled by a lossy channel with transmissivity $\tau_\text{d}$ followed by a perfect single-photon detector. We find that non-perfect efficiency impacts the success probability but has a small impact on the fidelity, as we'll see in this section. 

Typically, it is more feasible in an experiment to use on-off photon detectors. On-off detectors cannot discriminate between different numbers of photons but can only distinguish between vacuum and non-vacuum. Again, for input states with small amplitudes, the effect of on-off detection on the fidelity is small.

The reason that our generalised scissors are robust to these practical issues is due to the detection scheme. The detection scheme separates all the light into several modes and performs single-photon detection on those modes. In the high-fidelity regime, there is only a small number of photons in the device at any one time (if there was not, the device would not be working in a high fidelity regime), and so errors due to on-off detection or imperfect detectors are rare.

Another important practical aspect of the device is the resource mode. The efficiency of the resource is modelled by a lossy channel with transmissivity $\tau_\text{s}$ following preparation of the Fock-state resource, $|1\rangle$ for the $\oneinjectedscissor$ and $|3\rangle$ for the $\threeinjectedscissor$. The success probability and the fidelity are both negatively impacted, but for coherent states, not severely. For coherent state inputs, the $\threeinjectedscissor$ still performs ideal truncation and amplification up to two photons if two photons rather than three are injected into the device, of course with a reduced fidelity since the truncation is at second order not three. This surprising result allows the $\threeinjectedscissor$ to keep performing well even with loss on the resource mode. This result generalises to all scissors, i.e., the $n$-scissor ideally truncates and amplifies coherent states even if less than $n$ photons are injected into the device.

In \cref{fig:GEOF} we also include the effect of inefficient detectors and an inefficient resource ($\tau_\text{s}=\tau_\text{d}=0.7$). We find that under realistic conditions the 3-scissor can still distil entanglement above the deterministic bound under conditions for which this is impossible for the 1-scissor.

Generalised scissors are robust to inefficient single-photon detectors, only the success probability is affected. More experimentally feasible than single-photon detectors are on-off detectors. The on-off measurement operators are
\begin{align}
\hat{\Pi}_\text{off} &=  \ketbra{0}{0},\\
\hat{\Pi}_\text{on} &=  \Id - \ketbra{0}{0}.
\end{align}

With on-off detection, \cref{eq:cat4relayder} for example needs correcting; the output state is no longer pure and is given by the density operator
\begin{align}
\rho_\text{out} = \sum_{n,m,l=1}^\infty \ket{\psi_{n,m,n}}\bra{\psi_{n,m,n}},
\end{align}
where
\begin{align}
\ket{\psi_{n,m,n}} &= \mathcal{N}_\text{cat} \sum_{k=0}^3 i^k   e^{-\frac{|A|^2}{2}} e^{-\frac{|A'|^2}{2}} e^{-\frac{|C|^2}{2}} e^{-\frac{|C'|^2}{2}}  \; \frac{{A'}^n}{\sqrt{n!}} \frac{{C{}}^m}{\sqrt{m!}} \frac{{C'}^l}{\sqrt{l!}} \ket{g \alpha i^k}_B.
\end{align}

\begin{figure}
\centering
\includegraphics[width=0.6\linewidth]{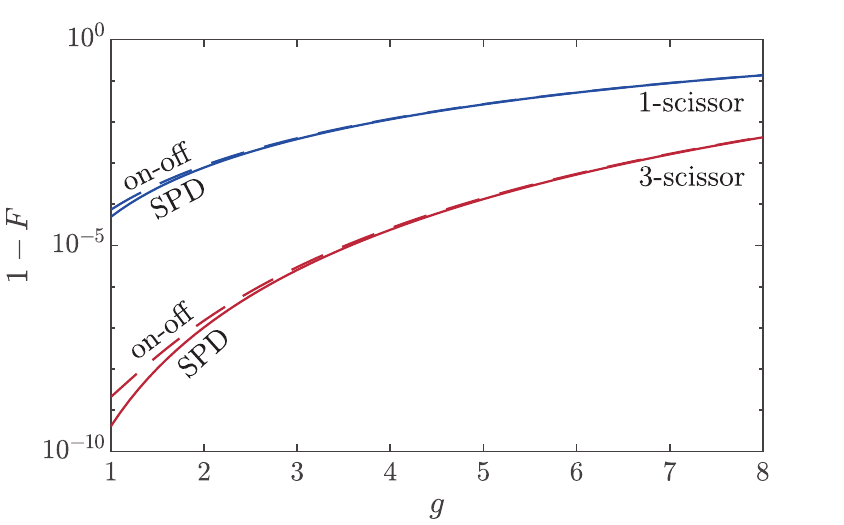}
\caption[a) Infidelity $1-F$ vs. the gain $g$ for the $\oneinjectedscissor$ and the $\threeinjectedscissor$ with perfect single-photon detectors (SPD) or on-off detectors.]{a) Infidelity $1-F$ vs. the gain $g$ for the $\oneinjectedscissor$ and the $\threeinjectedscissor$ with perfect single-photon detectors (SPD) or on-off detectors. The input state is a coherent state $\ket{\gamma}$ with magnitude $\gamma=0.1$. (There is very little increase in the probability of success for $\gamma = 0.1$ by using on-off detectors, thus, we have not included a similar plot showing the success probability.)\label{fig:onoff}}
\end{figure}

We plot the infidelity vs. gain in~\cref{fig:onoff} for single-photon detectors (SPD) and on-off detectors, (assuming perfect efficiency for all devices). The input state is a coherent state with magnitude $\gamma = 0.1$. In the high-fidelity regime of the scissors (low amplitude input states), this correction is small, and the scissors perform well with on-off detectors in place of SPDs.

\begin{figure}
\centering
\includegraphics[width=0.5\linewidth]{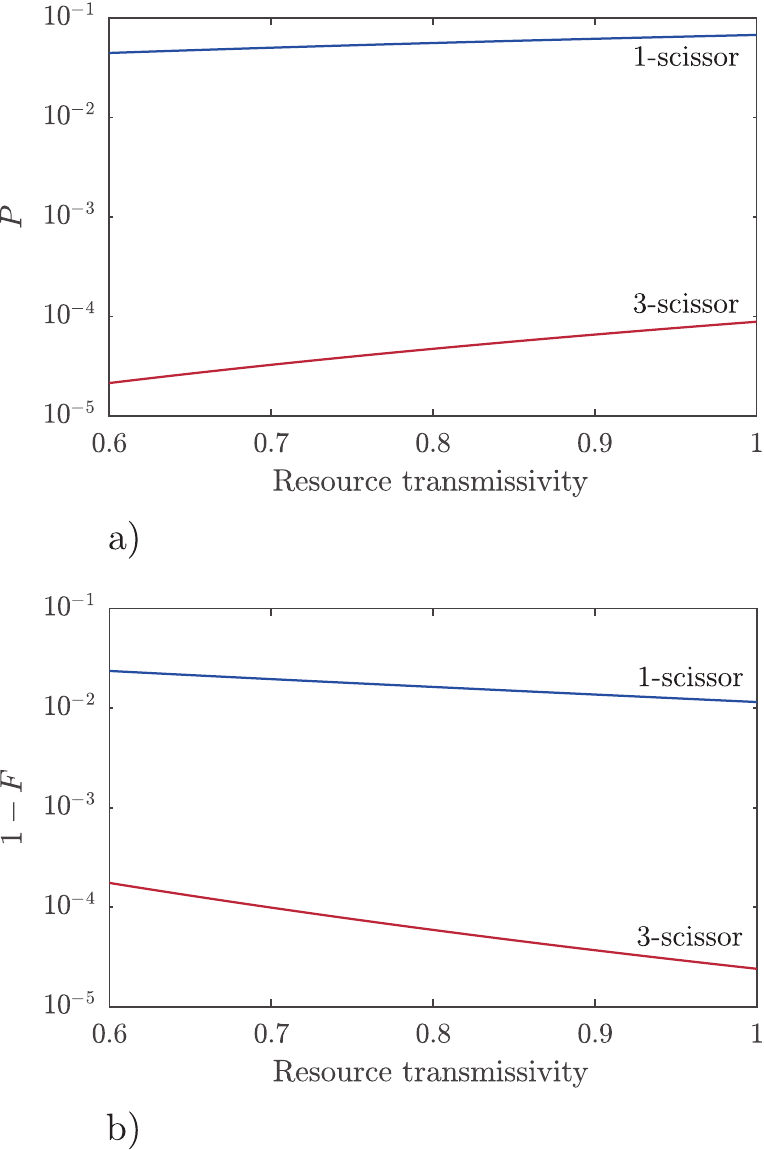}
\caption[a) Probability of success $P$ and b) infidelity $1-F$ vs. transmissivity $\tau_s$ of the resource mode for the $\oneinjectedscissor$ and the $\threeinjectedscissor$.]{a) Probability of success $P$ and b) infidelity $1-F$ vs. transmissivity $\tau_s$ of the resource mode for the $\oneinjectedscissor$ and the $\threeinjectedscissor$. The input state is a coherent state $\ket{\gamma}$ with magnitude $\gamma=0.1$. The gain is set to $g=4$.\label{fig:lossatresource}}
\end{figure}

Of greater concern for the practicality of generalised scissors is the efficiency of the resource Fock state. In~\cref{fig:lossatresource} we plot the success probability and infidelity of the output state with the target state $|g\gamma\rangle$ as a function of transmissivity $\tau_s$ of the resource mode. The input state is a coherent state $\gamma=0.1$ and the gain is fixed to $g=4$. For coherent state inputs, the scissors are surprisingly robust to resource inefficiencies. This can be explained by our result where less-than $n$ photons enter the device as a resource, derived in~\cref{sec:appendix2-scissor}.

\section{\label{sec:appendixEPR}Repeater-like operation of the quantum scissors}

In this section, we consider the quantum scissors with loss distributed into the device as per~\cref{Chap3fig:lossy_scissors} (middle configuration) compared with the usual protocol with the measurement at the end (end configuration).

\begin{figure}
    \begin{center}
        \includegraphics[width=0.6\linewidth]{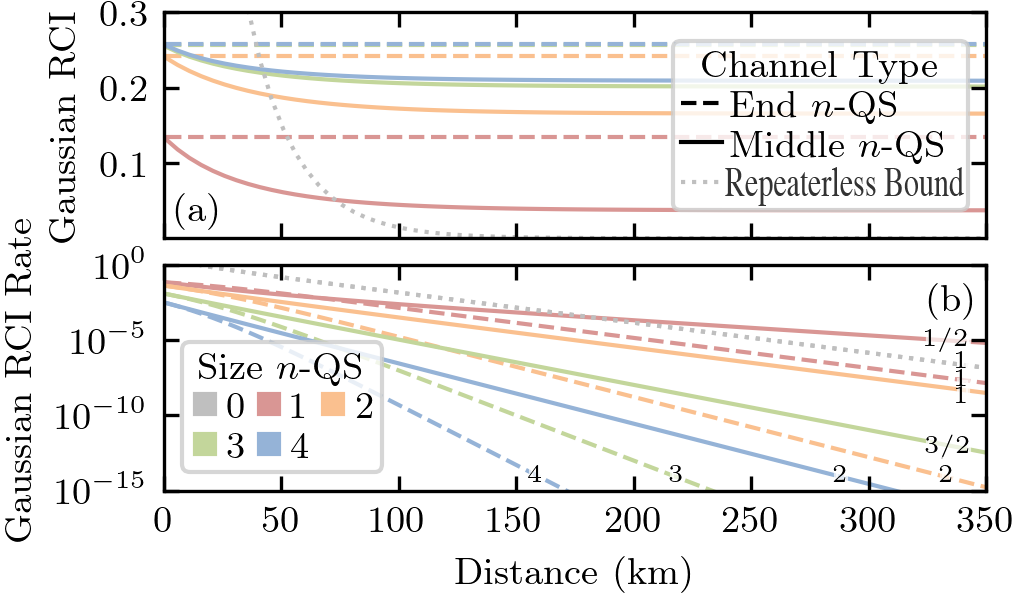}
        \caption[Gaussian reverse coherent information (GRCI) of a two-mode squeezed vacuum (TMSV) state with $\chi=0.25$ squeezing versus distance (km).]{\label{fig:GRCIRate} Gaussian reverse coherent information (GRCI) of a two-mode squeezed vacuum (TMSV) state with $\chi=0.25$ squeezing versus distance (km) for our $n$-scissor protocol at various $n$. We consider placing the QFT measurement either at the end of the channel (dashed lines) or in the middle of the channel (solid lines), with $g=1/\sqrt{\eta}$ for the end configurations and $g=1$ for the middle configurations. (a) The Gaussian reverse coherent information (GRCI) achieved by our $n$-scissor protocol at various $n$. (b) Average rate of quantum communication given by the probability $\times$ GRCI, where the small numbers refer to the long-distance scaling ($\eta^{1/2}$, $\eta^1$, $\eta^{3/2}$, etc.). This can be compared with the repeaterless bound (dotted grey line).}
    \end{center}
\end{figure}

We consider the reverse coherent information (RCI)~\cite{PhysRevLett.102.210501} as an achievable rate for entanglement distillation in bits per use of the channel; in particular, we calculate the Gaussian RCI from the covariance matrix. We set $g=1/\sqrt{\eta}$ for the end configuration and $g=1$ for the middle configuration. We can see loss tolerance in~\cref{fig:GRCIRate}a. In this plot, we can see the effect of the truncation on the entanglement at $d=0$, since $g=1$ for both cases. Recall that the input state $|\chi\rangle$ has the coefficients $\chi^k$, and since $\chi=0.25$ each additional term will contribute significantly less to the entanglement for small $\chi$.

We can also calculate the RCI multiplied by the success probability as shown in Fig.~\ref{fig:GRCIRate}b. This can then be compared to the repeaterless bound. The low-order scissors have better success probability, but these pay a penalty in terms of the quality of the output state since they do not keep higher-order terms.

Quantum repeaters can improve the rate-distance scaling of the higher-order scissors, which can be extremely useful. In other words, by dividing the lossy channel into additional shorter segments, a chain of the single-node $n$-scissor protocols (in series) can independently distil entanglement of the shorter links to improve the rate-distance scaling, as will be shown in fantastic detail for the $1$-scissor in~\cref{Chap:simple_repeater}.

\section{Chapter discussion and conclusion}

We have shown our generalised scissors 
amplify and truncate arbitrary input states without distorting the Fock coefficients, assuming perfect implementation. Considering more realistic devices, we find that in the working regime of high fidelity, imperfect single-photon detectors, or employing on-off detectors 
has little effect on the fidelity and only impacts the success probability. Of greater importance 
is the efficiency of the Fock-state resource. For coherent-state inputs, we find that the $n$-scissor device is naturally and surprisingly robust to non-ideal resource efficiency.  
Realistic devices perform well for entanglement distillation as well.

Another possible application of our scissors would be to engineer optical states~\cite{escher2005synthesis}. For example, making a slight change to our $n$-scissor device; in particular by accepting a different measurement outcome, a different state will be heralded. This heralded state may be potentially useful for instance, we speculate that it would be possible to generate truncated cat-like states in this way.

Generalised scissors belong to a class of protocols known as tele-amplification~\cite{neergaard-nielsen2013quantum}. Scissors are tele-amplification devices upon taking the amplitude of the entangled cat-state resource to zero (derivation presented in~\cref{sec:appendixscissors})~\cite{Guanzon2022cat_relay}.

The laws of quantum physics put absolute limits on the performance of probabilistic NLA~\cite{Pandey_2013}. A natural question to ask is how the scissors compare against these ultimate bounds. Within the high-fidelity region, NLAs have success probabilities that decrease exponentially with $N$ (the order of truncation), and this is an unavoidable consequence of attempting noiseless linear amplification~\cite{Pandey_2013}. Our scissors do not obtain the ultimate limits of NLA; one would require more complicated (probably highly nonlinear) devices, such as the proposal in Ref.~\cite{PhysRevA.89.023846}.

To conclude this chapter, we have introduced new quantum scissors which truncate and ideally amplify optical states up to $n\in\mathbb{N}^+$ using linear optical components. Compared to the use of multiple scissors in parallel we found that the new scissors are more efficient for noiseless linear amplification and more practical for experimental implementation. This device may be scaled up to $n$ numbers of photons at the cost of a diminishing probability of success. We expect that our generalised scissors will in some situations 
improve the performance of existing experiments in quantum communication and make theorised protocols realisable in the near future.

Finally, we showed that a distributed $n$-scissor quantum relay is loss-tolerant with fast rates, hence is useful as a building block for quantum repeater networks.


%% file: chapter_CVrepeater/chapter_CVrepeater.tex
\chapter{CV quantum repeater}
\label{Chap:CV_QR}	
\pagestyle{headings}

\noindent
The results of this chapter have been published in the following.\\

\begin{sloppypar}
\noindent
1.~\cite{PhysRevA.102.052425} J. Dias, \textbf{M. S. Winnel}, N. Hosseinidehaj, and T. C. Ralph, ``Quantum repeater for continuous-variable entanglement distribution,'' \href{http://dx.doi.org/10.1103/physreva.102.052425}{Phys. Rev. A \textbf{102}, 052425 (2020)}.\\
\end{sloppypar}

\noindent
We thank Josephine Dias for major contributions to the development of the results of this chapter and the preparation of the text, while the author of this thesis, \textbf{Matthew Winnel}, contributed by performing numerical calculations of the secret key rates and being involved in the discussions. The results of later chapters owe a lot to the insights and discoveries made in this chapter.\\

\noindent
In this chapter, we present a CV repeater scheme using NLA for entanglement distillation and optimal Gaussian entanglement swapping. Our scheme uses NLA for entanglement distillation. We show that using practical quantum scissors for NLA and under the assumption of ideal quantum memories and perfect sources and detectors, our scheme beats the repeaterless bound~\cite{Pirandola_2017} and scales like the fundamental repeater bounds~\cite{Pirandola_2019}. We reduce the number of physical resources in~\cref{Chap:simple_repeater} by commuting the loss into the quantum scissor.

\section{Introduction}

In this chapter, we present an improvement upon the protocol presented in Ref.~\cite{dias2017}. Like Refs.~\cite{dias2017, PhysRevResearch.2.013310}, our repeater uses the single-photon ($n=1$) quantum scissor discussed in~\cref{Chap:scissors} to distil CV entangled states. Unlike Ref.~\cite{PhysRevResearch.2.013310}, which uses non-deterministic non-Gaussian entanglement swapping, our CV quantum repeater uses  Gaussian entanglement swapping with post-selection. The post-selection is required to ensure the state is close to a Gaussian state, because the scissor truncates the state, making it non-Gaussian. Our scheme utilises a different Gaussian entanglement swapping setup than Ref.~\cite{dias2017}; thus, we can report an improvement in the attainable key rates. The CV repeater scheme presented in this chapter surpasses the fundamental repeaterless repeaterless bound \cite{Pirandola_2017} for a total distance of 322 km. The repeater considered in this chapter is similar to Ref.~\cite{Ghalaii_2020}, except here we consider a realistic NLA.

The results in this chapter do not model the effects of imperfect quantum memories, sources or detectors. Furthermore, the channel is assumed to be pure loss and without thermal noise. We improve the repeater in this chapter in~\cref{Chap:simple_repeater}, and there we indeed consider imperfect sources, with inefficiencies and dark counts, and additional thermal noise in the channel.

This chapter is arranged in the following way: in~\cref{Chap4sec:repeater} we explain the structure of our CV repeater and in~\cref{Chap4sec:results} we present results. Finally, in~\cref{Chap4sec:conc} we discuss and conclude.

In~\cref{Chap:simple_repeater}, we improve this repeater by reducing the required resources and removing the requirement for post-selection.

\section{CV repeater with NLA and dual homodyne \label{Chap4sec:repeater}}

We saw in~\cref{Chap2CVrepeatercomponents} that first-generation quantum repeaters are based on three core elements: entanglement distribution, entanglement swapping, and entanglement purification (or distillation). See Ref.~\cite{munro2015inside} for a review. In the following section, we will give an overview of how each of these elements will be implemented in our CV repeater in this chapter. 
\begin{figure}
\centering
\subfloat[]{%
  \includegraphics[scale=1.3]{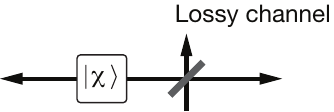} \label{Chap4fig:entdist}%
}

\subfloat[]{%
  \includegraphics[scale=1.3]{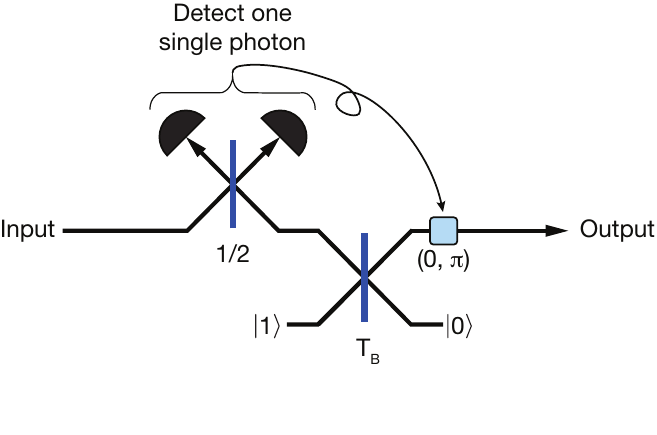}
  \label{Chap4fig:NLA}
  }

\subfloat[]{%
  \includegraphics[scale=1.3]{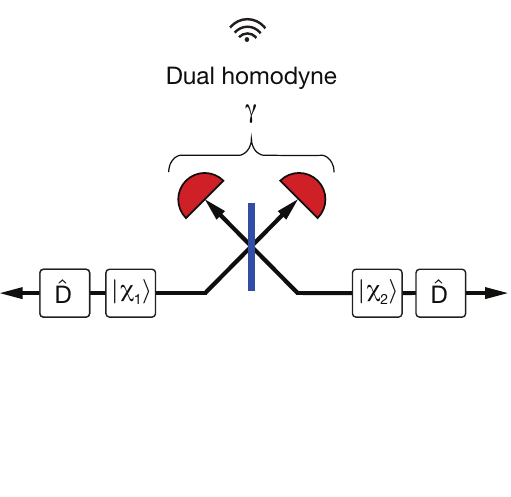}
  \label{Chap4fig:entswap1}
  }
\caption[Components of the CV quantum repeater]{Components of the CV quantum repeater \protect\subref{Chap4fig:entdist} Entanglement distribution in the CV quantum repeater. One mode of a TMSV state is sent through a lossy channel to a neighbouring repeater node. The other mode of the entangled state remains in the same node. \protect\subref{Chap4fig:NLA} The CV repeater uses the NLA to distil entangled TMSV states. The simplest linear optics construction of the NLA is pictured here consisting of a single-photon quantum scissor. The input is combined with an ancilla photon which has passed through a beamsplitter of tunable ratio $\xi$. This is related to the gain of the NLA via $g=\sqrt{(1-\xi)/\xi} $. The combined modes are detected and success is heralded when a single photon is detected at one output and none at the other. \protect\subref{Chap4fig:entswap1} Gaussian entanglement-swapping protocol from Ref.~\cite{PhysRevA.83.012319}. Modes of two independent TMSV states are combined and input into a dual-homodyne detection (dual HD). The results of the detection are sent in both directions to both output modes where displacements are performed accordingly.}
\label{Chap4fig:Components}
\end{figure}
\subsection{Entanglement distribution}
Beginning with entanglement distribution, the entangled resource states used in our protocol are the Gaussian two-mode squeezed vacuum (TMSV) state:
\begin{equation}
\ket{\chi}_{AB} = \sqrt{1-\chi^2} \sum \chi^n \ket{n}_{A} \ket{n}_{B} .
\label{Chap4eq:chi5}
\end{equation}
where $0<\chi<1$ is the two-mode squeezing parameter.
Distribution of these states \cref{Chap4eq:chi5}, is performed asymmetrically (see~\cref{Chap4fig:entdist}) with entangled states being generated at each node of the quantum repeater and then one mode of the entangled state is passed through a lossy channel through to the neighbouring node. One mode of each of these entangled states would be decohered by loss from transmission through the channel while the other mode remains untouched in the same node.

\subsection{Entanglement distillation}
In our CV repeater, entanglement distribution in the repeater links is followed immediately by distillation on the entangled mode that has passed through the lossy channel. Entanglement distillation is a necessary component in first-generation repeaters, needed to combat the decoherence effects from channel loss and entanglement-swapping operations. In the scheme of Refs.~\cite{dias2017, PhysRevResearch.2.013310} and in this work, the NLA~\cite{ralph2009nondeterministic} is used to distil the entangled states. We saw in~\cref{Chap:scissors} that when implemented with linear optics, the simplest NLA comprises of a single-photon quantum scissor~\cite{ralph2009nondeterministic}. The single-photon quantum scissor implementation of the NLA has been demonstrated experimentally~\cite{Kosis13,xiang2010heralded}, and more specifically entanglement distillation on TMSV states decohered by loss has been demonstrated with a similar device \cite{Ulanov15}.

\subsection{Entanglement swapping}
Following entanglement distribution and distillation, our CV repeater will use deterministic Gaussian entanglement swapping \cite{loock1999unconditional} to connect the entangled repeater links. We employ the optimal Gaussian entanglement-swapping protocol described in Ref.~\cite{PhysRevA.83.012319}. This involves sending classical signals to both ends of the channel and conducting displacements on both modes (see~\cref{Chap4fig:entswap1}). This is unlike other protocols (including CV teleportation) where classical communication and displacements are only performed on one mode. In this way, two pure Gaussian entangled states can be swapped and the resulting entangled state remains pure. In general, for any two Gaussian states, entanglement swapping in this way is optimal~\cite{PhysRevA.83.012319}. 

The use of the optimal Gaussian entanglement swapping scheme of Ref.~\cite{PhysRevA.83.012319} represents the main difference between this work and the work in Ref.~\cite{dias2017} which used CV teleportation. In this work, we also consider the use of post-selection based on the results of the dual HD in the swapping scheme. Qualitatively, this means that based on the results of the dual HD some results will be rejected and some will be accepted; thus, entanglement swapping in our repeater is not deterministic. Post-selection in our scheme is necessary because the truncation due to the single-photon quantum scissor deteriorates the raw key and adds non-Gaussianity. This effect is more pronounced for large measurement outcomes and thus we use post-selection to filter the measurement results, accepting results that are close to zero. 

\section{Results \label{Chap4sec:results}}
\subsection{Single-node repeater}
\begin{figure}
\centering
  \includegraphics[width=1\linewidth]{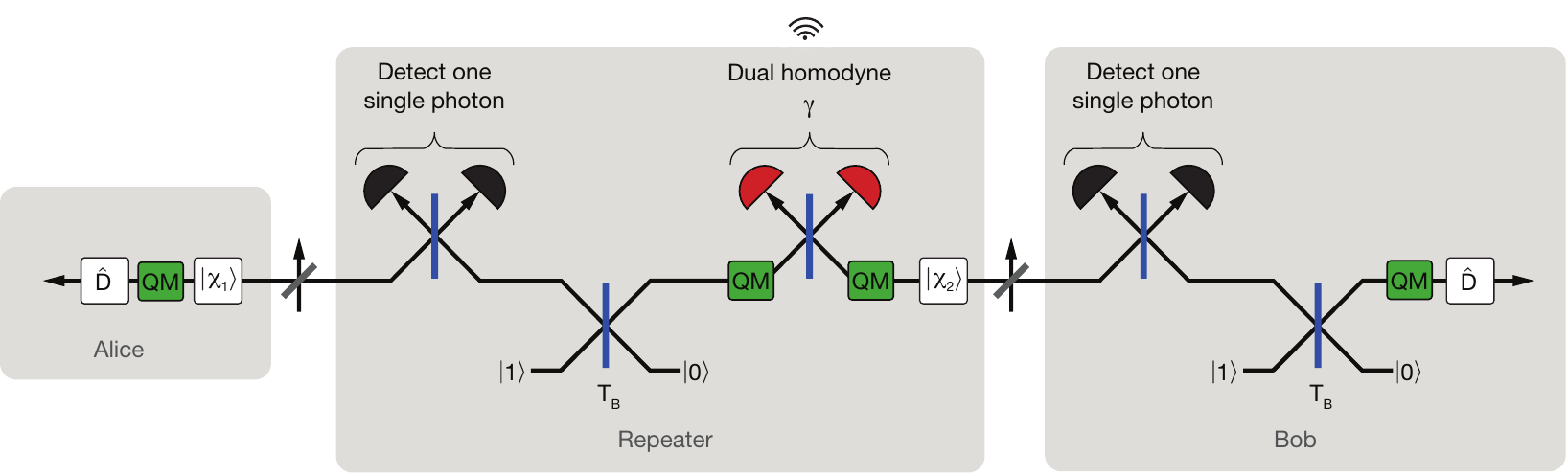}
\caption[Simplest implementation of our first-generation CV repeater with NLA and dual homodyne.]{Simplest implementation of our first-generation CV repeater with NLA and dual homodyne. The entangled resource states used are TMSV states (given by \cref{Chap4eq:chi5}). Entanglement distillation is performed by a quantum scissor. Once successful distillation has been heralded, the distilled state is stored in a quantum memory (QM) where it will wait for the neighbouring quantum scissor to succeed. Gaussian entanglement swapping is conducted by a dual homodyne of the two modes at the repeater node, if the outcome of the dual homodyne is within the accepted post-selection range around zero, this is followed by classical communication of the results of the detection being sent to Alice and Bob and both modes are then displaced accordingly ($\hat{D}$). This configuration is asymmetric as the two inputs to the dual homodyne are not the same. This setup requires one source to be placed with Alice and a quantum scissor to be placed at the repeater node.}
\label{Chap4fig:asymrep1}
\end{figure}
The simplest implementation of our CV-repeater protocol is shown in~\cref{Chap4fig:asymrep1}. It is formed using a single repeater node in the centre of the channel with a single-photon quantum scissor for NLA. Entanglement distribution is performed by sending one mode of a TMSV state \cref{Chap4eq:chi5} through the channel between the single repeater node and ends of the channel. The mode of the entangled state that had passed through the lossy channel is then distilled using the scissor.

\subsection{Directions of operation}

\begin{figure}
\centering
\subfloat[]{%
  \includegraphics[scale=1.3]{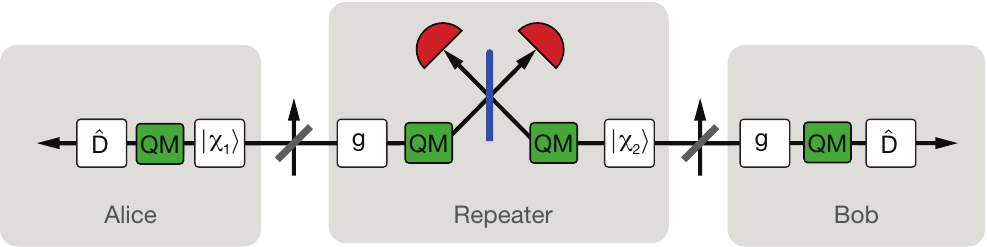} \label{Chap4fig:asymmetric}%
}

\subfloat[]{%
  \includegraphics[scale=1.3]{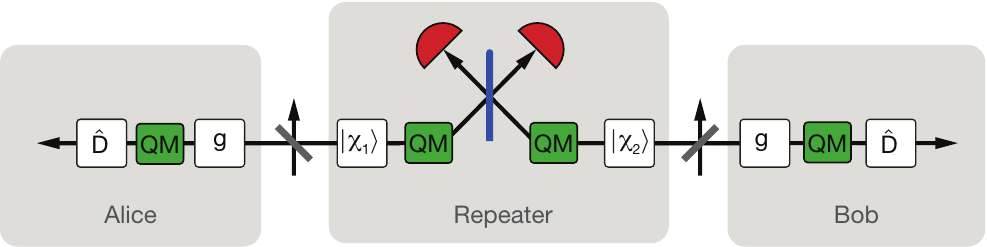}
  \label{Chap4fig:symmetric}
  }

\subfloat[]{%
  \includegraphics[scale=1.3]{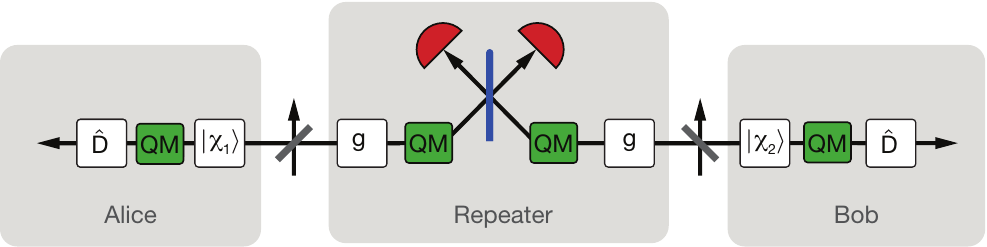}
  \label{Chap4fig:MDI}
  }
  
  \subfloat[]{%
  \includegraphics[scale=1.3]{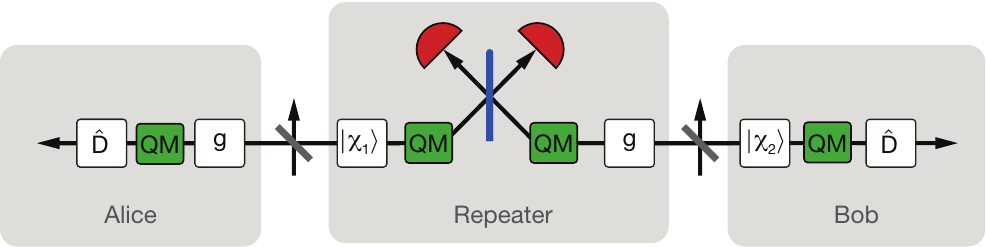} \label{Chap4fig:reverse}%
}
\caption[Different configurations of the CV repeater.]{Different configurations of the CV repeater. The sub-figures are ordered from highest performance to lowest performance. These configurations are \protect\subref{Chap4fig:asymmetric} asymmetric, \protect\subref{Chap4fig:symmetric} symmetric, \protect\subref{Chap4fig:MDI} MDI-like, and \protect\subref{Chap4fig:reverse} reverse asymmetric. The symbols are as follows. $\ket{\chi_1},\ket{\chi_2}$ are two-mode squeezed vacuum state sources, QM are quantum memories, $g$ is noiseless linear amplification, the measurement in the repeater is dual homodyne, and $\hat{D}$ are displacements conditioned on the dual-homodyne measurement outcome. The channels between nodes are lossy channels.}
\label{Chap4fig:configurations}
\end{figure}

\begin{figure}
\centering
\includegraphics[width=0.55\linewidth]{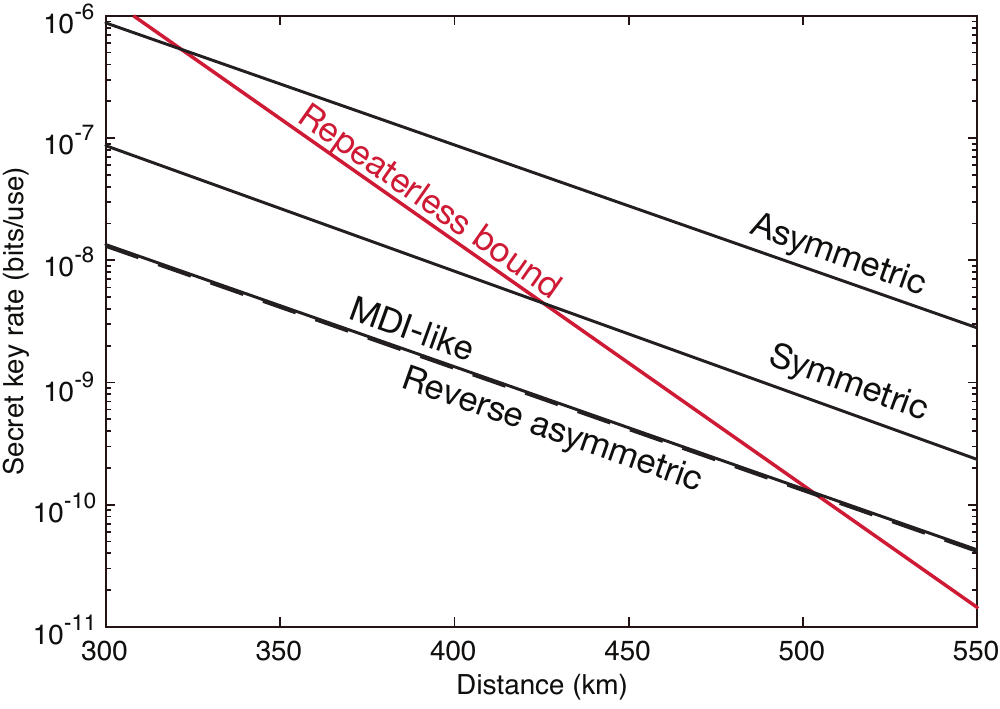}
\caption[Secret key rate for different configurations of the CV repeater.]{Secret key rate for different configurations of the CV repeater shown in~\cref{Chap4fig:configurations}. The key rate assumes a CV-QKD protocol where Alice and Bob perform homodyne detection and reverse reconciliation. Asymmetric is the optimal configuration with reverse reconciliation. The MDI-like configuration might be useful for measurement-device-independent protocols despite its sub-optimal performance.}
\label{Chap4fig:directionality}
\end{figure}

For a point-to-point protocol, reverse reconciliation works much better than direct reconciliation. This is because the eavesdropper gets less information about the key when Bob is the reference. So an important question for our CV repeater protocol is what configuration works best. The various configurations are shown in~\cref{Chap4fig:configurations}. The asymmetric configuration works best, where all quantum states are propagated in a direction from Alice towards Bob. The performance of each configuration is summarised in~\cref{Chap4fig:directionality}. Note that all protocols beat the repeaterless bound, which means they are all effective repeaters; however, there is a preferred direction, from Alice towards Bob. Having a preferred direction is a limitation since we would like to send quantum information in any direction we like in a network, and it is an important problem to solve especially for large CV quantum networks. We solve this problem for all our repeaters introduced in future chapters.

\subsection{Asymmetric repeater}

We focus on the asymmetric repeater configuration since it is the optimal configuration. While the quantum scissor operation is non-deterministic, both entangled states are independent at this point in the protocol. Therefore, both quantum scissors can operate independently and simultaneously. When a quantum scissor heralds successful operation, we assume ideal, high-quality quantum memories are available to store the distilled entanglement until the other quantum scissor is successful. After both entangled states have been distilled, they are then swapped by mixing the two modes at the repeater node and conducting a dual-homodyne detection. For the results to be accepted, the measurement outcome of the dual-HD results must fall within a certain radius around zero. If this is successful, the results of this detection are then sent to Alice and Bob and a displacement is performed on each mode based on the results of the detection, which completes the entanglement-swapping operation.   
\begin{figure}
\centering
\includegraphics[width=0.5\linewidth]{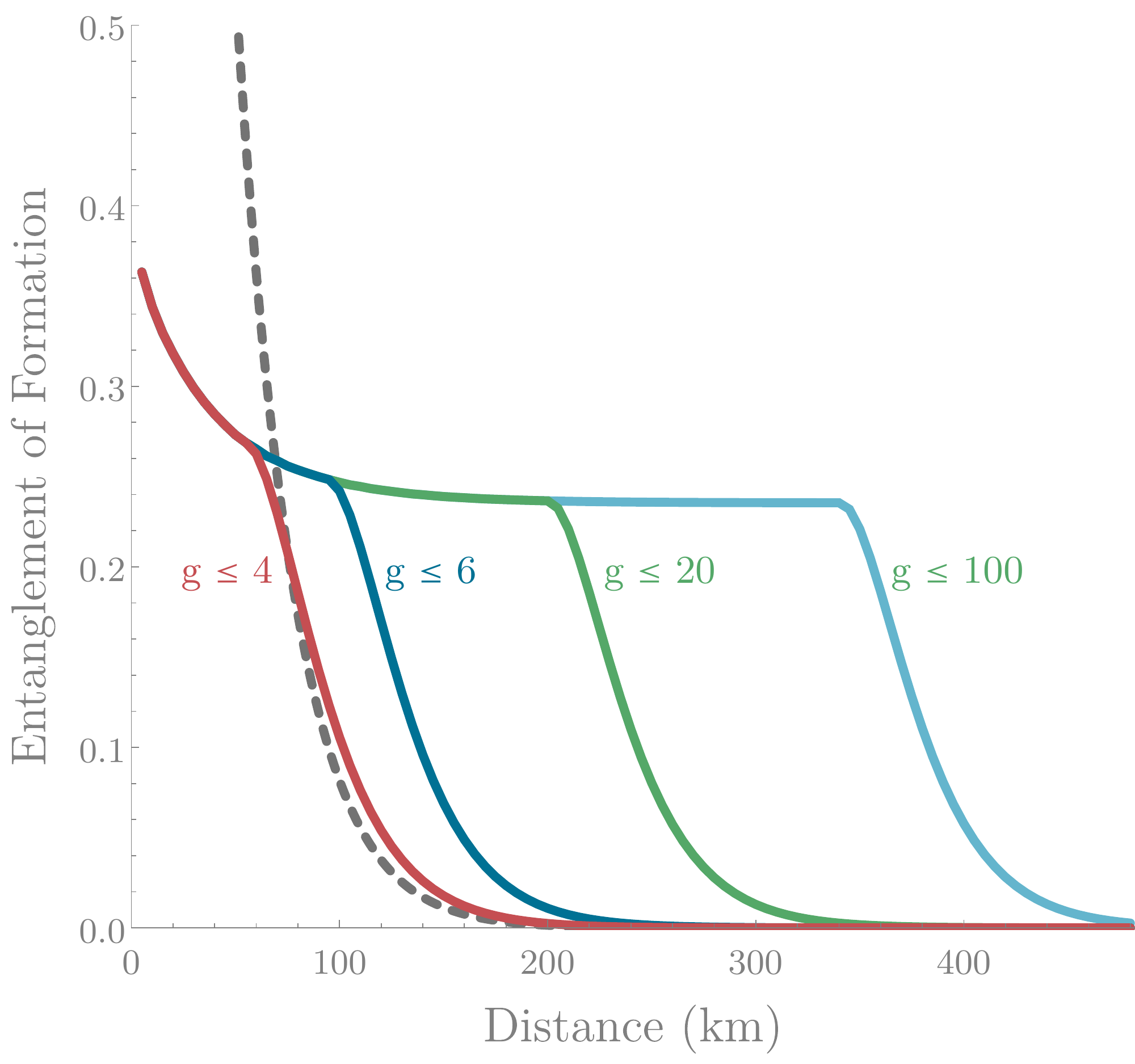}
\caption[Entanglement of formation of the CV repeater]{Entanglement of formation of the CV repeater. The solid, coloured lines show the entanglement of formation between ends of the channel using the single-node CV repeater with TMSV state sources of squeezing $\chi=0.3$. Each line shows the optimal EOF attainable when the NLA gain has been restricted to some maximum value (each line has been labelled with this maximum gain). The dashed, dark grey line is the entanglement of formation of an unphysical, infinitely-squeezed TMSV state (\cref{Chap4eq:chi5} with $\chi\to1$) transmitted through an optical fibre channel of the same distance. }
\label{Chap4fig:eof}
\end{figure}

Initially, we consider the maximum entanglement that can be distributed via our repeater by evaluating the entanglement of formation (EOF) \cite{PhysRevA.69.052320,akbarikourbolagh2015entanglement,marian2008entanglement} between the end stations when post-selection of the HD results lying very close to zero are accepted. This result is given in~\cref{Chap4fig:eof} where we show the entanglement of formation between end stations of our CV repeater using TMSV sources of fixed squeezing $\chi=0.3$. We compare this to the EOF of an unphysical, infinitely-squeezed TMSV state distributed through the same loss. Each solid line in~\cref{Chap4fig:eof} shows the highest EOF achievable for various maximum NLA gains. At shorter distances, EOF maximises for lower gains. It can be seen on~\cref{Chap4fig:eof} that there is a turning point on each solid line. This turning point marks the distance beyond which maximum EOF is achieved by the maximum allowable NLA gain. For maximum gains of 5 or higher, the EOF surpasses the direct transmission EOF at a distance of 70 km.  While the red line for $g\leq4$ does produce an improvement over the direct transmission EOF, this occurs for a total channel distance of 75 km. Additionally, for NLA gains of $g\leq 3$ we do not observe an improvement beyond the direct transmission EOF. Note that in~\cref{Chap4fig:eof} and throughout this thesis we have considered optical fibre with loss of 0.2 dB/km.

\begin{figure}
\centering
\includegraphics[width=0.5\linewidth]{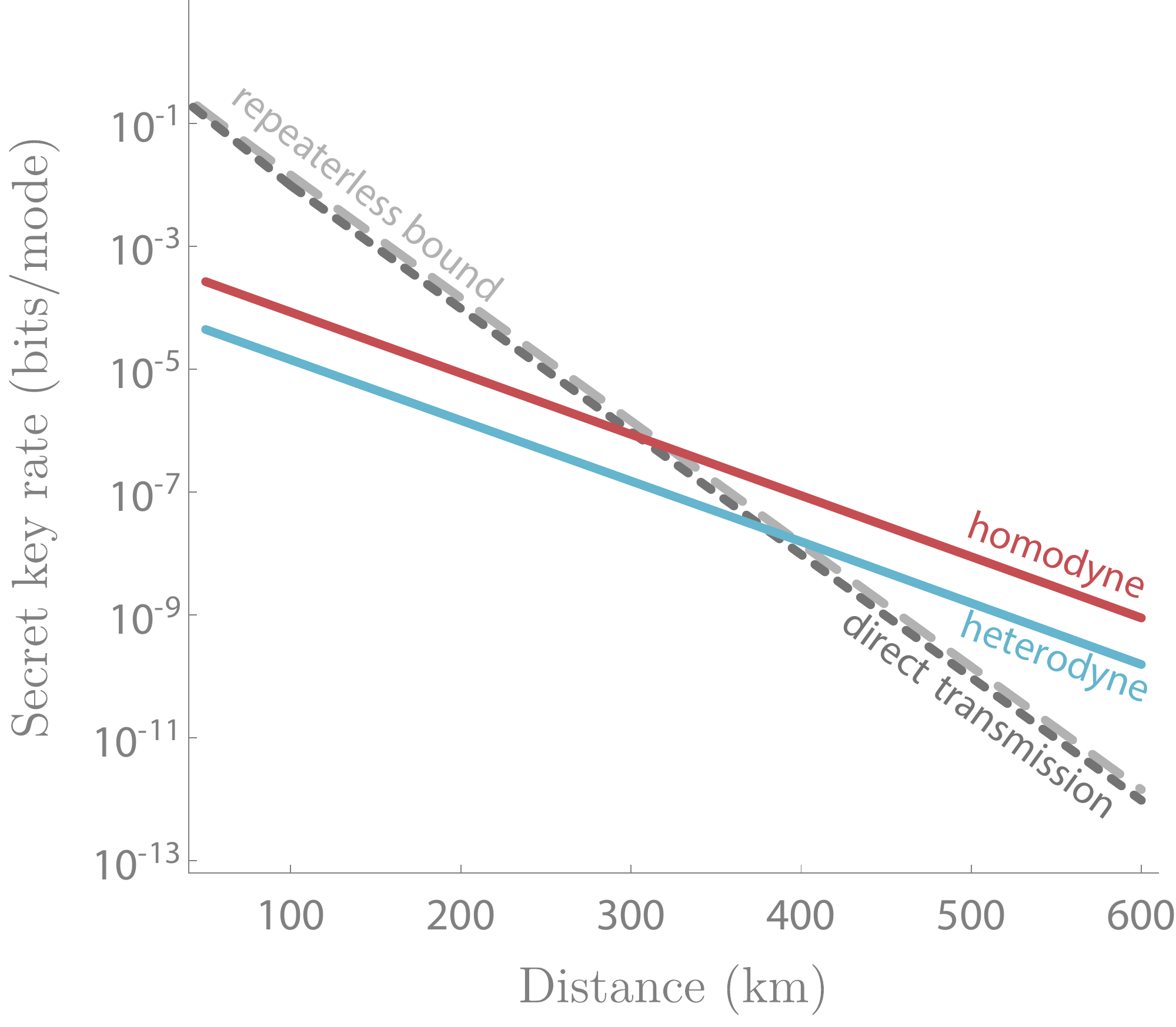}
\caption[Secret key rate of the single-node repeater]{Secret key rate of the single-node repeater shown in~\cref{Chap4fig:asymrep1} using single-photon quantum scissors. The blue line shows the key rate for a heterodyne-based QKD protocol and post-selection cut-off at $\gamma_{\text{max}}=0.4$.  The red line shows the same result except with a homodyne-based QKD protocol and a larger post-selection cut-off of $\gamma_{\text{max}}=0.5$. The dashed, dark grey line is the key rate (using a homodyne-based CV-QKD protocol) for direct transmission through the channel using optimised, finite squeezing ($\chi<1$). Reconciliation efficiency for the homodyne, heterodyne and direct transmission lines has been set to $95\%$.  The dashed, light-grey line is a fundamental upper bound on the maximum secret key rate that is achievable using direct transmission without a quantum repeater (repeaterless bound) \cite{Pirandola_2017}. }
\label{Chap4fig:result1}
\end{figure}

As a second figure of merit, which importantly incorporates the probability of success, we consider the secret key rate achievable by the CV quantum repeater. The secret key rate of the scheme shown in~\cref{Chap4fig:asymrep1} can surpass the absolute maximum secret key rate (repeaterless bound) for direct transmission (shown by the dashed, grey line on~\cref{Chap4fig:result1} \cite{Pirandola_2017}). The secret key rate presented in~\cref{Chap4fig:result1} is defined as:
\begin{equation}
\mathrm{Secret\, key\, rate} = K \times R_{\text{rep}}
\end{equation} 
where $K$ is the raw key rate given in~\cref{eq:DW} calculated from the covariance matrix of the output state, and $R_{\text{rep}}$ is the rate of successful operation of the entire repeater which depends on the success probability of the scissor and post-selection. The rate $R_{\text{rep}}$ for successful operation of the CV repeater for $2^n$ links is calculated via
\begin{equation}
R_{\text{rep}} = \frac{1}{Z_n\left(P_{\text{NLA}}\right) }\times   \prod_{i=0}^{n-1} \frac{1}{Z_i\left(P_{\text{PS}i}\right) } ,
\label{Chap4eq:reprate}
\end{equation}
where $P_{\text{PS}i} $ is the probability of successful post-selection in the various entanglement swapping rounds when $2^i$ swaps need to occur successfully for the repeater protocol to proceed. Here, $i=n-1$ corresponds to the first round of entanglement swapping with  $2^{n-1}$ swaps, and $i=0$ corresponds to the final swap. The function $Z_n\left(P\right)$ is the average number of steps to generate successful outcomes in $2^n$ probabilistic operations, each with success probability $P$ \cite{PhysRevA.83.012323}:
\begin{equation}
Z_n \left(P\right) =\sum^{2^n}_{j=1} \binom{2^n}{j} \frac{\left(-1\right)^{j+1}}{1-\left(1-P\right)^j}.
\end{equation}
For the single-node results in~\cref{Chap4fig:result1}, the repeater rate \cref{Chap4eq:reprate} is simply:
\begin{equation}
R_{\text{rep}}  =  \frac{1}{Z_1\left(P_{\text{NLA}}\right) } \times  \frac{1}{Z_0\left(P_{\text{PS0}}\right) }.
\end{equation}

Using the output entangled state of the single-node repeater, where we include the derivation in~\cref{app:derivation_single_node}, we calculate asymptotic key rates using~\cref{eq:DW} for two entanglement-based CV-QKD protocols, one where Alice and Bob both perform heterodyne detection \cite{Weedbrook_2004} and the other where they both perform homodyne detection \cite{PhysRevA.63.052311} to their entangled modes to obtain a raw key. These are shown on~\cref{Chap4fig:result1} by the blue and red lines respectively. The key rate shown is for reverse reconciliation, where Bob is the reference for reconciliation which is favourable in high-loss regimes. Optimisation of the normalised rate has been performed at each point over both gain of the quantum scissors and the strength of the TMSV state sources. Note that the success probability of the quantum scissor decreases as the gain is increased.  Optimal performance is achieved for squeezing of $0.31< \chi_{\mathrm{opt}}<0.36$. 

In~\cref{Chap4fig:result1}, the key rates for the homodyne-based CV-QKD protocol outperform the heterodyne protocol at all distances. This is because a positive key rate can be achieved using the homodyne-based CV-QKD protocol for larger post-selection cut-off regions. Larger cut-off regions correspond to a bigger post-selection success probability and thus increase the overall key rate. For the results in~\cref{Chap4fig:result1}, the heterodyne protocol uses a post-selection cut-off of $\gamma_{\text{max}}=0.4$ which was found to be roughly optimal. However, the homodyne result uses a larger cut-off of $\gamma_{\text{max}}=0.5$ and produces a higher key rate. 

\cref{Chap4fig:result1} shows that the repeaterless bound is beaten for a total channel distance of 322 km. The repeater can surpass the direct transmission key rate at 305 km. This represents a significant improvement upon the single-node operation of the CV  repeater in Ref.~\cite{dias2017} which uses CV teleportation and beats the repeaterless bound for distances above 500 km \cite{dias2020quantum}. While beating the repeaterless bound at 322 km is a significant improvement, we improve it again in a much simpler protocol in~\cref{Chap:simple_repeater}, and saturate the capacity in~\cref{Chap:purification}. We emphasise these distances are total channel distances, meaning the point at which the protocol beats direct transmission, 305 km, corresponds to 152.5 km of optic fibre between Alice and the node (and between Bob and the node).

\subsection{Multi-node repeater with nested swapping}
\begin{figure*}
\centering
\includegraphics[width=1\linewidth]{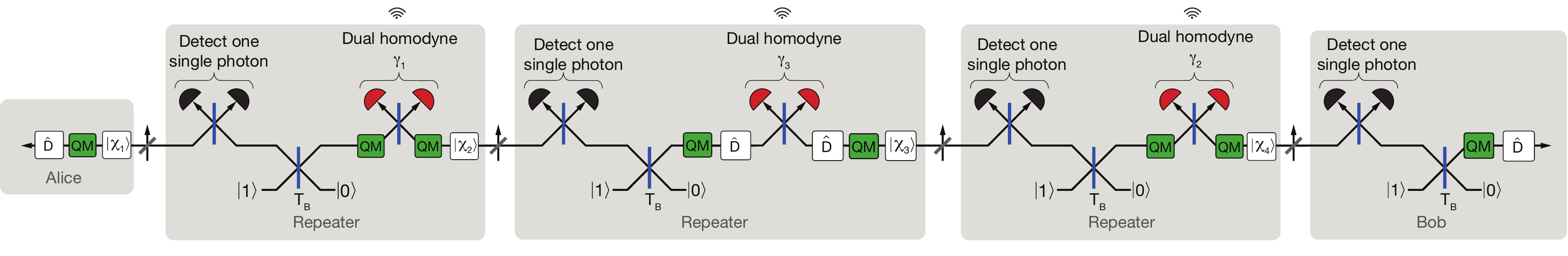}
\caption[Four links of the CV quantum repeater using one level of nested swapping]{Four links of the CV quantum repeater using one level of nested swapping. By comparison with~\cref{Chap4fig:asymrep1}, it can be seen that two independent implementations of the two-link (single node) CV repeater are connected via nesting within another Gaussian entanglement-swapping protocol (\cref{Chap4fig:entswap1}). Quantum memories are required to hold the distilled entangled states until all quantum scissors are successful. Then deterministic, nested entanglement swapping can proceed. }
\label{Chap4fig:nestedAsym}
\end{figure*}

\begin{figure}
\centering
\includegraphics[width=0.5\linewidth]{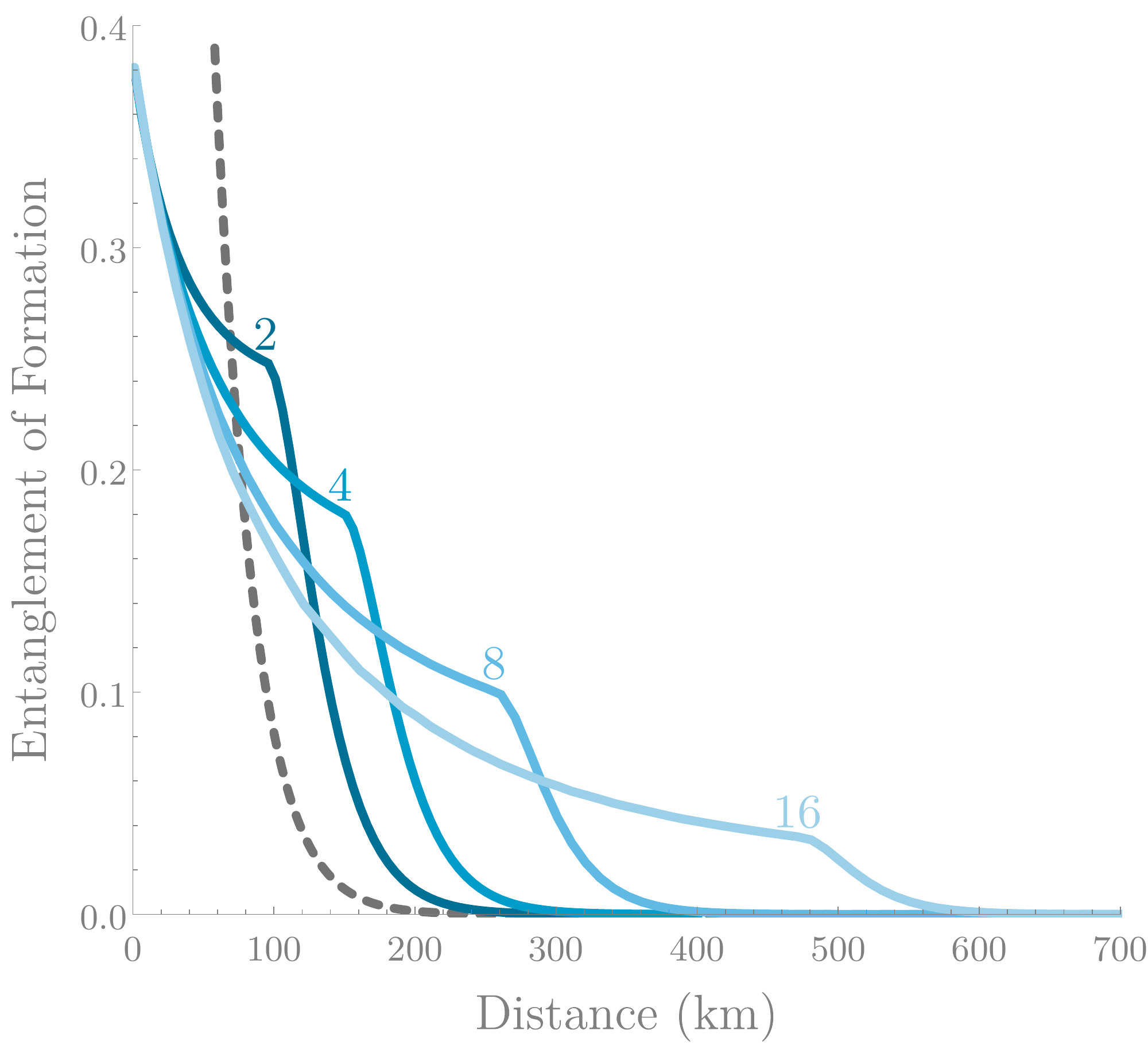}
\caption[Entanglement of formation of the CV quantum repeater for operation with two and more links]{Entanglement of formation of the CV quantum repeater for operation with two and more links. All solid blue lines show the EOF between end stations of the CV repeater for various numbers of repeater links (labelled). Like~\cref{Chap4fig:eof}, these results are achieved when post-selection of homodyne results lying close to 0 are accepted. The CV repeater uses an optimised NLA gain limited to $g\leq 6$ and TMSV sources of squeezing $\chi=0.3$. The dashed, dark grey line is the EOF of an infinitely-squeezed TMSV state distributed through an optical fibre channel of the same total distance.}
\label{Chap4fig:eofAll}
\end{figure}

To use this CV quantum repeater over long distances, more nodes along the channel are required as well as more entanglement-swapping operations to connect the entangled links.  To illustrate how this would proceed, see~\cref{Chap4fig:nestedAsym} with four links of the repeater connected via three repeater nodes. The protocol in~\cref{Chap4fig:nestedAsym} is just two copies of the asymmetric entanglement-swapping protocol in~\cref{Chap4fig:asymrep1} connected via another Gaussian entanglement swapping with post-selection. 

For even longer distances and more repeater nodes, nesting proceeds in this way, where the output of two identical and independent copies of the protocol in~\cref{Chap4fig:nestedAsym} would be connected within another entanglement-swapping operation. It is important to note that our repeater does not use nested entanglement distillation, meaning distillation occurs after entanglement distribution and not at any time after. Structuring the repeater in this way has an extremely favourable effect on the repeater rates, as it lowers the number of probabilistic operations occurring within the protocol.

Again, we initially study the entanglement that may be distributed in this way. Like~\cref{Chap4fig:eof}, the results in~\cref{Chap4fig:eofAll} show the entanglement that may be distributed between end stations of the CV repeater when  results lying very close to 0 are accepted. While we showed the effect of increasing gain on EOF in the results in~\cref{Chap4fig:eof}, for a fair multi-node comparison we restrict all NLA gains to the same maximum value in~\cref{Chap4fig:eofAll}; as an example, we use $g\leq 6$. As expected, increasing the maximum NLA gain results in larger distances over which entanglement may be distributed. However, even with NLA gain restricted to $g\leq 6$,~\cref{Chap4fig:eofAll} shows how our CV repeater may be used to distribute entanglement hundreds of kilometres beyond what is achievable using direct transmission with an unphysical, infinitely-squeezed source.

\begin{figure}
\centering
\includegraphics[width=0.5\linewidth]{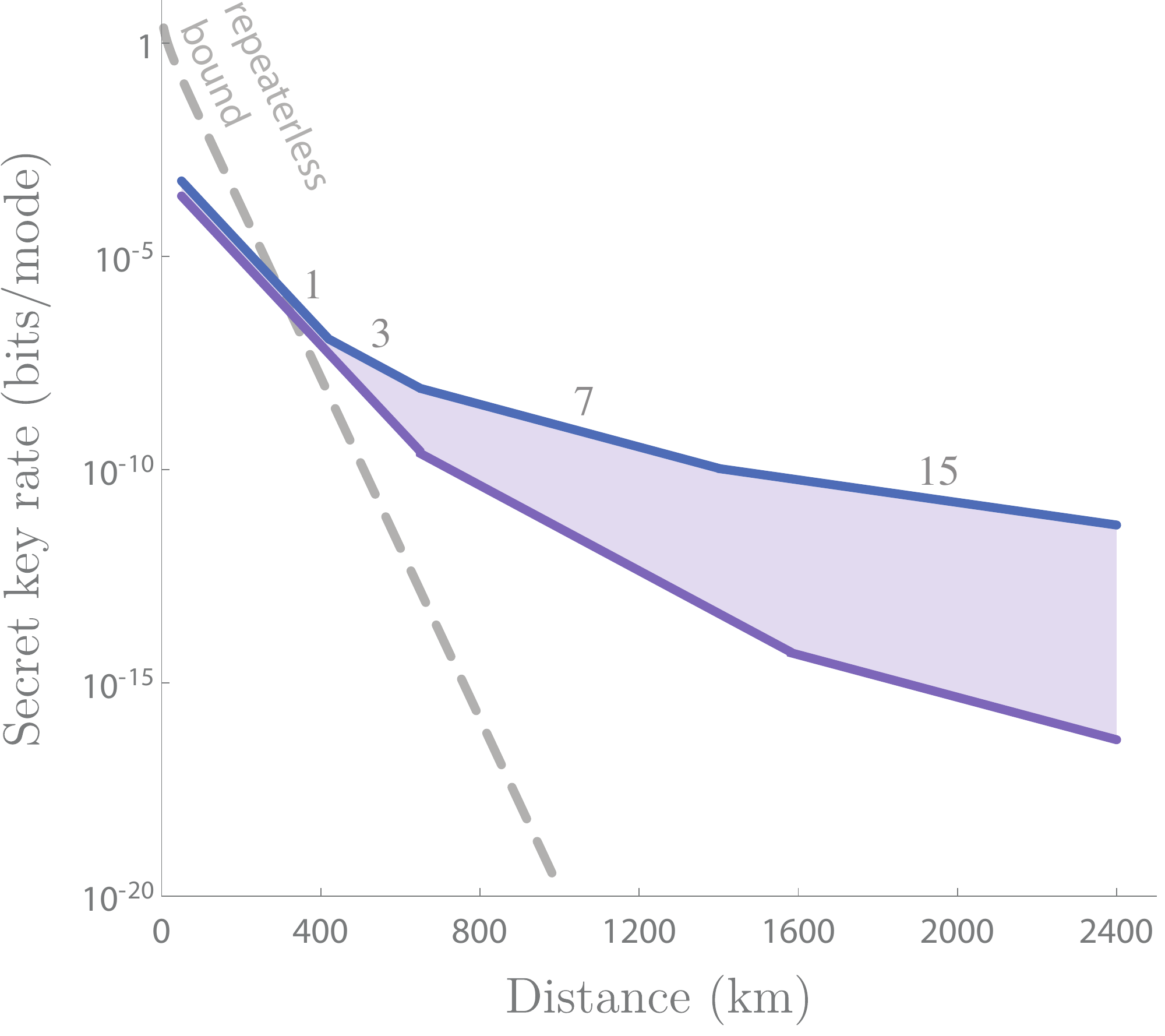}
\caption[Lower and upper bounds on key rates of the CV quantum repeater]{Lower and upper bounds on key rates of the CV quantum repeater. The solid, blue lines represent different numbers of repeater nodes (repeater links) used along the channel, each line is labelled with the number of repeaters. The post-selection cut-off has been set to $\gamma_{\text{max}}=0.5$. The solid purple line corresponds to the lower bound calculated via the method in Appendix \ref{app:lower}. The CV repeater rates shown are for a homodyne-based CV-QKD protocol and assume a reconciliation efficiency of $95\%$. The dashed, light-grey line is the repeaterless bound \cite{Pirandola_2017}.}
\label{Chap4fig:krL}
\end{figure}

We then consider how our CV repeater performance scales with distance for CV QKD. In~\cref{Chap4fig:krL}, we give an upper and lower bound on the secret key rate of our CV quantum repeater and show how it changes with more repeater nodes. Determining the actual output state of the multi-node CV repeater is intractable because it involves integrating over all dual homodyne outcomes $\gamma$. However, an upper bound on the raw secret key rate $K$ can be calculated from the ideal ($\gamma=0$) case, multiplied by the realistic rate of successful operation of the entire repeater $R_{\text{rep}}$, determined numerically. In the case of two links, where the upper bound and the exact numerical result can be compared, we find the two results are close. Given the favourable performance of our repeater with the homodyne-based CV-QKD protocol in~\cref{Chap4fig:result1}, we present results in~\cref{Chap4fig:krL} focusing only on performance with the homodyne-based protocol. The repeater rate $R_{\text{rep}}$ was obtained via \cref{Chap4eq:reprate} with post-selection probabilities calculated numerically and the post-selection cutoff is fixed at all instances to $\gamma_{\text{max}}=0.5$. For smaller post-selection cutoffs, the output state yields a higher raw key rate due to higher correlations; however, it comes at the expense of a lower probability of successful post-selection. We also provide a lower bound on the secret key rate of the CV quantum repeater.

We now detail how to perform the relevant calculations for multiple links to go beyond the simplest case of two links. We proceed by using two copies of the heralded two-mode entangled state which have been distributed along four initial segments of the channel:
\begin{equation}
\hat{\rho}_{ABMN} \left(\gamma_1, \gamma_2\right) = \hat{\rho}_{AB} \left(\gamma_1\right) \otimes  \hat{\rho}_{MN} \left(\gamma_2\right),
\label{Chap4eq:abmn}
\end{equation}
where both $\hat{\rho}_{AB} \left(\gamma_1\right) $ and $\hat{\rho}_{MN} \left(\gamma_2\right)$ correspond to the output state is conditioned on measurement outcomes $\gamma_1$ and $\gamma_2$ from dual HDs at nodes 1 and 3 respectively (see~\cref{Chap4fig:nestedAsym}). The two copies, $\hat{\rho}_{AB} \left(\gamma_1\right)$ and $  \hat{\rho}_{MN}\left(\gamma_2\right) $, are explicitly given in~\cref{app:CV_repeater} by~\cref{Chap4eq:rhoOut}.  Modes $B$ and $M$ are mixed at the central node and a dual HD is conducted on both modes, giving the total conditional output state:
\begin{align}
\hat{\rho}_{AN}\left(\gamma_1, \gamma_2, \gamma_3 \right) 
&= \mathrm{Tr}_{BM} \left[\hat{\rho}_{ABMN} \left(\gamma_1, \gamma_2\right) \otimes \ket{\gamma_3}_{BM}\tensor[_{BM}]{\bra{\gamma_3}}{}  \right].
\label{Chap4eq:an}
\end{align}
Finally, the output modes $A$ and $N$ are displaced by the measurement outcome $\gamma_3$ scaled by classical gains $\lambda_a$ on mode $A$ and $\lambda_n$ on mode $N$: 
\begin{align}
\hat{\rho}_{out} \left(\gamma_1, \gamma_2, \gamma_3 \right)  
 &=\hat{D}_N \left( \lambda_n \gamma_3 \right) \hat{D}_A \left( \lambda_a \gamma_3 \right) \hat{\rho}_{AN}\left(\gamma_1, \gamma_2, \gamma_3 \right) \hat{D}_N^\dagger \left( \lambda_n \gamma_3 \right) \hat{D}_A^\dagger \left( \lambda_a \gamma_3 \right).
\end{align}
We have outlined here the process for calculating the output state of four links of the CV quantum repeater, the output state of eight and higher links proceeds in the same way. 
\subsection{Lower bound\label{app:lower}}
Evaluating the performance of our CV repeater via the method outlined in the previous section is not tractable for the multi-node repeater. This is because the results obtained need to be integrated over each dual HD measurement outcome. While for two links, results can be obtained via numerical integration (and are given in~\cref{Chap4fig:result1}), for four and higher links we will model the performance of our CV repeater by averaging the output density matrix after each entanglement swapping step, that is, instead of~\cref{Chap4eq:abmn}, the density matrices are first averaged over the accepted post-selection region: 
\begin{equation}
\hat{\rho}_{AB}\to \int_0^{2\pi} \int^{\gamma_{\text{max}}}_{0} \hat{\rho}_{AB} \left(\gamma\right) |\gamma|\,\mathrm{d} \phi_{\gamma} \text{d} |\gamma|.
\label{Chap4eq:avrho}
\end{equation}
As previously described, in the four-link repeater scheme, two copies of the  state are used:
\begin{equation}
\hat{\rho}_{ABMN} = \hat{\rho}_{AB} \otimes  \hat{\rho}_{MN} .
\end{equation}
The two averaged output states are then combined and swapped:
\begin{align}
\begin{split}
\hat{\rho}_{AN} \left(\gamma_3 \right) 
&= \mathrm{Tr}_{BM} \left[\hat{\rho}_{ABMN} \otimes \ket{\gamma_3}_{BM}\tensor[_{BM}]{\bra{\gamma_3}}{}  \right].
\end{split}
\end{align}
This is followed by a displacement on modes $A$ and $N$:
\begin{align}
\begin{split}
\hat{\rho}_{out}\left(\gamma_3 \right)  
& \hspace{15pt}= \hat{D}_N \left( \lambda_n \gamma_3 \right) \hat{D}_A \left( \lambda_a \gamma_3 \right) \hat{\rho}_{AN}\left(\gamma_3 \right) \hat{D}_N^\dagger \left( \lambda_n \gamma_3 \right) \hat{D}_A^\dagger \left( \lambda_a \gamma_3 \right).
\label{Chap4eq:avgout}
\end{split}
\end{align}
 The success probability of the final entanglement swap at the central node is given by:
\begin{equation}
P_{PS} = \frac{\int_0^{2\pi} \int^{\gamma_{\text{max}}}_{0} \mathrm{Tr}\hat{\rho}_{AN} \left(\gamma_3 \right)|\gamma_3| \, \mathrm{d} \phi_{\gamma_3} \text{d} |\gamma_3| }{\int_0^{2\pi} \int^{\infty}_{0}\mathrm{Tr} \hat{\rho}_{AN}\left(\gamma_3 \right)  |\gamma_3|\,\mathrm{d} \phi_{\gamma_3} \text{d} |\gamma_3|}.
\label{Chap4eq:Pnest}
\end{equation}
Calculation of the covariance matrix proceeds by using the output state~\cref{Chap4eq:avgout} and numerically integrating to average over the accepted post-selection region. 

By averaging the density matrices before input into subsequent entanglement swapping, the calculations become tractable. However, as $\gamma$ is a classical parameter, this averaging will unavoidably lead to an overestimation in the noise present in the output state. Therefore, we present the results gained from this method as a lower bound to the key rates achievable by our CV quantum repeater. While this method may be used to estimate the probability of success of the nested entanglement-swapping operations~\cref{Chap4eq:Pnest} and thus can be used to estimate $R_{\mathrm{rep}}$, the raw key rate $K$ calculated from the covariance matrix of the output state~\cref{Chap4eq:avgout} will suffer from the overestimation of noise. 

For four links of the CV repeater, the repeater rate~\cref{Chap4eq:reprate} is given by:
\begin{equation}
R_{\text{rep}}  =  \frac{1}{Z_2\left(P_{\text{NLA}}\right) } \times  \frac{1}{Z_1\left(P_{\text{PS}1}\right)} \times  \frac{1}{Z_0\left(P_{\text{PS}0}\right) }
\end{equation}
where $ P_{\text{PS1}}$ is the probability of successful post-selection in the two base-level entanglement swaps, and $P_{\text{PS}0}$ is the probability of success of post-selection in the single higher-level entanglement swap.

\subsection{Upper bound\label{app:upper}}
An upper bound on the raw secret key rate $K$ can be determined from the ideal output state of the CV repeater protocol. That is, the output state achieved conditioned on the measurement outcome of $\gamma=0$, which results in no displacement. In this ideal case, the covariance matrices for the output states of the two, four, eight and sixteen link schemes are analytically solvable. We use the raw key rate $K$ calculated from the ideal ($\gamma=0$) case, multiplied by the realistic rate of successful operation of the entire repeater $R_{\text{rep}}$. This rate depends on the success probability of the NLA given in~\cref{Chap4eq:Psucc}, and the probabilities of success of post-selection calculated via the method explained in the previous section~\cref{Chap4eq:pps} and~\cref{Chap4eq:Pnest}.

We can compare the upper bound to the secret key rate of the post-selected output state. The upper bound and numerically integrated key rates are quite close for fewer numbers of links. This is because the covariance matrices of the ideal output state and the post-selected output state for $\gamma_{\text{max}}=0.5$ are close.

Finally, we can use our upper bound to give the region of estimated performance of our CV quantum repeater and this is shown in~\cref{Chap4fig:krL}. As previously noted, the upper bound uses the fixed post-selection cut-off of $\gamma_{\text{max}}=0.5$ at all swapping levels. However, the post-selection cut-off of the lower bound varies at each level. This is because calculating the lower bound requires small post-selection cut-offs for initial swaps (i.e. $\gamma$ close to 0) since we swap average density matrices~(\cref{Chap4eq:avrho}). Rough optimisation of the overall key rates including raw key rate and repeater rate yields the following post-selection cut-offs.  For the results in~\cref{Chap4fig:krL}, the two-link lower bound uses a cut-off of $\gamma_{\text{max}}=0.5$. The four-link lower bound uses a post-selection cut-off of $\gamma_{\text{max}}=0.2$ at the base level and $\gamma_{\text{max}}=0.45$ at the upper-level entanglement swap. Lastly, the eight-link lower bound uses cut-offs  $\gamma_{\text{max}}=0.06,\,0.15,\,0.4$ at the base-level, mid-level, and highest-level entanglement swaps respectively. Note that the region between upper and lower bound increases for longer distances due to the compounding effect of noise from averaging after multiple entanglement swapping rounds as well as the reduction in lower bound repeater rate due to the smaller post-selection cut-offs.

\section{Chapter summary \label{Chap4sec:conc}}

In this chapter, we have presented here a novel scheme for a CV repeater. We emphasise our approach here is different from that of Ref.~\cite{dias2017} as we focus on the distribution of CV entanglement, rather than preparing an improved channel. We have shown here that even with reasonably small NLA gains $g\leq6$, our repeater can distribute entanglement hundreds of kilometres beyond what is achievable with an unphysical, infinitely-squeezed TMSV state via direct transmission.  Additionally, we have shown that when these distributed entangled states are used for CV QKD, we can improve upon the rates achieved from a previous CV repeater in the literature \cite{dias2017}. In our view, this improvement is attributed to the use of the optimal Gaussian entanglement-swapping protocol described in Ref.~\cite{PhysRevA.83.012319} in conjunction with post-selection. Despite the entanglement swapping being non-deterministic due to the use of post-selection, we have found here that we can indeed achieve an improvement.

While our CV QKD analysis incorporates non-ideal reconciliation efficiency, it is idealised in all other senses. A remaining question to be answered would be how the performance of our CV repeater is affected by experimental inefficiencies including inefficient single-photon sources in the scissors,  inefficient homodyne detection and imperfect quantum memories. Specifically with inefficient single-photon sources, prior work has shown that this inefficiency causes a gain saturation effect thus limiting the actual achievable gain of the NLA \cite{Kosis13,xiang2010heralded} with maximum reported gains of $g^2=11\pm 1$ \cite{Ulanov15}. For distances larger than 130 km, the gain for optimal operation of our CV repeater is greater than this maximum reported gain. Further improvements in photon production and detection efficiency will be needed to obtain these higher gains; however, we note that single-photon source efficiency is constantly improving. Operation of this repeater may be further optimised by use of a different distillation protocol.

In this chapter, we presented a CV quantum repeater using quantum scissors for entanglement distillation and dual-homodyne detection for entanglement swapping. In the next chapter, we improve our CV repeater scheme by noting that quantum scissors are tolerant to loss and act as loss-tolerant quantum relays themselves, and, motivated by the fact that states are truncated by the scissors, the dual-homodyne measurement is removed, replacing all entanglement resources outside of Alice's station with single-photon entangled states and replacing the CV Bell-state measurement for entanglement swapping with a DV Bell-state measurement. This considerably simplifies the analysis since we do not have to evaluate integrals over continuous homodyne-measurement outcomes.

%% file: chapter_simple_repeater/chapter_simple_repeater.tex
\chapter{Quantum repeaters without quantum memories}
\label{Chap:simple_repeater}	
\pagestyle{headings}

\noindent
The results of this chapter appear in the following.\\

\noindent
1.~\cite{winnel2021overcoming} \textbf{M. S. Winnel}, J. J. Guanzon, N. Hosseinidehaj, and T. C. Ralph, ``Overcoming the repeaterless bound in continuous-variable quantum communication without quantum memories,'' \href{https://arxiv.org/abs/2105.03586}{arXiv:2105.03586 (2021)}.\\

\noindent
In this chapter, we introduce a CV repeater protocol that overcomes the repeaterless bound and scales like the single-repeater bound using just one quantum scissor (see~\cref{Chap:scissors}), combining the entanglement distillation and entanglement-swapping elements of other repeater proposals (such as the one introduced in~\cref{Chap:CV_QR}) into a single step, thus, removing the need for quantum memories. Implementing a standard CV-QKD protocol using our repeater, we predict key rates which surpass the repeaterless bound. Our protocol works well for non-ideal single-photon sources and non-ideal single-photon detectors and can tolerate some level of excess noise, making our protocol implementable with existing technology. We show that our scheme can be extended to longer repeater chains using quantum memories, using less physical resources than previous schemes, basically halving the required physical resources and operations as the scheme in~\cref{Chap:CV_QR}. Furthermore, for applications beyond QKD, our scheme generalises to higher order using higher-order quantum scissors (introduced in~\cref{Chap:scissors} and Refs.~\cite{PhysRevA.102.063715,PhysRevLett.128.160501}) and distils more entanglement at the cost of a reduced probability of success.


\section{Introduction}

We saw in earlier chapters that the repeaterless bound~\cite{Pirandola_2017} sets the fundamental rate-distance limit and cannot be surpassed without a quantum repeater. ``Twin-field'' (TF) QKD~\cite{Lucamarini_2018} can overcome the repeaterless bound and scales proportionally to the single-repeater bound without complex repeater components such as quantum memories. TF QKD is based on DV systems but it deviates significantly from standard DV-QKD protocols, and other applications for it have not been identified. In this chapter, we are concerned with CV systems~\cite{Weedbrook_2012} where the quantum information is encoded in an infinite-dimensional Hilbert space which is advantageous for QKD because Alice and Bob can use coherent states and efficient homodyne detection~\cite{Weedbrook_2004}. No simple CV protocol has been proposed that can beat the repeaterless bound. It is important to find one because the ideal performance of CV QKD is better than the ideal performance of DV QKD since the Hilbert space is larger~\cite{Pirandola_2020}, and CV approaches are required to saturate the single-repeater bound (we achieve this feat in~\cref{Chap:purification}).

In~\cref{Chap2CVrepeatercomponents,Chap:CV_QR}, we saw that first-generation quantum repeaters are based on three essential ingredients: entanglement distribution, entanglement distillation, and entanglement swapping. First, entangled states are distributed between neighbouring nodes. Second, entanglement is distilled non-deterministically that overcomes loss and noise in the links. Quantum memories are required to hold onto the quantum states while neighbouring links succeed in distilling their entanglement. Third, joint measurements are performed on some of the modes, heralding entanglement between more-distant stations. Additional rounds of distillation and swapping can entangle stations separated over greater distances. This approach requires quantum memories to overcome the repeaterless bound.

In this chapter, we introduce a CV repeater protocol which surpasses the repeaterless bound with a simple architecture and without quantum memories. This is possible by combining the entanglement distillation and entanglement swapping elements of previous CV proposals into a single step using just one linear-optical device called a quantum scissor, described in~\cref{Chap:scissors}~\cite{pegg1998optical,ralph2009nondeterministic,xiang2010heralded,barbieri2011nondeterministic,PhysRevA.102.063715}. Scissors non-deterministically perform noiseless linear amplification (NLA) whilst truncating all higher-order Fock numbers (hence the name scissor). Scissors have proved useful in many quantum-communication schemes, for instance, to enhance point-to-point QKD~\cite{Ghalaii_2020_scissors} and for quantum repeaters~\cite{PhysRevResearch.2.013310,PhysRevA.102.052425}. Previously, scissors have been used for the entanglement-distillation step only; however, our protocol highlights that quantum scissors can, in fact, operate as loss-tolerant quantum relays by themselves. That is, they simultaneously perform entanglement distillation and entanglement swapping and the repeaterless bound can be beaten without quantum memories. Furthermore, our protocol can be extended to higher order (since scissors can be extended to higher order) and distil large amounts of entanglement at the cost of a reduced probability of success.

We demonstrate how our CV repeater protocol can be used for CV QKD and compute asymptotic secret key rates secure against collective attacks for the simple no-switching protocol based on coherent states and heterodyne detection~\cite{Weedbrook_2004}. We give the eavesdropper (Eve) full control of the single-photon sources and single-photon detectors. Our protocol fixes the ``directional problem'' in CV QKD (as will be discussed later) which plagues, for example, the CV measurement-device-independent (MDI) protocol~\cite{Ma_2014,Zhang_2014,Ottaviani_2015,Pirandola2015}, discussed in~\cref{sec:MDI_intro}. Using quantum memories, we show how our protocol can be extended into longer repeater chains that scale like the corresponding repeater bounds.

The first-order version of our protocol with single-photon detection at the halfway central node has recently been demonstrated for entanglement distribution~\cite{PhysRevLett.125.110506}, and researchers are actively trying to build our simple repeater discussed in this chapter for CV QKD.


\begin{figure*}
\centering
\includegraphics[scale=1.3]{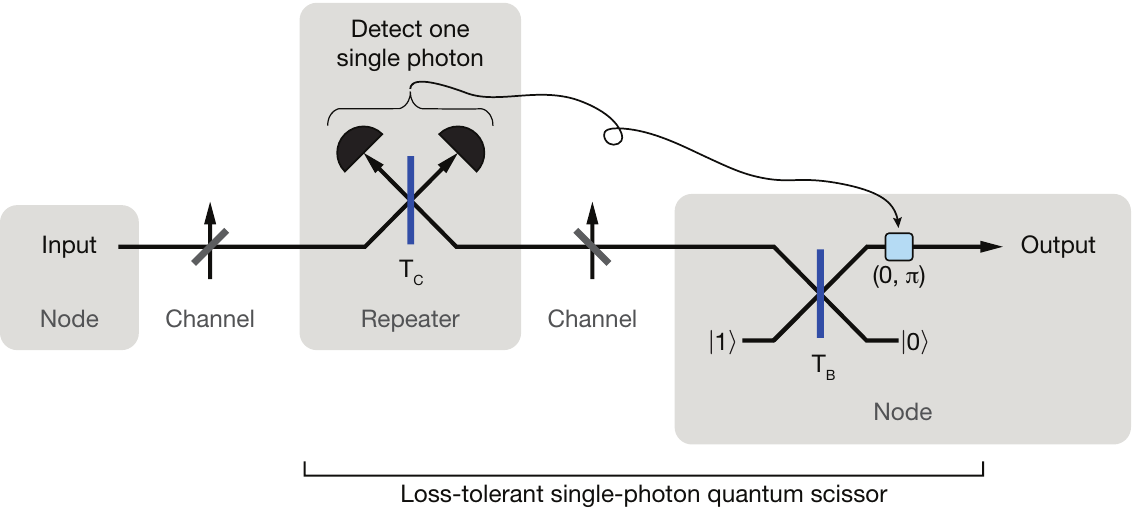}
\caption[Our simple CV repeater without quantum memories]{Our simple CV repeater without quantum memories simultaneously teleports and amplifies quantum information across a lossy channel via an untrusted node. The resource for the teleportation is single-photon entanglement, prepared at the receiver. The repeater node performs a Bell-state measurement, detecting a single photon, without determining which path the photon took. The receiver node and the repeater node together form a single-photon quantum scissor, described in~\cref{Chap:scissors}. The gain of the NLA can be selected by varying the beamsplitter transmissivities, $T_B$ and $T_C$, as well as the position of the repeater between the end nodes. A passive phase shift may be required as shown depending on which detector found a photon.}
\label{SIMfig:memoryless_repeater}
\end{figure*}


\section{Our simple CV repeater without memories}

\begin{figure}
\centering
\includegraphics[scale=1.3]{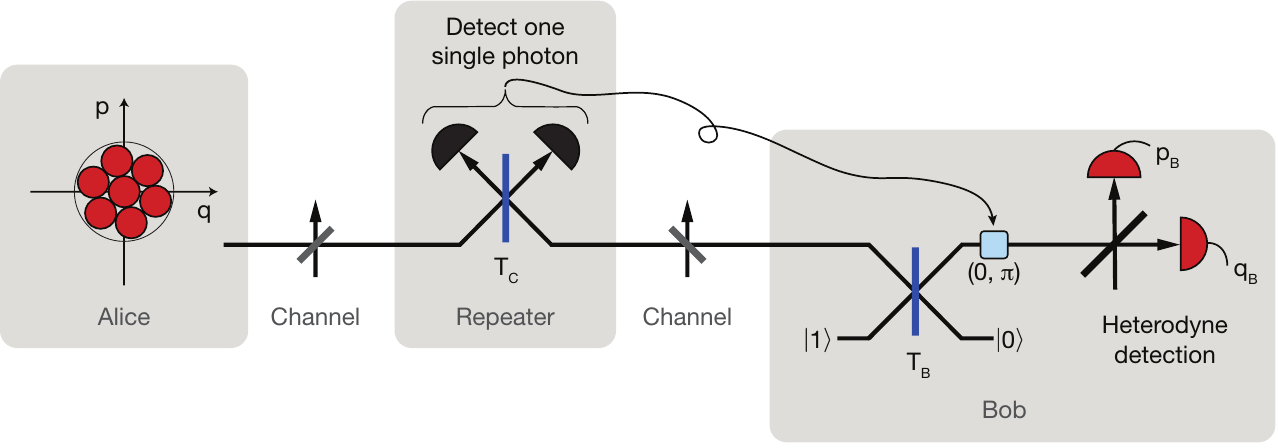}
\caption[No-switching CV-QKD protocol with a memoryless repeater]{No-switching CV-QKD protocol with a memoryless repeater showing the prepare-and-measure version. Alice chooses a coherent state at random from a two-dimensional Gaussian distribution and forwards it to Charlie down a thermal-noise channel. Meanwhile, Bob prepares a single-photon entangled state and sends one mode also towards Charlie. Charlie interferes the modes he receives and performs photon-number-resolving detection. A single click heralds strong correlations between Alice and Bob. Bob measures his remaining mode with heterodyne detection. The probability of success scales like the square root of the transmissivity of the total distance, thus, the repeater is effective at improving the rate-distance scaling. The gain of the NLA is tuned both by Charlie's exact location between Alice and Bob and the transmissivities of the beamsplitters, $T_C$ and $T_B$.}
\label{SIMfig:CV_QKD_repeater}
\end{figure}

\begin{figure}
\centering
\includegraphics[scale=1.3]{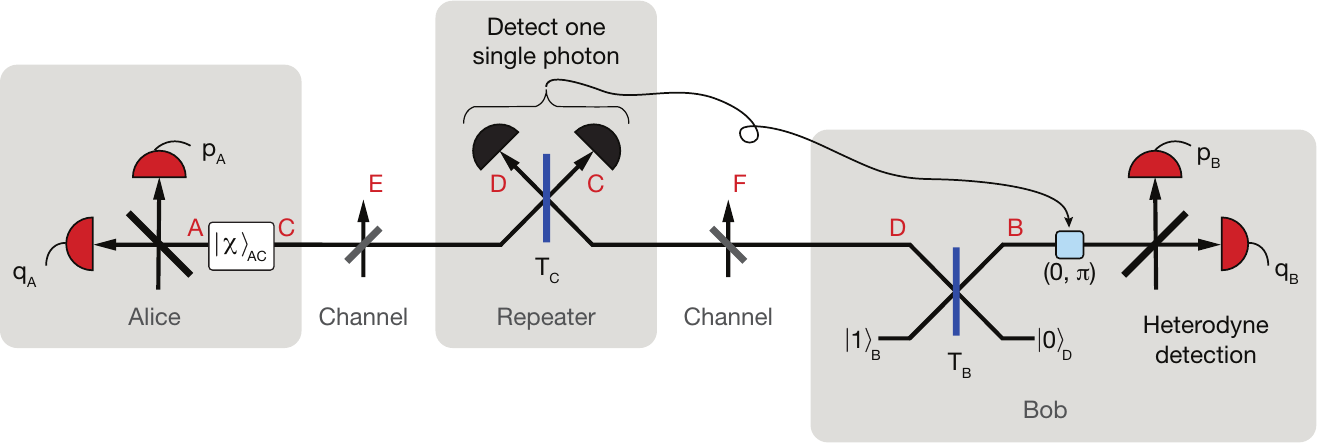}
\caption[Entanglement-based version of the no-switching CV-QKD protocol with a memoryless repeater]{Entanglement-based version of the no-switching CV-QKD protocol with a memoryless repeater. Alice prepares a TMSV state and performs heterodyne detection. The rest of the protocol is the same as the prepare-and-measure version shown in~\cref{SIMfig:CV_QKD_repeater}.}
\label{SIMfig:CV_QKD_repeater_EB_version}
\end{figure}

Our simple CV repeater protocol is shown in~\cref{SIMfig:memoryless_repeater}. The sender node transmits quantum information to the repeater, while the receiver node shares single-photon entanglement with the repeater. Conditioned on a single-photon detection in the repeater node, the quantum information is teleported and amplified to the receiver. The repeater node and the receiver node together consist of a single-photon quantum scissor, with the measurement distributed into the channel. In particular, when the quantum information is Gaussian, the entanglement and purity are especially preserved and the fidelity is high.

We consider our CV repeater protocol for practical CV QKD with a Gaussian distribution, as shown in~\cref{SIMfig:CV_QKD_repeater}. The standard one-way Gaussian CV-QKD protocols are based on a Gaussian modulation of squeezed states or coherent states, and homodyne or heterodyne detection. We focus here on the coherent-state protocol without switching~\cite{Weedbrook_2004} since it is the simplest to implement. 

In the prepare-and-measure (PM) version, shown in~\cref{SIMfig:CV_QKD_repeater}, Alice prepares coherent states selected at random from a Gaussian modulation of variance $V_A$ to send to Bob. The channel of total transmissivity $\eta$ is divided into two shorter links each with transmissivity $\eta_A$ and $\eta_B$, such that $\eta=\eta_A \eta_B$, with an untrusted intermediate station, Charlie, between the trusted parties, Alice and Bob. Bob's prepared entangled state and Charlie's station together form a single-photon quantum scissor~\cite{pegg1998optical,ralph2009nondeterministic,barbieri2011nondeterministic}. When Charlie registers just a single click at one of his detectors, strong correlations are heralded between Alice and Bob resulting in a virtually lossless communication channel. Bob performs heterodyne measurements on his final states. His data are correlated with the states Alice sent and after classical post-processing and privacy amplification, they share a secret random key.

The entanglement-based (EB) version is shown in~\cref{SIMfig:CV_QKD_repeater_EB_version} which is used for security analysis of the PM version or for entanglement distillation. Alice initially prepares a two-mode squeezed vacuum (TMSV) state, $\ket{\chi}_{AC} =  \sqrt{1{-}\chi^2} \sum_{n=0}^\infty \chi^n \ket{n}_A\ket{n}_C$, with two-mode-squeezing parameter $r$, variance $\nu=\cosh{2r}$, $\chi=\tanh{r}$, and mean photon number $\bar{n}=\sinh^2{r}$, and where $\ket{n}$ are Fock-number states. The modulation variance $V_A$ in the PM version is related to $\chi$ in the EB version by $\nu=V_A{+}1=\cosh{(2\tanh^{-1}{\chi})}$. The two versions are equivalent if Alice performs heterodyne detection on her mode.

Quantum scissors~\cite{PhysRevA.102.063715} perform noiseless linear amplification (and de-amplification) on arbitrary input states up to some Fock number $k$. They work by combining the input state with an entanglement-resource state and post-selecting on $k$ detectors registering a single click, teleporting and amplifying the input state onto the outgoing mode.  The transmissivity of Bob's beamsplitter, $T_B$, sets the gain. Let's now consider loss between Bob and Charlie. Remarkably, we note here that quantum scissors perform linear amplification sufficiently well despite the loss if the input states are selected from a classical mixture but now with a modified gain which depends on the loss. We say that the scissor is loss tolerant. This result was mentioned previously concerning more general scissor-like devices called ``tele-amplification''~\cite{neergaard-nielsen2013quantum}, and these devices are loss tolerant if the input states are restricted to a classical mixture of coherent states on a ring in phase space. A classical mixture is precisely what Alice sends towards Bob for Gaussian-modulated CV QKD. Thus, quantum scissors are suitable as repeaters for CV QKD. We call it a repeater since it actually ``repeats'' in the sense that it beats the repeaterless bound.


When placed halfway between Alice and Bob, the success probability of the repeater scales as the square root of the total channel transmissivity so the protocol can surpass the repeaterless bound~\cite{Pirandola_2017} which scales with total channel transmissivity~\cite{takeoka2014}. This is in contrast to CV-MDI QKD~\cite{Pirandola2015}, discussed in~\cref{sec:MDI_intro}, which cannot beat the repeaterless bound (the dual homodyne relay is not effective at ``repeating'' Gaussian states). The secret key rate for an ideal implementation of the coherent-state protocol using our CV repeater is plotted in~\cref{SIMfig:rate} and we beat the bound at 223 km, improving the distance reported for the CV repeater presented in~\cref{Chap:CV_QR} which was 322 km. That is, our improved repeater in this chapter enjoys a higher secret key rate using fewer resources.

We introduced the repeater in terms of teleportation and NLA; however, it can also be viewed as hybrid entanglement swapping where DV entanglement is used to fix the ``directional problem'' in CV QKD. The directional problem is that reverse reconciliation works much better than direct reconciliation because the reference should be the more noisy state (i.e., Bob). This means that the CV-MDI QKD protocol~\cite{Pirandola2015} works extremely asymmetrically. It also means that recent repeater proposals~\cite{PhysRevResearch.2.013310,PhysRevA.102.052425,Ghalaii_2020_scissors} are set up such that they always work better if the states are propagated down lossy channels in a direction from Alice towards Bob and not in the other direction. So, in this chapter (and indeed~\cref{Chap:purification}) we have fixed this problem.

The gain of the NLA is tuned by changing the exact location of the repeater between Alice and Bob (i.e., varying $\eta_A$ and $\eta_B$) as well as in the traditional way~\cite{ralph2009nondeterministic} of tuning the transmissivities of the beamsplitters in the scissor, $T_C$ and $T_B$. We in general let Charlie's beamsplitter be balanced, $T_C=1/2$. Then the gain of the repeater for pure loss is $g = \sqrt{\frac{ T_B}{\eta_B(1-T_B)}}$. This is the same gain relation for loss-tolerant tele-amplification, see Eq. 5 of Ref.~\cite{neergaard-nielsen2013quantum}. Putting this all together, in our CV repeater protocol, Alice's state experiences loss which is amplified by the scissor where the gain $g$ is tuned by the location of Charlie and the transmissivity of Bob's beamsplitter.

For pure loss, the global output state shared between Alice, Bob, and the environment, heralded by a single click at the mode-$D$ detector, is
\begin{multline}
|\psi\rangle_{ABEF} = \sqrt{\frac{1{-}\chi^2}{2}} \sum_{k=0}^\infty  (1{-}\eta_A)^\frac{k}{2} \biggl[    \chi^k \sqrt{\eta_B}  \sqrt{1{-}T_B}  \ket{k}_A\ket{0}_B\ket{k}_E\ket{0}_F \\  + \chi^{k{+}1} \sqrt{k{+}1} \sqrt{\eta_A}  \bigl(\sqrt{T_B}  \ket{k{+}1}_A\ket{1}_B\ket{k}_E\ket{0}_F  + \sqrt{1{-}\eta_B} \sqrt{1{-}T_B} \ket{k{+}1}_A\ket{0}_B\ket{k}_E\ket{1}_F \bigr)  \biggr],\label{SIMeq:output}
\end{multline}
where $k$ is the number of photons lost in Alice's link. For each $k$, the scissor operates ideally and noiselessly for the first two terms; however, for the third term, mode $D$ lost one photon to mode $F$ and this is an error. Thus, there are three types of noise on the global output state: loss on Alice's link, truncation noise from the scissor, and decoherence due to the error term when the scissor loses a photon.

Alice and Bob's final state $\hat{\rho}_{AB}$ is obtained by tracing over the environmental modes, $E$ and $F$. The total success probability is $P=2\braket{\psi}{\psi}$, where the factor of two is because there are two successful click patterns, one of which heralds a passive $\pi$-phase shift on the output state and is easily corrected~\cite{PhysRevA.102.063715}, Charlie simply tells Bob which detector fired.  We present the derivation of~\cref{SIMeq:output} next.


\subsection{Ideal implementation for pure loss}

In this subsection, we describe how to calculate the final output state of the entanglement-based (EB) version of our single-repeater CV-QKD protocol for overcoming the repeaterless bound, shown in~\cref{SIMfig:CV_QKD_repeater_EB_version}.

We first consider the ideal implementation of our protocol shown in~\cref{SIMfig:CV_QKD_repeater_EB_version}, consisting of a perfect single-photon source and perfect single-photon detectors. Initially, Alice and Bob each prepare an entangled resource state. Alice prepares a TMSV state $\ket{\chi}_{AC}$, while Bob prepares a single photon mixed with vacuum on a beamsplitter with transmissivity $T_B$. The beamsplitter transformation is
\begin{align}
    \hat{B}(\theta) &= e^{{\theta}(\hat{a}^\dagger \hat{b} - \hat{b} \hat{a}^\dagger)},
\end{align}
where $\hat{a}$ and $\hat{b}$ are the annihilation operators of the two modes, and the transmissivity is determined by $T_B = \cos^2{\theta}$. Hence, Alice and Bob's initial state is
\begin{align}
    |\psi_0\rangle_{ACDB} &= \ket{\chi}_{AC} ( \sqrt{T_B} \ket{0}_D\ket{1}_B + \sqrt{1{-}T_B} \ket{1}_D\ket{0}_B ).
\end{align}

Modes $C$ and $D$ are individually propagated through different lossy channels with transmissivity $\eta_A$ and $\eta_B$, respectively. The lossy channels are modelled by introducing environmental modes $E$ and $F$, initially in the vacuum state (or a thermal state to model excess noise), and mixed on a beamsplitter. After the pure-loss links, the state becomes
\begin{multline}
|\psi_0\rangle_{ACEDBF} \to  \sqrt{1{-}\chi^2} \sum_{n=0}^\infty \chi^n \sum_{k=0}^n \sqrt{{n \choose k}}
(1{-}\eta_A)^{\frac{k}{2}} \eta_A^{\frac{n-k}{2}}|n\rangle_{A} |n-k\rangle_{C} |k\rangle_E\\
 \biggl[ \sqrt{T_B} \ket{0}_D\ket{1}_B \ket{0}_F +  \sqrt{1{-}T_B} \bigl ( \sqrt{\eta_B} \ket{1}_D\ket{0}_B  \ket{0}_F + \sqrt{1{-}\eta_B}\ket{0}_D\ket{0}_B \ket{1}_F \bigl )\biggr].
\end{multline}

Modes $C$ and $D$ are combined on a beamsplitter, which we assume with no loss of generality a transmissivity $T_C=1/2$, and measured with single-photon detectors. We post-select on the instances where the mode-$D$ detector registers a single photon and the mode-$C$ detector registers vacuum (i.e. ${}_{C}\langle 0| {}_{D}\langle 1|$). The output state is
\begin{multline}
|\psi\rangle_{ABEF} = \sqrt{\frac{1{-}\chi^2}{2}} \sum_{k=0}^\infty  (1{-}\eta_A)^\frac{k}{2} \biggl[    \chi^k \sqrt{\eta_B}  \sqrt{1{-}T_B}  \ket{k}_A\ket{0}_B\ket{k}_E\ket{0}_F \\ + \chi^{k{+}1} \sqrt{k{+}1} \sqrt{\eta_A}  \bigl(\sqrt{T_B}  \ket{k{+}1}_A\ket{1}_B\ket{k}_E\ket{0}_F + \sqrt{1{-}\eta_B} \sqrt{1{-}T_B} \ket{k{+}1}_A\ket{0}_B\ket{k}_E\ket{1}_F \bigr)  \biggr]. \label{eq:EPRdir}
\end{multline}

Alice and Bob's final state $\hat{\rho}_{AB}$ is obtained by tracing out the environmental modes, $E$ and $F$. The probability that this protocol succeeds is $P=2\braket{\psi}{\psi}$ where the factor of $2$ accounts for the other click pattern (i.e. ${}_{C}\langle 1| {}_{D}\langle 0|$), which heralds a passive $\pi$-phase shift on the output state, easily correctable by Bob.

In the symmetric configuration, the repeater is placed exactly in the centre between Alice and Bob. In this case, we expect symmetric loss $\eta_A=\eta_B=\sqrt{\eta}$, such that it's best to set $T_C=T_B=1/2$, which results in a final output state 
\begin{multline}
|\psi (\text{symmetric})\rangle_{ABEF} = \sqrt{\frac{1{-}\chi^2}{2}} {\eta}^\frac{1}{4} \sum_{k=0}^\infty  (1{-}\sqrt{\eta})^\frac{k}{2} 
\biggl[    \chi^k    \ket{k}_A\ket{0}_B\ket{k}_E\ket{0}_F \\ +   \chi^{k{+}1} \sqrt{k{+}1}  \bigl( \ket{k{+}1}_A\ket{1}_B\ket{k}_E\ket{0}_F + \sqrt{1{-}\sqrt{\eta}} \ket{k{+}1}_A\ket{0}_B\ket{k}_E\ket{1}_F \bigr)  \biggr].
\end{multline}
It is important to consider how the success probability $P$ depends on the transmissivity of the channel $\eta$. In the limit of large distances (i.e. small $\eta$ transmissivity) and small $\chi$ input states, the success probability scales as $P \propto \sqrt{\eta}/2$. Thus, the secret key rate also scales proportional to $\sqrt{\eta}$, and hence the protocol can overcome the repeaterless bound~\cite{Pirandola_2017} which scales like $\eta$~\cite{takeoka2014}.

Let us now consider the output state more carefully. The number of lost photons from mode $C$ to mode $E$ is given by $k$. There is also loss in the scissor itself, from mode $D$ to mode $F$. The first term corresponds to the instance where Bob's photon clicked the detector, while the second term is where Alice's photon clicked the detector. In particular, no photons were lost from mode $D$ for the first two terms, which represent the states in which the scissor operated perfectly. The third term is an error state, where mode $D$ lost one photon to mode $F$.

Thus, there are three types of errors: loss on Alice's link (mode $E$), truncation noise because of the quantum scissor, and decoherence because of the loss \textit{in} the scissor (mode $F$). The loss on Alice's link is overcome by increasing the gain $g$ of the scissor. The truncation noise is minimised by choosing the input variance $V_A$ and gain $g$ so that the output state is mostly a superposition of a single photon and vacuum. The decoherence term is unavoidable due to the loss in the scissor; however, as we stress in this chapter, using Bob as the reference for reconciliation, we still get a positive key rate.

\section{Computation of the secret key rate}

The asymptotic secret key rate $K$ is given by the raw key, $K_\text{raw}$, multiplied by the rate of successful operation of the repeater protocol, $R$. That is, $K= R K_\text{raw}$. The raw key is given by the asymptotic secret key rate formula, $K_\text{raw}=\beta I_{AB}{-}\chi_{EB}$~\cite{devetak2005distillation} (see~\cref{eq:DW} and surrounding text), where $I_{AB}$ is Alice and Bob's classical mutual information, $\beta$ is the reconciliation efficiency, and $\chi_{EB}$ is an upper bound on Eve's maximal information.

Eve's information is upper bounded by $\chi_{EB}$ which can be bounded as a function of the heralded covariance matrix, $\Gamma_{AB}$, in the EB version of the protocol. Since Alice's modulation is Gaussian, $\Gamma_{AB}$ is directly accessible by Alice and Bob if they perform the PM version of the experiment. From Alice and Bob's total data in the PM scheme, they can reconstruct the equivalent EB scheme. Then they use their reconstructed EB version to estimate $\Gamma_{AB}$, and use Gaussian optimality~\cite{PhysRevLett.96.080502} to upper-bound Eve's information $\chi_{EB}$. No assumption on the channel is required in an experiment for asymptotic security. 

However, since we do not have access to experimental data, we assume ambient conditions to simulate the parameters accessible to Alice and Bob in an experiment. We consider that the so-called ``excess noise'' is coming from the environment as input thermal noise ($\hat{\rho}_{\text{th}}$) with variance $V$. Consider Ref.~\cite{PhysRevLett.125.010502} which performed a long-distance CV-QKD experiment. They had an input excess noise $\xi=0.0081$ shot noise units (SNU) for 32.45 dB of loss (i.e., 162.25 km at 0.2dB/km). The distances involved in our system are much greater than 100 km, so we consider an amount of input thermal noise such that an equivalent amount of excess noise on the direct transmission system is $\xi=0.02$ SNU at 350 km (70 dB). This noise with variance $V$ we inject evenly into all links. That is, the input thermal noise variance is fixed and excess noise $\xi$ builds up over long distances. All key rates plotted in this chapter have this amount of thermal noise.

\begin{figure}
\centering
\includegraphics[width=0.65\linewidth]{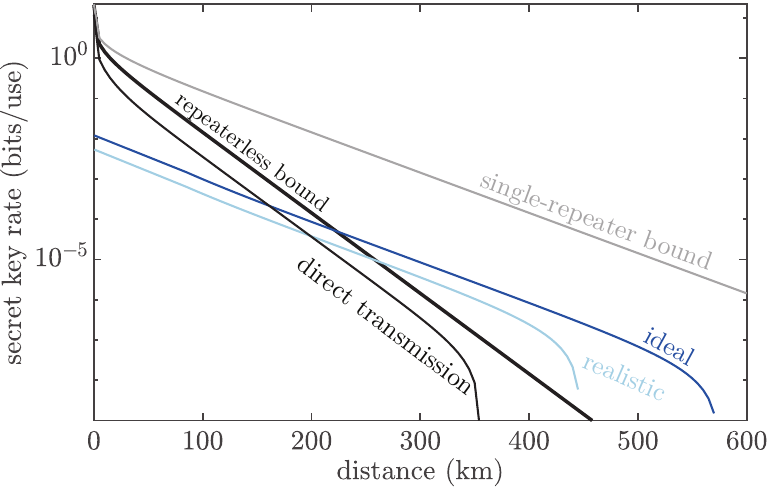}
\caption[Secret key rate of our CV-QKD protocol based on coherent states and heterodyne detection]{Secret key rate of our CV-QKD protocol based on coherent states and heterodyne detection versus the total distance between Alice and Bob assuming standard optical fibre (0.2 dB/km) and excess noise. The amount of excess noise is defined in the text. The strength of Alice's TMSV state is $\chi=0.4$ ($V_A=0.38$ in the PM version), and the beamsplitters and location of the relay are roughly optimised. The reconciliation efficiency is $\beta= 0.95$. For the realistic implementation (light blue), we assume $75\%$ single-photon source and single-photon detector efficiencies and $10^{-8}$ dark-count rate probability, whilst these are assumed perfect for ideal (dark blue). We plot direct transmission for optimised modulation variance and with excess noise. We also plot the point-to-point repeaterless bound~\cite{Pirandola_2017} and the single-repeater bound~\cite{Pirandola_2019}. Also note that the squeezed-state protocol and homodyne detection can increase the key rate, achieving similar rates as the best variants of TF QKD.}
\label{SIMfig:rate}
\end{figure}

The ideal secret key rate of our protocol based on coherent states and heterodyne detection is shown in~\cref{SIMfig:rate} for fixed thermal noise (the amount defined above). The modulation variance is $\chi=0.4$ and the relay is placed off-centre, slightly closer to Bob than Alice, which gives good key rates and is probably close to optimal (we cannot numerically compute key rates for arbitrarily large $\chi$). The reconciliation efficiency is $\beta=0.95$. For comparison, we plot the key rate for direct transmission for optimised modulation variance. The repeaterless~\cite{Pirandola_2017} and single-repeater~\cite{Pirandola_2019} bounds are also shown. These bounds are $-\log_2{(1-\eta^{1/N})}$ where $N$ is the number of links dividing the total distance, i.e., the number of repeaters is $N{-}1$ (for the repeaterless bound $N=1$ and for the single-repeater bound $N=2$). Our ideal protocol beats the repeaterless bound at 223 km and scales like the single-repeater bound. It beats direct transmission at 166 km.

Scissors are robust to non-ideal single-photon sources and detectors (and even on-off detectors) in the high fidelity regime of operation, i.e., low-energy input states~\cite{PhysRevA.102.063715}. These experimental imperfections generally decrease the success probability but do not greatly affect the fidelity. We plot the secret key rate for a realistic implementation in~\cref{SIMfig:rate} assuming $75\;\%$ single-photon source and single-photon detector efficiencies and $10^{-8}$ dark-count rate probability. The realistic curve beats the repeaterless bound at 260 km and direct transmission at 203 km.

Placing the repeater symmetrically between Alice and Bob, $\eta_A=\eta_B=\sqrt{\eta}$ and $T_C=T_B=0.5$, also gives good key rates. We refer to this setup as the symmetric configuration (here, $\chi = 0.1$ is about optimal). However, we find that by moving the repeater slightly closer to Bob and increasing the variance of the state prepared by Alice, and tuning the beamsplitters accordingly, the success probability is increased (at the expense of some decrease of the raw key), thus, improving the secret key rate overall. It is the asymmetric configuration that we plot in the figure noting that the symmetric configuration would decrease the key rate such that the repeaterless bound is beaten at about 50 km greater distance. Also, CV QKD based on squeezed states and homodyne detection~\cite{PhysRevA.63.052311} can increase the key a fair amount, obtaining similar rates achievable by the best TF-QKD variants.

\subsection{The asymptotic secret key rate formula}

The secret key rate is given by
\begin{align}
    K =  R  K_\text{raw},
\end{align}
where $R$ is the rate of successful operation of the repeater protocol and $K_\text{raw}$ is the raw key. For our single-repeater protocol, $R$ is simply the success probability, $P$, of the repeater detecting a single click, $R=P=2\braket{\psi}{\psi}$. The raw key, $K_\text{raw}$, is given by the asymptotic secret key rate formula $K_\text{raw}=\beta I_{AB}-\chi_{EB}$~\cite{devetak2005distillation}, where $I_{AB}$ is Alice and Bob's classical mutual information, $\beta$ is the reconciliation efficiency, and $\chi_{EB}$ is an upper bound on Eve's maximal information (the Holevo bound with Bob who is the reference side of the information reconciliation).

In what follows, we show that Alice and Bob's mutual information can be lower bounded and Eve's maximal information can be upper bounded using only the covariance matrix shared between Alice and Bob in the EB version, thereby giving a lower bound on the secret key rate. Since the output state $\hat{\rho}_{AB}$ is close to Gaussian, the exact key rate should be close to our lower bound. The output state $\hat{\rho}_{AB}$, due to the symmetry of the protocol, has a covariance matrix of the following form:
\begin{align}
   \Gamma_{AB} &= \begin{bmatrix}
        a & 0 & c & 0\\
        0 & a & 0 & -c\\
        c & 0 & b & 0\\
        0 & -c & 0 & b\\
    \end{bmatrix}.\label{eq:CM}
\end{align}

\subsection{Alice and Bob's mutual information}

The mutual information quantifies the amount of correlations between Alice and Bob, measured in bits of correlation. It is given by~\cite{Sanchez2007QuantumIW}
\begin{align}
    I_{AB} &= S(A)-S(A|B),
\end{align}
where $S(\cdot)$ is the Shannon entropy. The notation $S({A|B})$ means the conditional Shannon entropy of Alice's data conditioned on Bob's measurements.

The output state is close to a Gaussian state so we can approximate Alice and Bob's information $I_{AB}$ using the covariance matrix $\Gamma_{AB}$ shared between Alice and Bob. For the protocol based on coherent states and heterodyne detection, we have 
\begin{align}
    I_{AB} &\approx  \log_2 \left({ \frac{V_{A}}{V_{A|B}}}\right) = \log_2 \left({ \frac{1+a}{1+a-\frac{c^2}{1+b}}}\right),\label{eq:I_AB}
\end{align}
where $a,b,$ and $c$ are elements of the covariance matrix in~\cref{eq:CM}.
Numerical calculations confirm that this underestimates Alice and Bob's information. Hence, the key rate is secure using $\Gamma_{AB}$ to approximate $I_{AB}$.

\subsection{Upper bounding Eve's information}

Eve's maximal information with Bob is given by the Holevo quantity~\cite{holevo1973bounds}
\begin{equation}
\chi_{EB} = S(\hat{\rho}_E) - S(\hat{\rho}_{E|b}),
\end{equation}
where $S(\hat{\rho}_E)$ is the von Neumann entropy of Eve's state, and $S(\hat{\rho}_{E|b})$ is the von Neumann entropy of Eve's state conditioned on Bob's measurement. 

However, we cannot calculate $\chi_{EB}$ directly because we do not know Eve's optimal attack since the protocol is both non-Gaussian and non-deterministic. Since Alice's initial modulation is Gaussian, Alice and Bob have access to a covariance matrix $\Gamma_{AB}$ in the EB version from the parameters they observe in the PM version. To compute secret key rates, we can simulate $\Gamma_{AB}$ assuming a thermal-lossy channel. Given $\Gamma_{AB}$, estimated by Alice and Bob in the simulated experiment, we can then bound Eve's information using Gaussian optimality~\cite{PhysRevLett.96.080502} which says that for an arbitrary quantum state $\hat{\rho}_{AB}$ shared between Alice and Bob, the Gaussian state $\hat{\rho}^*_{AB}$ with the same covariance matrix $\Gamma_{AB}$ as for $\hat{\rho}_{AB}$ gives the maximal Holevo information. Thus, Eve's information can be upper bounded given the covariance matrix $\Gamma_{AB}$ shared between Alice and Bob, even if the state is non-Gaussian. Explicitly, we have
\begin{equation}
\chi_{EB} = S(AB) - S(A|B).
\end{equation}
$S(AB)$ and $S(A|B)$ can be calculated from the symplectic eigenvalues $\nu_k$ of the respective covariance matrix, $\Gamma_{AB}$ and $\Gamma_{A|b}$, via the relation $S(\cdot)=\sum_{k=1}^N \frac{\nu_k+1}{2}\log_2 \frac{\nu_k+1}{2} - \frac{\nu_k-1}{2}\log_2 \frac{\nu_k-1}{2},$ where $N$ is the number of modes~\cite{PhysRevA.59.1820}.

\begin{figure}
\centering
\includegraphics[width=0.65\linewidth]{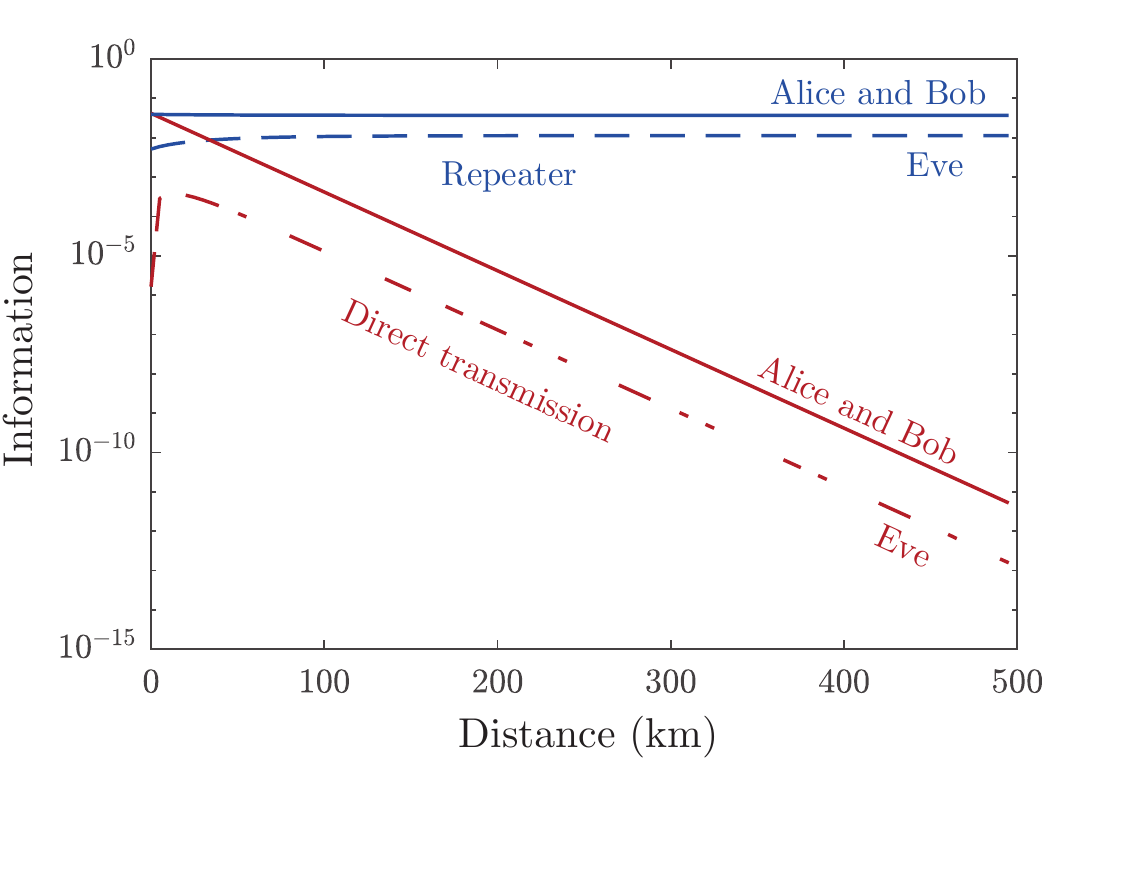}
\caption[Mutual information and Eve's information.]{Alice and Bob's mutual information and Eve's information for fixed $\chi=0.12$ as a function of distance in km for direct transmission (red) and with the simple CV repeater (blue). Notice that the repeater heralds entanglement between all parties, Alice and Bob and Eve, but since Alice and Bob's mutual information is greater than the information Eve has, the key rate is positive.}
\label{SIMfig:AliceBobEveinfo_chipoint12}
\end{figure}

To illustrate the workings of our simple repeater compared with using no repeater (i.e., direct transmission), in~\cref{SIMfig:AliceBobEveinfo_chipoint12}, we plot Alice and Bob's mutual information and Eve's information with and without a repeater. For direct transmission, Alice and Bob's mutual information, and Eve's maximal information, decrease exponentially with distance, but since the mutual information is greater than Eve's information, the key rate is positive. On the other hand, with the repeater, Alice and Bob's mutual information and Eve's information is approximately constant with distance, while the success probability of the protocol scales exponentially with distance. Since Alice and Bob's information is greater than Eve's information, the key rate is positive. The output state needs to be sufficiently Gaussian otherwise Eve's maximal information is greater than Alice and Bob's mutual information and our bound on the secret key rate is zero.

\subsection{Asymmetric configuration}

In~\cref{SIMfig:rate}, we tune the beamsplitters and $\chi$ so that the key rate is roughly as optimal as possible. Specifically, the energy of Alice's TMSV state is $\chi=0.4$ ($V_A=0.38$ in the PM version), the beamsplitter transmissivities in the scissor are $T_C=1/2$ and $T_B=2/3$, and the location of the repeater is moved slightly towards Bob such that $\sqrt{\eta_A}g=0.21$, where $g = \sqrt{\frac{ T_B}{\eta_B(1-T_B)}}$. This means the gain does not amplify back up to the energy of the input state; this is necessary because the modulation here is large and the scissor introduces too much truncation noise for larger gain. This configuration is approximately optimal. There is a trade-off between success probability and entanglement distillation.

\subsection{Excess noise and experimental imperfections}

In this subsection, we detail how we numerically incorporated excess noise and experimental imperfections in our calculations.

In CV QKD, noise on top of pure loss is called ``excess noise''. We model this additional noise by considering thermal-state inputs from the environment, instead of vacuum as in the pure loss case. The variance of the input thermal noise is chosen such that for direct transmission it is equivalent to ``excess noise'' $\xi= 0.02$ SNU at 70 dB loss (that is, 350 km at 0.2 dB/km). This amount of noise is probably what can be expected in future demonstrations, for instance, consider the long-distance CV-QKD experiment over 202.81 km from Ref.~\cite{PhysRevLett.125.010502}. At their longest distance, they have an excess noise of $\xi= 0.0081$ SNU at 32.45 dB loss (that is, 162.25 km at 0.2 dB/km).

We model the efficiency of the non-ideal single-photon source by placing a beamsplitter of transmissivity $\tau_\text{s}$ after the source (here, the transmissivity is the efficiency). We model the detection efficiency of the single-photon detectors by placing a beamsplitter of transmissivity $\tau_\text{d}$ before each detector. To model dark counts, we assume a thermal state of mean photon number $\bar{n}_d$ is incident on the auxiliary beamsplitter port and choose $\bar{n}_d$ such that the required dark-count rate is achieved. For each $\tau_\text{d}$ there is a $\bar{n}_d$ that gives a specific dark-count rate.

To compute secret key rates for noisy channels and to include experimental imperfections, we do not analytically calculate the output density matrix $\hat{\rho}_{AB}$. Rather, we do it numerically in Fock space.

\section{Repeater chain using quantum memories}

\begin{figure}
\includegraphics[width=1\linewidth]{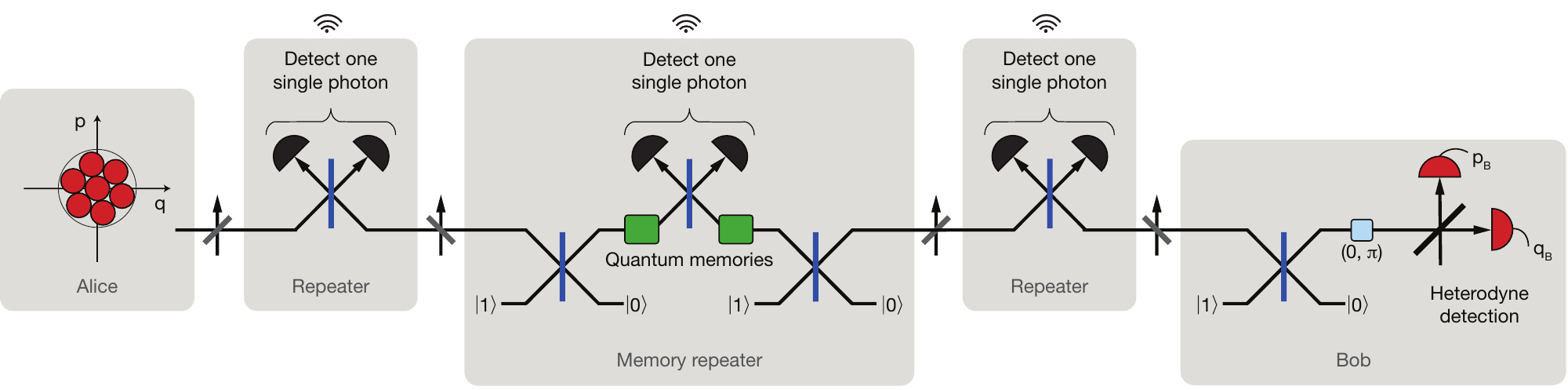}
\caption[Extending our CV-QKD repeater protocol into a three-repeater chain using quantum memories]{Extending our CV-QKD repeater protocol into a three-repeater chain using quantum memories (solid squares) and additional quantum scissors. Shown here is the entanglement-based version. At the lowest level, the quantum scissors (shown in blue) perform entanglement swapping and entanglement distillation (without the need for quantum memories) and are tolerant to loss in the links. At the higher level, quantum memories hold onto the state of the lower level and an additional quantum scissor (shown in red) (i.e., a partial Bell measurement) performs entanglement swapping. The secret key rate scales of order $\eta^{1/4}$, the same scaling as the fundamental three-repeater bound. This repeater protocol requires far fewer resources than previous CV schemes~\cite{PhysRevResearch.2.013310,PhysRevA.102.052425,Ghalaii_2020_scissors}. Note that quantum scissors have two successful click patterns, one of which heralds a $\pi$-phase shift on the output mode~\cite{PhysRevA.102.063715}. This phase shift is easily corrected via passive phase shifts but they are not shown in the figure for simplicity.}
\label{SIMfig:repeater}
\end{figure}

\begin{figure}
\centering
\includegraphics[width=0.6\linewidth]{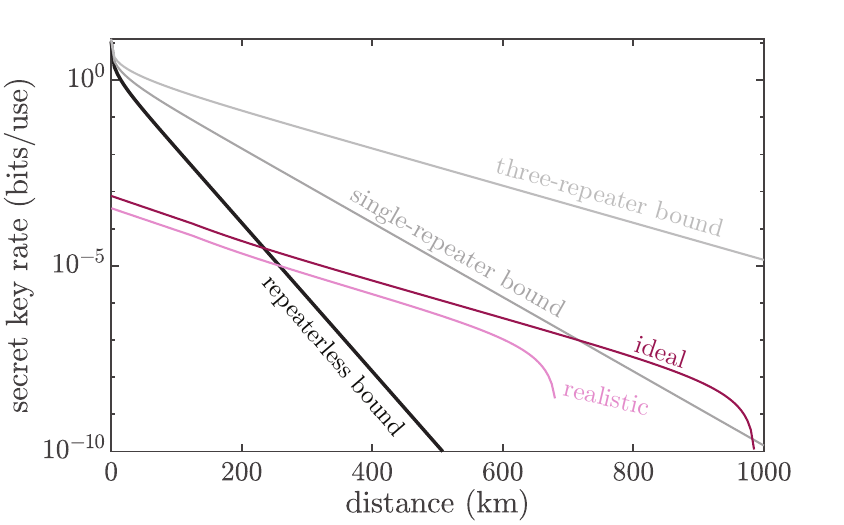}
\caption[Secret key rate versus the total distance between Alice and Bob with excess noise of our three-repeater chain]{Secret key rate versus the total distance between Alice and Bob with excess noise of our three-repeater chain for the no-switching protocol based on coherent states and heterodyne detection. The repeater chain is set up in a slightly asymmetric configuration with $\chi=0.2$ to roughly optimise the key rate. The realistic implementation (pink) assumes $75\;\%$ single-photon source and single-photon detector efficiencies and $10^{-8}$ dark-count rate probability. The quantum memories are assumed to be ideal.}
\label{SIMfig:rate_4links}
\end{figure}

We extend our single-node protocol to a three-repeater chain in~\cref{SIMfig:repeater} and we plot the secret key rate in~\cref{SIMfig:rate_4links}. It scales like the three-repeater bound, ${-}\log_2{(1{-}\eta^{1/4})}$~\cite{Pirandola_2019}.  We achieve the same scaling with half the number of resources as previous CV repeater proposals~\cite{PhysRevResearch.2.013310,PhysRevA.102.052425,Ghalaii_2020_scissors}, in particular, the one introduced in~\cref{Chap:CV_QR}. Our ideal three-repeater chain beats the repeaterless bound at 234 km and the single-repeater bound at 717 km.

For the realistic implementation, we assume $75\;\%$ single-photon source and single-photon detector efficiencies and $10^{-8}$ dark-count rate probability. The quantum memories are assumed ideal. Details for computing secret key rates of our three-repeater chain are presented next. The chain can straightforwardly be extended to include more repeaters in a longer repeater chain.

Now, we describe how to calculate secret key rates of the three-repeater chain shown in~\cref{SIMfig:rate_4links}. We use quantum memories to store the heralded state of our single-node protocol and then do entanglement swapping on two copies of the state using another quantum scissor. We calculate the output state $\hat{\rho}_{AB}$ numerically in Fock space.

The rate, $R$, of successful operation of the entire three-repeater protocol depends on the success probabilities of the higher and lower levels in the following way:
\begin{align}
R &= \frac{1}{Z_1(P_\text{lower level})} P_\text{higher level},
\end{align}
where $P_\text{higher level}$ is the success probability of the entanglement-swapping scissor (red), and $P_\text{lower level}$ is the minimum success probability of the two entanglement-distillation scissors (blue). The function $Z_n(P)$ is the average number of steps required to generate successful outcomes in $2^n$ probabilistic operations, each with success probability $P$~\cite{PhysRevA.83.012323}:
\begin{align}
    Z_n(P) &= \sum_{j=1}^{2^n} {n \choose j} \frac{(-1)^{j+1}}{1-(1-P)^j}.
\end{align}

For the same number of resources, our protocol scales like the square root of the scaling of previous schemes~\cite{PhysRevResearch.2.013310,PhysRevA.102.052425,Ghalaii_2020_scissors} (we can go twice as far with the same number of resources). That is, the scheme from Ref.~\cite{PhysRevA.102.052425} requires four memories, two scissors, and a Bell measurement to scale like the single-repeater bound and to beat the repeaterless bound at about 400 km, whereas, our protocol uses the same resources to achieve three-repeater scaling and we beat the bound at 234 km and is robust to some excess noise and experimental imperfections.

We tune the beamsplitters and $\chi$ so that the key is roughly optimised. That is, the two-mode-squeezing parameter is $\chi=0.2$, which is a bit too large for the protocol so the effective change in amplitude of the total channel is $0.42$, set by the location of the repeaters and the transmissivity of the primary gain beamsplitter (with transmissivity $0.73$), and all other beamsplitters are $50{:}50$; that is, the second two links are in symmetric configuration while the first two links are in an asymmetric configuration, and the central node is positioned in the exact centre between Alice and Bob.

\section{Entanglement distillation with higher-order scissors}

We introduced in~\cref{Chap:scissors}, that quantum scissors generalise to higher order at the cost of a reduced success probability~\cite{PhysRevA.102.063715}. For some applications, where entanglement distillation is more important than rate, for instance, for teleportation protocols and distributed quantum computing, higher-order quantum scissors may prove useful. When used as a repeater (i.e., in a similar setup as in~\cref{SIMfig:CV_QKD_repeater_EB_version}), higher-order scissors, for example, the three-photon scissor, cannot beat the repeaterless bound; however, it distils a large amount of entanglement, more than the single-photon scissor can achieve. The single-photon repeater scales like $ \sqrt{\eta}/2$ whilst the three-photon repeater scales like $\frac{3}{64} {\eta}^\frac{3}{2}$. The prefactor is a penalty due to using linear optics and waiting for the required number of clicks in the detector. Our scissors generalise to all positive integers~\cite{PhysRevLett.128.160501} but we focus on the first-order and third-order devices since they were discovered first~\cite{PhysRevA.102.063715}.

First, recall from Ref.~\cite{neergaard-nielsen2013quantum} that tele-amplification devices perform NLA on a restricted set of superposition states consisting of coherent states on a ring in phase space. The devices are loss tolerant meaning that they can faithfully teleport and amplify classical mixtures of coherent states. When used in this way, the device is referred to as a loss-tolerant quantum relay.

\begin{figure}
\centering
\includegraphics[width=0.45\linewidth]{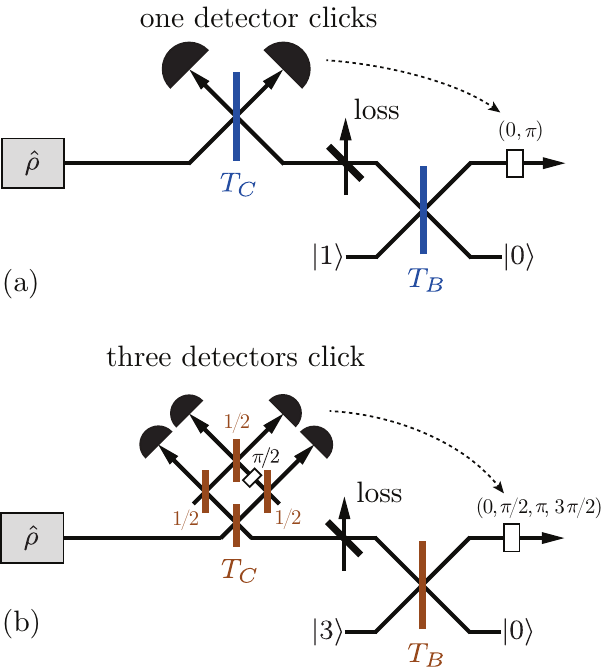}
\caption[Quantum scissors perform ideal truncation and noiseless linear amplification up to some Fock number]{Quantum scissors perform ideal truncation and noiseless linear amplification up to some Fock number. Shown here are two examples. (a) single-photon scissor~\cite{pegg1998optical} which truncates all terms greater than Fock-number one, and (b) three-photon scissor~\cite{PhysRevA.102.063715} which truncates all terms greater than Fock-number three. A passive phase-shift (white rectangle) correction is required on the outgoing mode, depending on which detectors fire~\cite{PhysRevA.102.063715}. Note there is an important $\pi/2$ phase shift in the three-scissor interferometer. An extremely useful feature is that the quantum scissors perform remarkably well for thermal input states if there is loss on the entanglement resource, as shown in the figure. We say that the scissor is loss tolerant.}
\label{fig:scissor}
\end{figure}

Quantum scissors perform simultaneous truncation and NLA up to Fock number $k$ (called a $k$-photon scissor or a $k$-scissor). The first and third-order quantum scissors are shown in~\cref{fig:scissor}. Tele-amplification devices are exactly quantum scissors as the cat amplitude goes to zero. Therefore, quantum scissors are approximately loss tolerant (strictly loss tolerant only on the vacuum state). Thus, quantum scissors can be used as loss-tolerant quantum repeaters for CV QKD. We call them repeaters rather than relays since they repeat effectively and can overcome the repeaterless bound. If the input states are restricted to classical mixtures of coherent states, the device performs pretty-good NLA despite loss on the entanglement mode, as shown in~\cref{fig:scissor}.

We want to be clear that scissors perform ideal NLA up to some Fock number but the NLA is not perfect if the scissor is used as a quantum repeater. It is always better in terms of fidelity to use a lossless quantum scissor (i.e., moving the repeater into Bob's station); however, sacrificing some fidelity by using the scissor as a repeater improves the scaling of the probability of success and increases the secret key rate.

\begin{figure}
\centering
\includegraphics[width=0.6\linewidth]{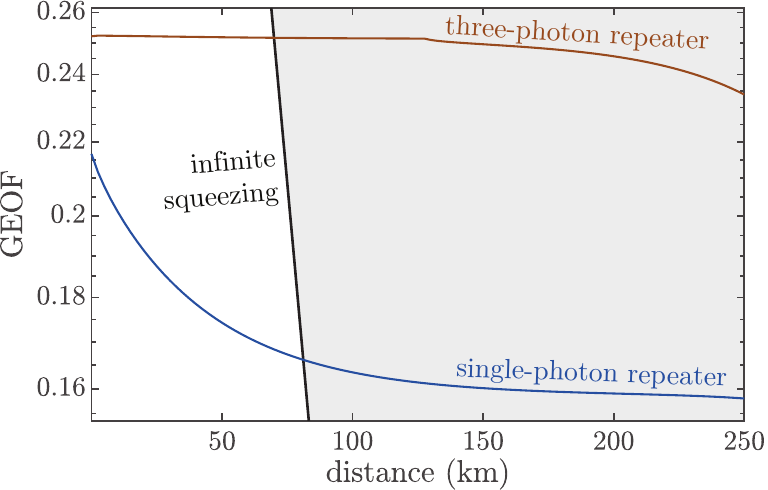}
\caption[Gaussian entanglement of formation (GEOF) versus distance for the repeater protocol]{Gaussian entanglement of formation (GEOF) versus distance for the repeater protocol like that shown in~\cref{SIMfig:CV_QKD_repeater_EB_version}, using a first-order and third-order device, in symmetric configuration ($T_C=T_B=0.5,\eta_A=\eta_B=\sqrt{\eta}$), $\chi=0.2$, and with excess noise. We also plot the GEOF for a TMSV state with infinite squeezing. The shaded region is unattainable by any deterministic protocol. It is clear from this figure that higher-order repeaters distil more entanglement and work for larger-energy input states, beyond what is possible for the first-order repeater.}
\label{SIMfig:GEOF_3scissor}
\end{figure}

As an entanglement measure, the entanglement of formation quantifies the minimum entanglement needed to prepare an entangled state from a classical one. We calculate the Gaussian entanglement of formation~\cite{PhysRevA.69.052320} as an entanglement measure using results from Ref.~\cite{PhysRevA.96.062338}. We calculate the Gaussian entanglement of formation (GEOF)~\cite{PhysRevA.69.052320,PhysRevA.96.062338} of the final state, $\hat{\rho}_{AB}$, to evaluate the performance of the first- and third-order repeater in symmetric configuration.  We plot this in~\cref{SIMfig:GEOF_3scissor} for $\chi=0.2$. For small $\chi$, the repeater introduces only a little non-Gaussian noise. We note in the figure that the entanglement is barely unchanged with distance.

\begin{figure}
\centering
\includegraphics[width=0.6\linewidth]{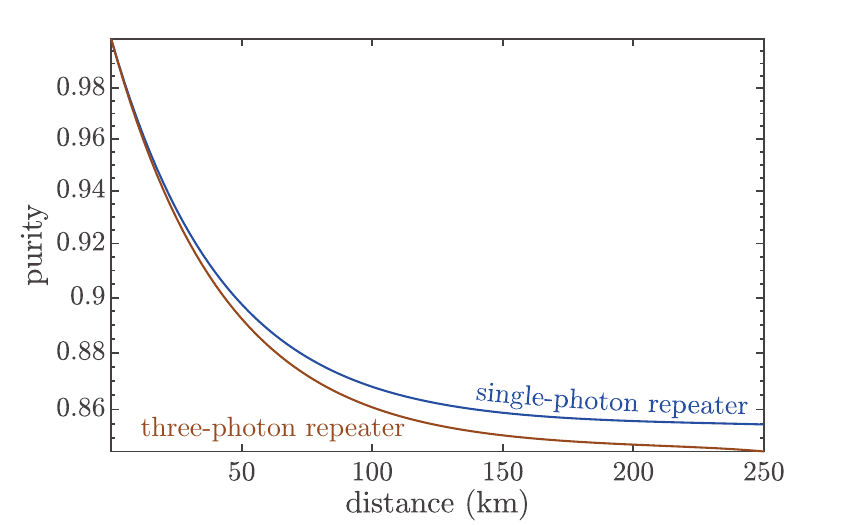}
\caption[Purity versus distance for our repeater protocol]{Purity versus distance for our repeater protocol shown in~\cref{SIMfig:CV_QKD_repeater_EB_version} in symmetric configuration ($T_C=T_B=0.5,\eta_A=\eta_B=\sqrt{\eta}$), two-mode-squeezing parameter $\chi=0.2$, and with excess noise. The purity for infinite two-mode-squeezing is not shown because it decreases exponentially with distance whereas, with the repeater, the purity stays high with distance.}
\label{fig:purity_3scissor}
\end{figure}

Another important figure of merit is purity. The purity of $\hat{\rho}_{AB}$ is defined as $\Tr({\hat{\rho}_{AB}^2})$. In~\cref{fig:purity_3scissor}, we plot the purity calculated numerically from the density matrix, $\hat{\rho}_{AB}$, as a function of distance. In the high-fidelity regime, the output state remains very pure as a function of distance, whereas the infinitely-squeezed TMSV state exponentially loses its purity with distance.

\section{CV-MDI protocols for beating the repeaterless bound}

\begin{figure}
\centering
\subfloat[]{%
  \includegraphics[width=1\linewidth]{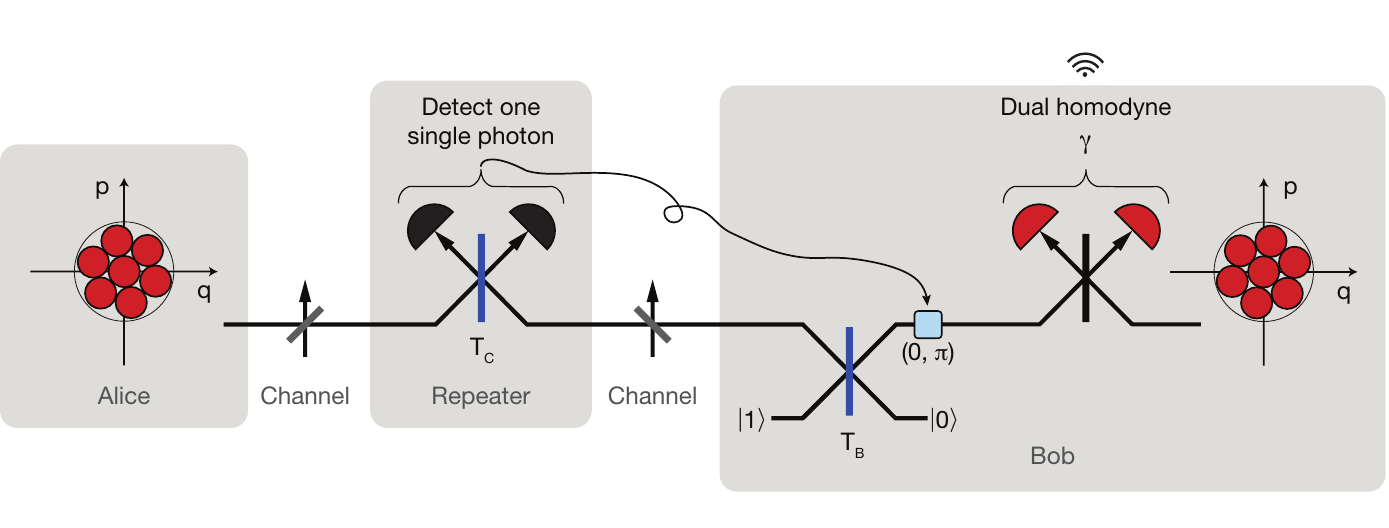}
  \label{SIMfig:MDI_repeater_protocol}
  }
  
  \subfloat[]{%
    \includegraphics[width=1\linewidth]{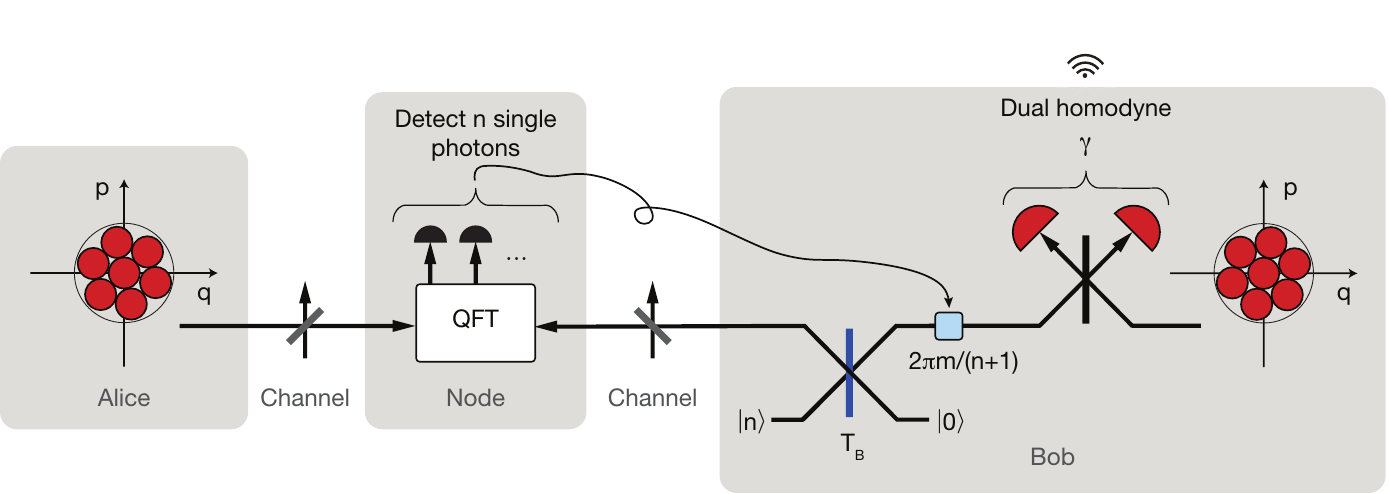}
  \label{SIMfig:MDI_repeater_protocol_higher_order}
  }
  
\subfloat[]{%
  \includegraphics[width=1\linewidth]{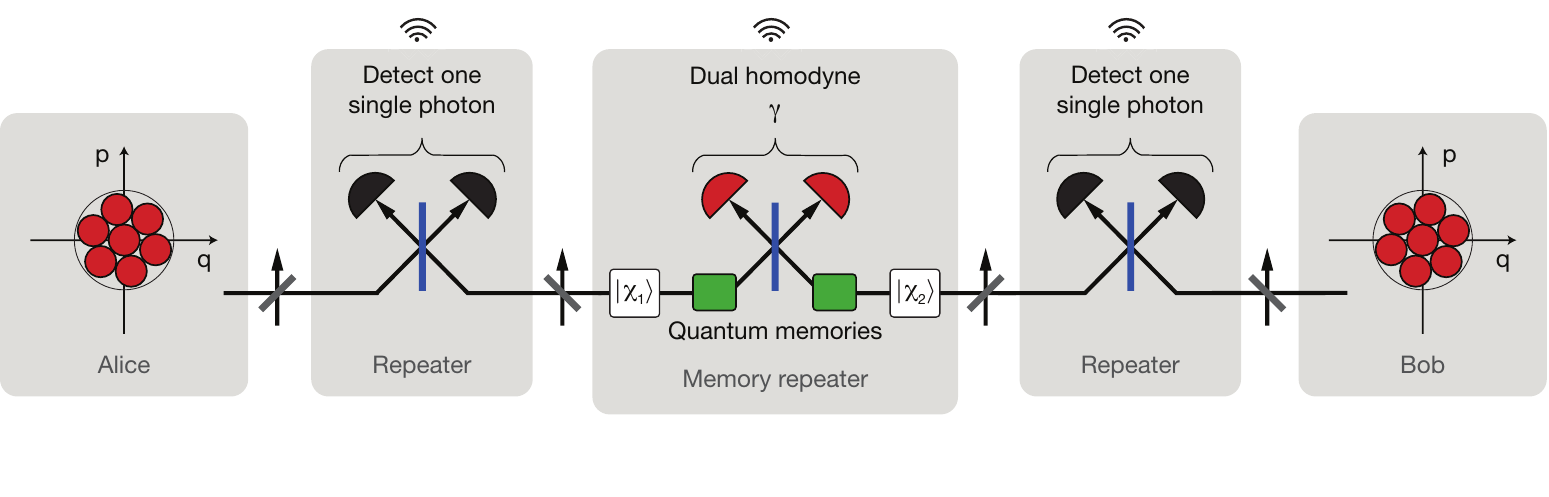}
  \label{SIMfig:MDI_three_repeater_chain}
  }
\caption[CV-MDI QKD repeaters]{CV-MDI-QKD repeaters. \protect\subref{SIMfig:MDI_repeater_protocol} CV-MDI QKD protocol with rate that scales like $\sqrt{\eta}$ and beats the repeaterless bound. \protect\subref{SIMfig:MDI_repeater_protocol_higher_order} CV-MDI QKD protocol to higher order with rate scaling $\eta^{n/2}$, where $n$ is the order of the NLA. \protect\subref{SIMfig:MDI_three_repeater_chain} CV-MDI QKD protocol with three-repeater chain and ideal quantum memories, with an ideal rate that scales like $\eta^{1/4}$ and beats the two-repeater bound.}
\label{SIMfig:MDI}
\end{figure}

Recall from~\cref{sec:MDI_intro} that CV measurement-device-independent (MDI) QKD is promising since it is immune to side-channel attacks, but unfortunately is strongly limited by the transmission distance. Our protocol may be useful for CV-MDI protocols, without this fundamental limitation. We present some interesting ideas in~\cref{SIMfig:MDI}. We have verified that estimating a bound on the secret key rate using the asymptotic secret key rate formula does give excellent positive rates which scale optimally. We leave detailed analyses of these preliminary results for future work.

\section{Optimising CV entanglement distillation}

To end this chapter, we provide an enjoyable, general method for solving a semidefinite program (\hypertarget{hyperlinklabel_SDP}{\hyperlink{hyperlinkbacklabel_SDP}{SDP}}) for optimising the success probability of general entanglement-distillation protocols, and of course their equivalent PM versions for QKD. Indeed, our SDP problem found our simple repeater protocol described in this chapter and Ref.~\cite{winnel2021overcoming}, demonstrating the usefulness of the method. We hoped that the SDP would allow us to achieve the ultimate capacity, the single-repeater bound, but this was not possible. We instead solve this problem in~\cref{Chap:purification} by other means.

\begin{figure}
\includegraphics[width=0.6\linewidth]{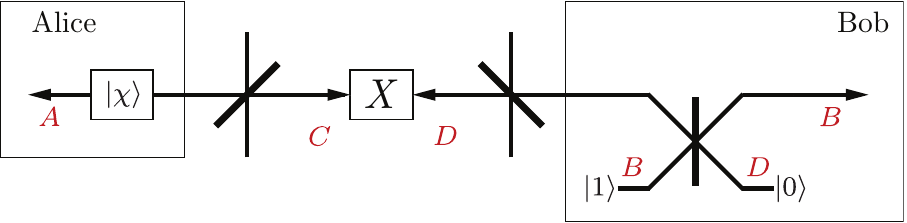}
\centering
\caption[Setup of the SDP problem]{Setup of the SDP problem for maximising the success probability of an unknown measurement $X$, given arbitrary, selected entanglement resources at Alice and Bob. Alice and Bob's entanglement resources shown are $\ket{\chi}$ and $(\ket{01}-\ket{10})/\sqrt{2}$. This choice gives very good rates and may be the optimal single-mode protocol in this configuration.}
\label{fig:SDP_problem}
\end{figure}

We now set up the SDP problem. Consider the general entanglement-swapping procedure shown in~\cref{fig:SDP_problem}. Alice and Bob share arbitrary entangled states with the repeater node across a lossy or noisy channel, resulting in the state $\rho_{ACDB}$, shared between Alice, Bob, and the repeater. The state shared between Alice and Bob after the unknown two-mode POVM $X$ is $\rho_{AB} \propto \tilde{\rho}_{AB} \equiv \tr_{CD}[(\mathds{1}\otimes X \otimes \mathds{1})  \rho_{ACDB}] $ (where $\tilde{\rho}_{AB}$ is the state before normalisation). Then, the SDP finds the maximum success probability for $X$ given that the target state has a similar covariance matrix to the output state $\rho_{AB}$, but $\rho_{AB}$ should not have an identical covariance matrix to the target state since that might require a very small probability of success. 

The SDP problem goes as follows. 
\begin{align*}
    &\text{maximise}\; P=\tr(\tilde{\rho}_{AB})\\
    &\text{where} \; \tilde{\rho}_{AB} = \tr_{CD}[(\mathds{1}\otimes X \otimes \mathds{1})  \rho_{ACDB}]\\
   \text{such that}\\
    1.\;\;\;\;\; &X \geq 0\\
    2.\;\;\;\;\;&Y \geq 0\\
    3.\;\;\;\;\;&X + Y = \mathds{1} \\
    4.\;\;\;\;\;& \tr[(\mathds{1}\otimes \hat{q}^2)\tilde{\rho}_{AB}] = P \; \tr[(\mathds{1}\otimes \hat{q}^2) \ketbra{\chi}{\chi}] \\
    5.\;\;\;\;\;& \tr[(\mathds{1}\otimes \hat{p}^2)\tilde{\rho}_{AB}] =  P \;\tr[(\mathds{1}\otimes \hat{p}^2) \ketbra{\chi}{\chi}]\\
    6.\;\;\;\;\;& \tr[(\hat{q}\otimes \hat{q})\tilde{\rho}_{AB}] =  P \; \tr[(\hat{q}\otimes \hat{q}) \ketbra{\chi}{\chi}]\\
    7.\;\;\;\;\;& \tr[(\hat{p}\otimes \hat{p})\tilde{\rho}_{AB}] =  P\;\tr[(\hat{p}\otimes \hat{p}) \ketbra{\chi}{\chi}],
\end{align*}
where the target state $\ket{\chi}$ may be any state, but it is usually the state Alice initially prepares (usually a TMSV state). Note there is truncation involved here; otherwise, it cannot be run numerically on a computer. We truncate at a Fock number chosen large enough so that our results are independent of the particular chosen cutoff.

In words, the SDP does the following: Maximise the success probability of the measurement in the relay, $\tr(\tilde{\rho}_{AB})$, with the constraints that $X$ is a physical POVM and the covariance matrix of the output state is the same as the covariance matrix of the target state, or at least part of the covariance matrix is the same (see the following paragraph).

Not all the constraints on the covariance matrix (i.e., constraints 4-7) are required. Further, additional constraints decrease the success probability because the SDP finds the ideal NLA transformation, which has zero success probability. Other combinations of constraints can also work. For entanglement resource $(\ket{00}+\ket{11})/\sqrt{2}$ at Bob, we find that we do not need constraints 4 and 5, for example. For higher-order entanglement consisting of more Fock components, we find that constraints 4, 5, and 6 but not 7 give excellent results. Using all the constraints computes very low success probabilities ($10^{-10}$), and this is because the SDP has no ``wriggle-room''; it is attempting ideal NLA which is unphysical and has zero success probability.

\begin{figure}
\includegraphics[width=0.6\linewidth]{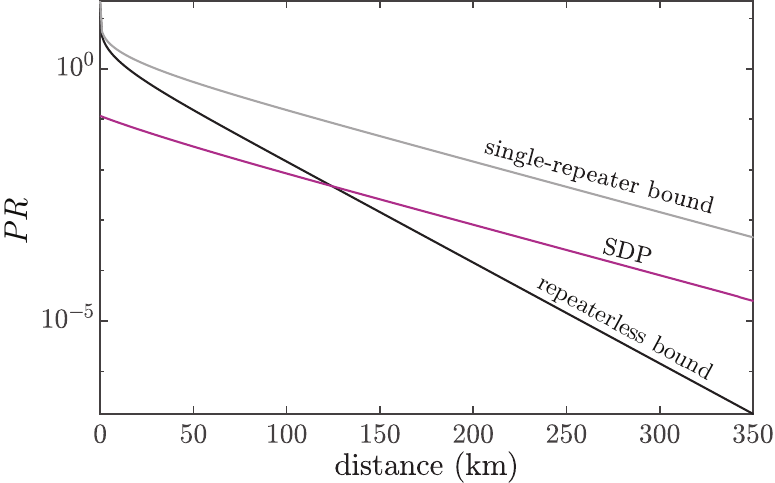}
\centering
\caption[Ideal performance of SDP-found output state ]{Ideal performance given by the success probability, $P$, multiplied by the RCI, $R$, of the SDP-found output state for entanglement resources $\ket{\chi}$ at Alice and $(\ket{01}+\ket{10})/\sqrt{2}$ at Bob. $\chi_\text{input}=\chi_\text{target} = 0.25$. We beat the bound at 124 km. Therefore, this is not quite optimal since by fine-tuning the parameters (beamsplitter transmissivities and modulation size) we can beat the repeaterless bound at about 119 km for this protocol.}
\label{fig:SDP}
\end{figure}

We plot the success probability multiplied by the RCI, $R$, of the output state found by the SDP for entanglement resources $\ket{\chi}$ at Alice, and single-photon entanglement at Bob. The initial and target squeezing were chosen to be $\chi=0.25$. The SDP finds our distributed quantum-scissor protocol, and it is unable to find a more optimal protocol. To improve these results, we need to consider operations on multiple modes, which becomes the subject of~\cref{Chap:purification}, where we achieve the capacity of the pure-loss channel.

\section{Discussion and Conclusion}
In this chapter, we introduced a practical CV quantum-communication protocol based on a quantum scissor which when used for CV QKD overcomes the repeaterless bound without the need for quantum memories. The scissor is used as a loss-tolerant quantum repeater. It is useful for QKD and entanglement distillation of TMSV states. Equivalently, the scheme can be thought of as hybrid entanglement swapping, using CV and DV entanglement. The DV entanglement fixes the directional problem in CV QKD. The single-photon measurement is a powerful non-Gaussian resource, projecting onto a state with strong correlations between Alice and Bob allowing the generation of a secret key or the distillation of entanglement.

The position of the repeater and the transmissivities of the beamsplitters together set the gain of the NLA. By placing the repeater symmetrically between Alice and Bob and using balanced beamsplitters, the success probability scales of order $\sqrt{\eta}$ and the gain of the NLA across the second link overcomes the loss in the first link. Thus, the state is sufficiently teleported and amplified from Charlie to Bob at a $\sqrt{\eta}$-scaling probability of success.

The protocol is tolerant to some excess noise and imperfections of the single-photon sources and single-photon detectors. The protocol can be extended into a chain of quantum repeaters using quantum memories. This would require just half the number of resources as recent CV quantum-repeater architectures~\cite{PhysRevResearch.2.013310,PhysRevA.102.052425,Ghalaii_2020_scissors}, for example, the one presented in the previous chapter.

One of the main results of this chapter is that we in general improve the scaling of many recent protocols based on quantum scissors to the square root of their old scaling and our protocol is quite robust to excess noise. This promotes the success probability of higher-order scissors into regimes of practical use.

One cannot input arbitrary quantum states into TF QKD, so it is not a quantum repeater in this sense. Our protocol can distribute quantum states long distances. It is also interesting to note that TF QKD requires the two links to have similar levels of loss~\cite{Zhong2021}, whereas, our protocol works both symmetrically and asymmetrically by optimising the gain and we always beat the repeaterless bound for an ideal implementation except in the completely asymmetric case when the repeater is inside Bob's station. One advantage of TF QKD is that the only measurement is performed in Charlie's station and so can be controlled by Eve. For our protocol, we have a CV measurement in Bob's station which we assume is trusted.

An interesting feature would be to generalise our memoryless protocol to reliably connect multiple trusted users. Future work would involve extending security analysis to the finite-size regime.

Achieving the single-repeater bound requires very large modulations. An open question is how to approach the single-repeater bound without quantum memories if it is possible; that is, how to increase the size of the modulation without introducing more truncation noise from the scissor. In~\cref{Chap:purification}, we introduce a protocol with rates that saturate the ultimate achievable rates, but for an ideal experiment and ideal quantum memories.


%% file: chapter_purification/chapter_purification.tex
\chapter{Ultimate-rate quantum repeaters}
\label{Chap:purification}	
\pagestyle{headings}

\noindent
The results of this chapter have been published in the following.\\

\noindent
1.~\cite{https://doi.org/10.48550/arxiv.2203.13924} \textbf{M. S. Winnel}, J. J. Guanzon, N. Hosseinidehaj, and T. C. Ralph, ``Achieving the ultimate end-to-end rates of lossy quantum communication networks,'' \href{https://www.nature.com/articles/s41534-022-00641-0}{npj Quantum Information \textbf{8}, 129 (2022)}.\\

\noindent
The field of quantum communications promises the faithful distribution of quantum information, quantum entanglement, and absolutely secret keys; however, the highest rates of these tasks are fundamentally limited by the transmission distance between quantum repeaters. \bblack{The ultimate end-to-end rates of quantum communication networks are known to be achievable by an optimal entanglement distillation protocol followed by teleportation. In this chapter, we give a practical design for this achievability. Our ultimate design is an iterative approach, where each purification step operates on shared entangled states and detects loss errors at the highest rates allowed by physics. As a simpler design, we show that the first round of iteration can purify completely at high rates. We propose an experimental implementation using linear optics and photon-number measurements} which is robust to inefficient operations and measurements, showcasing its near-term potential for real-world practical applications.

\section{Introduction}

\bblack{The goal of this chapter is to give a physical realisation for achieving the ultimate rates of quantum communications, i.e., the repeaterless (PLOB) bound and the ultimate end-to-end rates of quantum networks, which could pave the way for experimental implementations. While the highest achievable secret key rate for point-to-point CV QKD saturates the repeaterless bound~\cite{Pirandola_2020}, it does not provide a physical design for entanglement distillation.} Furthermore, \bblack{it is impossible to distil Gaussian entanglement using Gaussian operations only~\cite{PhysRevA.66.032316,PhysRevLett.89.137903}} so quantum repeaters must use non-Gaussian elements~\cite{PhysRevA.90.062316}.

Protocols based on infinite-dimensional systems are required to saturate the ultimate limits, i.e. CV protocols. However, previous practical quantum repeater designs are unable to distil entanglement at the ultimate rates~\cite{dias2017,PhysRevA.98.032335,PhysRevResearch.2.013310,Ghalaii_2020,PhysRevA.102.052425,winnel2021overcoming}.

Previous repeater techniques based on NLA or entanglement swapping fail to achieve the ultimate limits under pure loss since the output state is not pure and the success probability is zero for high-energy input states. For instance, the schemes based on noiseless linear amplification~\cite{Ghalaii_2020,PhysRevA.102.052425,winnel2021overcoming} have the same rate-distance scaling as the ultimate bounds but do not saturate them. We now explain why.

Noiseless linear amplification is a distinguished technique for the distillation of continuous-variable (CV) entanglement but it is unable to purify completely from loss. It is unable to return a loss-attenuated entangled state to its original loss-free entangled state.

To see this, consider pure loss acting on one mode of a CV two-mode squeezed vacuum (TMSV) state with the squeezing parameter $\chi\in [0,1]$. Pure loss is equivalent to interacting the data mode with vacuum on a beamsplitter with transmissivity $\eta\in[0,1]$. Consider ideal noiseless linear amplification~\cite{Pandey_2013} with gain $g\geq1$ applied to the lossy arm to correct the loss. The final state after ideal noiseless linear amplification of the lossy mode results in another {\textit{lossy mixed}} TMSV state but with different parameters:
\begin{align}
    \ket{\chi} \lossto \mathcal{L}_\eta (\ketbra{\chi}{\chi}) \NLAto \mathcal{L}_{\eta'} (\ketbra{\chi'}{\chi'}),
\end{align}
where $\mathcal{L}_\eta$ is the completely-positive trace-preserving (CPTP) map for the pure-loss channel with transmissivity $\eta$. The new parameters, $\chi'$ and $\eta'$, are related to the old parameters, $\chi$ and $\eta$, in the following way:~\cite{Blandino2012CV-QKD, PhysRevA.89.023846, PhysRevA.91.062305}
\begin{align}
    \chi' &= \sqrt{1-\eta+\eta g^2} \chi,\\
    \eta' &= \frac{g^2}{1-\eta+\eta g^2} \eta.
\end{align}
The effective loss is zero when $\eta'=1$ which for finite gain is only possible when $\eta=1$. This means that it is impossible to completely correct loss using noiseless linear amplification. Also, as $g\to\infty$, the success probability tends to zero.

Entanglement swapping is another technique that can distil entanglement; however, it can be thought of as noiseless linear amplification with some loss distributed across the measurement~\cite{winnel2021overcoming,PhysRevLett.128.160501}. Thus, entanglement swapping is also unable to completely purify CV entanglement. Other distillation techniques such as photon subtraction are also strictly limited in their purification abilities~\cite{subtraction2010}.

Noiseless linear amplification is nondeterministic. Ideal noiseless linear amplification is unphysical since the success probability is zero. Physical noiseless linear amplification is possible with a non-zero success probability by considering an energy cutoff. For instance, consider noiseless linear amplification which amplifies and truncates the input state at Fock number $d-1$ (where $d$ is the dimension of the output state). Then, the success probability scales like $\eta^{d-1}$. Our purification scheme uses the entanglement between multiple modes so larger states can be protected against loss with a much higher success probability than noiseless linear amplification. Our success probability depends on the number of photons in the code, not the dimension of the state that is encoded. Our single-shot purification scheme has the improved success probability $\eta^{k}$ where in general $k\leq d-1$. Furthermore, purification removes completely the effect of pure loss since errors are detected, while noiseless linear amplification is unable to do this.

To summarise, noiseless linear amplification and entanglement swapping (i.e., techniques for the first and second generations of quantum repeaters) fail to saturate the ultimate rate limits under pure loss for two reasons: 1. the output entangled states are strictly non-pure for any amount of loss and finite gain, and 2. the success probability depends on the energy cutoff and tends to zero for large input states.

The third generation of quantum repeaters~\cite{surfact_code_2010} uses quantum error correction~\cite{gottesman2009introduction} and is a purely one-way communication scheme. It promises high rates since it does not require back-and-forth classical signalling; however, here the ultimate rates are bounded by the unassisted quantum capacity of each pure-loss link~\cite{PhysRevLett.98.130501,PhysRevA.86.062306,PhysRevA.95.012339,Wilde_2018}, $  \log_2{( \frac{\eta}{1-\eta} )} <  C$. This means the third-generation rate is zero if $\eta \leq 0.5$, which translates to a maximum link distance of about 15 km for optical fibre with a loss rate of 0.2 dB/km. In contrast, the two-way assisted capacity of pure loss allows a nonzero achievable rate at all distances. It is interesting to note that the family of GKP codes~\cite{GKP2001} achieves the unassisted capacity of general Gaussian thermal-loss channels with added thermal noise, where pure loss is the zero-temperature case, up to at most a constant gap~\cite{GKP2019}. \bblack{Likewise, our main result is to give a practical protocol that achieves the two-way assisted capacity of the pure-loss channel.}

In summary, all three generations of quantum repeaters are unable to operate at rates that saturate the ultimate limits of quantum communications. Motivated by this reality, we introduce an iterative protocol to \bblack{purify completely} from pure loss and achieve the capacity of the channel. Our schemes are inspired by Refs.~\cite{Bennett_1996,PhysRevLett.84.4002}. \bblack{The idea is that neighbouring nodes locally perform photon-number measurements on copies of shared CV entanglement across the lossy channel, followed by \textit{two-way} classical communication to compare photon-number outcomes.} We show that the highest rates of our purification scheme, requiring two-way classical communication, achieve the fundamental limits of quantum communications for pure-loss channels. In contrast to quantum error correction, we describe purification as a quantum error detection scheme against loss. {We consider a much-simpler design with good rates requiring only \textit{one-way} classical communication and no iteration.}

In this chapter, the required measurements are quantum non-demolition measurements (\hypertarget{hyperlinklabel_QND}{\hyperlink{hyperlinkbacklabel_QND}{QND}}) of total photon number of multiple modes and can be implemented experimentally using \bblack{linear optics and photon-number measurements}. This implementation is naturally robust against the inefficiencies of the detectors and gates. Alternatively, these QND measurements can be implemented using high finesse cavities and cross–Kerr nonlinearities~\cite{PhysRevLett.84.4002}.

Now we present our results. First, we introduce our iterative protocol for the complete purification of high-dimensional entanglement, saturating the two-way assisted capacity of the bosonic pure-loss channel. Then, we \bblack{show that our} protocol without iteration \bblack{(i.e. single-shot)} still gives high rates which fall short of the ultimate limits by at worst a factor of $0.24$. Finally, we explain how to implement our protocol using \bblack{linear-optics and photon-number measurements}.

\section{Iterative purification} 

\bblack{Alice and Bob share copies of a state which is entangled in photon number, such that in a lossless situation they will always measure the same number of photons. Our purification technique, in a pure loss situation, is for Alice and Bob to each locally count the total number of photons contained in multiple copies of the shared entangled states, and then compare the results. If they locally find a different total number of photons, this means photons were lost. Alice and Bob then iteratively perform total photon number measurements over smaller subsets of states until their outcomes are the same, and hence distil pure entanglement.} We prove in Ref.~\cite{https://doi.org/10.48550/arxiv.2203.13924} that the highest average rates of the protocol achieve the capacity of the pure-loss channel.

\begin{figure}
\includegraphics[width=0.5\linewidth]{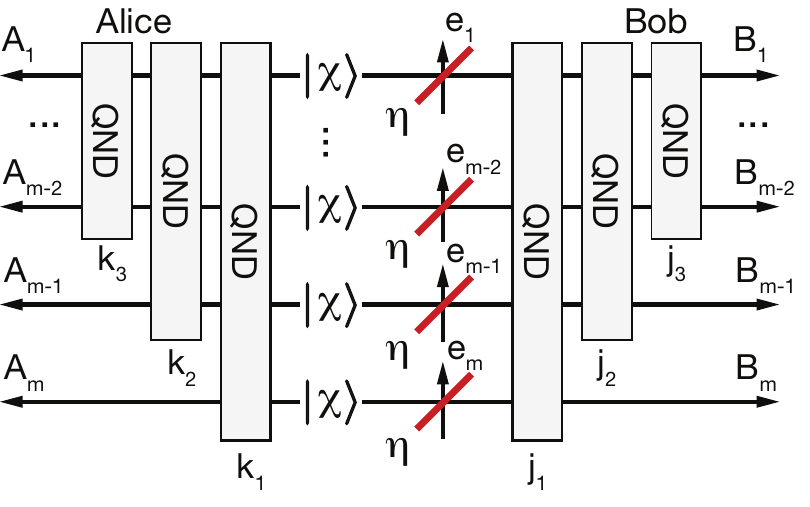}
\centering
\caption[Our iterative protocol for the complete purification of Gaussian continuous-variable quantum states.]{{Our iterative protocol for the complete purification of Gaussian continuous-variable quantum states}~\cite{PhysRevLett.84.4002}. Alice shares $m$ two-mode squeezed vacuum states with Bob across independent pure-loss channels with transmissivity $\eta$. \bblack{Iterative QND measurements of the total photon number at Alice's and Bob's sides followed by classical communication herald pure states whenever $k_n=j_n$, which means the $n$th round is successful and the protocol is complete. The highest average rate of quantum communication achieves the capacity (the repeaterless PLOB bound~\cite{Pirandola_2017}).} Alice's measurements encode the quantum information onto the rails, and Bob's measurements purify the entanglement and decode the quantum information.}
\label{PURfig:QND}
\end{figure}

We now consider our protocol in detail. The protocol is shown in~\cref{PURfig:QND}. Consider two neighbouring nodes in a network, Alice and Bob, separated by a repeaterless link. \bblack{Round one of our iterative protocol is identical to entanglement purification of} Gaussian CV quantum states from Ref.~\cite{PhysRevLett.84.4002}; however, our protocol includes an iterative procedure.  Alice prepares $m$ copies of a pure two-mode squeezed vacuum (TMSV) state, $\ket{\chi}=\sqrt{1-\chi^2} \sum_{n=0}^{\infty} \chi^n \ket{n}\ket{n}$ in the Fock photon-number basis, with squeezing parameter $\chi \in [0,1]$. \bblack{The unique entanglement measure, $E$, for a bipartite pure state, $\ket{\phi}$, is given by the von Neumann entropy, $S$, of the reduced state, i.e., $E =-\tr{(\rho_A \log_2{\rho_A})}$, where $\rho_A=\tr_B{\ketbra{\phi}{\phi}}$. This means Alice initially prepares $m E_\chi$ ebits of entanglement, where $E_\chi = G[(\lambda_k{-}1)/2]$, where $G(x)=(x{+}1)\log_2(x{+}1){-}x\log_2(x)$, $\lambda_k=2\bar{n}{+}1$, $\bar{n} = \sinh(r)^2$, and $r = \tanh^{-1}{\chi}$.}

Alice shares the second mode of each of the $m$ pairs with Bob across the link.  The error channel we consider is bosonic pure loss, modelled by mixing the data rails with vacuum modes of the environment, or a potential eavesdropper (Eve), on a beamsplitter with transmissivity $\eta$.

Alice encodes the quantum information into a quantum-error-detecting code so that Bob can detect errors on his side. To do this, she performs a QND measurement of the total photon number on the $m$ modes and obtains outcome $k_1$ \bblack{(where the subscript $1$ refers to the round one of iteration)}, and shares this information with Bob via classical communication. Alice's measurement projects the system before the channel onto a maximally-entangled state~\cite{PhysRevLett.84.4002}
\begin{align}
    \ket{\phi_{k_1,m}}_{AB} &\propto (1-\chi^2)^\frac{m}{2} \chi^{k_1} \stackrel{n_{1}+n_{2}+\cdots +n_{m}=k_1}{\mathrel{\mathop{\sum }\limits_{n_{1},n_{2},\cdots ,n_{m}}}}   \ket{ n_{1},n_{2},\cdots ,n_{m}}_{A} \ket{ n_{1},n_{2},\cdots ,n_{m}}_{B} \\
    &\propto (1{-}\chi^2)^\frac{m}{2} \chi^{k_1}    \sum_{\mu=0}^{d_{k_1,m}{-}1}   \ket{ \mu_{k_1,m}}_A \ket{ \mu_{k_1,m}}_{B},
\end{align} where $ \ket{\mu_{k_1,m}} = \ket{n_1^{(\mu)},n_2^{(\mu)},\cdots,n_m^{(\mu)}}$ can be viewed as orthogonal basis states which form a quantum-error-detecting code, each composed of $\sum_{i=1}^m n_i^{(\mu)}=k_1$ photons. We discuss the code in detail later. The pure maximally-entangled state $\ket{\phi_{k_1,m}}_{AB}$ has entanglement $E_{k_1,m}=\log_2{d_{k_1,m}}$ ebits with dimension $d_{k_1,m} = {k_1+m-1 \choose k_1}$, and the probability of Alice's measurement outcome, $k_1$, is $P_{k_1,m}^{\text{Alice}} = (1-\chi^2)^m \chi^{2 k_1} d_{k_1,m}$. The dimension, $d_{k_1,m}$, is the total number of ways $k_1$ identical photons can be arranged among the $m$ distinct rails.

Bob then performs a QND measurement of total photon number across the $m$ rails on his side to detect loss errors. If Bob obtains the outcome $j_1$ photons and knows that Alice sent $k_1$ photons, then both Alice and Bob know that $k_1-j_1$ photons were lost. Bob's QND measurement, with success probability $P_{k_1,j_1}^\text{Bob}=(1{-}\eta)^{k_1{-}j_1} \eta^{j_1} {k_1 \choose j_1}$, heralds a renormalised mixed state shared between Alice and Bob which does not depend on loss for any outcome $j_1$. That is, together Alice's and Bob's QND measurements remove all dependence on loss and have exchanged success probability for entanglement, while they learn that $k_1-j_1$ photons were lost to the environment. All outcomes besides zero at Alice and Bob herald useful entanglement. If $j_1\neq k_1$, they must do further rounds of purification (iteration) since the output state is not pure. If $j_1=k_1$, then the output state is strictly pure and purification is complete in a single round. For a simpler protocol, Alice and Bob may post-select on outcomes $j_1=k_1$ without further iteration. We show later in the chapter that this single-shot protocol still gives excellent rates.

Explicitly, the global output state heralded by outcomes $k_1$ and $j_1$ is
\begin{align}
&\ket{\phi_{k_1,j_1,m}}_{ ABe }  \propto  (1-\chi^2)^{\frac{m}{2}} \chi^{k_1} (1-\eta)^\frac{k_1-j_1}{2} \eta^\frac{j_1}{2}  \stackrel{n_{1}+n_{2}+\cdots +n_{m}=k_1}{\mathrel{\mathop{\sum }\limits_{n_{1},n_{2},\cdots ,n_{m}}}} \;  \nonumber \stackrel{l_1+l_2+\cdots+l_m=k_1-j_1,\;\; l_i\leq n_i \forall i}{\mathrel{\mathop{\sum }\limits_{l_{1},l_{2},\cdots ,l_{m}}}} \;  \nonumber \\ &\;\;\;\;\; \sqrt{{n_1 \choose l_1}  {{n_2 \choose l_2}} {{n_3 \choose l_3}} \cdots {{n_m \choose l_m}}} \;   \ket{ n_{1},n_{2},\cdots ,n_{m}} _{A} \ket{ n_{1}-l_{1},n_{2}-l_{2},\cdots ,n_{m}-l_{m}}_{
B }   \ket{ l_{1},l_{2},\cdots ,l_{m}}_{e },\label{PUReq:phi_jk}
\end{align}
where $A,B,e$ refer to the $m$-rail quantum systems owned by Alice, Bob, and the environment, respectively, as shown in~\cref{PURfig:QND}. The full derivation of this state is in~\cref{app:derivation_states}. The factor $(1-\eta)^\frac{k_1-j_1}{2} \eta^\frac{j_1}{2}$ is outside the sum; thus, we have the remarkable result that the renormalised output state shared between Alice and Bob does not depend on $\eta$. Therefore, the entanglement shared between Alice and Bob also has no dependence on $\eta$, which has been exchanged for probabilities.

\bblack{Additional rounds of purification can purify more entanglement after the initial round. One approach is for Alice and Bob to locally perform QND measurements as in round one but on $m-n+1$ rails, and obtain outcomes $k_n,j_n$, where $n$ refers to the round number. At round $n$, there is no entanglement shared between Alice and Bob on the last $n-1$ rails and the photon number of each of these rails is completely known. At round $n$, these last $n-1$ rails can be discarded while the $m-n+1$ rails should be kept.
}

\subsection{Exact rate of iterative entanglement purification for finite numbers of rails}

The rate of our purification protocol (in ebits per use) is maximised if Alice performs her first measurement offline (i.e., setting $P_{k_1,m}^{\text{Alice}}=1$), where she obtains outcome $k_1$. For finite $m$, there is a finite $k_1$ which optimises the rate. {However, for large squeezing $\chi\to1$ outcomes $k_1$ are dominated by $k_1\to\infty$ with unit probability. Therefore, the large squeezing limit $\chi\to\infty$ without $k_1$ pre-selection is equivalent to $k_1\to\infty$ with offline $k_1$ pre-selection.}

{Taking Alice’s first measurement to be done with result $k_1$ offline (which can be chosen in advance to optimise the rate or, for example, the practicality of the protocol)}, the rate for finite $m$ is
\begin{align}
    E_{k_1,m}(\eta) &= \frac{1}{m} \sum_{n=1}^{m-1} \sum_{j_1,k_2,j_2,\dots,k_n,j_n} P E, \label{PUReq:exact_rate_finite_m}
\end{align}
where the sum is constrained by
\begin{align}
    &k_1 \geq k_2 \geq k_3 \geq \dots \geq k_n,\\
    &j_1 \geq j_2 \geq j_3 \geq \dots \geq j_n,\\
    &k_s-k_{s+1} \geq j_s-j_{s+1} \; \forall s ,\\
    &k_s > j_s \; \forall s \neq n, \\
    &k_n=j_n,
\end{align}
where the probability of success for a particular combination of outcomes, $j_1,k_2,j_2,\dots,k_n,j_n$, for a given $k_1$ and $m$ is
\begin{align}
      P &= (1-\eta)^{k_1-j_1} \eta^{j_1}   \frac{ {{k_n+m-n} \choose k_n} {k_n \choose j_n} }{ {k_1+m-1 \choose k_1} }  \left[ \prod_{s=1}^{n-1} {{k_s-k_{s+1}} \choose {j_s-j_{s+1}}} \right] ,\label{PUReq:probability}
\end{align}
where a maximally-entangled state is generated with entanglement
\begin{align}
      E &= \log_2{\left[ { {{k_n+m-n} \choose k_n} } \right]}.\label{PUReq:entanglement}
\end{align}
The rate $E_{k_1,m}(\eta)$ for finite $m$ can be optimised over Alice's initial outcome $k_1$ prepared offline and the number of rails $m$ as a function of $\eta$. We numerically compute the rate for small $k_1$ and $m$.

The rate of our purification protocol (in ebits per use) is maximised if Alice performs her first measurement offline, where she obtains outcome $k_1$. For finite $m$, there is an optimal choice of $k_1$.

\begin{figure}
\centering
\includegraphics[width=0.5\linewidth]{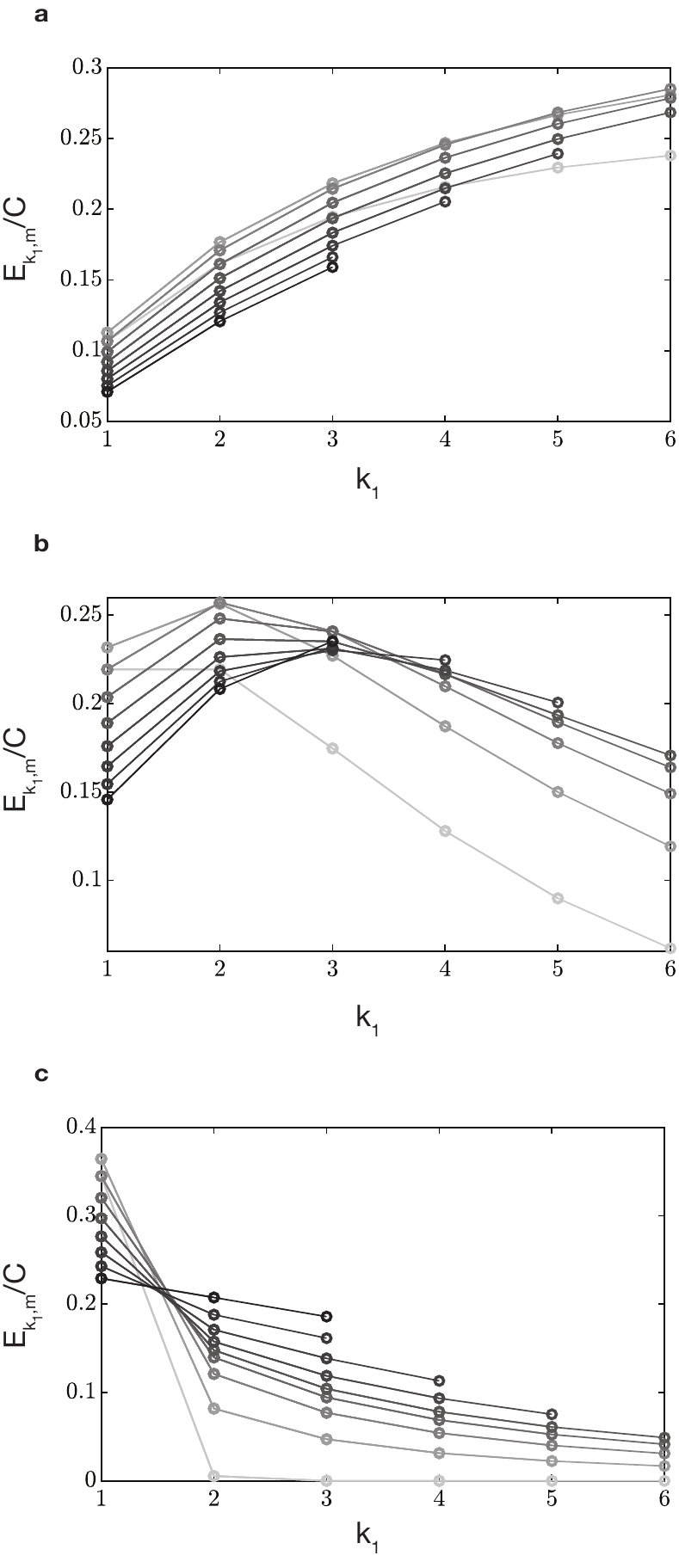}
\caption[Ratio of our finite $m$ rate, $E_{k_1,m}$ with the two-way assisted quantum capacity.]{Ratio of our finite $m$ rate, $E_{k_1,m}$ with the two-way assisted quantum capacity, $C$, (repeaterless PLOB bound)~\cite{Pirandola_2017} as a function of $k_1$ at Alice at a fixed distance of \textbf{a} 1 km, \textbf{b} 10 km, and \textbf{c} 100 km. The number of rails ranges from $m=2$ (light grey) to $m=9$ (black). The figure shows that for finite $m$ there is an optimal $k_1$ at a particular distance. Unfortunately, we cannot simulate large numbers of rails because we have to sum over many measurement outcomes. Analytically, we prove that $E_{k_1,m}/C$ $\to 1 $ as $m\to\infty$ and $k_1\to \infty$.\label{fig:finite_m_rate}}
\end{figure}

\bblack{
We numerically compute the rate for finite $m$ (and include up to $n=m-1$ rounds in the sum shown by~\cref{PUReq:exact_rate_finite_m}). We plot it in~\cref{fig:finite_m_rate}. Iteration becomes useful for $k_1>1$. Even for small $k_1$, the rates are quite close to the repeaterless bound.
}

We show next that the highest rate of our iterative protocol achieves the capacity, $C$, for $m\to\infty$ and $k_1\to\infty$ (i.e., $\chi\to1$). We show this without having to compute~\cref{PUReq:exact_rate_finite_m} directly which would be arduous.

\subsection{Optimality of our protocol} 

The RCI~\cite{PhysRevLett.102.050503,PhysRevLett.102.210501}, $R$, gives an achievable lower bound on the distillable entanglement, $E_{\text{D}}$, and on the optimal secret key rate. The RCI is defined in~\cref{sec:RCI}. We will show that our protocol is optimal for entanglement distillation as $m\to\infty$ and $\chi\to1$ and that no entanglement is lost {between rounds}. {We use the RCI as a benchmark to test the quality of our distillation procedure. The optimal distillation protocol implicit by the RCI is not required here since our scheme gives the same performance as the implicit optimal protocol round after round for large $m$.}

We require that the {weighted} average von Neumann entropy of the {reduced} pure states heralded after round one, $S_1$,  plus the average RCI of the failure states heralded after round one, $F_1$, equals the RCI of the state before round one. Note that the RCI equals the von Neumann entropy for pure states. We have
\begin{align}
    S_1+F_1 &= \frac{1}{m} \sum_{k_1=0}^\infty \sum_{j_1=0}^{k_1} P_{k_1,j_1,m} R_{k_1,j_1,m},\label{PUReq:sum}
\end{align}
in units of ebits per channel use, {where
\begin{align}
    S_1 &= \frac{1}{m} \sum_{k_1=0}^\infty  P_{k_1,j_1=k_1,m} R_{k_1,j_1=k_1,m},\\
    F_1 &= \frac{1}{m} \sum_{k_1=0}^\infty \sum_{j_1=0}^{k_1-1} P_{k_1,j_1,m} R_{k_1,j_1,m},
\end{align}
}where $P_{k_1,j_1,m}  {=} P_{k_1,m}^{\text{Alice}}  P_{k_1,j_1}^\text{Bob}$, where $P_{k_1,m}^{\text{Alice}}=(1{-}\chi^2)^m \chi^{2k_1} d_{k_1,m}$ and $P_{k_1,j_1}^\text{Bob}=(1{-}\eta)^{k_1{-}j_1} \eta^{j_1} {k_1 \choose j_1}$. 
{When the first round succeeds, the entanglement of the renormalised maximally-entangled pure-state shared between Alice and Bob heralded by outcomes $k_1=j_1$ is given by the von Neumann entropy of the reduced state:
\begin{align}
     E_{k_1=j_1,m} &= R_{k_1,j_1=k_1,m} =  \log_2{\left[ { {k_1+m-1 \choose k_1} } \right]},\label{PUReq:E_kj}
\end{align}
}which does not depend on the transmissivity, $\eta$, nor the amount of two-mode squeezing, $\chi$. When the first round fails, the RCI of the renormalised mixed state shared between Alice and Bob heralded by each pair of outcomes $k_1$ and $j_1$ is
\begin{align}
    R_{k_1,j_1,m} =  \log_2{\left[ \frac{ {k_1+m-1 \choose k_1} }{ {k_1-j_1+m-1 \choose k_1-j_1} } \right]},\label{PUReq:E_kj}
\end{align}
which also does not depend on the transmissivity, $\eta$, nor the amount of two-mode squeezing, $\chi$. See~\cref{app:derivation_states} for the derivation.

{The amount of squeezing is scaled to infinity $\chi\to1$, such that Alice initially measures a large amount of photons $k_1\to\infty$ and $m/k_1\to0$. Furthermore, from the fact that $P_{k_1,j_1}^\text{Bob}$ is a binomial distribution, Bob will most likely measure $j_1 \approx \eta k_1$ photons. Using these conditions, we checked numerically and prove in Ref.~\cite{https://doi.org/10.48550/arxiv.2203.13924} that~\cref{PUReq:sum} approaches}
\begin{align}
   \lim_{\chi\to1} (S_1+F_1) 
   &= - \frac{m-1}{m} \log_2{(1-\eta)} = \frac{m-1}{m} C,
\end{align}
\bblack{which ensures that round one of purification is optimal since there is no loss of rate after round one as $m\to\infty$, given that the average RCI at the end of the round equals the initial RCI of the protocol. Thus, there exists an optimal protocol to follow round one which can saturate the two-way assisted quantum capacity (the repeaterless bound) using our protocol as an initial step.} The entanglement is \textit{optimally} exchanged for success probability.

\bblack{Similarly, we proved numerically and in Ref.~\cite{https://doi.org/10.48550/arxiv.2203.13924} that round $n$ is optimal since there is no loss of rate at round $n$ as $m\to\infty$, up to the same factor, $\frac{m-1}{m}$. This factor comes from Alice's and Bob's measurements of photon number.}

To quantify this loss of entanglement, consider the protocol before the channel. We'll see that for finite $m$ some entanglement is immediately lost after Alice's QND measurement. {Ref.~\cite{PhysRevLett.84.4002} defined the entanglement ratio, denoted by $\Gamma_{1}$}, as the average entanglement heralded by Alice's QND measurement divided by the total initial entanglement $m E_\chi$, that is,
\begin{align}
    \Gamma_{1} &{\equiv} \frac{\sum_{k_1=0}^\infty P_{k_1,m}^{\text{Alice}} E_{k_1,m}}{m E_\chi}.
\end{align}
In the limit of a large number of rails $\lim_{m\to\infty} \Gamma_1 = 1$ for all $0<\chi\leq1$, which means asymptotically Alice's QND measurement {heralds no loss of} entanglement. However, for finite $m$ in the limit of large squeezing, $\lim_{\chi\to1} \Gamma_1 = (m-1)/m$. So, for a finite number of rails $m$ some entanglement is lost since $1/2 < (m-1)/m <1$. This means we must take $m\to\infty$ to get the highest rates.

{Similarly, to quantify the entanglement lost at round $n>1$, we define the entanglement ratio, $\Gamma_{k_{n-1}}$, as the weighted average entanglement heralded at round $n$ over all outcomes $k_n$ normalised by the weighted entanglement heralded by outcome $k_{n-1}$ at the previous round, $n-1$, that is,}
\begin{align}
   {\Gamma_{k_{n-1}}} &{\equiv} {\frac{\sum_{k_n} P_{k_n} E_{k_n} }{ P_{k_{n-1}} E_{k_{n-1}}}}  =   \frac{ \sum_{k_n} d_{k_n} \log_2{d_{k_{n}}}}{d_{k_{n-1}}\log_2{d_{k_{n-1}}}},\label{PUReq:ratio_n}
\end{align}
{where $d_{k_n} =  {{k_n+m-n} \choose k_n} $ and $d_{k_{n-1}} =  {{k_{n-1}+m-n+1} \choose k_{n-1}} $. $P_{k_n}$ ($P_{k_{n-1}}$) and $E_{k_{n}}$ ($E_{k_{n-1}}$) are defined in~\cref{PUReq:probability,PUReq:entanglement}, for a given $k_n$ ($k_{n-1}$), and of course we can take $j_i=k_i$ for all $i$ since here we consider no loss channel. Many of the factors cancel giving the simple expression in~\cref{PUReq:ratio_n}.} Curiously, for large numbers of rails the amount of entanglement lost at round $n$ is the same as for round one, i.e., $\lim_{m\to\infty} {\Gamma_{k_{n-1}}} = (m-1)/m$ for \textit{all} $k_{n-1}$. {This result ensures that ``encoding'' into $k_n$ photons is asymptotically optimal throughout the entire duration of our iterative procedure up to the factor $(m-1)/m$, which approaches unity for large $m$.}

\subsection{Achieving the capacity} 

The average {rate in case of success} of pure entanglement distilled at round $n$ in ebits per use of the channel is
\begin{align}
    S_n = \frac{1}{m} \sum_{k_1,j_1,\dots,k_n,j_n}  P E\; {\delta_{k_n,j_n}},
\end{align}
and the average {rate in case of failure} of entanglement distilled at round $n$ in ebits per use of the channel is lower bounded by the average RCI
\begin{align}
    F_n = \frac{1}{m} \sum_{k_1,j_1,\dots,k_n,j_n} P R \; {(1-\delta_{k_n,j_n})},
\end{align}
where $P$ is the probability from~\cref{PUReq:probability} and $R =  \log_2{\left[ \frac{ {k_n+m-n \choose k_n} }{ {k_n-j_n+m-n \choose k_n-j_n} } \right] }$ is the RCI of the heralded states. {Note $\delta_{k_n,j_n}$ is the Kronecker delta function.} For the {rate in case of success}, the RCI equals the von Neumann entropy since the states are pure, $R=E$. The entanglement {in case of failure} at round $n$ will be purified at a later round.

\bblack{Since our protocol is optimal at round $n$ for $\chi\to 1$ and $m\to\infty$, we have the following expressions:
\begin{align}
    S_1 + F_1  &=   \frac{m-1}{m} C \\
    \lim_{m\to\infty} (S_{n} + F_{n} )&= \lim_{m\to\infty} \left( \frac{m-1}{m} F_{n-1}\right),\label{equalities}
\end{align}
for $2 \leq n \leq m$. We solve this system of equations by addition, and we find that the average {rate in of success} of {purification using our iterative procedure for}
$m\to\infty$ and $\chi\to1$ is
\begin{align}
    E_\text{iteration} &= \lim_{m\to\infty}  \sum_{n=1}^m S_n\\ 
    &=  \lim_{m\to\infty} \frac{m-1}{m} C - \lim_{m\to\infty} \sum_{n=1}^{m-1}  \frac{F_n}{m}\\
    &= C.
\end{align}
We achieve the capacity (repeaterless PLOB bound) in the limit of a large number of rails, where $\lim_{m\to\infty} \sum_{n=1}^{m-1}  \frac{F_n}{m} \to 0$ and $\lim_{m\to\infty} (m-1)/m\to1$. See Ref.~\cite{https://doi.org/10.48550/arxiv.2203.13924} for the proof.}

\subsection{Achievable rates of repeater networks}

\begin{figure}
\includegraphics[width=1\linewidth]{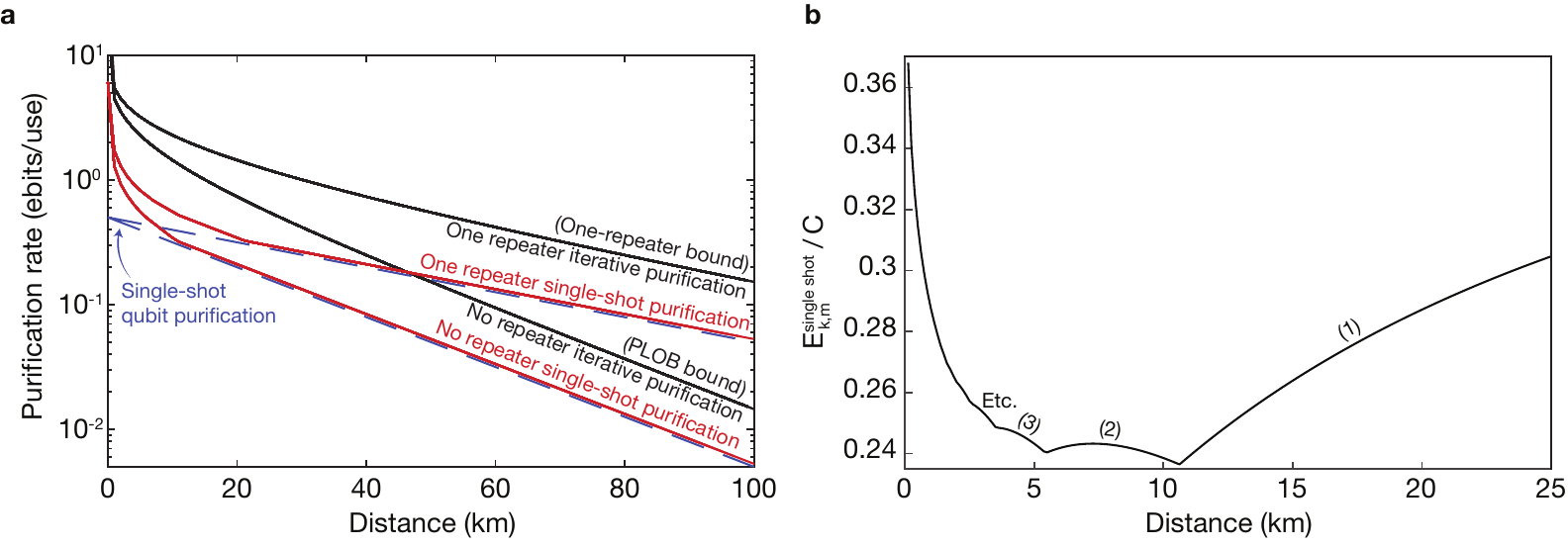}
\caption[Optimal rates of entanglement purification.]{{Optimal rates of entanglement purification.} a) We plot the highest rates of iterative purification (black), single-shot purification (red), and single-shot purification for qubits (blue dashed) as a function of the total end-to-end distance, for optical fibre with a loss rate of 0.2 dB/km. Purification is used to distribute and purify entanglement between nodes and entanglement swapping connects the end users. To illustrate, we plot rates without a repeater and with one repeater. Equivalently, these rates are secret key rates since ebits are specific types of secret bits~\cite{secure_key_2005}. Our highest rates coincide with the ultimate limits of quantum communications (black), \bblack{shown here the repeaterless bound~\cite{Pirandola_2017} and the one-repeater bound~\cite{Pirandola_2019}.} At all distances the output states are strictly pure. The scaling improves for increased numbers of repeaters (not shown). b) The ratio of the optimised rate of our single-shot purification protocol without a repeater with the repeaterless bound as a function of distance, showing how close it comes to saturating the bound (and therefore end-to-end quantum-repeater networks in general). The optimal number of photons, $k$, at each distance, is shown in rounded brackets. At shorter distances codes with more photons are optimal. The three-rail encoding ($m=3$) is optimal at most distances, with sometimes a larger $m$ optimal at distances less than about 1.6 km. At large distances, the ratio $E_{k,m}^\text{single shot}/C$ approaches $\ln{3}/3 \sim 0.366$. As the distance goes to zero, $E_{k,m}^\text{single shot}/C=1/2$.}
\label{PURfig:optimal_rate}
\end{figure}

Our protocol purifies completely at the repeaterless bound rate. Assuming ideal quantum memories are available, after teleportation (entanglement swapping) we achieve the ultimate end-to-end rates of quantum communication networks {by adopting the routing methods of Ref.~\cite{Pirandola_2019}. That is, the results can be extended beyond chains to consider more complex topologies and routing protocols~\cite{Pirandola_2019}.} We describe details about entanglement swapping in~\cref{Chap:math_background}. We plot the highest rates of iterative purification as a function of total distance with no repeater and one repeater in~\cref{PURfig:optimal_rate}a (black lines) which coincide with the capacities. We also plot rates in~\cref{PURfig:optimal_rate}a for single-shot purification for finite $m$ where Alice and Bob stop after the first round which is a much more practical design. We discuss in detail those single-shot rates next.

Note that for our iterative purification scheme, the dimension of the states will most likely be different so the nodes should link up the states with the same dimension. This is always possible if they have access to many copies held in quantum memories. Otherwise, if the states have different dimensions, then entanglement swapping is limited to the smallest dimension. Alternatively, the nodes could transfer all pure entanglement onto qubits before the swapping. Keeping the dimension the same (i.e., single-shot purification which we discuss next, rather than the iterative procedure) is more practical and then $d$-dimensional swapping is done on those states.

\section{Single-shot purification}
Purification is complete after a single round if no photons are lost and Alice and Bob detect the same number of photons, $j_1=k_1$. If Bob detects less photons than Alice, $j_1<k_1$, further purification is required; however, it is most practical to disregard those outcomes and keep only the outcome $j_1=k_1$ which purifies in one shot. Alice's measurement can be pre-selected and prepared offline; thus, improving the rate of a particular pair of outcomes. The single-shot rate for a given $m$ and $k\equiv k_1 = j_1$ can be quite close to the repeaterless bound rate. 

{Recall the optimal protocol implicit by the RCI is based on one-way classical communication~\cite{PhysRevLett.102.210501}, whereas, our iterative procedure requires two-way classical communication. Here, the single-shot protocol requires only one-way classical communication, like the implicit optimal protocol.} Alice and Bob agree on the quantum-error-detecting code ($k$ and $m$) in advance, so Alice does not need to send any classical information towards Bob. Bob only needs to send classical information towards Alice, telling her when the protocol succeeds. This greatly simplifies the required back-and-forth signalling in a quantum network.

\subsection{Highest achievable rate for single-shot purification against the bosonic pure-loss channel} 

The rate of single-shot purification is as follows. The probability that Alice obtains outcome $k$ is $P_{k,m}^{\text{Alice}}=(1-\chi^2)^m \chi^{2k} d_{k,m}$; however, for the single-shot protocol we can incorporate this step into Alice's state preparation and assume this is done offline such that $P_{k,m}^{\text{Alice}}=1$. This means the rate will not depend on $\chi$. When Bob obtains the outcome which matches Alice's, $k$, his probability is $P_{k}^\text{Bob} = \eta^k$ and the output state is pure with entanglement $E_{k,m} = \log_2{d_{k,m}}$. Thus, the achievable rate of single-shot entanglement purification in ebits per use of the channel is
\begin{align}
    E_{k,m}^\text{single shot} (\eta)&= \frac{{P_{k}^\text{Bob} E_{k,m}}}{m}  = \frac{\eta^k }{m} {\log_2{ {k+m-1 \choose k}}}.\label{PUReq:rate}
\end{align}
The rate is divided by the number of rails, $m$, to compare directly with the repeaterless bound~\cite{Pirandola_2017}. This is required because Alice and Bob exploit a quantum channel whose single use involves the simultaneous transmission of $m$ distinct systems in a generally entangled state.

This rate is for a perfect implementation without any additional losses, errors, or noise. The protocol heralds pure states in a single attempt, so if we have ideal quantum memories, after entanglement swapping (see~\cref{Chap:math_background} for details on entanglement swapping) we can distribute entanglement between ends of a network without any loss of rate, i.e., at the rate given by~\cref{PUReq:rate} where $\eta$ is the transmissivity of the most destructive link.

Our optimised rate over $k$ and $m$ is shown in~\cref{PURfig:optimal_rate}a, without and with one quantum repeater (with a repeater, we assume ideal quantum memories). The repeaterless bound can ultimately be broken at 46.3 km using single-shot purification between nodes and a single repeater for entanglement swapping. In~\cref{PURfig:optimal_rate}b, we show the ratio of our optimised single-shot rate with the repeaterless bound, showing that at long distances the protocol falls short of the repeaterless bound by just a factor of $\ln{3}/3 \sim 0.366$, and remarkably, our rate approaches $\frac{1}{2} C$ at short distances. At short distances, the rate is optimised for larger $k$ while at long distances $k=1$ is always optimal since the probability that photons arrive scales like $\eta^k$ at long distances. The optimal number of rails is about $m=3$ at most distances (including long distances) since increasing the number of rails decreases the rate like $1/m$. Larger $m$ is sometimes optimal at short distances.

The rate of our single-shot purification protocol is unable to saturate the ultimate limit because we postselect on Bob's outcomes, throwing away useful entanglement when $j\neq k$ and $j>0$. Keeping all measurement outcomes, the rate increases; however, it is less practical to do so.

\subsection{Quantum error detection} 

In this section, we describe our single-shot purification protocol as quantum error detection (\hypertarget{hyperlinklabel_QED}{\hyperlink{hyperlinkbacklabel_QED}{QED}}). Consider encoding an arbitrary finite-dimensional single-rail state with dimension $d_{k,m}$:
\begin{align}
    \ket{\psi_{{k,m}}} &=  \frac{1}{\sqrt{N_\psi}} \sum_{\mu=0}^{d_{k,m}-1} c_{\mu} \ket{\mu},\label{PUReq:psi}
\end{align}
where $N_\psi=\sum_\mu^{d_{k,m}-1} |c_\mu|^2$.

The code is a subspace with $d_{k,m}$ dimensions (\mbox{$d_{k,m}$-dimensional}), a subspace of the infinite-dimensional Hilbert space chosen to detect loss errors non-deterministically. It is represented by the projector onto the subspace
\begin{align}
    \mathcal{P}_{k,m} = \sum_{\mu=0}^{{d_{k,m}}-1} \ketbra{\mu_{k,m}}{\mu_{k,m}},
\end{align}
where $ \ket{ \mu_{k,m}}= \ket{n_1^{(\mu)}, n_2^{(\mu)}, n_3^{(\mu)},\cdots,n_m^{(\mu)}}$ are the orthogonal basis states (code words) which make up the code subspace (code space), where $\mu$ is the logical label. Note $n_i^{(\mu)}$ is the number of photons in the $i$th rail, which depends on the logical label $\mu$, as well as $k$ and $m$ where $\sum_{i=1}^m n_i^{(\mu)}=k$. There are $d_{k,m}$ code words, i.e., for a given code the set of code words is $\{ \ket{ \mu_{k,m}} \} = \{ \ket{0_{k,m}}, \ket{1_{k,m}},\ket{2_{k,m}},\cdots \}$. Quantum information up to dimension $d_{k,m}$ can faithfully be transmitted to Bob, conditioned that he detects no errors. Photon loss (and photon gains) will result in states outside of code space, which we can distinguish as an error.

The set of logical states forming the \mbox{$d_{k,m}$-dimensional} basis of the code consists of all possible ways $k$ identical photons can be arranged among the $m$ distinct rails. For example, $\ket{\mu_{k=2,m=3}} \in \{\ket{0,0,2},\ket{0,2,0},\ket{2,0,0},\ket{0,1,1},\ket{1,0,1},\ket{1,1,0}\}$. The code space is a subspace of the full Hilbert space of the $m$ rails which introduces the redundancy required for error detection, that is, $k$ photons and $m$ rails can encode \mbox{$d_{k,m}$-dimensional} states. The dimension grows rapidly with $k$ and $m$. For example, with just $k=4$ photons and $m=5$ rails, we can efficiently encode 70-dimensional states, i.e., truncated at Fock number $\ket{69}$, with success probability $P_{j=k=4}^\text{Bob}=\eta^k=\eta^4$. This is a great advantage over noiseless linear amplification, for example, if we choose the gain of the amplifier to be $g=1/\sqrt{\eta}$ to overcome the loss then the success probability is  $P_\text{NLA}=1/g^{2(d_{k,m}-1)} = \eta^{(d_{k,m}-1)}=\eta^{69}$, totally impractical. Furthermore, noiseless linear amplification fails to purify completely and cannot completely overcome the loss.

The encoding step is
\begin{align}
    \mathcal{S}&= \sum_{\mu=0}^{d_{k,m}-1}  \ket{\mu_{k,m}} \bra{\mu},
\end{align}
which maps Fock states, $\ket{\mu}$, from a single mode to the code words, $\ket{\mu_{k,m}}=\ket{n_1^{(\mu)},n_2^{(\mu)},\cdots,n_m^{(\mu)}}$. The combined operation of encoding, loss, and decoding is a completely-positive trace-non-increasing map
\begin{align}
    \mathcal{E} &= \mathcal{S}^{-1} \circ  \mathcal{L}^{\otimes m} \circ \mathcal{S}=\mathcal{E} = \eta^{k/2} \sum_{\mu=0}^{d_{k,m}-1} \ketbra{\mu}{\mu},\label{PUReq:map}
\end{align}
where $ \mathcal{L}^{\otimes m}$ is the map for independent applications of the pure-loss channel on the $m$ rails, the decoding step, $\mathcal{S}^{-1}$, performs a QND measurement of the total photon number, and if $k$ photons arrive, then it successfully decodes to single rail. The decoding step succeeds only if no photons are lost. The combined operation, $\mathcal{E}$, is a scaled identity map up to the $(d_{k,m}{-}1)$th Fock state; thus, the protocol succeeds with success probability $P_{k}^\text{Bob} = \tr{(\rho_{AB})} = \eta^k$ with unit fidelity.

The input state may be entangled with another mode. For example, we may consider encoding one arm of an arbitrary entangled state, $\ket{\psi_{k,m}} \propto  \sum_{\mu=0}^{d_{k,m}-1} c_\mu \ket{\mu}_A\ket{\mu}_B$. In this case, the operation, $\mathcal{E}$, acts on Bob's mode only and Alice leaves her mode alone. The final state shared between Alice and Bob is $\rho_{AB} = (\Id \otimes \mathcal{E}  ) \rho_{\text{in}}$, where $\rho_{\text{in}}$ is the initial state and $\Id$ is the identity on mode $A$. Note there is still useful entanglement if photons are lost. The entanglement-distribution rate of single-shot error detection for a maximally-entangled initial state $\ket{\phi_{k,m}}_{AB}=\frac{1}{\sqrt{{d_{k,m}-1}}} \sum_{\mu=0}^{d_{k,m}-1} \ket{\mu}_A\ket{\mu}_B$ is given by~\cref{PUReq:rate}, showing that error detection and purification indeed are equivalent.

One might consider using purification to distribute Gaussian entanglement for long-distance CV QKD between trusted end users of a network, performing the entanglement-based CV-QKD protocol based on homodyne detection~\cite{PhysRevA.63.052311,gottesman2001secure} or heterodyne detection~\cite{Weedbrook_2004}. Consider the initial data state to be a truncated TMSV state with dimension $d_{k,m}$ given by
\begin{align}
    \ket{\chi_{k,m}} &=  \sqrt{\frac{\chi^2-1}{\chi^{2{d_{k,m}}}-1}} \sum_{\mu=0}^{d_{k,m}-1} \chi^\mu \ket{\mu}_A \ket{\mu}_B.\label{PUReq:chi}
\end{align}
The state is Gaussian except for the hard truncation in Fock space. The protocol works as follows. First, truncated TMSV states, $\ket{\chi_{k,m}}$, are distributed using our single-shot purification scheme between all repeater nodes and held in quantum memories. Once successful, CV entanglement swapping is used to entangle the end users who use the entanglement to perform CV QKD. While the purification scheme can be complex (depending on the chosen protocol size), the CV entanglement swapping is simple. It works by performing dual-homodyne measurements on some of the modes, followed by conditional displacements, swapping the entanglement~\cite{PhysRevA.83.012319}, see~\cref{Chap:math_background} for details on how to compute the secret key rate.

\subsection{A simple example: the qubit code}

The simplest nontrivial code uses a single photon, $k=1$, in two rails, $m=2$, (i.e., unary dual-rail) and protects qubit systems, $d_{k=1,m=2}=2$, from loss. It is equivalent to the original purification protocol from Ref.~\cite{Bennett_1996}, but in this context, we use it to purify entanglement completely from pure loss to first order in Fock space and we can protect arbitrary single-rail qubit states from loss. The code words can be defined $\ket{{0}_{k=1,m=2}} \equiv \ket{0,1}$ and $\ket{{1}_{k=1,m=2}} \equiv \ket{1,0}$. The projector onto the code space is $\mathcal{P} =\ketbra{0,1}{0,1} + \ketbra{1,0}{1,0}$. The encoding step is
\begin{align}
    \mathcal{S} &= \ketbra{{0}_{k=1,m=2}}{0} + \ketbra{{1}_{k=1,m=2}}{1}\\
    &= \ketbra{0,1}{0} + \ketbra{1,0}{1},
\end{align}
which maps the vacuum component of the data mode onto a single photon of the second rail and the single-photon component of the data mode onto a single photon of the first rail. If either of these photons is lost, the protocol fails. The success probability is $P_{j=k=1}^\text{Bob}=\eta$. The maximum single-shot rate in ebits per use is $D_{k{=}1,m{=}2}^\text{single shot}(\eta)=\eta/2$, as shown by the dashed line in~\cref{PURfig:optimal_rate}a (with and without a repeater).

\section{Physical implementation}

Entanglement purification (iterative and single shot) requires joint QND measurements on multiple rails. These measurements can be performed via controlled-SUM (\hypertarget{hyperlinklabel_CSUM}{\hyperlink{hyperlinkbacklabel_CSUM}{CSUM}}) quantum gates and photon number-resolving measurements. Another technique is to use high finesse cavities and cross–Kerr nonlinearities~\cite{PhysRevLett.84.4002}. 

\bblack{More simply, our scheme can be implemented using beamsplitters and photon-number-resolving detectors; however, it also requires an entangled resource state, $\ket{\Omega_{k,m}}$. This is a common technique in linear optics~\cite{Yan:21,Yan2021}. See Ref.~\cite{Kok2007} for a review of quantum information processing using linear optics. The resource state can also be generated using linear optics and photon-number measurements. That is, we have a simple method of implementing our protocol, though at some cost to the success probability.}

\begin{figure}
\includegraphics[width=0.5\linewidth]{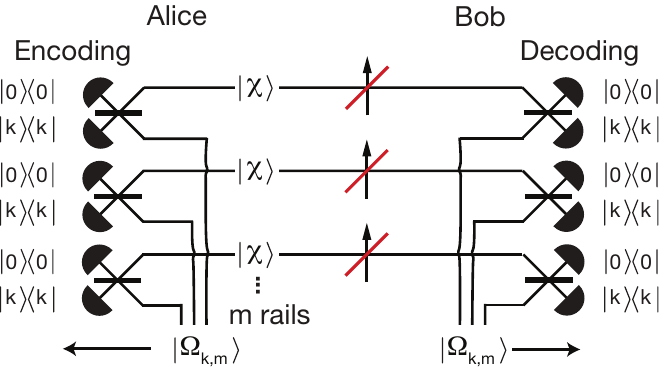}
\centering
\caption[Implementation using linear optics and number measurements.]{\bblack{{Implementation using linear optics and number measurements.}} The aim of the protocol is for Alice to share quantum information contained in a \mbox{$d_{k,m}$-dimensional} pure state, $\ket{\psi_{k,m}}$, which can be entangled with another system which Alice keeps, with Bob separated by a pure-loss channel. As shown in this figure, the aim is to distribute a pure maximally-entangled state between Alice and Bob. Alice shares $m$ entangled states, $\ket{\chi}$, with Bob. Encoding/decoding simply requires beamsplitters, photon detectors, and $(m+1)$-mode entangled resource states, $\ket{\Omega_{k,m}}$. We can assume Alice's encoding (the state heralded after Alice's measurement) is prepared offline so that the success probability of the protocol is the probability of Bob's side only. There is; however, a success-probability penalty for using linear optics, noting crucially that this penalty does not depend on the loss. The protocol succeeds when Bob detects the same number of photons Alice sent, $k$. If Bob detects less photons, multiple rounds of error detection are required or Alice and Bob can simply disregard those states.}
\label{PURfig:implementation}
\end{figure}

We focus on the single-shot linear-optics protocol, shown in~\cref{PURfig:implementation}. Alice shares $m$ entangled states, $\ket{\chi}$, with Bob\bblack{. Note each rail could have a different $\chi$, distilling different amounts of entanglement between Alice and Bob at the end of the protocol; however, maximal entanglement is heralded when all rails have the same value of $\chi$.} \bblack{Alice and Bob each prepare locally multimode resource states, $\ket{\Omega_{k,m}}$, consisting of $(m+1)$ modes. We assume they do this offline so it does not affect the quantum communication rate:}
\begin{align}
    \ket{\Omega_{k,m}} \propto \sum_{\mu=0}^{d_{k,m}-1} f_\mu \ket{\mu} \ket{{\mu_{\widetilde{k,m}}}},\label{PUReq:omega}
\end{align}
where $\ket{\mu}$ are Fock states, $\ket{{\mu_{\widetilde{k,m}}}}$ are ``anticorrelated'' code words. Writing the usual code words in the Fock basis as $\ket{{\mu_{k,m}}}= \ket{n_1^{(\mu)}, n_2^{(\mu)}, n_3^{(\mu)},\cdots,n_m^{(\mu)}}$, where $n_i^{(\mu)}$ is the number of photons in the $i$th rail, and where $\sum_{i=1}^m n_i^{(\mu)}=k$, recalling that the code space was defined as the set of all states with this property, then $\ket{{\mu_{\widetilde{k,m}}}} =  \ket{k{-}n_1^{(\mu)}, k{-}n_2^{(\mu)}, k{-}n_3^{(\mu)},\cdots,k{-}n_m^{(\mu)}}$. The coefficients, $f_n$, are
\begin{align}
     f_n &= \left [ {k \choose n_1^{(\mu)}} {k \choose n_2^{(\mu)}} \cdots {k \choose n_m^{(\mu)}} \right]^{-1/2}.\label{PUReq:fn}
\end{align} 

\bblack{The last $m$ modes of  \ket{\Omega_{k,m}} are fed into the beamsplitters with the distributed entanglement and are measured by photon-number detectors. The first mode is kept locally by the user and remains at the end of the protocol, as shown in ~\cref{PURfig:implementation}.}

Detecting $k$ photons means there were no loss events since $n_i^{(\mu)} + k - n_i^{(\mu)} = k$. That is, all photons are accounted for in the circuit and the output state is strictly pure. There may be useful entanglement for measurement outcomes other than $\ket{0,k}\bra{0,k}$ at each pair of detectors, and further purification (iteration) can increase the rate, but we do not consider it due to practicality.

Adjusting the amount of entanglement prepared for each rail, adjusting the loss on each rail, or selecting different coefficients in the resource state, $f_n$, results in a different output state, which may be useful for certain tasks. For example, if the entangled states prepared for each rail are identical and have the same amount of squeezing and $f_n$ is chosen as in~\cref{PUReq:fn}, then a maximally-entangled state will be heralded between Alice and Bob at the output. For another example, consider the dual-rail case ($m=2$). If the second rail is maximally entangled and $f_n$ is chosen as in~\cref{PUReq:fn}, then the output state is the initial state of the top rail. This tuning of the circuit parameters is useful, for instance, for CV QKD where the target state is a truncated TMSV state.

The linear-optics scheme detects all errors and outputs a pure state. There is; however, an additional success probability penalty using linear optics because of Bob's decoding measurement. We assume Alice prepares offline. Then, the success probability is
\begin{equation}
    P_{k,m}^\text{Bob linear optics} = \frac{ \eta^k}{2^{m(k-1)}\sum_{n=0}^{d-1} f_n^2},\label{PUReq:P}
\end{equation}
which depends on $m$. Compare this with the ideal purification protocol where $P_{k}^\text{Bob}=\eta^k$ which does not depend on $m$. The penalty paid for using linear optics is mainly due to the exponential $2^{m(k-1)}$ factor, which is painful for anything other than $k=1$, where $P_{k=1}^\text{Bob linear optics}=\eta/m$.

\subsection{Derivation of the linear-optics circuit}

In this section, we consider how to implement our protocol in an experimentally tractable platform. In this regard, we show that a linear-optical network, coupled with an entangled resource state, implements the required decoding action at Bob's side. We also derive the success probability of this decoding action.

First, consider our ideal single-shot error-detection protocol. Alice prepares the two-mode entangled state $\ket{\psi_{d_{k,m}}}_{A,A1}\propto  \sum_{\mu=0}^{d_{k,m}-1} c_\mu \ket{\mu}_A \ket{\mu}_{A1}$ and encodes the second mode $(A1)$ into our quantum-error-detecting code to send to Bob. Recall that the encoding step is
\begin{align}
    \mathcal{S}_{A1}&= \sum_{\mu=0}^{d_{k,m}-1}  \ket{\mu_{k,m}}_{1,2,3,\dots,m} \bra{\mu}_{A1},
\end{align}
where $\ket{\mu_{k,m}}_{1,2,3,\dots,m}$ are the $m$-rail code words of the code. 
Alice applies it to mode $A1$ of her initial state resulting in
\begin{align}
    (\Id_A \otimes \mathcal{S}_{A1}) \ket{\psi_{d_{k,m}}}_{A,A1}  &\propto  \sum_{\mu=0}^{d_{k,m}-1} c_\mu \ket{\mu}_A \ket{\mu_{k,m}}_{1,2,3,\dots,m}.
\end{align}
After the lossy channels, if Bob detects no loss errors then he successfully decodes, and Alice and Bob share the initial state $\ket{\psi_{d_{k,m}}}_{A,B}$ at the end of the protocol.

Now let us consider the linear-optics circuit, focusing on Bob's side since Alice can encode in a similar way to how Bob decodes since it is simply the complex conjugate transformation. Alice's encoding is assumed to be done offline and we assume that she has perfectly encoded the quantum information into the $m$-rail code, $(\Id_A \otimes \mathcal{S}_{A1}) \ket{\psi_{d_{k,m}}}_{A,A1}$, and then Bob performs the following action:
\begin{figure}[H]
\centering
\includegraphics[width=0.5\linewidth]{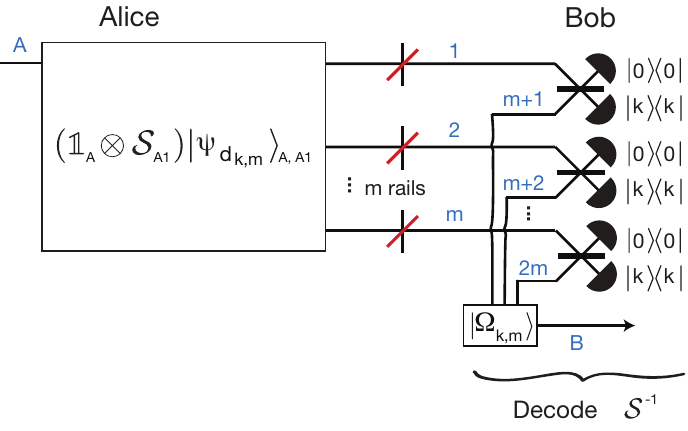}
\end{figure}
This decoding action can be written operationally as follows
\bblack{\begin{align}
    \ket{\psi_{d_{k,m}}} &\propto M_{1,m+1} M_{2,m+2} \cdots M_{m,2m} (\Id_A \otimes \mathcal{S}_{A1}) \ket{\psi_{d_{k,m}}}_{A,A1} \otimes \ket{\Omega_{k,m}}_{m+1,m+2,\dots,2m,B}\\
    &\propto M_{1,m+1} M_{2,m+2} \cdots M_{m,2m}  \sum_{\mu=0}^{d_{k,m}-1} c_\mu \ket{\mu}_A \ket{\mu_{k,m}}_{1,2,\dots,m} \ket{\Omega_{k,m}}_{m+1,m+2,\dots,2m,B}\;,\label{eq:lin_optics_state}
\end{align}
where $M_{u,v}$ are the required measurements on pairs of modes $u$ (sent from Alice through the channel) and $v$ (from Bob's resource state), and $\ket{\Omega_{k,m}}$ is the resource state at Bob's side. The measurements, $M_{u,v}$, consist of mixing modes $u$ and $v$ on a balanced beamsplitter and post-selecting $k$ photons at one output and vacuum at the other
\begin{align}
    M_{u,v} &\equiv \frac{1}{\sqrt{2^k}} \sum_{m=0}^k \sqrt{ k \choose m } \bra{m}_u \bra{k-m}_v.
\end{align}}
This expression can be derived by considering how $|0\rangle|k\rangle$ evolves through a 50:50 beamsplitter, then taking the complex conjugate (see Eq. S2 of~\cite{PhysRevLett.128.160501}). Recall that the code words in the Fock basis are $\ket{{\mu_{k,m}}}= \ket{n_1^{(\mu)}, n_2^{(\mu)}, n_3^{(\mu)},\cdots,n_m^{(\mu)}}$. Then we define our entanglement resource state as
\begin{align}
    \ket{\Omega_{k,m}}_{m+1,m+2,m+3,\dots,2m,B} \propto \sum_{\mu=0}^{d_{k,m}-1} f_\mu \ket{{\mu_{\widetilde{k,m}}}}_{m+1,m+2,m+3,\dots,2m} \ket{\mu}_{B},
\end{align}
where $\ket{{\mu_{\widetilde{k,m}}}} =  \ket{k{-}n_1^{(\mu)}, k{-}n_2^{(\mu)}, k{-}n_3^{(\mu)},\cdots,k{-}n_m^{(\mu)}}$, and $f_\mu$ is some free scaling factor which we will determine later.

Now, writing~\cref{eq:lin_optics_state} in full we get
\begin{align}
    \ket{\psi_{d_{k,m}}} &\propto \sum_{m_1=0}^k \sqrt{k \choose m_1} \bra{m_1} \bra{k-m_1}\; \sum_{m_2=0}^k \sqrt{k \choose m_2} \bra{m_2} \bra{k-m_2}\; \sum_{m_3=0}^k \sqrt{k \choose m_3} \bra{m_3} \bra{k-m_3}\; \cdots \nonumber \\ \;\;\;\;\; & \;\;\;\;\;\;\sum_{\mu=0}^{d_{k,m}-1}\sum_{\mu'=0}^{d_{k,m}-1} c_\mu f_{\mu'}  \ket{\mu}_A \ket{n_1^{(\mu)}, n_2^{(\mu)}, n_3^{(\mu)},\cdots,n_m^{(\mu)}} \ket{k{-}n_1^{(\mu')}, k{-}n_2^{(\mu')}, k{-}n_3^{(\mu')},\cdots,k{-}n_m^{(\mu')}} \ket{\mu'}_B .\label{eq:lin_optics_state_full}
\end{align}

\bblack{Here, modes labelled $A$ and $B$ are left alone, while the measurement operators act on pairs of modes, one from the channel and one from Bob's resource state.}

The only non-zero terms in~\cref{eq:lin_optics_state_full} are for $m_i=n_i^{(\mu)}$ and $k-m_i=k-n_i^{(\mu')}$, which means that $\mu=\mu'$. Thus, we have shown that these measurement actions and resource entangled state can perform the required decoding action
\begin{align}
    \ket{\psi_{d_{k,m}}}_{A,B} &\propto   \sum_{\mu=0}^{d_{k,m}-1} c_{\mu}f_{\mu}  \sqrt{  {k \choose n_1} {k \choose n_2} \cdots {k \choose n_m}  } \ket{\mu}_A \ket{\mu}_B \\
    &\propto \sum_{\mu=0}^{d_{k,m}-1} c_\mu \ket{\mu}_A \ket{\mu}_B,
\end{align}
in which we get back our original entangled state (but with one arm now at Bob). However, this requires that we set our resource state to have the following factors
\begin{align}
    f_{\mu} &= \left[ {k \choose n_1} {k \choose n_2} \cdots{k \choose n_m} \right]^{-1/2}.
\end{align}

Finally, we note that the success probability of this action, Bob's measurements after loss characterised by transmissivity $\eta$, is given by
\begin{equation}
    P_{k,m}^\text{Bob linear optics} = \frac{\eta^{k}}{2^{m(k-1)}S_{k,m}}\;.\label{eq:lin_optics_P}
\end{equation}
We have an $1/2^{m(k-1)}$ factor from the $m$ beamsplitter measurements $M$, as well as a factor from the normalised resource state
\begin{align}
    S_{k,m} := \left[ \sum_{\mu=0}^{d_{k,m}-1} f_\mu^2 \right].
\end{align}
The final rate of entanglement purification depends on the probability of Bob's side but not Alice's side since her encoding step is assumed to have been prepared offline:
\begin{align}
    E_{k,m}^\text{single shot linear optics} (\eta)&= \frac{1}{m} P_{k,m}^\text{Bob linear optics} E_{k,m}\\
    &= \frac{1}{m} P_{k,m}^\text{Bob linear optics} {\log_2{ {{k+m-1} \choose k}}},
\end{align}
which for $k=1$ is
\begin{align}
    E_{k=1,m}^\text{single shot linear optics} (\eta)&= \frac{\eta }{m^2} {\log_2{ {m \choose k}}}.
\end{align}

The linear-optics circuit leads naturally to a controlled-SUM gate using linear optics \bblack{and number measurements}, albeit with distorted coefficients (this distortion ultimately has no effect in our protocol since we immediately measure the state).

Since Alice's encoded state is prepared offline, it is useful to consider more generally that she encodes an arbitrary state into the code: $(\Id \otimes \mathcal{S}) \ket{\psi_{k,m}}$.

\subsection{Preparation of resource states} 

For our \bblack{linear-optics and number measurement} circuit, the resource state, $\ket{\Omega_{k,m}}$, and Alice's encoded state she prepares offline, are multi-mode entangled states. Once these states are prepared, our scheme requires just beamsplitters and photon detectors. One practical way to prepare these states is to use a Gaussian Boson Sampler (GBS)~\cite{PhysRevLett.119.170501} and post selecting on a specific photon-number-resolving measurement click pattern on some of the modes of the output~\cite{PhysRevA.100.052301,PhysRevA.100.022341}. This allows our scheme to be implemented entirely using linear-optics \bblack{and number measurements}. Using the GBS method for the simplest scheme with $k=1$ and $m=2$, we have found the resource state, $\ket{\Omega_{k{=}1,m{=}2}} = (\ket{0,1,0} + \ket{1,0,1}) /\sqrt{2}$, can be prepared with high fidelity, $F>0.999$, with success probability $\approx 10^{-6}$. This was found by optimising the parameters of a GBS network via a machine learning algorithm called ``basin hopping''~\cite{PhysRevA.100.012326}. In~\cite{Josh_github}, we have provided our code that implements this algorithm, as well as the parameter set we found that generates this resource state. Alternatively, adaptive phase measurements~\cite{Lund_2005} can be used to prepare the needed resource states directly from dual-rail Bell pairs or GHZ-like states which may have a higher probability of success.

\subsection{Experimental imperfections} 

Our linear-optics circuit is robust to loss. The quality of the state Alice sends can be managed since her encoding is done offline. In any case, we are interested in the noise introduced by the protocol, not the noise in the initial state we are trying to transmit. The measurement and detection scheme at Bob's side is such that all photons are accounted for. This means if Alice's encoding is perfect, the protocol can correctly identify if any photons are lost in the channel or at the detectors. 

\bblack{Next, we perform a numerical simulation incorporating inefficient detectors with dark counts and a thermal-noise channel. Our protocol is robust to practical values of these imperfections.}

Our purification technique is robust to additional losses such as using inefficient detectors. If we consider that Alice has done her encoding offline and has a perfect state preparation, then additional losses at Bob's side, at the resource states and the detectors, affect the success probability of Bob's side but not the purity of the output state. Therefore, the circuit is robust to both losses in the channel and at the resource state and detectors on Bob's side. All photons must arrive at Bob's detectors for the device to succeed, otherwise, it fails.

To show this, recall that the output state of the circuit after Bob's measurement is given by~\cref{eq:lin_optics_state_full} and we must have that $m_i=n_i^{(\mu)}$ and $k-m_i=k-n_i^{(\mu')}$, which means that $\mu=\mu'$. Additional losses anywhere on the $m$ rails (i.e., in the channel, at the detectors, or resource state) will decrease the success probability since it is less likely that $m_i=n_i^{(\mu)}$ and $k-m_i=k-n_i^{(\mu')}$ for all $i$, but it will not decrease the purity when the protocol succeeds. Losses on the modes Alice and Bob keep do affect the purity of course.

Interestingly, if the losses are not the same on all the rails, the output state is still pure, but the amount of entanglement is increased or decreased depending on the ratio of the losses on each rail.

While the linear-optics circuit is robust to losses at Bob's detectors and resource states, unfortunately, dark counts, thermal noise, and the use of bucket detectors will in general decrease the purity and fidelity of the output state; however, numerical simulations suggest that practical values still result in good rates. We analyse these effects next.

Here, we analyse the implementation of our single-shot protocol for $k=1$ and $m=2$ using linear optics and photon-number measurements under a thermal-loss channel and with inefficient detectors and dark counts. These imperfections mostly affect the high-loss regime. The linear-optics probability is dominated by the $2^{m(k-1)}$ factor which favours $k=1$. $k=1$ is usually optimal, furthermore, $k=1$ is always optimal in the high-loss regime. So, we only consider $k=1$.

Recall that the rate of the $k=1$ protocol using linear optics and photon-number measurements for a pure loss channel and no imperfections is
\begin{align}
    E_{k=1,m}^\text{single shot linear optics} (\eta)&= \frac{\eta }{m^2} {\log_2{ {m \choose k}}}.
\end{align}

\begin{figure}
\centering
\includegraphics[width=0.6\linewidth]{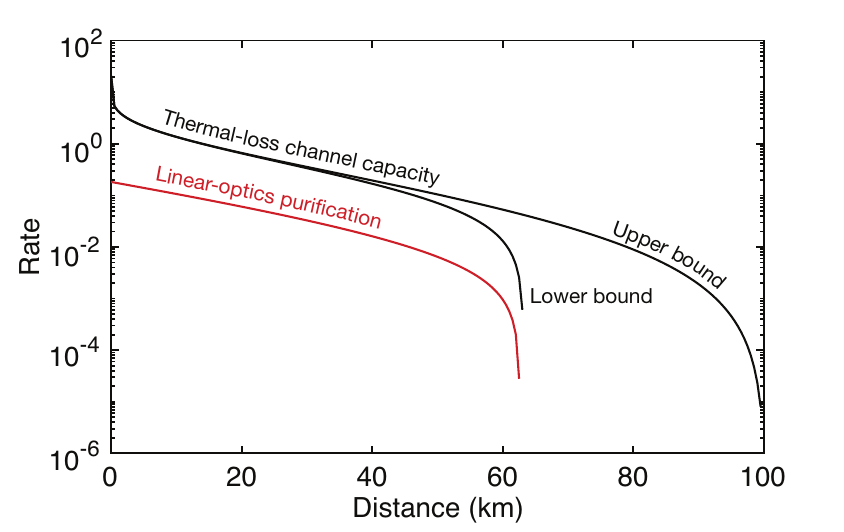}
\caption[Linear-optics purification with imperfections.]{\bblack{{Linear-optics purification with imperfections.} Rate vs. distance for the dual-rail qubit protocol using linear-optics and photon-number measurements. The linear-optics purification rate is given by the RCI. We assume thermal noise in the channel with mean photon number $\bar{n}=0.01$, and imperfections at the detectors with $\eta_\text{efficiency}=0.5$ and dark-count rate $r_\text{dark} = 10^{-6}$. The upper and lower bounds on the channel capacity of the thermal-loss channel are also shown which give the ultimate limits of repeaterless quantum communications~\cite{Pirandola_2017}.}}\label{fig:dark_counts}
\end{figure}

With inefficient detectors, dark counts, and thermal noise, we perform a numerical calculation of the RCI  and the results are shown in~\cref{fig:dark_counts}. We simulate the thermal-loss channel using the Kraus-operator representation for the channel~\cite{PhysRevA.84.042311,PhysRevA.96.062306} with mean photon number $\bar{n}=0.01$. We model dark counts and detector inefficiencies (additional losses) by a thermal-loss channel before each detector at Bob's side with efficiency $\eta_\text{efficiency}=0.5$ and dark-count rate $r_\text{dark}=10^{-6}$. The protocol seems to be tolerant to these imperfections, especially at short distances. Eventually, the thermal noise in the channel overwhelms the protocol. The upper and lower bounds on the channel capacity are also shown~\cite{Pirandola_2017}.

\subsection{The directionality issue} 

Often in quantum communications, it is best if quantum states propagate in one preferred direction (from Alice towards Bob) when reverse reconciliation is used. This is a persistent problem in CV quantum communications. For instance, CV measurement-device-independent protocols~\cite{Ma_2014,Zhang_2014,Ottaviani_2015,Pirandola2015} work only in an extremely asymmetric configuration, with the node (ineffective as a repeater) positioned close to one of the trusted parties. This directionality problem is also present in the repeater protocols considered in~\cref{Chap:CV_QR} and Refs.~\cite{PhysRevA.102.052425,Ghalaii_2020}, where they work best in one direction, which for reverse reconciliation is again from Alice towards Bob. Our purification schemes allow states to propagate in any direction in a quantum network since the output states are pure. Using purification for CV-QKD works the same for both direct and reverse reconciliations.  The users are identical and can be swapped (i.e., there is no difference between reverse reconciliation and direct reconciliation). This is an important requirement for large CV networks. We have no directionality problem. Note that the memoryless CV repeater protocol introduced in~\cref{Chap:simple_repeater} (i.e., Ref.~\cite{winnel2021overcoming}) also fixes the directionality problem.

\section{Chapter summary}\label{PURsec:discussion}

\bblack{We have presented a physical protocol that achieves the two-way assisted quantum capacity of the pure-loss channel~\cite{Pirandola_2017}.} Error \textit{correction} requires short distances between neighbouring repeater nodes, while in contrast, we showed simple error \textit{detection} can saturate the repeaterless bound. An open question is what protocol can saturate the fundamental limits for added thermal noise and close the gap between the theoretical upper and lower bounds~\cite{Pirandola_2017}. Furthermore, perhaps error detection can saturate the ultimate limits for other noise models, \bblack{for instance, dephasing}~\cite{Lami2023}.

\bblack{Our protocol is an optimal one and can be performed using CSUM quantum gates and measurements. However, it is unknown whether other gates and measurements may also achieve the capacity, and if they can do so more efficiently, i.e., approaching the repeaterless bound more quickly with the number of iteration rounds, $n$.}

Our ultimate protocol is experimentally challenging since it requires iterative purification steps. However, we show that by simplifying our protocol to just a single round of purification, we can still achieve excellent rates. This is superior to other nondeterministic techniques such as noiseless linear amplification where it is impossible to remove all effects of loss.

Our purification protocol can be implemented using \bblack{linear-optics and number measurements}. This does introduce some probability penalty (which does not depend on the loss). However, having an all-optical design is experimentally convenient.

One limitation of our results is the requirement for high-performance quantum memories. However, purification outputs pure states which is extremely beneficial. Firstly, pure states can handle a higher amount of decoherence coming from nonideal quantum memories. Secondly, distributing pure states will prevent how much thermal noise builds up across a network during entanglement swapping. Finally, purity may be beneficial for both point-to-point and repeater-assisted CV QKD, improving the signal-to-noise ratio and decreasing Eve's knowledge of the key, speeding up the classical post-processing part of the protocol. 

Thus, purification is inherently a useful technique for overcoming the unavoidable losses in quantum communication networks. Remarkably, purification can be employed using linear optics \bblack{and number measurements} and is compatible with existing DV and CV infrastructure. Finally, we remark that since the lossy entanglement is completely purified, it can be used for exotic tasks, such as device-independent quantum key distribution and demonstrating Bell nonlocalities, over much longer distances than previously imagined. In the next chapter, we consider repeaters with quantum error correction, removing the need for classical signalling.

%% file: Chapter7/Chapter7.tex
    
    \chapter{Third-generation CV quantum repeaters}
    \label{Chap:QEC}	
    \pagestyle{headings}

    \noindent
    At the time of writing, the results presented in this chapter have not been published or appear elsewhere.
    
    $\;$

    \noindent
    Third-generation quantum repeaters use quantum error correction to protect against noise in the channel and promise very-high rates because they are a one-way scheme. In this chapter, we introduce an approach for the third generation of continuous-variable quantum repeaters. We use the cat code to extend the distance of continuous-variable quantum key distribution with a discrete modulation. We assume an optimal recovery against photon loss for the cat code and calculate asymptotic secret key rates, bounding Eve's attack via numerical optimisation since the protocol is non-Gaussian. We consider teleportation for the encoding and decoding of the quantum information. Our results can be extended to other bosonic codes.

    \section{Introduction}

    Quantum repeaters are classified into three generations based on their approach to protecting the quantum information from noise~\cite{Muralidharan2016,munro2015inside}. The first and second generations use entanglement swapping and purification techniques to herald a highly-entangled state between the two ends of a long-distance channel. This entanglement can be used for communication protocols such as quantum key distribution (QKD)~\cite{Gisin_2002,RevModPhys.81.1301,Pirandola_2020} or teleportation~\cite{PhysRevLett.70.1895}. The first two generations of quantum repeaters are limited by the two-way classical signalling required for their operation. The third generation works differently. The approach uses quantum error correction (QEC)~\cite{gottesman2009introduction} to overcome loss and noise in the channel. It needs only one-way signalling; thus, promises very high rates. This is similar to how classical repeaters work.

    Third-generation repeaters for DV quantum information where loss and noise are overcome using CV techniques have been explored previously and we refer to this as ``DV repeater protocols using CV techniques''. In Refs.~\cite{Li_2017,PhysRevA.94.042332}, for instance, they use the single-mode cat code~\cite{PhysRevA.59.2631,PhysRevLett.111.120501,Mirrahimi_2014,PhysRevA.97.032346,PhysRevX.10.011058}. The cat code is well known for being the first QEC code to surpass the break-even point in a QEC experiment~\cite{ofek2016demonstrating}.

    Third generation approaches for CV quantum information have not been explored before and this is because it is impossible to protect arbitrary CV states from noise using a single-mode error-correcting code. There is no room for redundancy. There are generalisations from qubit codes to CV codes~\cite{PhysRevLett.80.4084} and there exist quantum erasure-correcting codes (which are possible for instance if you can probe the losses from the data mode into the environment~\cite{PhysRevLett.101.130503}). There are multi-mode codes~\cite{Aoki2009} and there are also error-mitigation techniques~\cite{PhysRevLett.100.030503,dias2018quantum}. Recently, it was shown that by encoding an oscillator into many oscillators~\cite{Noh_2020} one can suppress noise on arbitrary CV states.

    The third generation of quantum repeaters works by directly encoding the quantum information into a quantum-error-correcting code which is robust to noise in the channel. Unlike the first generation, the third-generation approaches do not need quantum memories.

    In this chapter, we explore the third generation of quantum repeaters, where both the quantum communication protocol is CV and the quantum repeaters and error correction are CV. We introduced quantum error \textit{detection} for CV states in~\cref{Chap:purification}, and in this chapter, we introduce an approach for CV repeaters with CV quantum error \textit{correction}.

    In usual one-way CV QKD protocols, Alice encodes quantum information into the quadratures of light. To do this, she prepares a Gaussian modulation of coherent states~\cite{Grosshans_2002,Weedbrook_2004} (or squeezed states~\cite{PhysRevA.63.052311,gottesman2001secure}) and sends them to Bob who ultimately measures them with homodyne or heterodyne detection. Alice may instead prepare a discrete modulation where she selects states at random from a finite set. Discrete-modulation protocols are appealing because good reconciliation procedures can be found even at long distances (at very low signal-to-noise ratios)~\cite{leverrier2011continuousvariable}. Important for us is that a discrete modulation forms a qudit. The qudit can be completely encoded into bosonic codes which protect the quantum information against noise in the channel. Also, there are ways to deal with the non-Gaussianity in the security analysis of deterministic discrete-modulation CV QKD~\cite{Liao_long_distance,PhysRevX.9.021059,PhysRevX.9.041064,papanastasiou2020continuousvariable,kaur2020asymptotic,matsuura2020finitesize}.

    The main result of this chapter is the use of single-mode codes to increase the secret key rate and secure distance of CV-QKD protocols with a discrete modulation. We are interested in single-mode codes (or codes with a small number of modes) because they are highly efficient for quantum communication, cleverly protecting the quantum information using fewer modes and fewer ``moving parts''~\cite{PhysRevA.97.032346}. The discrete-modulation protocol that we consider is CV in the sense that Alice prepares a CV state encoding information into the quadratures of light, Bob performs a CV measurement, for example, heterdyne detection, and we bound Eve's information by calculating the Holevo bound via the covariance matrix. The QEC is CV in the sense that it exploits the infinite-dimensional Hilbert space to protect the quantum information from noise and the error correction is performed entirely in a single mode or a small number of modes.
    
    We choose to encode the ensemble prepared by Alice into a cat code~\cite{PhysRevA.59.2631,PhysRevLett.111.120501,Mirrahimi_2014,PhysRevA.97.032346,PhysRevX.10.011058} but our work may be extended to other bosonic codes, for instance, binomial codes~\cite{Michael_2016} and GKP codes~\cite{GKP2001}. We focus on a two-state CV-QKD protocol for motivation and extend the idea to a four-state protocol since it performs better than the two-state protocol.
    
    One interesting point is that the finite set prepared by Alice is itself encoded in a cat code~\cite{PhysRevA.97.032346,PhysRevX.10.011058}. This trivial cat code is unable to detect single-loss events which is why the ensemble needs to be encoded into a larger cat code. Quantum teleportation between cats of different sizes provides a method for encoding and decoding the quantum information, and possibly to perform the error correction (called tele-correction). These points motivate us to investigate cat codes in this chapter.

    \subsection{Overview of the main results of this chapter}
    
    We consider a two-state and a four-state discrete-modulation protocol and protect the quantum information using cat codes. We calculate lower bounds on the secret key rate in the asymptotic limit of infinitely-long keys. Our results are secure against collective attacks using the security method from Ref.~\cite{PhysRevX.9.021059} which uses numerical optimisation to bound the covariance matrix (without this optimisation, the secret key rates are secure against Gaussian collective attacks).

    We assume an explicit recovery map for the cat code which performs optimally against photon loss~\cite{matt_honours} and pretty well for a small amount of added thermal noise. This recovery can be modified in the presence of a large amount of thermal noise to correct gain events.
    
    Recall that the ultimate rate-loss trade-off of point-to-point quantum communication is given by the repeaterless bound~\cite{Pirandola_2017} and we show that our protocol can beat the repeaterless bound for sufficiently high coupling efficiency and high-quality error recovery.
    
    We describe that the encoding and decoding steps may be performed via teleportation using linear optics, photon counting, and a non-Gaussian entanglement resource. The required entanglement resource is a Bell state where the entangled modes are encoded in different cat codes. Our teleportation protocol is motivated by~Ref.~\cite{Neergaard-Nielsen2013}.


    \section{The error-corrected repeater protocol for CV QKD}
    
\begin{figure}
    \centering
    \includegraphics[width=0.8\textwidth]{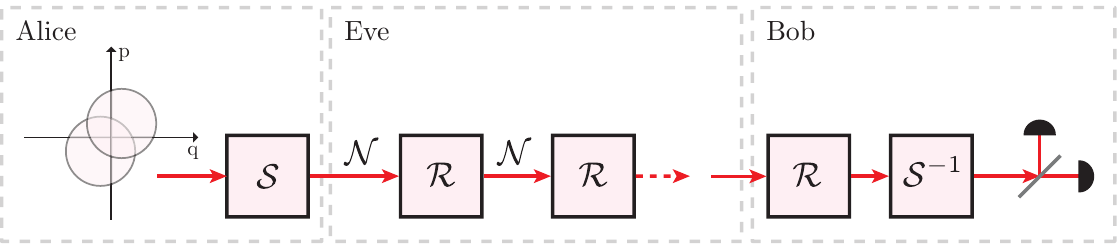}
    \caption[The CV-QKD repeater protocol consisting of a discrete-modulation of coherent states and quantum error correction]{The CV QKD repeater protocol consisting of a discrete-modulation of coherent states prepared by Alice, followed by encoding $\mathcal{S}$ into a bosonic code, alternating error channel $\mathcal{N}$ and recovery $\mathcal{R}$ along the entire distance, and finally decoding $\mathcal{S}^{-1}$ and heterodyne detection at Bob.\label{Chap7fig:twostateprotocol}}
    \end{figure}

    Our discrete-modulation protocol with quantum repeaters is shown in~\cref{Chap7fig:twostateprotocol}. Alice chooses a coherent state at random from a finite set, such as two states or four states (which is approximately Gaussian for small amplitudes). To extend the secure distance and speed up the secret key rate, Alice encodes the quantum information into a bosonic code. Repeater stations along the channel recover the state from noise. Bob decodes after a final recovery operation and performs heterodyne detection.
    
    The combined quantum channel between Alice and Bob is described by a completely-positive trace-preserving (CPTP) map $\mathcal{E}$~\cite{weedbrook2012gaussian} consisting of encoding $\mathcal{S}$, alternating noise $\mathcal{N}$ and recoveries $\mathcal{R}$, and decoding $\mathcal{S}^{-1}$. The combined channel $\mathcal{E}$ is
    \begin{align}
        \mathcal{E} &= \mathcal{S}^{-1} \circ (\mathcal{R} \circ \mathcal{N})^{N_\text{links}} \circ  \mathcal{S},\label{Chap7eq:map}
    \end{align}
    where $N_\text{links}$ is the number of links. The number of repeater stations is $N_\text{links}{-}1$ and there is a final recovery at Bob's station before decoding.

    \subsection{Error-correcting codes}
    
    A single-mode qubit code is a two-dimensional subspace of the infinite-dimensional Hilbert space. It is defined by the projector onto the code subspace
    \begin{align}
        \mathcal{P}_\text{code} &= \ketbra{0_\text{code}}{0_\text{code}} + \ketbra{1_\text{code}}{1_\text{code}},
    \end{align}
    where $\ket{0_\text{code}}$ and $\ket{1_\text{code}}$ form any orthogonal basis of the code subspace~\cite{PhysRevA.97.032346}.
    
    The cat codes belong to the class of single-mode codes based on phase-space rotation symmetry~\cite{PhysRevX.10.011058}. The codewords can be written as a superposition of discrete-rotated coherent states
    \begin{subequations}
    \begin{align}
        \ket{0_{\text{cat}(\beta,N)}} &= \frac{1}{\sqrt{\mathcal{N}_0}} \sum_{m=0}^{2N-1} e^{i\frac{m\pi}{N}\hat{n}} \ket{\beta},\\
        \ket{1_{\text{cat}(\beta,N)}} &= \frac{1}{\sqrt{\mathcal{N}_1}} \sum_{m=0}^{2N-1} (-1)^m e^{i\frac{m\pi}{N}\hat{n}} \ket{\beta},
    \end{align}\label{Chap7eq:catcode}%
    \end{subequations}
    where $\beta$ is the amplitude of the coherent states in superposition and $N$ is the degree of discrete rotation symmetry. $\mathcal{N}_0$ and $\mathcal{N}_1$ are the normalisation constants. An order-$N$ cat code consists of a superposition of $2N$ coherent states around a circle in phase space. In Fock space, the code words form a ladder of occupation number. The spacing of the ladder protects the quantum information from loss since small shifts up and down move the state out of the code subspace. Up to $N-1$ loss events are detectable errors for an order-$N$ cat code. Shifted codes can also be considered by shifting the unshifted Fock-ladder up or down $s$ rungs. The shifted cat codes have occupation on Fock states $n=s$ and $n=N+s$ modulo $2N$ for $s = \{0,1,...,N{-}1\}$. $N$ and $s$ define a code which can be optimised in any given situation. In this chapter, we do not focus too much on the shifting parameter, $s$. 

    \subsection{The two-state constellation and encoding}
    
    The prepare-and-measure (PM) version is the version of the protocol Alice and Bob implement in practice. In this section, we assume the modulation is along the $q$-quadrature axis. 
    
    Modulating in the $q$-direction, Alice selects a coherent state at random from the set $\{\ket{\alpha}, \ket{{-}\alpha}\}$ where $\alpha$ is a real number. That is, Alice prepares~\cite{leverrier2011continuousvariable}
    \begin{align}
        \rho_\text{PM} &= \frac{1}{2} ( \ketbra{\alpha}{\alpha} + \ketbra{{-}\alpha}{{-}\alpha} )\\
        &= c_0 \ketbra{\phi_0}{\phi_0} + c_1 \ketbra{\phi_1}{\phi_1},\label{Chap7eq:PMeigenstates}
    \end{align}
    where $c_0 = e^{-\alpha^2} \cosh{\alpha^2}$, $c_1 = e^{-\alpha^2}\sinh{\alpha^2}$ and
    \begin{align}
    \ket{\phi_0} &= \ket{0_{\text{cat}(\alpha,1)}} = \frac{\ket{\alpha}+\ket{{-}\alpha}}{\sqrt{2+2e^{-2|\alpha|^2}}}\\
    \ket{\phi_1} &= \ket{1_{\text{cat}(\alpha,1)}} = \frac{\ket{\alpha}-\ket{{-}\alpha}}{\sqrt{2-2e^{-2|\alpha|^2}}},
    \end{align}
    and $\ket{\phi_0}$ and $\ket{\phi_1}$ are orthogonal eigenstates of Alice's ensemble $\rho_\text{PM}$. These eigenstates are codewords of the order-$1$ cat code defined in~\cref{Chap7eq:catcode}. This code does not protect against pure loss since there is zero spacing between the code words in Fock space. Alice maps the quantum information into a bosonic code (with codewords $\ket{0_\text{code}}$ and $\ket{1_\text{code}}$), which protects against loss, for example, a higher-order cat code. The encoded state she sends to Bob is
    \begin{align}
    S\rho S^\dagger &= c_0 \ketbra{0_\text{code}}{0_\text{code}} + c_1 \ketbra{1_\text{code}}{1_\text{code}},
    \end{align}
    where one choice for the encoding step is $S = \ketbra{0_\text{code}}{\phi_0} + \ketbra{1_\text{code}}{\phi_1}$.

    Repeater stations between Alice and Bob recover the quantum information from errors. Bob decodes at the end to retrieve Alice's original state $\rho_{\text{PM}}$ with high fidelity.
    
    Security analysis is performed in the equivalent entanglement-based (EB) version. In the EB version, Alice prepares a pure bipartite state $\ket{\Phi_2}$. A projective measurement on the first mode $\{\ketbra{\psi_0}{\psi_0},\ketbra{\psi_1}{\psi_1}\}$ projects the second mode onto coherent states $\ket{\alpha}$ or $\ket{{-}\alpha}$, each with probability $1/2$. The purification which maximises the correlation between the two modes is~\cite{leverrier2011continuousvariable}:
    \begin{align}
        \ket{\Phi_2} &= \frac{1}{\sqrt{2}} (\ket{\psi_0}\ket{\alpha} + \ket{\psi_1}\ket{{-}\alpha} )\\
        &= \sqrt{c_0} \ket{\phi_0}\ket{\phi_0} + \sqrt{c_1} \ket{\phi_1}\ket{\phi_1},
    \end{align}
    where
    \begin{align}
        \ket{\psi_0} &= \frac{1}{\sqrt{2}} (\ket{\phi_0} + \ket{\phi_1}) \\
        \ket{\psi_1} &= \frac{1}{\sqrt{2}} (\ket{\phi_0} - \ket{\phi_1}).
    \end{align}
    Alice encodes the second mode into a bosonic code, for example:
    \begin{align}
        S \ket{\Phi_2} &= \sqrt{c_0} \ket{\phi_0}\ket{0_\text{code}} + \sqrt{c_1} \ket{\phi_1}\ket{1_\text{code}}.
    \end{align}
    Repeater stations are placed along the noisy quantum channel which recover the quantum information from errors. Bob performs one final recovery and decodes. Finally, Alice and Bob share a highly-entangled state $\rho_{AB}$.

    \subsection{Four-state protocol: Alice's modulation and encoding into a CV code\label{Chap7app:four_state_protocol}}
    
    In the PM version of the four-state protocol, Alice sends one of four random coherent states chosen from the set $\{\ket{\alpha},\ket{i\alpha}, \ket{{-}\alpha},\ket{{-i}\alpha}\}$ where $\alpha$ is a real number. That is, Alice sends a mixture $\rho_{\text{PM}}$ given by~\cite{PhysRevX.9.021059}
    \begin{align}
    \rho_{\text{PM}} &= \frac{1}{4} \sum_{k=0}^{3} \ketbra{i^k \alpha}{i^k \alpha}\\
    &= \sum_{m=0}^{3} c_m \ketbra{\phi_m}{\phi_m}
    \end{align}
    where $\ket{\phi_m}$ are four-component cat states
    \begin{align}
        \ket{\phi_m} &= \frac{1}{4 \sqrt{c_m}} \sum_{k=0}^{3} i^{-mk} \ket{i^k \alpha},
    \end{align}
    with,
    \begin{align}
        c_0 &= \frac{1}{2} e^{-{\alpha^2}}(\cosh{\alpha^2}+\cos{\alpha^2})\\
        c_1 &= \frac{1}{2} e^{-{\alpha^2}}(\sinh{\alpha^2}+\sin{\alpha^2})\\
        c_2 &= \frac{1}{2} e^{-{\alpha^2}}(\cosh{\alpha^2}-\cos{\alpha^2})\\
        c_3 &= \frac{1}{2} e^{-{\alpha^2}}(\sinh{\alpha^2}-\sin{\alpha^2}).
    \end{align}
    
    In the EB version, Alice prepares the initial state
    \begin{align}
    |\Phi_{\text{4}} \rangle &\propto (\Id \otimes \sqrt{\rho_{\text{PM}}}) \sum_{n=0}^\infty |n\rangle |n\rangle\label{eq:PHI4}\\
    &= \sum_{m=0}^3 \sqrt{c_m} \ket{\phi_m} \ket{\phi_m}\\
    &= \frac{1}{2} \sum_{k=0}^3 \ket{\psi_k}\ket{i^k \alpha},\label{eq:psik}
    \end{align}
    where $|n\rangle$ is a Fock state with $n$ photons, and $\ket{\psi_k} =  \frac{1}{2} \sum_{m=0}^3 e^{-ikm \frac{\pi}{2} }\ket{\phi_m}$.
    
    One possible encoding map is
    \begin{align}
        S &= \ket{0_\text{code}}\ket{0_\text{code}}\bra{\phi_0} + \ket{0_\text{code}}\ket{1_\text{code}}\bra{\phi_1} +\ket{1_\text{code}}\ket{0_\text{code}}\bra{\phi_2} +\ket{1_\text{code}}\ket{1_\text{code}}\bra{\phi_3}.
    \end{align}

    Applying this encoding step onto the second mode maps the qudit onto two modes and gives
    \begin{align}
        S \ket{\Phi_4} &= \sqrt{c_0} \ket{\phi_0}\ket{0_\text{code}}\ket{0_\text{code}} + \sqrt{c_1} \ket{\phi_1}\ket{0_\text{code}}\ket{1_\text{code}} \\ & \nonumber \;\;\;\;\;\;\; + \sqrt{c_2}  \ket{\phi_2}\ket{1_\text{code}}\ket{0_\text{code}} + \sqrt{c_3} \ket{\phi_3}\ket{1_\text{code}}\ket{1_\text{code}}.
    \end{align}
    Alice keeps the first mode and sends the two encoded modes along a noisy channel. Successive error recovery is performed on the two modes, recovering the qubits from errors. After final recoveries at Bob's station, Bob decodes $\mathcal{S}^{-1}$ so that the final state shared by Alice and Bob has high fidelity with the initial entangled state prepared by Alice in the EB version. Bob performs heterodyne detection and Alice and Bob extract a secret key using classical post-processing.

    
    \subsection{Noise model}

    We do not know Eve's optimal attack because our protocol is non-Gaussian, and Alice and Bob do not have access to a covariance matrix in the EB version, but we can bound Eve's maximal information given the parameters accessible to Alice and Bob in an experiment, for example, by bounding the covariance matrix. To obtain numerical results we simulate the experimental parameters by modelling the fibre as a lossy thermal-noise channel $\mathcal{N}$ of transmissivity $T$ and excess noise $\xi$. The channel is equivalent to combining a thermal state with mean photon number $\bar{n}$ and variance $V=2\bar{n}+1=\frac{T \xi}{ 1-T} + 1$ with the input mode on a beamsplitter of transmissivity $T$. See~\cref{sec:thermal_noise} for details about the thermal-loss channel.

    The total distance $L_\text{tot}$ measured in km is divided into $N_\text{links}$ links, each with length $L_0$, transmissivity $T_0 = 10^{-0.02L_0}$ (assuming optical fibre with attenuation $0.2$ dB/km) and constant excess noise with thermal-noise variance $V=2\bar{n}+1$. We model the non-ideal coupling efficiency of the repeaters by additional pure loss with transmissivity $\eta_\text{c}$. In summary, the total noise channel for each link can be decomposed into pure loss with transmissivity $\tau \eta_\text{c}$, followed by quantum-limited amplification with gain $G$, where $\tau = T_0/G$ and $G=1+(1-T_0)\bar{n}$.

    
    \subsection{Recovery channel\label{Chap7sec:recovery_channel}}
    
    Third-generation repeaters require high-quality error correction and low operational errors. The optimal performance of a bosonic code can be determined in terms of the entanglement fidelity by solving a semidefinite program (SDP)~\cite{PhysRevA.97.032346}. Here, we use an explicit form for the recovery which gives the same optimal entanglement fidelity as found by an SDP. That is, we numerically verified that it performs optimally for the cat codes against photon loss~\cite{matt_honours}. We refer to this recovery as the truncated Cafaro-van Loock (CL) recovery since it appears without truncation in Ref.~\cite{PhysRevA.89.022316}.

    The Kraus operators of the truncated CL recovery are
     \begin{align}
         R_l &= \frac{\ketbra{0_\text{cat}}{0_\text{cat}}A_l^\dagger}{\sqrt{\bra{0_\text{cat}}A_l^\dagger A_l \ket{0_\text{cat}}}} + \frac{\ketbra{1_\text{cat}}{1_\text{cat}}A_l^\dagger}{\sqrt{\bra{1_\text{cat}}A_l^\dagger A_l \ket{1_\text{cat}}}}\\
              R_N &= \sqrt{\Id - \sum_{l=0}^{N-1} (R_l^\dagger R_l)}\label{Chap7eq:CL},
     \end{align}
    with $l = 0,1,2,...,N{-}1$ and where $A_l$ are the Kraus operators of the pure-loss channel. The reason we truncate the number of Kraus operators to $N-1$ is because a cat code of order $N$ can correct up to $N-1$ loss events. For example, the $N{=}2$ cat code has a two-fold rotation symmetry and the code words consist of superpositions of four coherent states in a circle in phase space. It is known as the single loss code because there is a spacing of one between the code words in Fock space and, therefore, it is robust against single loss events. The optimal, truncated CL recovery for the single-loss cat code consists of three Kraus operators, $R_0,R_1,R_2$.
    
    The Pauli channel from Ref.~\cite{Li_2017} is a good approximation to the optimal channel for moderate to large amplitude $\beta$. It transforms density matrices as
    \begin{align}
        \rho \to (1{-}\epsilon_X{-}\epsilon_Z) \rho + \epsilon_X X\rho X + \epsilon_Z Z\rho Z,
    \end{align}
    where $\epsilon_X$ and $\epsilon_Z$ are Pauli-$X$ (bit-flip) and Pauli-$Z$ (dephasing) error rates respectively (see Ref.~\cite{Li_2017} for details). $X$ and $Z$ are the Pauli matrices acting on the logical qubit. To physically implement the recovery, one can use the circuit presented in Ref.~\cite{Li_2017}.


    \subsection{Covariance matrices for the two-state protocol\label{Chap7sec:outputCM}}
    
    Before presenting numerical results, we discuss the effect of error correction on the EB covariance matrix shared between Alice and Bob for the two-state protocol. We rotate the protocol 45 degrees so that the modulation direction is diagonal to the quadrature axes (so that the quadrature variances are the same). The initial EB state Alice prepares (before encoding into the cat code) has covariance matrix
    \begin{align}
    \Gamma_0 &= \begin{bmatrix} 
    V_A \Id  & Z \mathbb{Z} \\
    Z \mathbb{Z} & V_B \Id
    \end{bmatrix},
    \end{align}
    where
    \begin{align}
    V_A &= \tr[(\Id + 2a^\dagger a)\rho]  = 1+2\alpha^2\\
    V_B &= \tr[(\Id + 2b^\dagger b)\rho] = V_A\\
    Z &= \tr [(ab+a^\dagger b^\dagger)\rho]\\
    &= 2 \Re \{ \tr [(a b) \rho] \} = 2 \alpha^2 \left( \frac{c_0^\frac{3}{2}}{c_1^\frac{1}{2}} + \frac{c_1^\frac{3}{2}}{c_0^\frac{1}{2}} \right),
    \end{align}
    where $a$ and $b$ are the annihilation operators on Alice and Bob's modes respectively and $c_0$ and $c_1$ were defined in~\cref{Chap7eq:PMeigenstates}.
    
    For direct transmission (i.e., no encoding and no error correction), after a lossy thermal channel with transmissivity $T$ and excess noise $\xi$, the covariance matrix is
    \begin{align}
    \Gamma_\text{direct} &= \begin{bmatrix}
    (V_A+1) \Id  & \sqrt{T} Z \mathbb{Z} \\
    \sqrt{T} Z  \mathbb{Z} & (T V_B + 1 + T \xi) \Id
    \end{bmatrix}.
    \end{align}

    The output covariance matrix for our error-corrected protocol is very different. The error channel is not loss and excess noise, like in direct transmission. The combined error channel after a single round of error correction assuming perfect encoding and decoding, a pure-loss noise channel, and optimal recovery is approximated by the Pauli channel from Ref.~\cite{Li_2017}. After one round of error correction for the two-state protocol given by the Pauli approximation ($L_\text{tot} = L_0$ and $N_\text{links}=1$), the output covariance matrix is
    \begin{align}
    \Gamma_\text{Pauli} &= \begin{bmatrix} 
    V_A \Id  & Z \mathbb{Z} \\
    Z \mathbb{Z} & V_B \Id
    \end{bmatrix},
    \end{align}
    where
    \begin{align}
    V_A &= 1+2\alpha^2\\
    V_B &= (1{-}\epsilon_X) (1+2\alpha^2) + \epsilon_X \left[ 1+2\alpha^2\left( \frac{c_0^2}{c_1} + \frac{c_1^2}{c_0} \right) \right]
    \end{align}
    \begin{align}
        Z = 2 \alpha^2 \left[   (1{-}\epsilon_X{-}\epsilon_Z)  \left( \frac{c_0^\frac{3}{2}}{c_1^\frac{1}{2}} + \frac{c_1^\frac{3}{2}}{c_0^\frac{1}{2}} \right)  - \epsilon_Z   \left( \frac{c_0^\frac{3}{2}}{c_1^\frac{1}{2}} + \frac{c_1^\frac{3}{2}}{c_0^\frac{1}{2}} \right) + \epsilon_X (2\sqrt{c_0 c_1})  \right].
    \end{align}

    The Pauli-$X$ bit-flip error increases Bob’s variance. This can be understood by considering the limit of small $\alpha$; Alice’s state is Fock zero and one, so a bit-flip error changes the mean photon number and thus the variance. The Pauli-$Z$ dephasing error decreases the covariance because of the minus sign. The Pauli-$X$ error also decreases the covariance because $2\sqrt{c_0 c_1}  \leq \frac{c_0^\frac{3}{2}}{c_1^\frac{1}{2}} + \frac{c_1^\frac{3}{2}}{c_0^\frac{1}{2}} $ for $c_0,c_1>0$.
    The dephasing and bit-flip errors are small after only a short distance so the output covariance matrix is close to the input covariance matrix. After many repeaters; however, the errors eventually build up.

    In an experiment, Alice and Bob implement the PM version of the protocol so the EB covariance matrix is not directly accessible. The parameters accessible in the PM protocol are the variances $v_q$ and $v_p$ and the covariances $c_q$ and $c_p$ of the $q$ and $p$ quadratures. The security analysis is performed in the equivalent EB version so the covariance matrix of the state $\rho_{AB}$ shared between Alice and Bob in the EB version must be estimated using experimental data. 
    
    In a usual direct-transmission Gaussian CV-QKD protocol, the measured parameters in the PM version allow one to estimate the transmissivity $T$ and the excess noise $\xi$ of the channel. For the direct-transmission discrete-modulation protocol (without repeaters), the state is non-Gaussian and the EB output state $\rho_{AB}$ is unknown because Eve's optimal attack i.e., the quantum channel, is unknown. Eve’s attack can be bounded by minimising the covariance $Z$ between Alice and Bob over all channels permissible by the PM parameters ($v_p,v_q,c_q,c_p$)~\cite{PhysRevX.9.021059}. In our protocol with repeaters, it is possible to estimate the covariance matrix of the state shared between Alice and Bob in the EB version from the PM parameters since Bob's variance is the same in the EB and PM versions, $V_B=v_B$, and Eve's information can be bounded by minimising the covariance $Z$ between Alice and Bob, by going over all channels compatible with the parameters observed in an experiment. 
    
    For direct transmission, the channel is parameterised by $T$ and $\xi$. With repeaters, the channel is a Pauli channel with parameters $\epsilon_X$ and $\epsilon_Z$. These Pauli error rates can be estimated by Alice and Bob from the measured parameters, thus, building up the whole covariance matrix.
    
    Because we do not have access to simulated data, we assume a lossy thermal-noise channel for the links between repeater stations to simulate data in an actual experiment (with additional coupling loss to model the inefficiency). Then we bound the covariance matrix via numerical optimisation (using an SDP).


     
    \section{A lower bound on the asymptotic secret key rate}
    
    The final state $\rho_{AB}$ in the EB version is
    \begin{equation}
    \rho_{AB} = (\Id \otimes \mathcal{E}) (|\Phi\rangle\langle\Phi|),
    \end{equation}
    where $\ket{\Phi}$ is the initial state prepared by Alice ($\ket{\Phi}=\ket{\Phi_2}$ for the two-state protocol and $\ket{\Phi}=\ket{\Phi_4}$ for the four-state protocol).
    
    A lower bound on the asymptotic secret key rate in the case of reverse reconciliation for collective attacks is given by the Devetak-Winter rate~\cite{devetak2005distillation} 
    \begin{align}
    K = \beta_\text{rec} I_{AB} - \chi_{EB},
    \end{align}
    also presented in~\cref{eq:DW}, where $I_{AB}$ is the classical mutual information between Alice and Bob, $\beta_\text{rec}$ is the reconciliation efficiency, and $\chi_{EB}$ is the Holevo information~\cite{holevo1973bounds}. Throughout this chapter, we set the reconciliation efficiency to $95\%$. In the following sections, we need to estimate Alice and Bob's mutual information $I_{AB}$ and upper-bound Eve's information $\chi_{EB}$.

    \subsection{Bounding Eve's information}
    
    Since the final state $\rho_{AB}$ is non-Gaussian, calculating the final covariance matrix shared between Alice and Bob is non-trivial. The optimality of Gaussian states~\cite{PhysRevLett.96.080502} allows us to calculate the supremum of the Holevo bound $\chi_{EB}$ over all states $\rho_{AB}$ with a fixed covariance matrix, but what is missing is a way to calculate the covariance matrix. For this, we use the method from Ref.~\cite{PhysRevX.9.021059}. Their method is to perform an optimisation over all states compatible with parameters observed in the PM version. 
    
    Gaussian channels with transmissivity $T$ and excess noise $\xi$ provide a realistic model for typical experiments, so we assume a Gaussian channel to determine numerical values for parameters observed in an experiment. The optimisation problem is set out in~\cref{Chap7app:SDP}.
    
    The Holevo information is upper bounded by the same quantity computed for a Gaussian state with the same covariance matrix as $\rho_{AB}$~\cite{PhysRevLett.102.180504}
    \begin{align}
    \chi_{EB} &= S(\rho_{AB}) - S(\rho_{A|b}),\label{Chap7eq:Holevo}
    \end{align}
    where $S(\rho_{AB})$ is the von Neumann entropy of Alice and Bob's state, and $S(\rho_{A|b})$ is the von Neumann entropy of Alice's state conditioned on Bob's measurement result, i.e., we assume Eve holds a purification of the state.

    \subsection{Alice and Bob's mutual information}

    Since the ensemble is approximately Gaussian, we can estimate Alice and Bob's mutual information $I_{AB}$ from the covariance matrix shared between Alice and Bob. Since the protocol is based on coherent states and heterodyne detection, we use the following~\cite{Sanchez2007QuantumIW}
    \begin{align}
        I_{AB} &\approx \log_2 \left({ \frac{V_{B} + 1}{V_{B|A^\text{het}} + 1}}\right),\label{Chap7eq:I_AB}
    \end{align}
    where $V_{B|A^\text{het}}$ is the covariance of Bob's mode conditioned on a heterodyne measurement on Alice's mode (heterodyne because Alice is sending coherent states and the ensemble is approximately Gaussian). We note that for direct transmission, the approximation (\cref{Chap7eq:I_AB}) is slightly less than the additive white Gaussian noise channel, therefore, we believe it to be okay to use~\cref{Chap7eq:I_AB} to approximate Alice and Bob's mutual information. Note also that for small $\alpha$, the discrete-modulation protocols are approximately Gaussian.
    
    For direct transmission Alice and Bob's mutual information $I_{AB}$ is given by a binary-input additive white Gaussian noise (AWGN) channel and can be approximated by the capacity of an AWGN channel~\cite{PhysRevX.9.021059}. So we have
    \begin{align}
        I_{AB} &\approx 2 \times C_\text{AWGN}(s)\\
        &\approx 2 \times \frac{1}{2} \text{log}_2 \left(1 + s\right)\\
        &\approx \text{log}_2 \left(1 + \frac{2T\alpha^2}{2+T\xi}\right),
    \end{align}
    where $s = \frac{2T\alpha^2}{2+T\xi}$ is the signal to noise ratio.

    \section{Bounding the covariance matrix via an SDP\label{Chap7app:SDP}}
    
    \subsection{Four-state protocol SDP}
    
    We introduce parameters accessible in an experiment, the variance $v_{B}$ and the covariance in each quadrature $c_p$ and $c_q$ of our simulated experiment. That is, Alice and Bob obtain parameters from the PM scheme and then we bound $\rho_{AB}$ for any general quantum channel $\mathcal{E}$.
    
    After sending the initial state $|\Phi_4\rangle$ (see~\cref{eq:PHI4}) down the channel $\mathcal{E}_{\text{G}}$ with Gaussian lossy thermal-noise channel for each link, with transmissivity $T$ and excess noise $\xi$, we obtain
    \begin{equation}
    \rho_{AB}^{\text{G}} = (\Id \otimes \mathcal{E}_{\text{G}}) (|\Phi_4\rangle\langle\Phi_4|).
    \end{equation}
    
    Therefore, we define the parameters $v_{B}$, $c_q$, and $c_p$ that Alice and Bob would obtain during a typical experiment as follows
    \begin{align*}
    v_{B} &= \frac{1}{2}\tr[(\Id  \otimes (\Id  + 2 b^\dagger b)) \rho_{AB}^{\text{G}}]\\
    c_q &= \tr[\left( (|\psi_0\rangle \langle \psi_0|-|\psi_2\rangle\langle \psi_2|) \otimes \hat{q} \right)\rho_{AB}^{\text{G}}] \\
    c_p &=\tr[\left(  (|\psi_1\rangle \langle \psi_1| - |\psi_3\rangle\langle \psi_3|) \otimes \hat{p}\right)\rho_{AB}^{\text{G}}],
    \end{align*}
    where $\ket{\psi_k}$ are given by~\cref{eq:psik}, and $\Pi = \sum_{k=0}^3 
    \ketbra{\psi_k}{\psi_k}$ is the orthogonal projector onto the space spanned by the four coherent states (see~\cite{PhysRevX.9.021059}).
    
    We minimise $Z = \tr[(a\otimes b+a^\dagger \otimes b^\dagger) \rho_{AB}]$ over all final states $\rho_{AB}$ compatible with the parameters observed in the experiment, thus, finding an upper bound for Eve and a lower bound on the secret key rate.
    
    We require the unknown state $\rho_{AB}$ to be positive semidefinite, to have trace 1, and to be compatible with $v_{B}$, $c_q$, and $c_p$ (by satisfying the corresponding linear constraints). Letting the unknown density matrix $\rho_{AB}$ be designated by $X$, then our SDP is formulated as follows:
    \begin{align}\label{Chap7eqn:sdp}
    \min & \, Z = {\tr}[(a\otimes b+a^\dagger \otimes b^\dagger)X]\\
    \text{such that} &\left\{
    \begin{array}{l}
    {\tr}(B_0 X) ={\tr}(B_0 \rho_{AB}^{\text{G}}) \nonumber = v_{B}\\
     {\tr}( B_q X) = {\tr}(B_q \rho_{AB}^{\text{G}}) \nonumber = c_q\\
      {\tr}( B_p X) = {\tr}(B_p \rho_{AB}^{\text{G}}) \nonumber = c_p\\
     {\tr_B}( X ) = {\tr_B}( \rho_{AB}^{\text{G}} ) \\
      X  \succeq 0, \nonumber
    \end{array}
    \right.
    \end{align}
    where
    \begin{align*}
    B_0 &= \Id  \otimes (\Id  + 2 b^\dagger b)\\
    B_q &=  (|\psi_0\rangle \langle \psi_0|-|\psi_2\rangle\langle \psi_2|) \otimes \hat{q} \\
    B_p &= (|\psi_1\rangle \langle \psi_1|-|\psi_3\rangle\langle \psi_3|) \otimes \hat{p},
    \end{align*}
    where $X$ is the unknown density matrix, $a$ and $b$ are the annihilation operators acting on Alice's and Bob's modes respectively. The optimum of this program is $Z^* ={\tr}[(a\otimes b+a^\dagger \otimes b^\dagger)X^*]$. We then calculate Eve's information $\chi_{EB}$ for the Gaussian state $\rho_{AB}^*$ with the same covariance matrix as $\rho_{AB}$. $\chi_{EB}$ is easily calculable from the symplectic eigenvalues~\cref{Chap7eq:Holevo} of the covariance matrix for $\rho_{AB}^*$.
    
\begin{figure}
    \centering
    \includegraphics[width=0.6\textwidth]{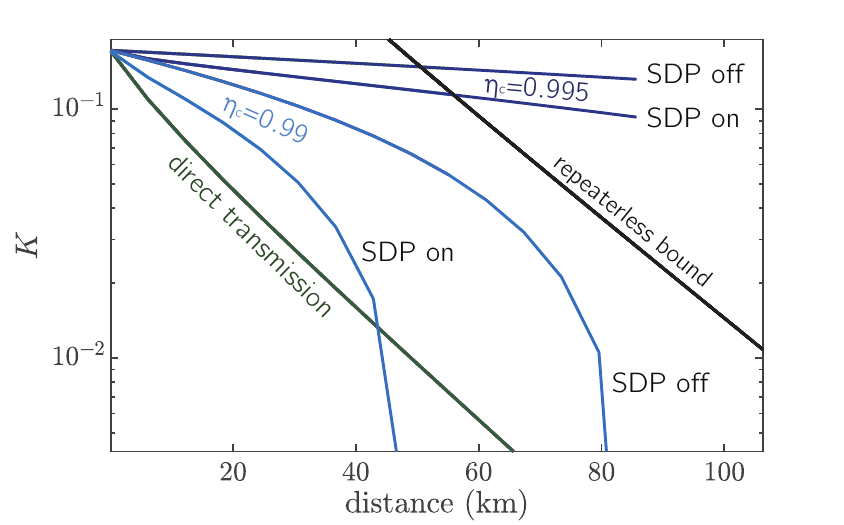}
    \caption[Secret key rate $K$ as a function of distance for the four-state protocol with repeaters (blue) or without repeaters (green)]{Secret key rate $K$ as a function of distance for the four-state protocol with repeaters (blue) or without repeaters (green). This plot shows the impact that running the SDP has on the key rate. We show results for two different coupling efficiencies $\eta_\text{c}=0.99$ and $\eta_\text{c}=0.995$. Also shown is the repeaterless bound. Note that for direct transmission whether the SDP is turned on or not gives identical results (green curve) since the channel is pure loss and pure loss is the only channel compatible with the PM parameters observed in the PM version. With repeaters, the error channel is approximated by a Pauli channel which adds non-Gaussianity so running the SDP decreases the key rate.\label{Chap7fig:SDP}}
    \end{figure}

    In~\cref{Chap7fig:SDP}, we plot the secret key rate as a function of distance for the four-state protocol with repeaters (blue) and for direct transmission (green), to show the effect of using the SDP. The repeaterless bound is also shown.
    
    \subsection{Two-state protocol SDP}

    We can also bound Eve's information for the two-state protocol; however, it gives loose bounds and poor results. We define the modulation direction to be at 45 degrees to the $q$ and $p$ quadrature directions. The program is
    \begin{align}\label{Chap7eqn:sdp}
    \min & \, Z = {\tr}[(a\otimes b+a^\dagger \otimes b^\dagger)X]\\
    \text{such that} &\left\{
    \begin{array}{l}
    {\tr}(B_0 X) ={\tr}(B_0 \rho_{AB}^{\text{G}}) \nonumber = v_{B}\\
     {\tr}( B_q X) = {\tr}(B_q \rho_{AB}^{\text{G}}) \nonumber = c_q\\
      {\tr}( B_p X) = {\tr}(B_p \rho_{AB}^{\text{G}}) \nonumber = c_p\\
     {\tr_B}( X ) = {\tr_B}( \rho_{AB}^{\text{G}} ) \\
      X  \succeq 0, \nonumber
    \end{array}
    \right.
    \end{align}
    where
    \begin{align*}
    B_0 &= \Id  \otimes (\Id  + 2 b^\dagger b)\\
    B_q &=  (|\psi_0\rangle \langle \psi_0|-|\psi_1\rangle\langle \psi_1|) \otimes \hat{q} \\
    B_p &= (|\psi_0\rangle \langle \psi_0|-|\psi_1\rangle\langle \psi_1|) \otimes \hat{p}.
    \end{align*}


    \section{Numerical results}
    
    We now present numerical lower bounds on the secret key rate. We assume that the encoding and decoding steps are perfect and that the recovery operation is the truncated CL recovery in \cref{Chap7eq:CL}, which performs optimally in the case of pure loss. We assume that the recoveries introduce additional pure loss due to a non-ideal coupling efficiency $\eta_\text{c}$. There is an optimal modulation $\alpha$, repeater spacing $L_0$, choice of cat code (with amplitude $\beta$, order of rotation-symmetry $N$, and Fock-shift parameter $s$) and encoding map for each choice of the channel parameters; total distance $L_\text{tot}$ and excess noise $\xi$ with thermal-noise variance $V$ (added constantly along the entire distance). We optimise these parameters assuming $\bar{n}_\text{code}\leq10$ and $N\leq5$ because lower-order cat codes are presumably more practical.

    Since the encoding and decoding steps are perfect and in the case of pure loss the recovery operation is optimal, our results provide the ultimate capabilities of the cat code for discrete-modulated CV QKD. The bound on the secret key rate is not tight because we underestimate Alice and Bob's information and overestimate Eve's information so our results lower bound the ultimate capabilities of the cat code for CV QKD.

    \subsection{Two-state protocol}
    
\begin{figure}
    \centering
    \includegraphics[width=0.6\textwidth]{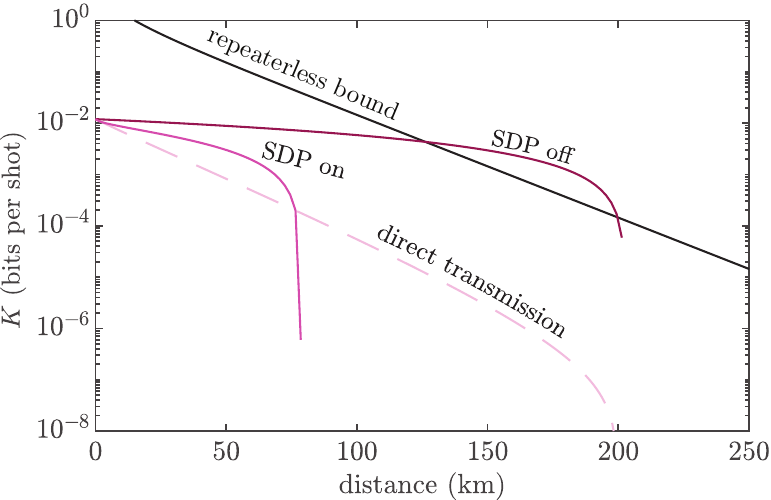}
    \caption[Asymptotic secret key rate $K$ vs. distance (km) for the two-state protocol]{Asymptotic secret key rate $K$ vs. distance (km) for the two-state protocol in units of bits per state prepared by Alice. The error-corrected protocol (using a cat code) is secure against Gaussian collective attacks (SDP off) or collective attacks (SDP on). We assume the encoding and decoding steps are perfect, the noise channel is pure loss with additional coupling efficiency $\eta_\text{c}=0.995$, and the recovery is optimal. We optimise over all cat codes (assuming $\bar{n}_\text{cat}\leq10$ and $N\leq5$) and repeater spacing $L_0$. Also plotted are direct transmission and the repeaterless bound. With SDP off, we beat the repeaterless bound at 126 km. \label{Chap7fig:twostateprotocolresults1}}
    \end{figure}

\begin{figure}
    \centering
    \includegraphics[width=0.6\textwidth]{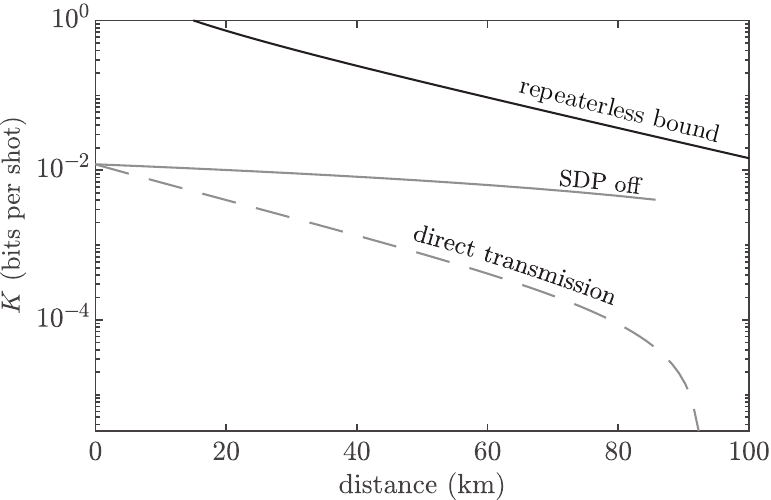}
    \caption[Same as~\cref{Chap7fig:twostateprotocolresults1} but with additional excess noise]{Same as~\cref{Chap7fig:twostateprotocolresults1} but with additional excess noise added in all the links with variance $V=1.00001$ (equivalent to $\xi=0.001$ at 100km of a direct-transmission lossy thermal-noise channel.\label{Chap7fig:twostateprotocolresults2}}
    \end{figure}

    The secret key rate for the two-state protocol is presented in~\cref{Chap7fig:twostateprotocolresults1,Chap7fig:twostateprotocolresults2} for pure loss and excess noise respectively. The coupling efficiency is $\eta_\text{c}=0.995$. The discrete-modulation amplitude is $\alpha=0.13$ which is about optimal. For direct transmission, the optimal amplitude $\alpha$ should be a function of distance but $\alpha = 0.13$ gives good performance for all distances so we keep it fixed. The behaviour of optimal $\alpha$ for the error-corrected protocol is nontrivial and depends on many things such as the security method, choice of bosonic code, etc. In~\cref{Chap7fig:twostateprotocolresults1,Chap7fig:twostateprotocolresults2} we roughly optimise the key assuming $\bar{n}_\text{cat}\leq10$ and $N\leq5$ and we get the unshifted order-$5$ cat code ($\beta{=}3.941,N=5,s=0$) and repeater spacing $L_0=0.0447$ km.
    
    We plot key rates secure against Gaussian collective attacks (SDP off) and collective attacks (SDP on) where we lower bound the key rate using the method from~\cite{PhysRevX.9.021059}). Also plotted are the repeaterless bound (black solid) and direct transmission (dashed).
    
    To calculate these numerical results, we used an artificially-truncated Fock space for $\rho_{AB}$ with dimension size $d_A d_B$, where $d_A$ and $d_B$ are the sizes of Alice and Bob's truncated Hilbert spaces. 
    
    In~\cref{Chap7fig:twostateprotocolresults2}, pure loss is a lot more dominant than thermal noise (considering that there is also coupling loss), therefore, the truncated CL recovery performs near-optimally in this very lossy regime. For a large amount of thermal noise, it becomes advantageous to modify the recovery.

    \subsection{Extension to a four-state protocol}
    
    The discrete-modulation protocol with repeaters can be extended to a four-state protocol. The PM and EB versions are outlined in~\cref{Chap7app:four_state_protocol}. We found that encoding the qudit into a single-mode code does not perform well; a better solution is to encode the qudit into two single-mode qubit codes. Having a qudit stored in a single mode requires a large mean photon number and leads to large probabilities of photon-loss events and so the qudit cat code performs quite poorly. Encoding the qudit in two single-mode codes requires a smaller mean photon number and the loss events are rarer, thus, spreading the quantum information across two modes better protects against loss.

    To simplify the calculation for the four-state protocol, the error channel is approximated by a Pauli channel~\cite{Li_2017}. We argue that this is a good approximation because the Pauli channel is a pretty-good approximation in the case of the two-state protocol. Working in Fock-space becomes intractable for the four-state protocol since Bob owns two modes and the truncated state for $\rho_{AB}$ has dimension size $d_A d_B^2$. For the two-state protocol, the Pauli channel agrees well with the numerical calculations (for moderate $\beta$), so we approximate the results for the four-state protocol using the Pauli channel.
    
\begin{figure}
    \centering
    \includegraphics[width=0.6\textwidth]{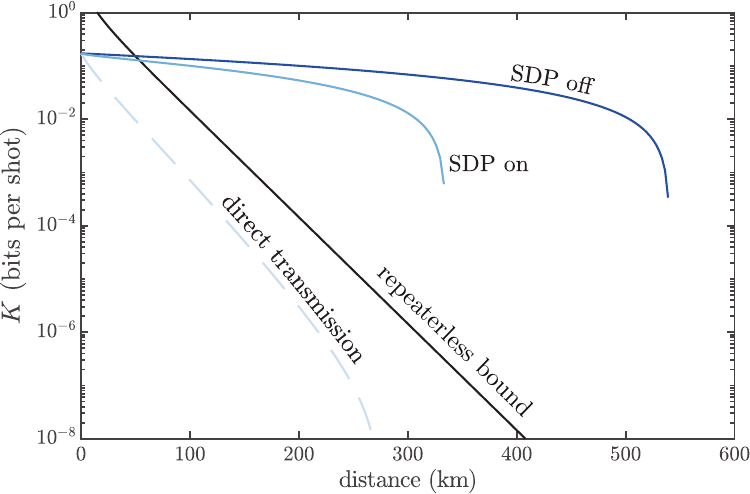}
    \caption[Asymptotic secret key rate $K$ vs. distance (km) for the four-state protocol and pure-loss channel and coupling efficiency $\eta_\text{c}=0.995$]{Asymptotic secret key rate $K$ in units of bits per state prepared by Alice vs. distance (km) for the four-state protocol and pure-loss channel and coupling efficiency $\eta_\text{c}=0.995$ (all other parameters and assumptions are the same as~\cref{Chap7fig:twostateprotocolresults1}). We beat the repeaterless bound at 50 km (SDP on) and 54 km (SDP off), and we achieve a better maximum distance than direct transmission.\label{Chap7fig:fourstateprotocolresults}}
    \end{figure}

    The secret key rate for the four-state protocol for pure loss is presented in~\cref{Chap7fig:fourstateprotocolresults}. We plot key rates secure against Gaussian collective attacks (SDP off) and collective attacks (SDP on). The coupling efficiency is $\eta_\text{c}=0.995$. Roughly optimal is the unshifted order-$5$ cat code with $\beta{=}4.566$ and repeater spacing $L_0=0.036$ km.

    \subsection{Dependence on the coupling efficiency}
    
\begin{figure}
    \centering
    \includegraphics[width=0.6\textwidth]{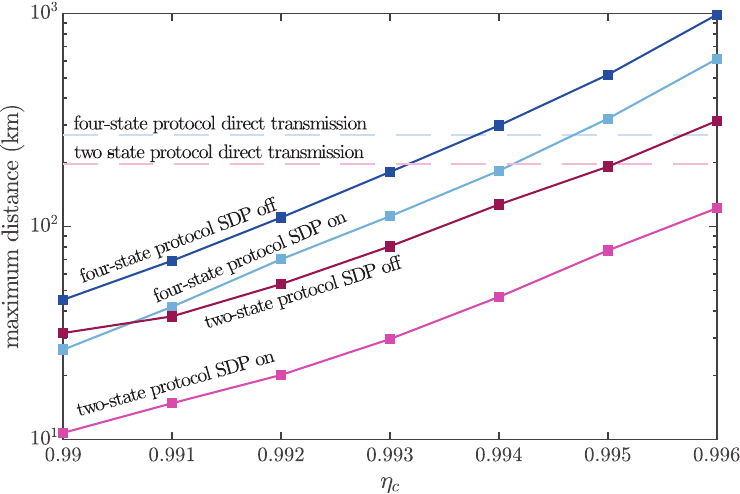}
    \caption[Lower bound on the maximum attainable distance as a function of coupling efficiency $\eta_\text{c}$ for the two-state and four-state protocols and pure-loss links]{Lower bound on the maximum attainable distance as a function of coupling efficiency $\eta_\text{c}$ for the two-state and four-state protocols and pure-loss links, secure against Gaussian collective attacks (SDP off) or collective attacks (SDP on). The repeater protocols perform better than direct transmission whenever the maximum distance is above the dashed lines.\label{Chap7fig:maxdistance}}
    \end{figure}

    The key rate strongly depends on the coupling efficiency $\eta_\text{c}$. The protocol performs well if $\eta_\text{c}$ is sufficiently high. \cref{Chap7fig:maxdistance} shows the maximum distance achievable as a function of the coupling efficiency $\eta_\text{c}$ for the two-state and the four-state protocols, secure against Gaussian or arbitrary attacks. The dashed lines are for direct transmission. Achieving a maximum distance above the dashed line means that the error-corrected protocol can achieve greater distances than direct transmission.
    
    \section{Physically implementing the encoding and decoding steps}
    
\begin{figure}
    \centering
    \includegraphics[width=0.5\textwidth]{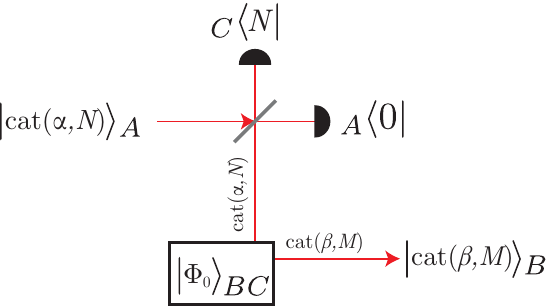}
    \caption[A device for teleporting a cat-encoded qubit from an order-$N$ cat code at the input (mode $A$) to an order-$M$ cat code at the output (mode $B$)]{A device for teleporting a cat-encoded qubit from an order-$N$ cat code at the input (mode $A$) to an order-$M$ cat code at the output (mode $B$). The entanglement resource state is a two-mode Bell state where each of the entangled qubit modes is encoded in different cat codes as shown (modes $B$ and $C$). The beamsplitter has transmissivity $1/2$ and the two ports count photons.\label{Chap7fig:catteleporter}}
    \end{figure}

    The encoding and decoding steps transfer the quantum information between cat codes of different sizes and orders of rotation symmetry. For the two-state protocol for instance, Alice initially encodes the qubit from cat($\alpha$,1) to cat($\beta$,$N$) and finally, Bob decodes back to cat($\alpha$,1) and measures the quadratures. Transferring the quantum information between different codes may be performed via teleportation. One example of such a teleportation device is shown in~\cref{Chap7fig:catteleporter}. It requires a highly non-Gaussian entangled resource state, beamsplitters, and photon counting. Another possibility for teleportation is to couple different cat states via the controlled-Phase interaction of Ref.~\cite{PhysRevX.10.011058}.
    
    Let's assume the initial code is cat$(\alpha,N)$ and the output code is cat$(\beta,M)$. The initial state of the linear-optical teleportation device is
    \begin{align}
        \ket{\psi}_{ABC} &=  (c_0 \ket{0_{\text{cat}(\alpha,N)}}_A + c_1 \ket{1_{\text{cat}(\alpha,N)}}_A) \ket{\Phi_0}_{BC},
    \end{align}
    where
    \begin{align}
        \ket{\Phi_0}_{BC} &= \frac{\ket{0_{\text{cat}(\beta,M)}}_B \ket{0_{\text{cat}(\alpha,N)}}_C +  \ket{1_{\text{cat}(\beta,M)}}_B \ket{1_{\text{cat}(\alpha,N)}}_C }{\sqrt{2}},
    \end{align}
    where $\ket{0_{\text{cat}(\alpha,N)}}$ and $\ket{1_{\text{cat}(\alpha,N)}}$ are the logical states of an order-$N$ cat code with amplitude $\alpha$. We pass modes $A$ and $C$ through a $50:50$ beamsplitter and measure the output modes with single photon detectors. We accept measurement outcomes: vacuum at mode $A$ and Fock state $\bra{N}$ at mode $C$. 
    \begin{align}
    &\propto (c_0 \ket{1_M} + c_1 \ket{0_M} ),\label{Chap7eq:teleportation}
    \end{align}
    where the success probability of the device is given by the norm of the output state which depends on the size and order of the input cat code to the device ($\alpha,N$).
    
    
    \section{Discussion and conclusion}

    Our numerical results show that the cat code can in principle speed up the rate of CV-QKD protocols. We focused on discrete-modulation CV QKD, where the states prepared by Alice are chosen from a finite set, forming a low-dimensional system which can be redundantly encoded into a bosonic code. The protocol is useful for three reasons: 1. The low-dimensional subspace of the protocol allows a proper encoding into a single or small number of modes such that the information is robust to errors. 2. Good reconciliation techniques exist for discrete-modulation CV QKD which make them preferable to continuous Gaussian-modulated CV QKD. 3. Non-Gaussianity can be dealt with in the security analysis~\cite{Liao_long_distance,PhysRevX.9.021059,PhysRevX.9.041064,papanastasiou2020continuousvariable,kaur2020asymptotic,matsuura2020finitesize}.
    
    Since a discrete modulation is a good approximation to a Gaussian for small modulation variances our work may inspire ideas for protecting arbitrary CV input states from noise. The entanglement-based version of a Gaussian protocol is equivalent to distributing a CV TMSV state $\ket{\chi}$ so our error-corrected protocol may inspire ideas for entanglement distribution as well.
    
    Our repeaters cannot extend the distance of the discrete-modulation protocol to arbitrarily far distances because eventually the errors build up and the secret key rate drops to zero. Fault tolerance is not a requirement in QKD. One just needs to be able to faithfully transmit qubits~\cite{DiVincenzo_2000} and prove security. One way to achieve fault-tolerance would be to add a level of DV encoding on top of the bosonic code so that the weaknesses of the base-level bosonic code can be protected by the overarching DV code.

    A fundamental limitation of the security method used here is that the bound on the secret key is not tight. Future work is to improve the tightness of estimating Alice and Bob's information and bounding Eve's information. Finite-size effects are also left for future work.
    
    An interesting direction would be to investigate how other techniques (for example, CV entanglement swapping or CV teleportation) and encodings (for example, coupled with other modes, concatenation with higher-level codes, etc.) may increase the rate and distance. One can also investigate other bosonic codes.
    
    We have introduced a new approach for the third generation of quantum repeaters in CV. We investigated discrete-modulation CV QKD using the single-mode cat code for the error correction. We calculated secret key rates in the asymptotic limit of infinitely long keys, valid against arbitrary collective attacks, bounding Eve's information by solving an SDP. Finally, we introduced teleportation for the encoding and decoding steps.


%% file: Conclusion/Conclusion.tex
\chapter{Conclusions and future work}
\label{Chap:Conclusion}


In this thesis, we introduced capacity-achieving quantum repeaters for continuous variables. Our quantum repeaters give excellent rates and allow for the transmission of large amounts of entanglement. We introduced quantum repeaters with the highest possible rates allowed by quantum physics. 

Some of our simplest quantum repeaters can be implemented in the lab in the near future. These low-hanging fruit experiments are expected to form the first multi-node quantum communication network. While they can perform simple, exotic tasks in quantum communications, they will also become building blocks for a global-scale quantum internet. Our architectures are useful for near-term demonstrations, but also stepping stones to more-advanced technologies. 

We provide ideas for possible future research directions in the concluding paragraphs of each chapter of this thesis. Please consult those specific pages for more details and unanswered questions. We summarise our recommendations for further research as follows.

First, there is a theoretical gap between the best achievable upper and lower bounds of quantum communication rates under the thermal-loss channel~\cite{Pirandola_2017}. Furthermore, it is unknown what practical repeater design can achieve the highest end-to-end rates for entanglement distillation under the thermal-loss channel~\cite{Pirandola_2019}. It is also unknown what practical repeater design can achieve the highest rates for other channels, such as dephasing. Another theoretical problem regards the lack of security proofs in CV QKD for non-Gaussian protocols that are non-deterministic.

One of the caveats of this thesis is that we assumed ideal quantum memories with unit efficiencies and infinite lifetimes. An important task is to consider all our protocols with non-ideal quantum memories, for instance, taking a similar approach to Ref.~\cite{Ghalaii_2020}. We expect that the rates will be excellent as long as the efficiencies of the memories are high and the additional noise introduced is not too great.

It would be interesting to consider how our repeaters combine into more complicated networks beyond chains, using the results from~\cite{Pirandola_2019}; that is, to reliably connect multiple users in a network where the repeaters are untrusted. Furthermore, what benefits are there from combining our different architectures into the same network? Some repeater designs may be better suited for different communication tasks and physical platforms.

All our protocols are useful for improving the performance of CV quantum channels. One could use our repeaters to do a similar analysis like Ref.~\cite{dias2018quantum}. Also, a recent interesting study performed a comparison of our repeater protocols in Ref.~\cite{https://doi.org/10.48550/arxiv.2207.13284}. Comparisons of this kind are useful for considering the trade-offs in an experiment. It would be beneficial to do a similar analysis with a focus on the efficiency and lifetime requirements of the quantum memories.

A non-deterministic error-corrected repeater, i.e., a hybrid approach, could enjoy benefits from both first-generation and third-generation approaches, i.e., combining the repeaters from~\cref{Chap:scissors,Chap:CV_QR,Chap:simple_repeater,Chap:purification} with the repeaters from~\cref{Chap:QEC}. The ultimate rate for a scheme would lie between the two-way assisted capacity and the unassisted capacity.

We wonder if it is possible to remove the hard truncation cutoff introduced by the quantum scissors used throughout~\cref{Chap:scissors,Chap:CV_QR,Chap:simple_repeater}. The truncation is a strongly non-Gaussian effect which can limit the performance of the scissors. In particular, it limits repeaters to the high-fidelity regime, i.e., for distributing CV entanglement without too much squeezing or large photon numbers.

An interesting research direction would be to dream up a repeater protocol that achieves the single-repeater bound without requiring quantum memories. The rate of our simple repeater, introduced in~\cref{Chap:simple_repeater}, scales excellently like the single-repeater bound, but does not saturate it.

With regards to the minimum-leakage protocol from~\cref{Chap:other}, one may consider a switching minimum-leakage protocol, rather than heralding. Another problem is to perform a finite-size security analysis for the minimum-leakage protocol. Since it cannot beat the repeaterless bound, it may be interesting to turn the minimum-leakage protocol into an effective quantum repeater.

In~\cref{Chap:CV_QR}, we introduced our CV repeater assuming pure-loss channels and perfect sources, devices, and quantum memories. Further work is required to consider our CV repeater in a thermal-loss channel, and with additional experimental imperfections.

Our iterative purification protocol, considered in~\cref{Chap:purification}, requires two-way classical communication, while the optimal protocol explicit via the RCI requires only one-way classical communication. We wonder if there is a practical protocol that can achieve the repeaterless bound using just one-way classical communication. Also, our purification protocol heralds pure states under the pure-loss channel. Highly pure states are useful in large quantum networks, and for more powerful demonstrations in quantum communications, such as violating a Bell inequality or testing quantum foundations.

While the internet is essentially a primitive telephone system, the global quantum internet will be more encompassing, connecting powerful machines manipulating and processing quantum information. Quantum repeaters will be vital for the quantum internet, protecting fragile quantum information from unwanted interactions with the environment.

Quantum physics has provided an understanding of physical phenomena, incredible medical breakthroughs, and large amounts of computing power. We should imagine that the future is filled with equally remarkable feats possible by processing information directly with the quantum systems themselves. 

If you are reading this thesis on a device capable of processing quantum information, then it is a very remarkable time! Your device may be connected to other devices in a global quantum internet, requiring highly-effective, high-rate quantum repeaters. In practice, limited experimental resources pose critical constraints on repeater performance, and a large experimental effort must be undertaken shortly to achieve the first scalable multi-node quantum communication network.


%% file: Appendix_minimum_leakage/Appendix_minimum_leakage.tex

\chapter{Supplementary material for the minimum-leakage protocol}

\section{\label{Chap8app:sec:appendix:heralding} Covariance matrix heralding protocol}

The final covariance matrix shared between Alice and Bob for our heralding protocol can be calculated from the entanglement-based scheme shown in~\cref{Chap8fig:heralding_EB} of the main text of this thesis. All the elements are Gaussian so it is a simple matter of applying the appropriate symplectic transformations.

The initial state of modes of $A1,B$ and $A2,A3$ are two TMSV sources.
\begin{align*}
\Gamma_{A1A2BA3\text{ initial state}} &= \left[ \begin{smallmatrix}   
                                 \mu &                 0 &                0 &                 0 & \sqrt{\mu^2-1} &                 0 &                0 &                 0\\
                0 &                \mu &                0 &                 0 &                0 & -\sqrt{\mu^2-1} &                0 &                 0\\
                0 &                 0 &               \mu &                 0 &                0 &                 0 & \sqrt{\mu^2-1} &                 0\\
                0 &                 0 &                0 &                \mu &                0 &                 0 &                0 & -\sqrt{\mu^2-1}\\
 \sqrt{\mu^2-1} &                 0 &                0 &                 0 &               \mu &                 0 &                0 &                 0\\
                0 & -\sqrt{\mu^2-1} &                0 &                 0 &                0 &                \mu &                0 &                 0\\
                0 &                 0 & \sqrt{\mu^2-1} &                 0 &                0 &                 0 &               \mu &                 0\\
                0 &                 0 &                0 & -\sqrt{\mu^2-1} &                0 &                 0 &                0 &                \mu\\
  \end{smallmatrix} \right] .
\end{align*}

Mode B is squeezed with squeezing $r$ and mode A3 with squeezing $-r$ and the covariance matrix becomes
\begin{align*}
    \Gamma_{A1A2BA3 \text{ after squeezing}} &= S \; \Gamma_{A1A2BA3\text{ initial state}} \; S^T,
\end{align*}
where
\begin{align*}
    S &= \begin{bmatrix}   
                  1 &   0 & 0 &  0 &  0 & 0 & 0 & 0 \\
                  0  &   1 & 0 &  0 &  0 & 0 & 0 & 0 \\
                  0  &   0 & 1 &  0 &  0 & 0 & 0 & 0 \\
                  0  &   0 & 0 &  1 &  0 & 0 & 0 & 0 \\
                  0  &   0 & 0 &  0 &  e^{-r} & 0 & 0 & 0 \\
                  0  &   0 & 0 &  0 &  0 & e^{r} & 0 & 0 \\
                  0  &   0 & 0 &  0 &  0 & 0 & e^{r} & 0 \\
                  0  &   0 & 0 & 0 &  0 & 0 & 0 & e^{-r} \\
  \end{bmatrix}.
\end{align*}
That is, after the squeezing we have
{\small{
\begin{align*}
\hspace*{-0.5cm} \Gamma_{A1A2BA3 \text{ after squeezing}} &= \left [ \begin{smallmatrix}  
                                              \mu &                        0 &                       0 &                         0 & e^{-r} \sqrt{\mu^2-1} &                        0 &                       0 &                         0\\
                        0 &                       \mu &                       0 &                         0 &                        0 & -e^{r} \sqrt{\mu^2-1} &                       0 &                         0\\
                        0 &                        0 &                      \mu &                         0 &                        0 &                        0 & e^{r} \sqrt{\mu^2-1} &                         0\\
                        0 &                        0 &                       0 &                        \mu &                        0 &                        0 &                       0 & -e^{-r} \sqrt{\mu^2-1}\\
 e^{-r} \sqrt{\mu^2-1} &                        0 &                       0 &                         0 &             \mu e^{-2r} &                        0 &                       0 &                         0\\
                        0 & -e^{r} \sqrt{\mu^2-1} &                       0 &                         0 &                        0 &              \mu e^{2r} &                       0 &                         0\\
                        0 &                        0 & e^{r} \sqrt{\mu^2-1} &                         0 &                        0 &                        0 &             \mu e^{2r} &                         0\\
                        0 &                        0 &                       0 & -e^{-r} \sqrt{\mu^2-1} &                        0 &                        0 &                       0 &              \mu e^{-2r}\\
  \end{smallmatrix} \right ].
\end{align*}
}}

Modes B and A3 are mixed on a balanced beam splitter and the transformed covariance matrix is
\begin{align*}
    \Gamma_{A1A2BA3 \text{ after beam splitter}} &= B \; \Gamma_{A1A2BA3 \text{ after squeezing}} \; B^T,
\end{align*}
where
\begin{align*}
    B &= \begin{bmatrix}   
                  1 &   0 & 0 &  0 &  0 & 0 & 0 & 0 \\
                  0  &   1 & 0 &  0 &  0 & 0 & 0 & 0 \\
                  0  &   0 & 1 &  0 &  0 & 0 & 0 & 0 \\
                  0  &   0 & 0 &  1 &  0 & 0 & 0 & 0 \\
                  0  &   0 & 0 &  0 &  \sqrt{0.5} & 0 & \sqrt{0.5} & 0 \\
                  0  &   0 & 0 &  0 &  0 & \sqrt{0.5} & 0 & \sqrt{0.5} \\
                  0  &   0 & 0 &  0 &  -\sqrt{0.5} & 0 & \sqrt{0.5} & 0 \\
                  0  &   0 & 0 & 0 &  0 & -\sqrt{0.5} & 0 & \sqrt{0.5} \\
  \end{bmatrix}.
\end{align*}

That is, we have
{\small{
\begin{align*}
\hspace*{-0.5cm} \Gamma_{A1A2BA3 \text{ after beam splitter}} &= \left [ \begin{smallmatrix}  
    \mu &                                    0 &                                   0 &                                     0 & \frac{e^{-r} \sqrt{\mu^2-1}}{\sqrt{2}} &                                     0 & -\frac{e^{-r} \sqrt{\mu^2-1}}{\sqrt{2}} &                                     0\\
                                     0 &                                   \mu &                                   0 &                                     0 &                                    0 &  -\frac{e^{r} \sqrt{\mu^2-1}}{\sqrt{2}} &                                     0 &   \frac{e^{r} \sqrt{\mu^2-1}}{\sqrt{2}}\\
                                     0 &                                    0 &                                  \mu &                                     0 &  \frac{e^{r} \sqrt{\mu^2-1}}{\sqrt{2}} &                                     0 &   \frac{e^{r} \sqrt{\mu^2-1}}{\sqrt{2}} &                                     0\\
                                     0 &                                    0 &                                   0 &                                    \mu &                                    0 & -\frac{e^{-r} \sqrt{\mu^2-1}}{\sqrt{2}} &                                     0 & -\frac{e^{-r} \sqrt{\mu^2-1}}{\sqrt{2}}\\
  \frac{e^{-r} \sqrt{\mu^2-1}}{\sqrt{2}} &                                    0 & \frac{e^{r} \sqrt{\mu^2-1}}{\sqrt{2}} &                                     0 &                         \mu \cosh{2r} &                                     0 &                          \mu \sinh{2r} &                                     0\\
                                     0 & -\frac{e^{r} \sqrt{\mu^2-1}}{\sqrt{2}} &                                   0 & -\frac{e^{-r} \sqrt{\mu^2-1}}{\sqrt{2}} &                                    0 &                          \mu \cosh{2r} &                                     0 &                         -\mu \sinh{2r}\\
 -\frac{e^{-r} \sqrt{\mu^2-1}}{\sqrt{2}} &                                    0 & \frac{e^{r} \sqrt{\mu^2-1}}{\sqrt{2}} &                                     0 &                         \mu \sinh{2r} &                                     0 &                          \mu \cosh{2r} &                                     0\\
                                     0 &  \frac{e^{r} \sqrt{\mu^2-1}}{\sqrt{2}} &                                   0 & -\frac{e^{-r} \sqrt{\mu^2-1}}{\sqrt{2}} &                                    0 &                         -\mu \sinh{2r} &                                     0 &                          \mu \cosh{2r}\\
  \end{smallmatrix} \right ].
\end{align*}
}}

Homodyne detection is performed on mode A3, assumed to be in the $x$ quadrature. The heralded covariance matrix is
\begin{align*}
\Gamma_{A1A2B|\text{hom. }A3_x} &= \Gamma_{A1A2B} - \sigma (\Pi \;\Gamma_{A3} \; \Pi)^{-1} \sigma^T,
\end{align*}
where $\Gamma_{A1A2B}$, $\Gamma_{A3}$ and $\sigma$ are submatrices given by
 \begin{align*}
\Gamma_{A1A2BA3 \text{ after beam splitter}} &= \begin{bmatrix}   
    \Gamma_{A1A2B} & \sigma \\
    \sigma^T & \Gamma_{A3}\\
  \end{bmatrix}.
\end{align*}

That is, we have
 \begin{align*}
\Gamma_{A1A2B|\text{hom. }A3_x} &= \left[  \begin{smallmatrix}  
               \frac{e^{4r} \mu^2 + 1}{\mu (e^{4r} + 1)} &                                    0 &        \frac{e^{2r} (\mu^2 - 1)}{\mu (e^{4r} + 1)} &                                     0 & \frac{\sqrt{2} e^{3r} \sqrt{\mu^2-1}}{e^{4r} + 1} &                                     0\\
                                                  0 &                                   \mu &                                                0 &                                     0 &                                                  0 &  -\frac{e^{r} \sqrt{\mu^2-1}}{\sqrt{2}}\\
          \frac{e^{2r} (\mu^2 - 1)}{\mu (e^{4r} + 1)} &                                    0 &            \frac{\mu^2 + e^{4r}}{\mu (e^{4r} + 1)} &                                     0 &   {\sqrt{2} e^{r} \sqrt{\mu^2-1}}{e^{4r} + 1} &                                     0\\
                                                  0 &                                    0 &                                                0 &                                    \mu &                                                  0 & -\frac{e^{-r} \sqrt{\mu^2-1}}{\sqrt{2}}\\
 \frac{\sqrt{2} e^{3r} \sqrt{\mu^2-1}}{e^{4r} + 1} &                                    0 & \frac{\sqrt{2} e^{r} \sqrt{\mu^2-1}}{e^{4r} + 1} &                                     0 &                     \frac{2 \mu e^{2r}}{e^{4r} + 1} &                                     0\\
                                                  0 & -\frac{e^{r} \sqrt{\mu^2-1}}{\sqrt{2}} &                                                0 & -\frac{e^{-r} \sqrt{\mu^2-1}}{\sqrt{2}} &                                                  0 &                          \mu \cosh{2r}\\
  \end{smallmatrix}  \right].
\end{align*}

Finally, the covariance matrix after the symmetric channel with transmissivity $T$ and excess noise $\xi$ is given by
\begin{align*}
\Gamma_{A1A2B \text{ after channel}|\text{hom. }A3_x} &= \begin{bmatrix}   
                  a_1 &    0 &              c_1 &    0 &         c_2 &         0 \\
  0 &    {\mu} &       0 &          0 &         0 &            c_3 \\
               c_1 &      0 &   a_2 &     0 &      c_4 &     0 \\
 0 &      0 &    0 &   {\mu} &     0 &   c_5 \\
 c_2 &     0 & c_4 &     0 & b_1 &    0 \\
  0 & c_3 &   0 & c_5 &   0 & b_2 \\
  \end{bmatrix},
\end{align*}
where
\begin{align*}
a_1 &= \frac{e^{4r}{\mu}^2 + 1}{{\mu}(e^{4r} + 1)}\\
a_2 &= \frac{{\mu}^2 + e^{4r}}{{\mu}(e^{4r} + 1)}\\
b_1 &= \frac{e^{4r} - T + T\xi - Te^{4r} + T\xi e^{4r} + 2T{\mu}e^{2r} + 1}{e^{4r} + 1}\\
b_2 &= T\xi - T + \frac{1}{2}T({\mu}e^{-2r} + {\mu}e^{2r}) + 1\\
c_1 &= \frac{e^{2r}({\mu}^2 - 1)}{{\mu}(e^{4r} + 1)} \\
c_2 &= \frac{\sqrt{2T}e^{3r}\sqrt{\mu^2-1}}{e^{4r} + 1}\\
c_3 &= \frac{-\sqrt{2T}e^{r}\sqrt{\mu^2-1}}{2} \\
c_4 &= \frac{\sqrt{2T}e^{r}\sqrt{\mu^2-1}}{e^{4r} + 1}\\
c_5 &= \frac{-\sqrt{2T}e^{-r}\sqrt{\mu^2-1}}{2}.\\
\end{align*}

The Holevo bound can be calculated from the symplectic eigenvalues of the covariance matrix via~\cref{eq:HOLEVO_pur} of the main text, i.e., $\chi_{EB} = S(\rho_{A1A2B}) - S(\rho_{A1A2|B})$.

Explicitly, we have that the Holevo quantity is
\begin{equation}
\chi_{EB} = S(\Gamma_{A1A2B \text{ after channel}|\text{hom. }A3_x}) - S(\Gamma_{A1A2 \text{ after channel}|\text{hom. }A3_x\text{ and }B_x})
\end{equation}
where $S(\cdot)$ is the von Neumann entropy which for Gaussian states can be determined numerically from the symplectic eigenvalues $\nu_k$ of the covariance matrix.

$\Gamma_{A1A2 \text{ after channel}|\text{hom. }A3_x\text{ and }B_x}$ is the covariance matrix after the channel of Alice's modes, $A1$ and $A2$, conditioned on Alice's $x$ measurement of mode $A3$ and Bob's $x$ measurement of $B$:
\begin{align*}
\Gamma_{A1A2 \text{ after channel}|\text{hom. }A3_x\text{ and }B_x} &= \Gamma_{A1A2 \text{ after channel}|\text{hom. }A3_x } - \kappa (\Pi \;\Gamma_{B \text{ after channel}|\text{hom. }A3_x} \; \Pi)^{-1} \kappa^T,
\end{align*}
where $\Gamma_{A1A2 \text{ after channel}|\text{hom. }A3_x }$, $\Gamma_{B \text{ after channel}|\text{hom. }A3_x}$ and $\kappa$ are submatrices given by
 \begin{align*}
\Gamma_{A1A2B \text{ after channel}|\text{hom. }A3_x} &= \begin{bmatrix}   
    \Gamma_{A1A2 \text{ after channel}|\text{hom. }A3_x } & \kappa \\
    \kappa^T & \Gamma_{B \text{ after channel}|\text{hom. }A3_x}\\
  \end{bmatrix}.
\end{align*}

Note that to eliminate Eve's information, Alice's heralded measurement (homodyne mode $A3$) and Bob's measurement (homodyne mode $B$) should both be performed in the same quadrature. In this thesis, for instance, we always assume modes $A3$ and $B$ of this protocol are measured in the $x$ quadrature. Alice measures $A1$ and $A2$ in $x$ and $p$ respectively in order to switch to the equivalent PM version of the protocol.

%% file: Appendix_scissors/Appendix_scissors.tex

\chapter{Supplementary material for our quantum scissors}

\section{\label{sec:appendixPsuc}Success probabilities}

In this section we write down the success probabilities. The probability of success is given by the squared norm of the unnormalized state which depends on the input state. 

For a single-photon quantum scissor ($\oneinjectedscissor$), the success probability is
\begin{align}
P_{\oneinjectedscissor} &=  2 \; \frac{1}{2(g^2+1)} (|c_0|^2+|g c_1|^2).
\end{align}
The squared norm has been multiplied by 2 because we assume that the phase shift on the single photon component given by the reverse click pattern can be corrected with a phase-shifter.

For a $\threeinjectedscissor$, the success probability is
\begin{align}\label{eq:PROB}
P_{\threeinjectedscissor} &= 4 \; \frac{3}{32} \left(\frac{1}{g^2+1}\right)^{3}  \; (|c_0|^2+|g c_1|^2 +|g^2 c_2|^2+|g^3 c_3|^2),
\end{align}
where we multiply by a factor of 4 which accounts for the other three measurement outcomes, which lead to a heralded phase shift of the state which can be passively corrected for with a phase-shifter.

The success probability for $N$ $\oneinjectedscissors$ in parallel is
\begin{align}
P_{\Nscissors} &=  \left(\frac{1}{g^2+1}\right)^{N}  \left( \sum_{n=0}^{N} \left| \frac{N!}{(N-n)!N^n} g^n c_n \right|^2 \right),
\end{align}
assuming again that phase flips in each $\oneinjectedscissor$ are corrected for by feeding forward. The total success probability decreases with $N$. For large $N$, the device approaches an ideal NLA and has a vanishing success probability.

\section{\label{sec:appendixmultiplescissors}NLA based on multiple single photon scissors in parallel}

The scheme for NLA from Ref.~\cite{ralph2009nondeterministic} performs the transformation
\begin{align}
\hat{T}_{\Nscissors} \ket{\psi} = \left(\frac{1}{g^2+1}\right)^\frac{N}{2} \sum_{n=0}^{N} \frac{N!}{(N-n)!N^n} g^n c_n \ket{n}.
\end{align}
NLA based on multiple single photon scissors perform distorted amplification for practical numbers of scissors $N$. It is assumed that for NLA based on multiple single photon scissors, phase flips on the one-photon component for each individual scissor are corrected by feeding forward (thus doubling the success probability of each scissor in the NLA), i.e., for $\twoscissors$ in parallel
\begin{align}
\hat{T}_{\twoscissors} \ket{\psi} &= \frac{1}{g^2+1}   (c_0 |0\rangle + g c_1 |1\rangle + \frac{1}{2} g^2 c_2 |2\rangle ),
\end{align}
\begin{align}
P_{\twoscissors} &=  \left(\frac{1}{g^2+1}\right)^{2} (|c_0|^2 + |g c_1|^2 + \frac{1}{4}|g^2 c_2|^2),
\end{align}
and for $\threescissors$
\begin{align}\label{AAA}
\hat{T}_{\threescissors} \ket{\psi} &= \left(\frac{1}{g^2+1}\right)^\frac{3}{2}   \; (c_0 |0\rangle + g c_1 |1\rangle + \frac{2}{3} g^2 c_2 |2\rangle +  \frac{2}{9} g^3 c_3 |3\rangle ),
\end{align}
\begin{align}
P_{\threescissors} &=  \left(\frac{1}{g^2+1}\right)^{3}  \; (|c_0|^2 + |g c_1|^2 + \frac{4}{9}|g^2 c_2|^2 + \frac{4}{81}|g^3 c_3|^2).
\end{align}

%% file: Appendix_CV_QR/Appendix_CV_QR.tex

\chapter{Supplementary material for our CV repeater}\label{app:CV_repeater}

\section{Derivation of the output state of the single-node CV repeater}\label{app:derivation_single_node}

In this appendix, we outline how to calculate the entangled output state of the single-node repeater protocol (\cref{Chap4fig:asymrep1} in the main text). Initial entanglement distribution begins with generating two independent two-mode Gaussian squeezed vacuum states of form: 
\begin{equation}
\ket{\chi}_{AC}= \sqrt{1-\chi^2} \sum_{n=0}^\infty \chi^n \ket{n}_A\ket{n}_C
\end{equation}
The sources are placed with Alice and the repeater node, with one mode of each entangled state distributed to the repeater node and Bob respectively. This is modelled by a pure-loss channel of transmission $\eta$.  This transforms mode $C$ as: 
\begin{align}
\begin{split}
\hat{U}_{BS}  \left[ \ket{n}_C\ket{0}_D \right]  = \sum_{p =0}^n \sqrt{\binom{n}{p}} \eta^{p/2} \left(1-\eta\right)^{(n-p)/2} \ket{p}_C\ket{n-p}_D ,
\end{split}
\end{align}
where mode $D$ is an environment mode. The state becomes:
\begin{align}
\begin{split}
\ket{\chi}_{AC}  \to  \sqrt{1-\chi^2} \sum_{n=0}^\infty \sum_{p =0}^n  \chi^n \ket{n}_A
& \sqrt{\binom{n}{p}} \eta^{p/2} \left(1-\eta\right)^{(n-p)/2} \ket{p}_C \ket{n-p}_D .
\end{split}
\end{align}
Entanglement distillation proceeds by acting an NLA on mode $C$ with gain $g$. The action of the NLA with the single-photon quantum scissors can be described by the following operation \cite{dias2017}:
\begin{equation}
\hat{T}_1 = \hat{\Pi}_1 g^{\hat{n}},
\end{equation}
where the truncation operator $\hat{\Pi}_1$ is defined as:
\begin{equation}
\hat{\Pi}_1 = {\frac{1}{\sqrt{g^2+1}}} \left(\ket{0}\bra{0}+\ket{1}\bra{1}\right).
\end{equation}
After this operation, the $\ket{2}$ and higher-order photon terms in mode $C$ are truncated by a single-photon quantum scissor and the state becomes: 
\begin{align}
\begin{split}
&\ket{\psi}_{ACD}=  \sqrt{\frac{1-\chi^2}{g^2+1}} 
\left(  \sum_{n=0}^\infty  \chi^n \left(1-\eta\right)^{n/2} \ket{0}_C \ket{n}_A   \ket{n}_D  
\right. 
\\
& \left.+ g  \sqrt{\eta} \sum_{n=1}^\infty  \chi^n  \sqrt{n} \left(1-\eta\right)^{(n-1)/2} \ket{1}_C \ket{n}_A  \ket{n-1}_D \right).
\end{split}
\label{Chap4eq:EPRlossQS}
\end{align}
The probability of success of this individual NLA can be found via the norm of the un-normalised state~\cref{Chap4eq:EPRlossQS} which is:
\begin{equation}
P_{NLA} =  \frac{\left(1-\chi ^2\right) \left(\chi ^2 \left(\eta  g^2+\eta -1\right)+1\right)}{\left(g^2+1\right) \left((\eta -1) \chi ^2+1\right)^2}.
\label{Chap4eq:Psucc}
\end{equation}
The final step in this single-node repeater protocol is the entanglement swapping operation. We use a second copy of the state~\cref{Chap4eq:EPRlossQS}, with modes $F$ and $B$  distributed between the repeater node and Bob respectively, given by:
\begin{align}
\begin{split}
& \ket{\psi}_{BFE}= \sqrt{\frac{1-\chi^2}{g^2+1}} 
\left(  \sum_{m=0}^\infty  \chi^m \left(1-\eta\right)^{m/2} \ket{0}_B  \ket{m}_F   \ket{m}_E \right.
\\
& \left. + g  \sqrt{\eta} \sum_{m=1}^\infty  \chi^m  \sqrt{m} \left(1-\eta\right)^{(m-1)/2} \ket{1}_B \ket{m}_F   \ket{m-1}_E \right),
\label{Chap4eq:EPRlossQS2}
\end{split}
\end{align}
where mode $E$ is an environment mode. With these two entangled states~\cref{Chap4eq:EPRlossQS} and~\cref{Chap4eq:EPRlossQS2}, modes $F$ and $C$ are combined at the repeater node and a dual homodyne detection is performed. To model this dual homodyne detection,  we project modes $F$ and $C$ onto the eigenstate \cite{hofmann2000fidelity, ide2001continuous}: 
\begin{equation}
 \ket{\gamma}_{FC} = \frac{1}{\sqrt{\pi}}\sum_{n=0}^\infty \hat{D}_C(\gamma)\ket{n}_C\ket{n}_F,
\label{Chap4eq:proj}
 \end{equation}
where $\gamma$ corresponds to the measurement outcome of the dual homodyne detection. The state after swapping can be found via:
\begin{equation}
\ket{\psi_{\mathrm{swap}}}_{ABDE}= \bra{\gamma}_{FC} \left[\ket{\psi}_{ACD}\otimes \ket{\psi}_{BFE} \right].
\label{Chap4eq:symSwap} 
\end{equation}
From~\cref{Chap4eq:symSwap} after corrective displacements on modes $A$ and $B$ , we find the following un-normalized entangled state shared between Alice and Bob (including environment modes $D$ and $E$) and conditioned on the measurement outcome of $\gamma$:
\begin{align}
\begin{split}
&\ket{\psi_{out}}_{ABDE}= \frac{1}{\sqrt{\pi}}  \frac{1-\chi^2}{g^2+1}  e^{-\left|\gamma\right|^2/2}  \hat{D}_A\left(\lambda_a \gamma\right) \hat{D}_B \left(\lambda_b \gamma\right)
\\
&   \bigg[  \sum_{n=0}^\infty \sum_{m=0}^\infty \chi^n \left(1-\eta\right)^{n/2}   \chi^m \left(1-\eta\right)^{m/2}  \frac{\left(-\gamma\right)^m}{\sqrt{m!}}  \ket{n}_D   \ket{n}_A    \ket{0}_B \ket{m}_E 
\\
&  +  g  \sqrt{\eta}  \sum_{n=0}^\infty  \sum_{m=1}^\infty   \chi^n \left(1-\eta\right)^{n/2}   \chi^m  \sqrt{m} \left(1-\eta\right)^{(m-1)/2}  \frac{\left(-\gamma\right)^m}{\sqrt{m!}}   \ket{n}_D \ket{n}_A \ket{1}_B \ket{m-1}_E
\\
&  + g  \sqrt{\eta} \sum_{n=1}^\infty    \sum_{m=0}^\infty  \chi^n  \sqrt{n} \left(1-\eta\right)^{(n-1)/2}  \chi^m \left(1-\eta\right)^{m/2} \gamma^* \frac{\left(-\gamma\right)^m}{\sqrt{m!}}   \ket{n-1}_D  \ket{n}_A  \ket{0}_B \ket{m}_E 
\\
&  + g  \sqrt{\eta} \sum_{n=1}^\infty    \sum_{m=1}^\infty  \chi^n  \sqrt{n} \left(1-\eta\right)^{(n-1)/2}  \chi^m \left(1-\eta\right)^{m/2}  \sqrt{m}\frac{\left(-\gamma\right)^{m-1}}{\sqrt{\left(m-1\right)!}}  \ket{n-1}_D  \ket{n}_A  \ket{0}_B \ket{m}_E 
\\
&  + g^2  \eta \sum_{n=1}^\infty \sum_{m=1}^\infty  \chi^n  \sqrt{n} \left(1-\eta\right)^{(n-1)/2}   \chi^m  \sqrt{m} \left(1-\eta\right)^{(m-1)/2}  \gamma^* \frac{\left(-\gamma\right)^m}{\sqrt{m!}}\ket{n-1}_D \ket{n}_A \ket{1}_B \ket{m-1}_E
\\
&  + g^2  \eta \sum_{n=1}^\infty \sum_{m=1}^\infty  \chi^n  \sqrt{n} \left(1-\eta\right)^{(n-1)/2}   \chi^m  \sqrt{m} \left(1-\eta\right)^{(m-1)/2}   \sqrt{m} \frac{\left(-\gamma\right)^{m-1}}{\sqrt{\left(m-1\right)!}} \ket{n-1}_D \ket{n}_A \ket{1}_B \ket{m-1}_E,
  \bigg]
  \label{Chap4eq:rhoAll}
\end{split}
\end{align}
where $\lambda_a$ and $\lambda_b$ correspond to the classical gains applied to scale the displacements on modes $A$ and $B$ respectively. The density matrix of the output state shared between Alice and Bob can be found via:
\begin{equation}
\hat{\rho}_{AB} \left(\gamma \right)= \mathrm{Tr}_{DE}\left[ \ket{\psi_{out}}_{ABDE}\bra{\psi_{out}}_{ABDE}\right].
  \label{Chap4eq:rhoOut}
\end{equation}

To find the probability of successful post-selection $P_{\text{PS}}$, we use the following:
\begin{equation}
P_{\text{PS}} = \frac{\int_0^{2\pi} \int^{\gamma_{\text{max}}}_{0} \mathrm{Tr} \hat{\rho}_{AB} \left(\gamma\right)|\gamma| \, \mathrm{d}\phi_\gamma  \text{d}|\gamma|}{\int_0^{2\pi} \int_{0}^\infty \mathrm{Tr} \hat{\rho}_{AB} \left(\gamma\right)|\gamma|\,\text{d}\phi_{\gamma} \mathrm{d} |\gamma| }.
\label{Chap4eq:pps}
\end{equation} 

From the entangled output state~\cref{Chap4eq:rhoOut} shared between Alice and Bob, we are now in a position to calculate the secret key rate assuming collective attacks given by \cite{garcia-patron2006unconditional}:
\begin{equation}
K= \beta I_{AB} - \chi_E,
\label{Chap4eq:skr}
\end{equation}
where $I_{AB}$ is the mutual information shared between Alice and Bob, $\chi_{EB}$ is the Holevo bound representing the maximum amount of quantum information accessed by Eve, and $0\leq \beta\leq 1$ is the reconciliation efficiency.  
We calculate the key rate from the covariance matrix of the entangled  output state shared between Alice and Bob. The covariance matrix elements were obtained using~\cref{Chap4eq:rhoAll} and~\cref{Chap4eq:rhoOut} and averaged over the accepted post-selection region. To be more specific, we accept results $\gamma$ of the dual HD which fall in a circular region centered on the origin to some maximum radius $\gamma_{\text{max}}$. Averaging was performed via numerical integration of each covariance matrix element.
A two-mode Gaussian state has covariance matrix in standard form:
\begin{equation}
V = \begin{bmatrix}
a\mathds{1}  &  c \mathbb{Z} \\
c \mathbb{Z} & b \mathds{1}
\end{bmatrix} .
\end{equation}
Even though the output entangled state is slightly non-Gaussian due to the quantum scissors operation and thus cannot be fully characterized by its covariance matrix, it is valid to use Gaussian key rate calculations as it overestimates Eve's information \cite{PhysRevLett.96.080502,navascues2006optimality,garcia-patron2006unconditional}.

\section{Mutual information and Eve's information}

We calculate the mutual information shared between Alice and Bob $I_{AB}$ for an entanglement-based protocol where Alice and Bob both conduct heterodyne detection on their entangled modes by \cite{Weedbrook_2004,Sanchez2007QuantumIW}:
\begin{equation}
I_{AB}^{\text{het}} = \log_2 \left( \frac{1+a}{1+a- \frac{c^2}{1+b}} \right) 
\end{equation}
and for an entanglement based protocol where Alice and Bob conduct homodyne detection \cite{PhysRevA.63.052311, Sanchez2007QuantumIW}:
\begin{equation}
I_{AB}^{\text{hom}} =\frac{1}{2} \log_2 \left( \frac{a}{a-\frac{c^2}{b}} \right) 
\end{equation}
We illustrate here how we calculate Eve's information given Bob as the reference for reconciliation which is the case for reverse reconciliation, giving $I_E=\chi_{BE}$ and representing the mutual information between Bob and Eve (for direct reconciliation where Alice is the reference we would have $I_E=\chi_{AE}$).  Eve's information $\chi_{BE}$ can be calculated via:
\begin{equation}
I_{BE} = S(E) -S(E|B) 
\label{Chap4eq:IEB}
\end{equation}
 where $S(E)$ is the Von-Neumann entropy of Eve's state before measurement and $S(E|B)$ is the Von-Neumann entropy of Eve's state conditioned on Bob's measurement outcome. $S(E)$ can be found by using the fact that Eve purifies Alice and Bob's system, giving $S(E) =S(AB)$ which is defined as:
 \begin{equation}
 S(AB) = G\left( \frac{\nu_1-1}{2} \right) +G\left(\frac{\nu_2-1}{2} \right)
 \end{equation}
 where $\nu_1$ and $\nu_2$ are the symplectic eigenvalues of the covariance matrix $V$ and 
 \begin{equation}
 G(x) = \left(1+x\right) \log_2 \left(1+x\right)- x \log_2 x . 
\end{equation}  
The symplectic eigenvalues $\nu_1$ and $\nu_2$ can be found via:
\begin{equation}
\nu_{1,2} = \sqrt{\frac{\Delta\pm \sqrt{\Delta^2-4 \det V}}{2}}
\end{equation}
where $\Delta= a^2+b^2 -2c^2$. The Von-Neumann entropy of the conditional state $S(E|B)$ is a function of the symplectic eigenvalue of the conditional covariance matrix, $\nu_3=a-\frac{c^2}{1+b}$:
\begin{equation}
S(E|B) = G \left(\frac{\nu_3-1}{2} \right) .
\end{equation}

%% file: Appendix_purification/Appendix_purification.tex



\chapter{Supplementary material for our purification protocol}\label{app:derivation_states}

In this Appendix, we derive the purified states heralded by Alice's and Bob's QND measurements.

\section{Heralded states after round one of purification} 

In this section, we derive the heralded states (\cref{PUReq:phi_jk} of the main text) during round one.

Alice prepares $m$ copies of infinite-dimensional TMSV states, $\ket{\chi}$, where $\chi \in [0,1]$ is the squeezing parameter. Then, she performs a mode-blind QND measurement on her side, and obtains an outcome of $k_1$ total photons. This measurement projects the $m$ TMSV states onto a maximally-entangled state, $\ket{\phi_{k_1,m}}_{AB}$, with dimension $d_{k_1,m} = {k_1+m-1 \choose k_1}$, entanglement $E_{k,m} = \log_2{d_{k,m}}$, and success probability $P_{k_1,m}^{\text{Alice}} = (1-\chi^2)^m \chi^{2 k_1} d_{k_1,m}$. After Alice obtains outcome $k_1$, Alice has the following maximally-entangled state:
\begin{align}
\ket{\phi_{k_1,m}}_{AB}=(1-\chi^2)^\frac{m}{2} \chi^{k_1}  \stackrel{n_{1}+n_{2}+\cdots +n_{m}=k_1}{\mathrel{\mathop{\sum }\limits_{n_{1},n_{2},\cdots ,n_{m}}}}   \ket{ n_{1},n_{2},\cdots ,n_{m}} _{A} \ket{ n_{1},n_{2},\cdots ,n_{m}}_{B },
\end{align}
where $A$ refers to the quantum system Alice keeps and $B$ refers to the quantum system which will be sent to Bob.

Alice propagates Bob's modes ($B_1,B_2,...,B_m$) across independent pure-loss channels. The pure loss is equivalent to introducing $m$ vacuum modes and mixing those with the $m$ data rails on beamsplitters of transmissivity $\eta$. We label the beamsplitter transformations $T_i$ which interact modes $B_i$ and $e_i$. The state after this interaction is
{\small{
\begin{align}\label{eq:after_channel}
T_1 T_2 T_3 \cdots T_m \ket{\phi_{k_1,m}}_{AB} \ket{0}_{e}  &= (1-\chi^2)^\frac{m}{2} \chi^{ k_1} \stackrel{n_{1}+n_{2}+\cdots +n_{m}=k_1}{\mathrel{\mathop{\sum }\limits_{n_{1},n_{2},\cdots ,n_{m}}}}   \stackrel{n_1}{\mathrel{\mathop{\sum }\limits_{l_1=0}}} \; \stackrel{n_2}{\mathrel{\mathop{\sum }\limits_{l_2=0}}} \; \stackrel{n_3}{\mathrel{\mathop{\sum }\limits_{l_3=0}}} \cdots \stackrel{n_m}{\mathrel{\mathop{\sum }\limits_{l_m=0}}} \; \sqrt{{n_1 \choose l_1}  {{n_2 \choose l_2}} {{n_3 \choose l_3}} \cdots {{n_m \choose l_m}}}  \; \\ &\;\;\;\;\;\; (1-\eta)^\frac{l_1+l_2+\cdots+l_m}{2} \eta^\frac{n_1+n_2+\cdots+n_m - (l_1+l_2+\cdots+l_m)}{2} \nonumber \\
&\;\;\;\;\;\;\;\;\;  \nonumber \ket{ n_{1},n_{2},\cdots ,n_{m}} _{A }  \ket{ n_{1}-l_{1},n_{2}-l_{2},\cdots ,n_{m}-l_{m}}_{B } \ket{ l_{1},l_{2},\cdots ,l_{m}}_{e },
\end{align}
}}
where $l_i$ is the number of photons lost from mode $B_i$ to $e_i$.

Finally, Bob performs a QND measurement and obtains an outcome of $j_1$ total photons. The terms that survive in~\cref{eq:after_channel} are those components where Bob's modes have total photons which add up to $j_1$, that is, $n_1-l_1+n_2-l_2+\cdots+n_m-l_m=j_1$. The total number of lost photons is the difference between Alice and Bob's outcomes, $l_1+l_2+\cdots+l_m = k_1-j_1$. Thus, the unnormalised final state shared between Alice and Bob is
\begin{align}
\ket{\phi_{k_1,j_1,m}}_{ABe }  &= (1-\chi^2)^\frac{m}{2} \chi^{ k_1} (1-\eta)^\frac{k_1-j_1}{2} \eta^\frac{j_1}{2} \stackrel{n_{1}+n_{2}+\cdots +n_{m}=k_1}{\mathrel{\mathop{\sum }\limits_{n_{1},n_{2},\cdots ,n_{m}}}} \;  \nonumber  \nonumber \\&\;\;\;\;\;\; \stackrel{l_1+l_2+\cdots+l_m=k_1-j_1,\; l_i\leq n_i \forall i}{\mathrel{\mathop{\sum }\limits_{l_{1},l_{2},\cdots ,l_{m}}}} \;  \nonumber \; \sqrt{{n_1 \choose l_1}  {{n_2 \choose l_2}} {{n_3 \choose l_3}} \cdots {{n_m \choose l_m}}} \; \\
&\;\;\;\;\;\;\;\;\;   \ket{ n_{1},n_{2},\cdots ,n_{m}} _{A}  \ket{ n_{1}-l_{1},n_{2}-l_{2},\cdots ,n_{m}-l_{m}}_{B } \ket{ l_{1},l_{2},\cdots ,l_{m}}_{e  },
\end{align}
which is~\cref{PUReq:phi_jk} of the main text as required. 

We can calculate the success probability (i.e., the probability to successfully get this output state) as follows
{\small{
\begin{align}
    P_{k_1,j_1,m} &= \langle \phi_{k_1,j_1,m} | \phi_{k_1,j_1,m} \rangle \\ 
    &= (1-\chi^2)^m \chi^{2 k_1} (1-\eta)^{k_1-j_1} \eta^{j_1} \stackrel{n_{1}+n_{2}+\cdots +n_{m}=k_1}{\mathrel{\mathop{\sum }\limits_{n_{1},n_{2},\cdots ,n_{m}}}} \;  \stackrel{l_1+l_2+\cdots+l_m=k_1-j_1,\; l_i\leq n_i \forall i}{\mathrel{\mathop{\sum }\limits_{l_{1},l_{2},\cdots ,l_{m}}}} \; \; {{n_1 \choose l_1}  {{n_2 \choose l_2}} {{n_3 \choose l_3}} \cdots {{n_m \choose l_m}}} \; \\
    &= (1-\chi^2)^m \chi^{2 k_1} (1-\eta)^{k_1-j_1} \eta^{j_1} \stackrel{n_{1}+n_{2}+\cdots +n_{m}=k_1}{\mathrel{\mathop{\sum }\limits_{n_{1},n_{2},\cdots ,n_{m}}}} {k_1 \choose j_1} \\ 
    &= (1-\chi^2)^m \chi^{2k_1} d_{k,m} (1-\eta)^{k_1-j_1} \eta^{j_1} {k_1 \choose j_1}\\
    &= P_{k_1,m}^{\text{Alice}} P_{k_1,j_1}^\text{Bob},
\end{align}
}}
We used the generalized Vandermonde identity to simplify the inner summation to ${k_1 \choose j_1}$; this does not depend on the outer summation, which has $d_{k_1,m}={k_1+m-1 \choose k_1}$ terms. We can assign $P_{k_1,m}^{\text{Alice}}=(1-\chi^2)^m \chi^{2k_1} d_{k,m}$ and $P_{k_1,j_1}^\text{Bob}=(1-\eta)^{k_1-j_1} \eta^{j_1} {k_1 \choose j_1}$, where we can interpret $P_{k_1,j_1,m}=P_{k_1,m}^{\text{Alice}} P_{k_1,j_1}^\text{Bob}$ as a joint probability distribution. Remarkably, while this success probability $P_{k_1,j_1,m}$ depends on the transmissivity $\eta$, the renormalised output state 
\begin{align}
    \ket{\phi_{k_1,j_1,m}'} = \ket{\phi_{k_1,j_1,m}}/\sqrt{P_{k_1,j_1,m}},
\end{align}
shared between Alice and Bob does not depends on $\eta$.

\section{Round two of purification}

Recall that in the first round, Alice obtains outcome $k_1$ and Bob obtains outcome $j_1$. They share the following state:
\begin{align}
\ket{\phi_{k_1,j_1,m}}_{ABe }  &= (1-\chi^2)^\frac{m}{2} \chi^{ k} (1-\eta)^\frac{k_1-j_1}{2} \eta^\frac{j_1}{2}  \nonumber \\&\;\;\;\;\;\; \stackrel{n_{1}+n_{2}+\cdots +n_{m}=k_1}{\mathrel{\mathop{\sum }\limits_{n_{1},n_{2},\cdots ,n_{m}}}} \;  \stackrel{l_1+l_2+\cdots+l_m=k_1-j_1,\; l_i\leq n_i \forall i}{\mathrel{\mathop{\sum }\limits_{l_{1},l_{2},\cdots ,l_{m}}}} \;   \nonumber \; \sqrt{{n_1 \choose l_1}  {{n_2 \choose l_2}} {{n_3 \choose l_3}} \cdots {{n_m \choose l_m}}} \; \\
&\;\;\;\;\;\;\;\;\;   \ket{ n_{1},n_{2},\cdots ,n_{m}} _{A }  \ket{ n_{1}-l_{1},n_{2}-l_{2},\cdots ,n_{m}-l_{m}}_{B } \ket{ l_{1},l_{2},\cdots ,l_{m}}_{e}.
\end{align}
Let us firstly consider the case when $j_1=k_1$, then $l_1+l_2+\cdots+l_m=0$ thus the output state is:
\begin{align}
\ket{\phi_{k_1,k_1,m}}_{ABe}  &= (1-\chi^2)^\frac{m}{2} \chi^{ k_1} \eta^\frac{k_1}{2} \stackrel{n_{1}+n_{2}+\cdots +n_{m}=k_1}{\mathrel{\mathop{\sum }\limits_{n_{1},n_{2},\cdots ,n_{m}}}}   \ket{ n_{1},n_{2},\cdots ,n_{m}} _{A }  \ket{ n_{1},n_{2},\cdots ,n_{m}}_{B } \ket{0,0,\cdots,0}_{e}.
\end{align}
We can see that the environment modes are all vacuum states and are separable from the state. When $j_1=k_1$ the state is purified in a single shot, and we do not need a second round.

Now, consider the case when $j_1<k_1$, we will need further rounds of purification. In the second round, one option for further purification is to measure the total photon number on $m-1$ rails instead of all $m$ rails. This means they learn if some errors happened in the $m$th rail, or in the first $m-1$ rails. If Alice obtains outcome $k_2$ and Bob obtains outcome $j_2$, they share the state:
{\small{
\begin{align}
\ket{\phi_{k_1,j_1,k_{2}, j_{2},m}}_{ABe }  &= (1-\chi^2)^\frac{m}{2} \chi^{ k_1} (1-\eta)^\frac{k_1-j_1}{2} \eta^\frac{j_1}{2} \nonumber \\
&\;\;\; \stackrel{n_{1}+n_{2}+\cdots +n_{m}=k_1,\; n_{1}+n_{2}+\cdots +n_{m-1}=k_2}{\mathrel{\mathop{\sum }\limits_{n_{1},n_{2},\cdots ,n_{m}}}} \;\;\; \nonumber \stackrel{l_1+l_2+\cdots+l_m=k_1-j_1,\; l_1+l_2+\cdots+l_{m-1}=k_2-j_2,\; l_i\leq n_i \forall i}{\mathrel{\mathop{\sum }\limits_{l_{1},l_{2},\cdots ,l_{m}}}} \;  \nonumber \;\\
&\;\;\sqrt{{n_1 \choose l_1}  {{n_2 \choose l_2}} {{n_3 \choose l_3}} \cdots {{n_m \choose l_m}}} \;   \ket{ n_{1},n_{2},\cdots ,n_{m}} _{A }  \ket{ n_{1}-l_{1},n_{2}-l_{2},\cdots ,n_{m}-l_{m}}_{B } \ket{ l_{1},l_{2},\cdots ,l_{m}}_{ e}.
\end{align}
}}

If $j_1<k_1$ and $j_2=k_2$, then no errors occurred in the first $m-1$ rails and the  output state is {\small{$\propto \stackrel{n_{1}+n_{2}+\cdots +n_{m-1}=k_2,\; n_{m}=k_1-k_2}{\mathrel{\mathop{\sum }\limits_{n_{1},n_{2},\cdots ,n_{m}}}}  \ket{ n_{1},n_{2},\cdots ,n_{m-1},k-k_2} _{A } \ket{ n_{1},n_{2},\cdots ,n_{m-1},j-j_2}_{B } \ket{ 0,0,\cdots ,0, k_1-j_1}_{e }$}}, so the $m$th rail is in a definite quantum state with no entanglement so this rail should be discarded. The rest of the $m-1$ modes are not entangled with the environment, so Alice and Bob's state of these rails is pure.

Finally, if $j_1<k_1,j_2<k_2$, we need still further purification rounds. After $m-1$ iterative rounds, we show next that we achieve the capacity as $\chi\to1,m\to\infty$.

%% file: bib_file2.tex
%

%% file: MainThesis.bbl
\begin{thebibliography}{196}
\bibitem [{Dav()}]{Dave_Winnel}%
  \BibitemOpen
  \href@noop {} {}\bibinfo {note} {{S}pecial thanks to {D}ave {W}innel for
  illustrating our simple quantum repeater experiment for continuous-variable
  quantum key distribution}\BibitemShut {NoStop}%
\end{thebibliography}
